# Ultrafast long-range energy transport via light-matter coupling in organic semiconductor films

[1]Raj Pandya, [1]Richard Y. S. Chen, [1]Qifei Gu, [1]Jooyoung Sung, [1]Christoph Schnedermann, [1]Oluwafemi S. Ojambati, [1]Rohit Chikkaraddy, [1]Jeffrey Gorman, [2]Gianni Jacucci, [2]Olimpia D. Onelli, [3]Tom Willhammar, [4]Duncan N. Johnstone, [4]Sean M. Collins, [4]Paul A. Midgley, [1]Florian Auras, [1]Tomi Baikie, [5]Rahul Jayaprakash, [6]Fabrice Mathevet, [7]Richard Soucek, [8]Matthew Du, [2]Silvia Vignolini, [5]David G Lidzey, [1]Jeremy J. Baumberg, [1]Richard H. Friend, [7]Thierry Barisien, [7]Laurent Legrand, [7]Alex W. Chin, [5]Andrew J. Musser, [8]Joel Yuen-Zhou, [9,10]Semion K. Saikin, [11]Philipp Kukura, [1]Akshay Rao*

[1]Cavendish Laboratory, University of Cambridge, J.J. Thomson Avenue, CB3 0HE, Cambridge, United Kingdom

[2]Department of Chemistry, University of Cambridge, Lensfield Road, Cambridge CB2 1EW, United Kingdom

[3]Department of Materials and Environmental Chemistry, Stockholm University, Stockholm, Sweden

[4]Department of Materials Science and Metallurgy, University of Cambridge, 27 Charles Babbage Road, CB3 0FS, Cambridge, United Kingdom

[5]Department of Physics & Astronomy, University of Sheffield, S3 7RH, Sheffield, United Kingdom

[6]Sorbonne Université, CNRS, Institut Parisien de Chimie Moléculaire (IPCM) UMR 8232, Chimie des Polymères, 4 Place Jussieu, 75005 Paris, France

[7]Institut des NanoSciences de Paris, UMR 7588/CNRS, Université Pierre et Marie Curie (Paris 6), Campus Boucicaut, 140 rue de Lourmel, 75015 Paris, France

[8]Department of Chemistry and Biochemistry, University of California San Diego, La Jolla, California 92093, United States

[9]Department of Chemistry and Chemical Biology, Harvard University, 12 Oxford Street, Cambridge, Massachusetts 02138, United States

[10]Kebotix Inc., 501 Massachusetts Avenue, Cambridge, Massachusetts 02139, United States

[11]Physical and Theoretical Chemistry Laboratory, Department of Chemistry, University of Oxford, South Parks Road, Oxford OX1 3QZ, UK.

*correspondence: ar525@cam.ac.uk

The efficient transport of energy in the form of spin-singlet excitons lies at the heart of natural light harvesting for photosynthesis and optoelectronic devices based on synthetic organic semiconductors. For optoelectronic applications, energy transport over long length scales would be highly desirable, but most organic semiconductors exhibit singlet exciton diffusion lengths between 5 – 50 nm [1,2], with only a handful of systems being suggested to exceed this[3–8]. This is because the primary mechanism for exciton transport is Förster resonance energy transfer (FRET), which involves site-to-site hopping within a disordered energy landscape, leading to short diffusion lengths. While there have been reports of higher diffusion lengths in highly ordered one-dimensional organic nanowires, these materials have proved challenging to integrate into optoelectronic devices. Hence, what is required is a new methodology that allows for long range energy transport within thin films of organic semiconductors, which form the basis of current optoelectronic devices. Here, we show that long-range and ultrafast transport of energy can be achieved at room temperature in a range of chemically diverse organic semiconductor thin films through light-matter coupling to form exciton-polaritons. These effects occur despite the absence of an external cavity, metallic or plasmonic structures. We directly visualize energy transport via femtosecond transient absorption microscopy with sub-10 fs temporal and sub-10 nm spatial precision and find energy transport lengths of up to ~270 nm at effective velocities of up to ~5 $\times 10^6$ m $s^{-1}$. Evidence of light-matter coupling in the films is provided via peak splittings in the reflectivity spectra, measurement of the polariton dispersion and emission from collective polariton states. The formation of these exciton-polariton states in organic semiconductor thin-films is a general phenomenon, independent of underlying materials chemistry, with the principal requirements being a high oscillator strength per unit volume and low disorder. These results and design rules will enable a new generation of organic optoelectronic and light harvesting devices based on robust cavity-free exciton-polaritons[9–11].

Exciton-polaritons (EPs) are quasiparticles formed by the hybridization of optically absorbing excitons with light modes. EPs occur in systems with significant light-matter coupling, based on either a high oscillator strength or the precise confinement of optical fields (*e.g.* cavities, nanophotonic systems), with ideally a combination of both [12–14]. As organic semiconductors sustain stable excitons at room-temperature, these materials are being actively studied for room temperature polaritonic devices [15–17]. Recent studies have almost exclusively focussed on cavity-based systems, where molecules are confined between metallic or dielectric mirrors [18,19] or in a plasmonic gap [12,20]. Early reports, however have also shown that in some densely packed organic crystals and aggregates, it is possible to form 'intrinsic' polaritons without an external structure to confine the light field [21–26], removing the need for complex sample fabrication. Yet, polaritonic effects in excitation energy transport have been largely unexplored.

Here, we study energy transport in model conjugated polymer, J-aggregate and H-aggregate thin film systems. Employing femtosecond transient absorption microscopy (fs – TAM) we are able to directly visualise energy transport on sub-10 nm length scales with sub-10 fs temporal resolution. Remarkably, we find that photoexcitations in these systems can move up to ~270 nm in ~50 fs with an effective velocity of ~5 $\times 10^6$ m $s^{-1}$. We show that these properties are enabled by the formation of exciton-polaritons (EPs) despite the absence of any cavity structure. In addition, we demonstrate that the magnitude of the energy migration is intricately linked to the sample morphology. These experimental results establish a new paradigm for energy transport in organic semiconductor thin films, based on room temperature polaritonic effects that transcend the underlying molecular building blocks.

To study energy transport in organic semiconductors thin films we choose three model systems: polydiacetylene (PDA), a conjugated polymer [27,28], in the form of aligned 2-5 µm chains; PIC, a molecular semiconductor that can self-assemble to form nanotubular J-aggregates based on the brickwork packing of cyanine dyes [29–31], with length 1-5 µm and tube diameter of 2-5 nm; and a perylene dimiide (PDI) which assembles to form ~100 µm long 'nanobelts' consisting of H-aggregated PDI molecules [32,33] (Figure 1a-c). In each case we fabricated thin films of the samples (see

SM **SI, S1** for film fabrication details). Figures 1a-c depict the molecular structures, packing motifs and electron microscopy images for the molecules, nanostructures and their thin films respectively. The systems have disparate chemical building blocks and packing motifs but all of them have high oscillator strengths (see SM structural characterisation **SI, S2**).

Figure 1d shows a schematic of the fs-TAM experiment. Here, a broadband sub-10 fs pump pulse (520 – 650nm) is focussed to the diffraction limit (FWHM ~270 nm) by a high numerical aperture (~1.1) objective onto the sample, with a wide field, counter-propagating probe pulse (8 fs, 680 – 790 nm, FWHM ~15 μm) focussed with a concave mirror, used to monitor the normalized change in image transmission with pump on ($T_{on}$) and off ($T_{off}$) as function of time delay between the pulses ($\frac{\Delta T}{T} = \frac{T_{on}-T_{off}}{T_{on}}$). By subtracting the extent of the spatial profile at a time *t,* from that at zero time delay between pump and probe ($t_0$) the propagation of population can be monitored (further details see **SI, S3**). The limit to localisation precision is determined by how well different spatial profiles can be resolved (right panel Figure 1d); based on the signal to noise ratio of the measurement this typically amounts to ~10 nm, and hence allows us to monitor spatiotemporal dynamics well below the diffraction limit.

Although the individual 1D nanostructures are highly ordered, their thin films are inhomogeneous (Figure 1a-c), as is common in organic semiconductors. Consequently, measurements were carried out on ~40 locations across multiple films per system, resulting in an improved statistical picture of how transport relates to morphology. This approach allows us to move beyond normal time-resolved studies where data is reported for a single ensemble measurement, which by definition cannot capture the underlying heterogeneity of the system.

Representative fs-TAM images obtained for PDA (blue), PIC (red) and PDI (green) following photoexcitation are shown in Figures 1e-g. We probe the energy transport by monitoring the exciton stimulated emission (SE) bands of PDA (670 nm) and PIC (600 nm), and the exciton photoinduced absorption (PIA) of PDI (720 nm), but note that the reported behaviour is independent of the exact probe wavelength within the band (**SI, S4**). In all three cases, there is a large and rapid expansion of the initial excitation spot. The 1D nature of the samples means that the growth is highly anisotropic with the excitation spot being a near symmetric Gaussian at $t_0$ but rapidly elongating in the wire direction (in the case of PDA and PIC this will be the average of wires within the pump spot, in PDI single wires are excited; see **SI, S2**). To quantify the extent of spatial transport, we extract the FWHM of each image along the propagation axis and convert it to a Gaussian standard deviation σ (see **SI, S3** includes discussion of propagation in orthogonal direction). The $t_0$ frame is identified from the σ value expected based on the diffraction limit of our pulses (σ ~135 nm corresponding to a FWHM ~310 nm) and by fitting the rise of the signal with the instrument response (**SI, S5**). In PDA the expansion is largest with σ increasing from roughly the diffraction limit of 134 ± 5 nm at t = 0 fs to 297 ± 5 nm at t = 90 ± 3 fs. For PIC σ increases from 143 ± 5 nm to 201 ± 5 nm in t = 100 ± 5 fs nm and in PDI the expansion is smallest, with a σ of 149 ± 8 nm at $t_0$ and σ of 206 ± 8 nm at t = 350 ± 3 fs. The large initial (sub-100 fs) expansion in all three samples suggests that excitations can travel surprisingly fast and far in these model organic semiconductors, beyond the sub-5 nm range reported on these timescales for FRET mediated systems [1,34].

To analyse the data further we plot the mean square displacement $MSD = \sigma(t)^2 - \sigma(t_0)^2$ as a function of time in Figures 2a-c. The behaviour is biphasic in nature for all three samples, with an initial ultrafast (sub-50 fs) phase labelled $R_1$, followed by a slower expansion over the following 200 fs, labelled $R_2$. The first phase ($R_1$) is particularly intriguing. Here the propagation velocity, $v = \frac{\sqrt{MSD}}{t}$ ($t$ is the time for which for the propagation in a given phase lasts), for all three systems is in the range of 3.5 ×10$^6$ – 5.0 ×10$^6$ m s$^{-1}$ (PDI<PIC<PDA), with transport distances ($\sqrt{MSD}$) between 50 – 200 nm (**SI, S6**). The speed of transport is two-orders of magnitude larger than that found in organic semiconductors [35–37] and above that reported for excitons in low temperature epitaxial GaAs [38] and free electrons in metallic films [39]. For $R_2$, $v$ typically lies in the range of 0.05 – 0.1 $\times 10^6$ m $s^{-1}$ with a large spread in the values for PDA, PDI and PIC depending on the sample location; in the majority of cases the transport length in $R_2$ is smaller than $R_1$, typically between 20 – 100 nm.

The slope of the $MSD$ plot (solid straight lines in Figure 2a-c) can be used to extract a diffusion coefficient, D ($MSD = \mathrm{D}t^{\alpha}$, $t$ is propagation time in $R_1$ or $R_2$ , $\alpha$ = 1), for each sample in the $R_1$ and $R_2$ regions. Although diffusion relates to strictly incoherent processes, the diffusion coefficient is commonly used to characterize transport in organic semiconductors and is hence extracted here for comparison (**SI, S3** further discussion). Repeating fs-TAM measurements on many different sample locations allows us to build up a histogram of diffusion coefficients, shown in Figure 2d-f. For all three systems, in $R_1$, and to a weaker extent in $R_2$, the order of magnitude of the diffusion coefficient exhibits a quasi-Gaussian distribution with a significantly smaller spread for $R_1$ compared to $R_2$ (**SI, S7**). This log-normally distributed behaviour can be rationalized based on the random selection of sample locations for measurement and distribution of disorder within the self-assembled films. The smaller spread observed for $R_1$ suggests however, that in this regime the diffusion coefficient (and the transport velocity) is related to an intrinsic property of the material and is not greatly perturbed by disorder. In $R_2$ the diffusion coefficient is significantly more sensitive to disorder, leading to a larger spread. A second salient feature of Figure 2d-f is that for both $R_1$ and $R_2$, the average diffusion coefficient ($\langle D \rangle$) follows the trend PDA>PIC>PDI: PDA $\langle D \rangle_{R_1}$= 4840 cm$^2$ s$^{-1}$, $\langle D \rangle_{R_2}$= 303 cm$^2$ s$^{-1}$; PIC $\langle D \rangle_{R_1}$= 3810 cm$^2$ s$^{-1}$, $\langle D \rangle_{R_2}$= 120 cm$^2$ s$^{-1}$ ; PDI $\langle D \rangle_{R_1}$= 2710 cm$^2$ s$^{-1}$, $\langle D \rangle_{R_2}$= 37 cm$^2$ s$^{-1}$. For both regimes this variation likely stems from the lower disorder in PDA as compared to PIC and PDI respectively (see **SM SI, S2** for further discussion).

In order to compare the experimentally obtained velocities and diffusion coefficients with theoretically expected values for excitonic transport, we first build a microscopic model of the electronic structure of the PDI system using a time-dependent density functional theory approach[40]. The choice of PDI is guided by the fact that it is the only system whose crystal structure can be accurately and completely solved (**SI, S2)**. Full details of the model and the parameters used are given in the SM (see **SI, S8** for details). The model accurately reproduces the absorption spectra of the PDI upon crystal packing, with the large blue shift caused by the H-aggregation in the system. With the electronic structure computed, the exciton transport parameters are calculated using the Haken-Reineker-Strobl (HRS) model[41,42] (for details see **SI, S8**), which provides a reasonable balance between the complexity of the model and the precision of the results. We consider two regimes, an early time regime in which the excitons will move ballistically at a velocity given by the dispersion of the exciton band and at later times we consider incoherent exciton diffusion. The ballistic velocity of an exciton we obtain from modelling, $v \approx 0.2 \times 10^6$ m s$^{-1}$, rests solely on the intermolecular coupling, and is about an order of magnitude smaller than that obtained in the $R_1$ region from experiments. This implies that the transport observed in the $R_1$ region cannot arise from ballistic propagation of excitons.

The exciton diffusion coefficient will depend on the exciton dephasing rate which cannot be directly extracted from experiments. In general, however, it is expected to be comparable to the homogeneous linewidth of the excitonic band. We find that in order to obtain a diffusion coefficient comparable to the one measured for the $R_2$ range, the dephasing rate should be on the order of ~10 meV, which is smaller than the absorption linewidth of around 50 meV. These estimates do not include structural disorder in the crystal lattice, which would additionally slow down exciton transport. Hence, transport within the $R_2$ regime is at the limit of what can be described using a pure excitonic model. Indeed, for comparable organic systems, theoretical calculations predict exciton velocities (within the $R_2$ region) between $0.02 \times 10^6 - 0.08 \times 10^6$ m s$^{-1}$ [43–46], placing the experimental value obtained here at the upper end of this range.

In Figure 3a-b we show the results of microscopic reflectivity measurements (sample illumination/collection area ~3 μm$^2$; **SI, S9)** for PDA and PIC, where in each material we see the strong absorption transition splits into two peaks in the reflectivity spectra. In PDA the zero-phonon transition (1.965 eV) is split by typically 17.5 ± 0.5 meV, whereas in PIC the J-aggregate exciton transition (2.132 eV) is split by 39.6 ± 0.6 meV. Additionally, in PIC we find this feature has an angle dependence distinct from the intrinsic Fabry-Pérot modes, revealing the state has partial photonic character (see **SI, S10**). Specular reflection measurements are commonly used to identify the formation of exciton-polaritons (EPs), where the strong coupling of light and exciton absorption at the exciton resonance results in the splitting in the reflectivity spectrum into upper and lower polariton bands[47]. The specular reflection hence suggests that these films support EPs, despite the absence of any cavities or nanophotonic systems.

Measuring the reflectivity at ~70 different sample locations (**SI, S9)**, as for fs-TAM, demonstrates the magnitude of the splitting does not vary significantly from site-to-site (within our resolution), but is only resolvable in ~30% of the regions examined. We attribute these variations to nanoscale inhomogeneities, *e.g.* from sample thickness or bundling of nanotubes (**SI, S2, S10**), which give rise to regions of increased oscillator strength. Transfer matrix modelling suggests that only regions with some such mechanism of locally enhanced oscillator strength can give rise to the splittings we detect. Furthermore, the model shows that the Rabi splitting scales with the square root of absorbance, in-line with our observations of a smaller splitting at the first vibrational peak of PDA (16.5 ± 0.5 meV), confirming the assignment to an exciton-polariton (see **SI, S10** for further discussion and modelling).

Based on these experimental results we propose that the exciton states in the aggregates are admixed with mildly confined photonic modes supported by the dielectric constant mismatch between the molecules and the surrounding environment. According to our computational simulations, we estimate (**SI, S11**) that the aggregates occupy as much as 10-30% of the mode volumes of these photonic modes, leading to modest collective light-matter couplings (see below) that create EPs separated by Rabi splittings that are smaller than the exciton inhomogeneous broadening. Hence, in contrast to molecular EP systems in cavities, a clear energetic separation between EPs and the dark-states (*i.e.*, exciton states which, in the absence of disorder, do not admix with photons) does not occur, and the latter, which we hereafter call subradiant, are endowed with small fractions of photonic character. While a detailed description of the emergent excitation energy transport pathways in these systems is beyond the scope of this work, we conjecture that the two regimes, $R_1$ and $R_2$, detected in our experiments correspond to transport mediated by EPs and by the quasi-localized subradiant states, respectively. At early times, the excitation will move ballistically according to a group velocity associated with the formation of a coherent EP wavepacket. Rapid population transfer from these EPs into the subradiant states, however, is entropically favourable due to the large density of states of the latter and this transfer should be irreversible given the room-temperature conditions and the small Rabi splittings featured by these systems. The kinetics of this transfer is mediated by vibrational relaxation and should exhibit a similar timescale to that in cavity polariton systems, on the order of ~50 fs [10]. Once population is collected in these subradiant states, transport of energy should be diffusive, yet with an interchromophoric coupling that is mildly enhanced by the admixture with photons due to disorder. These hypotheses are consistent with our experimental observation of two regimes of transport: a fast one (dominated by polaritons, which technically yields ballistic transport) that decays within 50 fs, and a slower one (dominated by subradiant states) that follows the former. Crucially, because exciton-polaritons are formed from interaction of the optical dipoles with the vacuum electromagnetic field, the polaritonic states are present and accessible even in the absence of excitation[48]. In the low-excitation-density regime, these states and the transport they mediate should be independent of photon flux, which we indeed observe (**SI, S6**) [49]. We emphasize that because fs-TAM inherently probes population, what we are measuring here is the transport of excitations within the semiconductor material, rather than photonic transport within a waveguide[33,50–52]. This difference is crucial since transport of the former is what is needed for optoelectronic applications (see **SI, S12** for further discussion). Furthermore, control measurements on the isolated PIC/PDI dye molecules, unaligned partially polymerized PDA and films of CdSe nanocrystals do not show such $R_1$ (polaritonic) behaviour (**SI, S13**) and hence show that the observations are directly related to the nanostructure of the materials.

In PDI, a multitude of homogeneously broadened vibronic peaks precludes observation of light-matter coupling via reflection measurements. The lack of the characteristic peak splitting may also be attributed to the absence of confined photon modes that are resonant with the highly absorbing molecular transitions (**SI**, **S11**). We track instead the intensity of emitted photoluminescence (PL) as a function of driving optical field (laser fluence). In the presence of strong-coupling an oscillatory light-matter response is expected due to periodic exchange of energy between the ground (excitation only) and excited states (ground state and photon), driven at the Rabi frequency, $\widetilde{\Omega}_R = \mu_{12}|E_0|/\hbar$ ($\mu_{12}$ is the transition dipole moment, $|E_0|$ is the magnitude of the driving optical electric field)[53–55]. For a given pulse duration, Rabi flopping is expected in the power dependence of the emission due to changing $\widetilde{\Omega}_R$ (here from 2.6 THz to 4.4 THz ) with the amplitude of the incident field $E_0$. As shown in Figure 3c, for isolated PDI wires, the PL intensity initially increases linearly with the applied field before beginning to oscillate at high optical fields. Importantly such oscillatory behaviour is not observed in bundles of PDI wires (likely due to disorder effects of the ensemble) or isolated monomers in solution (**SI, S14**). Modelling the oscillation frequency[56] (right panel Figure 3c), yields a damping time of 50 ± 10 fs *i.e.* time for which well-defined superpositions of the ground and excited states exist, in good agreement with the timescale of ultrafast transport in $R_1$. Although pump-induced oscillations are not necessarily indicative of strong-coupling[54], in the case of PDI wires we suggest that the formation of polaritons is required to overcome any rapid competing decay processes. This in turn leads to cooperative interaction and emission from the densely packed molecules that make up the nanobelt (~5 $\times 10^7$ within our collection volume).

In summary, we have shown that in a range of organic semiconductor films, we observe the transport of energy over hundreds of nanometres on ~50 fs timescales. We propose that this is due to the formation of exciton-polaritons, despite the absence of any external cavity structure. This opens the possibility of exploiting cavity quantum electrodynamic (QED) phenomena, typically associated with low temperature physics and epitaxial sample fabrication in a remarkably simple manner, allowing for ultralong polariton assisted energy propagation at room temperature in inhomogeneous organic semiconductor films. Key requirements for this new transport mechanism are: a high material oscillator strength, typically achievable via the coupling of ordered dipoles; a refractive index mismatch between the organic medium and its surroundings; and low disorder to minimize scattering of the wavefunction. We envisage that the polaritonic transport mechanism described here could be used to improve exciton transport in organic optoelectronic devices[10]. For low-threshold lasers and LEDs such phenomena could be used to generate emission with narrower linewidths due to the formation of EPs prior to Frenkel excitons[57]. Exploitation of traditional polaritonic effects in both cases has been limited due to the obstacles of outcoupling imposed by a physical microcavity[58]. Other applications could include novel chemical reactions[10], superfluidity[11] and nanoscale circuitry[9,59]. Although we have focussed our attention on quasi-1D systems the results presented are generalizable to any organic system fulfilling the above criteria, irrespective of dimensionality. These results call for the development of a new generation of molecular systems specifically designed to harness cavity-free polariton assisted energy transport.

**Figure 1**

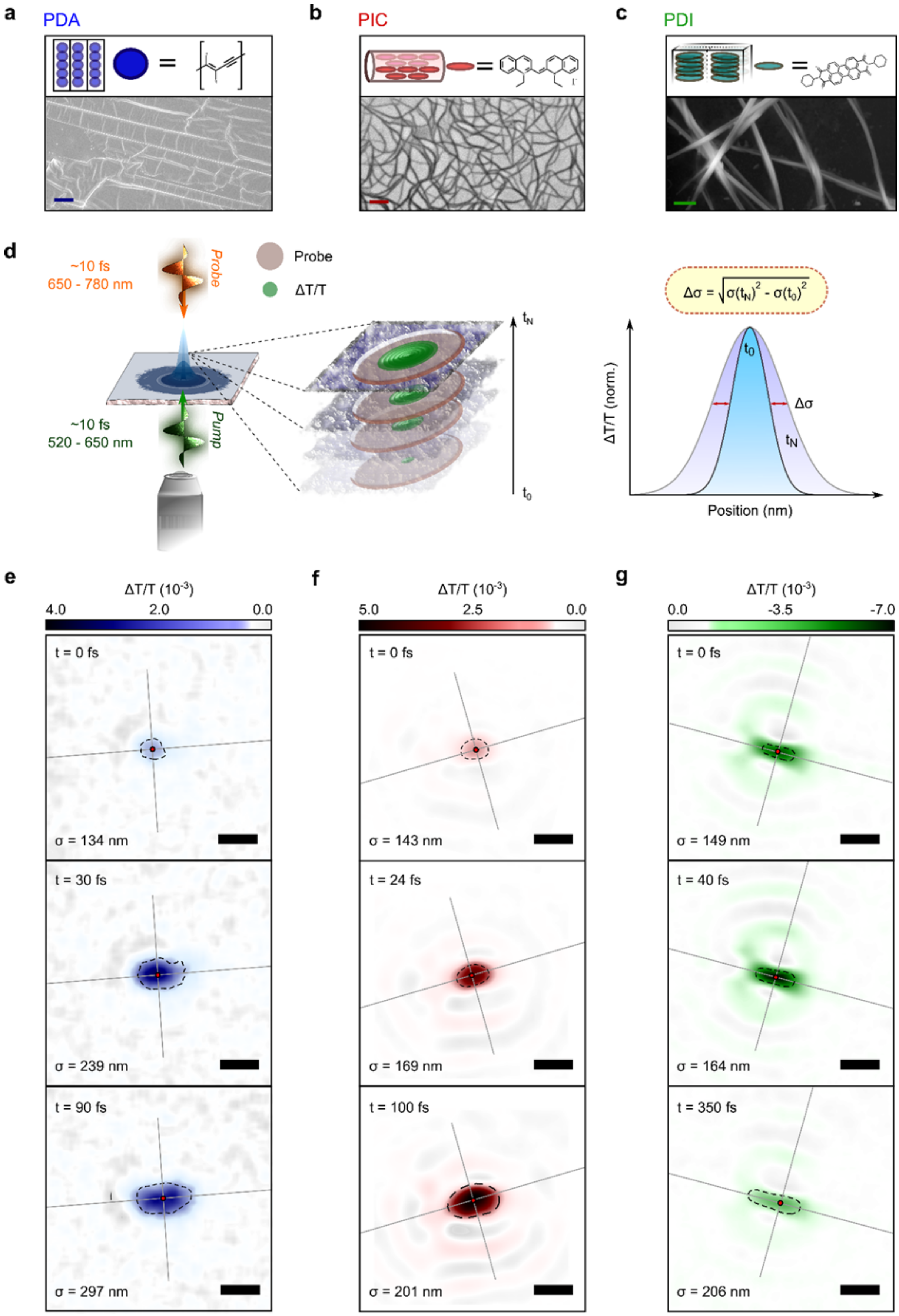

a
PDA
b
PIC
c
PDI
d
~10 fs
650 - 780 nm
Probe
~10 fs
520 - 650 nm
Pump
Probe
ΔT/T
t_N
t_0
Δσ = √(σ(t_N)^2 - σ(t_0)^2)
ΔT/T (norm.)
t0
Δσ
tN
Position (nm)
e
ΔT/T (10^-3)
4.0
2.0
0.0
t = 0 fs
σ = 134 nm
t = 30 fs
σ = 239 nm
t = 90 fs
σ = 297 nm
f
ΔT/T (10^-3)
5.0
2.5
0.0
t = 0 fs
σ = 143 nm
t = 24 fs
σ = 169 nm
t = 100 fs
σ = 201 nm
g
ΔT/T (10^-3)
0.0
-3.5
-7.0
t = 0 fs
σ = 149 nm
t = 40 fs
σ = 164 nm
t = 350 fs
σ = 206 nm

**Figure 1: Structural characterization of quasi-one dimensional organic semiconductors studied and representative fs-TAM images. a.** Cartoon and scanning electron-microscopy (SEM) image of PDA. The polymer chains consist of a single-double-single-triple bonded back bone and are near perfectly aligned in 100 $\mu m^2$ domains. Scale bar is 10 μm. **b.** Molecular packing and transmission electron-microscopy image of PIC nanotubes. The cyanine based monomers pack in a brickwork pattern to form nanotubes. They are embedded in a rigid sucrose-trehalose matrix and tend to be highly bundled in thin films; scale bar 100 nm **c.** PDI molecules pi-face stack in a quasi-H-aggregate fashion and in four distinct blocks (black outlines) to form ~100 μm long, 50 nm wide nanobelts; scale bar 250 nm. **d.** Schematic of fs-TAM. A diffraction limited pump pulse (green) is focused onto the sample by a high numerical aperture microscope objective. A probe pulse (yellow) is focussed from the top onto the sample in the wide-field. The transmitted probe is collected by the objective and imaged onto a digital camera (**SI, S1**). Comparison of the spatial extent of the signal (stack) at different time delays allows us to dynamically track changes in the population with ~10 nm precision (right hand graph). **e-g.** fs-TAM images of PDA, PIC and PDI at selected time delays following photoexcitation. The pump pulse covers the entire (excitonic) absorption of the systems, with probing carried out at 670 nm, 600 nm and 720 nm respectively. In all images in Figure 1e-g the scale bar is 500 nm and the dotted line indicates the radial Gaussian standard deviation (σ; numerical value bottom left) from the excitation centre of mass (red circle). Gray lines indicate the principle transport axes, along and orthogonal to the average wire direction (see **SI, S3** for respective line cuts).

## Figure 2

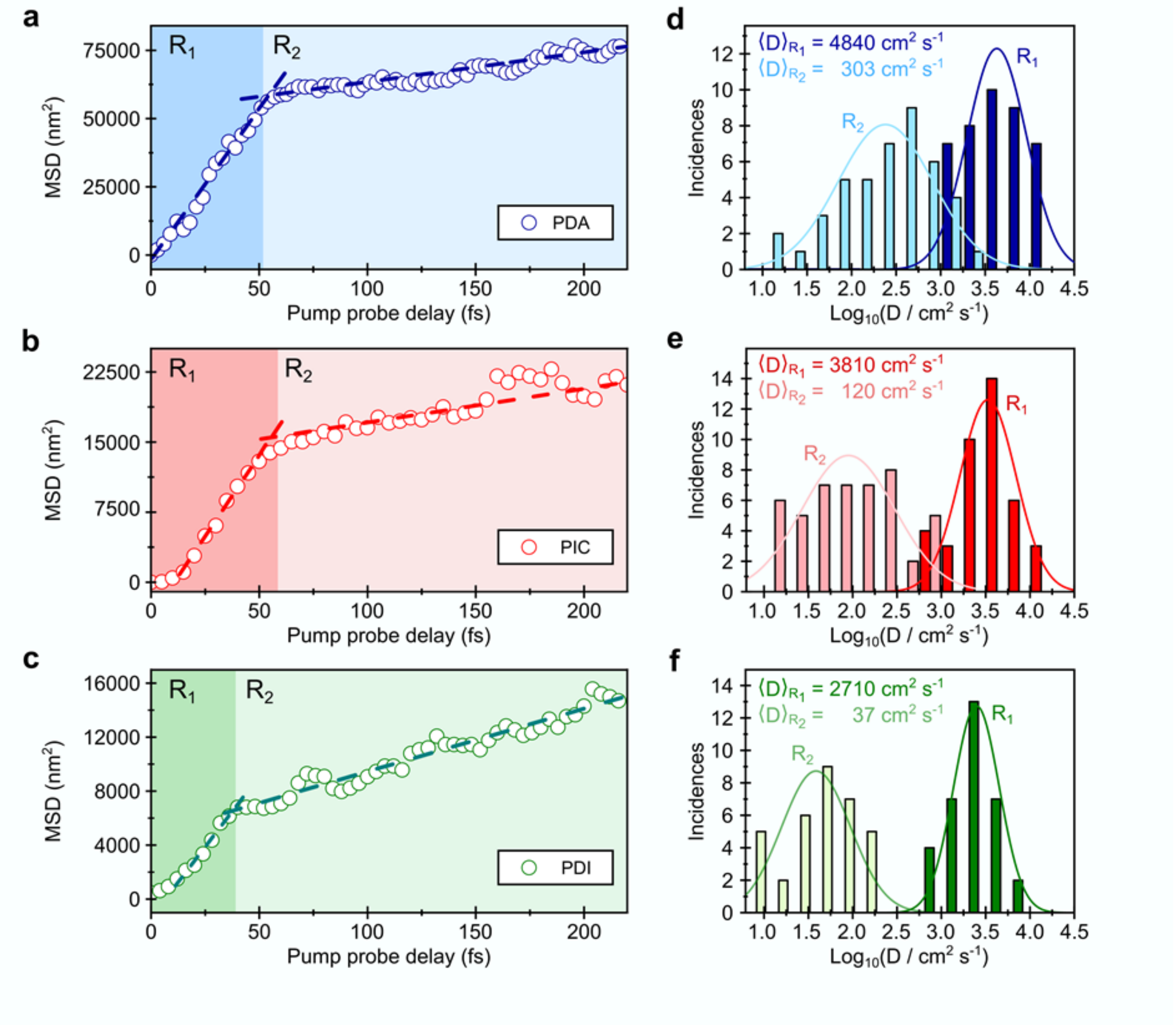

**Figure 2: Dynamics of energy propagation and distribution in transport lengths. a-c.** Representative plots of MSD ($\sigma(t)^2 - \sigma(t_0)^2$) as a function of time for PDA (**a** blue), PIC (**b** red) and PDI (**c** green)**.** The transport is divided into two regimes $R_1$ and $R_2$. Dashed lines show a straight line fit to the two regions from which a diffusion constant (D) can be estimated. **d-f.** fs-TAM measurements are repeated on many different sample locations. Histograms show the frequency of $Log_{10}$(D) (logarithm of diffusion coefficient, D) for PDA, PIC and PDI respectively. The data is divided between $R_1$ (dark shading) and $R_2$ (light shading). Solid line shows a normal distribution fit to the data. Mean diffusion coefficients obtained in both phases are indicated for comparison.

## Figure 3

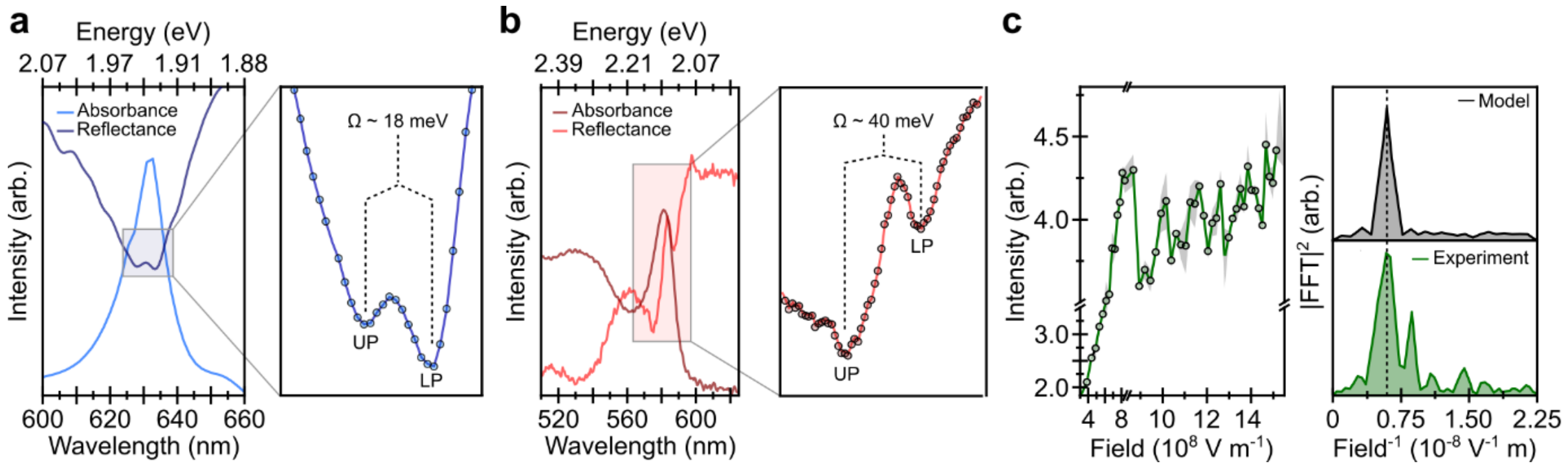

**Figure 3: Experimental evidence for light-matter coupling in organic nanostructures. a.** Absorption (light blue) and specular reflection (dark blue) spectra of PDA. Zooming into the zero-phonon peak at 1.965 eV (right) shows it is split by ~18 meV (Ω) into two branches, an upper (UP) and lower polariton (LP). **b.** Absorption (maroon) and specular reflection (red) spectra of PIC. The excitonic peak at 2.132 eV is split by ~40 meV due to the formation of excition-polaritons (right). **c.** Emission intensity as a function of optical field (laser fluence) for PDI. After increasing linearly, the PL intensity then begins to oscillate at a frequency of 0.71 $V^{-1}$ m after 8 $\times 10^{8}$ V $m^{-1;}$ gray shading shows uncertainty obtained from repeat measurements. Modelling the data[56] allows extraction of the damping time for Rabi flopping (**SI, S14**). Strong agreement between the model and experiment is highlighted by the Fourier transform of the oscillation (right). The extra frequencies in the experimental data arise from the detuning of the laser pulse (590 nm, FWHM ~10 nm) from the excitonic transition ($\lambda_{max}$ = 512 nm). In the case of the individual PDI chromophores or densely packed nanobelts no such behaviour is observed (**SI, S14**).

**Acknowledgements**

We thank the EPSRC (UK) and Winton Program for Physics of Sustainability for financial support. R.P. thanks O. Zadvorna, T. H. Thomas, A. Tanoh, A. Cheminal, W.M. Deacon and A. Alvertis (Cambridge) for assistance with experiments and useful advice. The authors also thank D. Beljonne (Mons) and S. Pannir-Sivajothi (UCSD) for fruitful discussions. This project has received funding from the European Research Council (ERC) under the European Union's Horizon 2020 research and innovation programme (grant agreements No 670405 and No 758826). The work of G.J, O.D.O. and S.V. was supported by the European Research Council (ERC-2014-STG H2020 639088). O.S.O acknowledges the support of a Rubicon fellowship from the Netherlands Organisation for Scientific Research. R.C. acknowledges support from Trinity College, University of Cambridge. T.W. acknowledges a grant from the Swedish research council (VR, 2014-06948) as well as financial support from the Knut and Alice Wallenberg Foundation through the project grant 3DEM-NATUR (no. 2012.0112). SMC acknowledges the Henslow Research Fellowship at Girton College, Cambridge. P.A.M thanks the EPSRC for financial support under grant number EP/R025517/1. C.S. acknowledges financial support by the Royal Commission for the Exhibition of 1851. We acknowledge access and support in the use of the electron Physical Science Imaging Centre (EM20527) at the Diamond Light Source. A.W.C., T.B., L.L., R. S. and F. M. acknowledge the CNRS (France) for financial support. A.J.M., R.J. and D.G.L. acknowledge support from EPSRC (UK) grant EP/M025330/1. M. D. and J.Y.-Z. were supported by the US Department of Energy, Office of Science, Basic Energy Sciences, CPIMS Program under Early Career Research Program award DE-SC0019188.

**Author contributions**

R.P., R.Y.S.C. and Q.G. prepared the PIC and PDI samples, performed the fs-TAM measurements and analysed the data under the guidance of J.S.. O.S.O. and R.C. performed Rabi flopping and reflectivity measurements and interpreted the data under the supervision of J.J.B.. G.J. and O.D.O. performed microscopic reflectivity measurements under the supervision of S.V.. J.G. and F.A. synthesized and characterized the PDI monomers under the supervision of R.H.F.. T.B. (Cambridge) performed KPFM-FM measurements and numerical simulations of the fs-TAM data. T.W., D.N.J. and S.M.C. structurally characterized the PDI nanobelts and solved the crystal structure under the supervision of P.A.M.. T.B., L.L., R. S. and F. M. synthesised the PDA samples. A.J.M., D.L. and R.J. performed transfer matrix simulations and interpreted the data. J.Y.Z. and M.D. performed MEEP calculations. S.K. developed the exciton diffusion model, performed numerical simulations and along with J.Y.Z. interpreted the data. A.R. and P.K. conceived the fs-TAM setup developed by C.S. and J.S.. R.P. and A.R. conceived the project, interpreted the data and wrote the manuscript with input from all authors.

**Supporting information**

Experimental methods, sample preparation and characterization, data analysis methods, reflectivity statistics, fs-TAM control measurements, transfer matrix and waveguiding modelling, exciton diffusion modelling, MEEP calculations.

**Notes**

The raw data associated with this manuscript is available at [url to be added in proof]. The authors declare no competing financial interests.

## References

1. Menke, S. M. & Holmes, R. J. Exciton diffusion in organic photovoltaic cells. *Energy Environ. Sci.* **7**, 499–512 (2014).

2. Mikhnenko, O. V., Blom, P. W. M. & Nguyen, T. Q. Exciton diffusion in organic semiconductors. *Energy Environ. Sci.* **8**, 1867–1888 (2015).

3. Bolinger, J. C., Traub, M. C., Adachi, T. & Barbara, P. F. Ultralong-range polaron-induced quenching of excitons in isolated conjugated polymers. *Science (80-. ).* **331**, 565–567 (2011).

4. Vogelsang, J., Adachi, T., Brazard, J., Vanden Bout, D. A. & Barbara, P. F. Self-assembly of highly ordered conjugated polymer aggregates with long-range energy transfer. *Nat. Mater.* **10**, 942–946 (2011).

5. Haedler, A. T. *et al.* Long-range energy transport in single supramolecular nanofibres at room temperature. *Nature* **523**, 196–199 (2015).

6. Lin, H. *et al.* Collective fluorescence blinking in linear J-aggregates assisted by long-distance exciton migration. *Nano Lett.* **10**, 620–626 (2010).

7. Jin, X. H. *et al.* Long-range exciton transport in conjugated polymer nanofibers prepared by seeded growth. *Science (80-. ).* **360**, 897–900 (2018).

8. Caram, J. R. *et al.* Room-Temperature Micron-Scale Exciton Migration in a Stabilized Emissive Molecular Aggregate. *Nano Lett.* **16**, 6808–6815 (2016).

9. Gao, W., Li, X., Bamba, M. & Kono, J. Continuous transition between weak and ultrastrong coupling through exceptional points in carbon nanotube microcavity exciton-polaritons. *Nat. Photonics* **12**, 363–367 (2018).

10. Ribeiro, R. F., Martínez-Martínez, L. A., Du, M., Campos-Gonzalez-Angulo, J. & Yuen-Zhou, J. Polariton chemistry: controlling molecular dynamics with optical cavities. *Chem. Sci.* **9**, 6325–6339 (2018).

11. Lerario, G. *et al.* Room-temperature superfluidity in a polariton condensate. *Nat. Phys.* **13**, 837–841 (2017).

12. Chikkaraddy, R. *et al.* Single-molecule strong coupling at room temperature in plasmonic nanocavities. *Nature* **535**, 127–130 (2016).

13. Lidzey, D. G. *et al.* Strong exciton-photon coupling in an organic semiconductor microcavity. *Nature* **395**, 53–55 (1998).

14. Snoke, D. Spontaneous Bose coherence of excitons and polaritons. *Science (80 ).* **298**, 1368–1372 (2002).

15. Herrera, F. & Spano, F. C. Cavity-Controlled Chemistry in Molecular Ensembles. *Phys. Rev. Lett.* **116**, 238301 (2016).

16. Schäfer, C., Ruggenthaler, M., Appel, H. & Rubio, A. Modification of excitation and charge transfer in cavity quantum-electrodynamical chemistry. *Proc. Natl. Acad. Sci.* **116**, 4883–4892 (2019).

17. Thomas, A. *et al.* Tilting a ground-state reactivity landscape by vibrational strong coupling. *Science (80 ).* **363**, 615–619 (2019).

18. Andrew, P. & Barnes, W. L. Energy transfer across a metal film mediated by surface plasmon polaritons. *Science (80 ).* **306**, 1002–1005 (2004).

19. Kéna-Cohen, S. & Forrest, S. R. Room-temperature polariton lasing in an organic single-crystal microcavity. *Nat. Photonics* **4**, 371–375 (2010).

20. Halas, N. J., Lal, S., Chang, W.-S., Link, S. & Nordlander, P. Plasmons in Strongly Coupled Metallic Nanostructures. *Chem. Rev.* **111**, 3913–3961 (2011).

21. Brillante, A., Philpott, M. R. & Pockrand, I. Experimental and theoretical study of exciton surface polaritons on organic crystals. I. (010) Face of TCNQ° single crystals. *J. Chem. Phys.* **70**, 5739–5746 (1979).

22. Dähne, L., Biller, E. & Baumgärtel, H. Polariton-induced color tuning of thin dye layers. *Angew. Chemie - Int. Ed.* **37**, 646–649 (1998).

23. Rose, T. S., Righini, R. & Fayer, M. D. Picosecond transient grating measurements of singlet exciton transport in anthracene single crystals. *Chem. Phys. Lett.* **106**, 13–19 (1984).

24. Weiser, G., Fuhs, W. & Hesse, H. J. Study of polariton resonances in a cyanine dye crystal. *Chem. Phys.* **52**, 183–191 (1980).

25. Tischler, J. R. *et al.* Solid state cavity QED: Strong coupling in organic thin films. *Org. Electron. physics, Mater. Appl.* **8**, 94–113 (2007).

26. Gentile, M. J., Núñez-Sánchez, S. & Barnes, W. L. Optical field-enhancement and subwavelength field-confinement using excitonic nanostructures. *Nano Lett.* **14**, 2339–2344 (2014).

27. Dubin, F. *et al.* Macroscopic coherence of a single exciton state in an organic quantum wire. *Nat. Phys.* **2**, 32 (2006).

28. Schott, M. The colors of polydiacetylenes: A commentary. *J. Phys. Chem. B* **110**, 15864–15868 (2006).

29. Jelley, E. E. Spectral absorption and fluorescence of dyes in the molecular state. *Nature* **138**, 1009–1010 (1936).

30. Bücher, H. & Kuhn, H. Scheibe aggregate formation of cyanine dyes in monolayers. *Chem. Phys. Lett.* **6**, 183–185 (1970).

31. Lebedenko, A. N., Guralchuk, G. Y., Sorokin, A. V., Yeflmova, S. L. & Malyukin, Y. V. Pseudoisocyanine J-aggregate to optical waveguiding crystallite transition: Microscopic and microspectroscopic exploration. *J. Phys. Chem. B* **110**, 17772–17775 (2006).

32. Che, Y., Yang, X., Balakrishnan, K., Zuo, J. & Zang, L. Highly polarized and self-waveguided emission from single-crystalline organic nanobelts. *Chem. Mater.* **21**, 2930–2934 (2009).

33. Chaudhuri, D. *et al.* Enhancing long-range exciton guiding in molecular nanowires by H-aggregation lifetime engineering. *Nano Lett.* **11**, 488–492 (2011).

34. Pandey, A. K. Highly efficient spin-conversion effect leading to energy up-converted electroluminescence in singlet fission photovoltaics. *Sci. Rep.* **5**, 7787 (2015).

35. Wan, Y., Stradomska, A., Knoester, J. & Huang, L. Direct Imaging of Exciton Transport in Tubular Porphyrin Aggregates by Ultrafast Microscopy. *J. Am. Chem. Soc.* **139**, 7287–7293 (2017).

36. Solomon, L. A. *et al.* Tailorable Exciton Transport in Doped Peptide-Amphiphile Assemblies. *ACS Nano* **11**, 9112–9118 (2017).

37. Tamai, Y., Ohkita, H., Benten, H. & Ito, S. Exciton Diffusion in Conjugated Polymers: From Fundamental Understanding to Improvement in Photovoltaic Conversion Efficiency. *J. Phys. Chem. Lett.* **6**, 3417–3428 (2015).

38. Snoke, D., Denev, S., Liu, Y., Pfeiffer, L. & West, K. Long-range transport in excitonic dark states in coupled quantum wells. *Nature* **418**, 757–757 (2002).

39. Griffiths, D. J. & Inglefield, *Introduction to Electrodynamics* (Addison Wesley, 1999).

40. Valleau, S., Saikin, S. K., Yung, M. H. & Guzik, A. A. Exciton transport in thin-film cyanine dye J-aggregates. *J. Chem. Phys.* **137**, 034109 (2012).

41. Saikin, S. K., Eisfeld, A., Valleau, S. & Aspuru-Guzik, A. Photonics meets excitonics: Natural and artificial molecular aggregates. *Nanophotonics* **2**, 21–38 (2013).

42. Haken, H. & Strobl, G. An exactly solvable model for coherent and incoherent exciton motion. *Zeitschrift für Phys.* **262**, 135–148 (1973).

43. Hestand, N. J. *et al.* Polarized absorption in crystalline pentacene: Theory vs experiment. *J. Phys. Chem. C* **119**, 22137–22147 (2015).

44. Bergman, A., Levine, M. & Jortner, J. Collision ionization of singlet excitons in molecular crystals. *Phys. Rev. Lett.* **18**, 593 (1967).

45. Meng, R. *et al.* Exciton transport in π-conjugated polymers with conjugation defects. *Phys. Chem. Chem. Phys.* **19**, 24971 (2017).

46. Tuszyński, J. A., Jørgensen, M. F. & Möbius, D. Mechanisms of exciton energy transfer in Scheibe aggregates. *Phys. Rev. E* **59**, 4374 (1999).

47. Berman, P. *Cavity Quantum Electrodynamics*. (1994).

48. Ebbesen, T. W. Hybrid Light-Matter States in a Molecular and Material Science Perspective. *Acc. Chem. Res.* **49**, 2403–2412 (2016).

49. Orgiu, E. *et al.* Conductivity in organic semiconductors hybridized with the vacuum field. *Nat. Mater.* **14**, 1123–1129 (2015).

50. van Vugt, L. K., Piccione, B., Cho, C.-H., Nukala, P. & Agarwal, R. One-dimensional polaritons with size-tunable and enhanced coupling strengths in semiconductor nanowires. *Proc. Natl. Acad. Sci.* **108**, 10050–10055 (2011).

51. Liao, Q. *et al.* An organic nanowire waveguide exciton-polariton sub-microlaser and its photonic application. *J. Mater. Chem. C* **2**, 2773–2778 (2014).

52. Takazawa, K., Inoue, J. I., Mitsuishi, K. & Takamasu, T. Fraction of a millimeter propagation of exciton polaritons in photoexcited nanofibers of organic dye. *Phys. Rev. Lett.* **105**, 067401 (2010).

53. Ojambati, O. S. *et al.* Quantum electrodynamics at room temperature coupling a single vibrating molecule with a plasmonic nanocavity. *Nat. Commun.* **10**, (2019).

54. Gerhardt, I. *et al.* Coherent state preparation and observation of Rabi oscillations in a single molecule. *Phys. Rev. A* **79**, 011402 (2009).

55. Lounis, B., Jelezko, F. & Orrit, M. Single molecules driven by strong resonant fields: Hyper-raman and subharmonic resonances. *Phys. Rev. Lett.* **78**, 3673–3676 (1997).

56. Fox, M. *Quantum Optics: An Introduction* (Oxford University Press, 2006)

57. Wang, M., Li, J., Hisa, Y.-C., Liu, X. & Hu, B. Investigating underlying mechanism in spectral narrowing phenomenon induced by microcavity in organic light emitting diodes. *Nat. Commun.* **10**, 1614 (2019).

58. Tischler, J. R., Bradley, M. S., Bulović, V., Song, J. H. & Nurmikko, A. Strong coupling in a microcavity LED. *Phys. Rev. Lett.* **95**, 036401 (2005).

59. Feist, J. & Garcia-Vidal, F. J. Extraordinary exciton conductance induced by strong coupling. *Phys. Rev. Lett.* **114**, 196402 (2015).

*Supplementary Material*

Table of Contents

## S1: Experimental Methods

**Femtosecond Transient Absorption Microscopy (fs-TAM)** Our fs-TAM setup is schematically depicted in Figure **SI1**. Pulses were delivered by a Yb:KGW amplifier (Pharos, LightConversion, 1030 nm, 5 W, 200 kHz) that seeded two broadband white light (WL) stages . The probe WL was generated in a 3 mm YAG crystal and adjusted to cover the wavelength range from 650-950 nm by a fused-silica prism-based spectral filter. In contrast, the pump WL was generated in a 3 mm sapphire crystal to extend the WL in the high frequency to 500 nm, and a short-pass filtered at 650 nm (Thorlabs, FESH650).

The pump pulses were focussed onto the sample using a single-lens oil immersion objective (100×, numerical aperture 1.1 NA) to a diffraction-limited spot of ~270 nm (FWHM, full bandwidth). In contrast, the counter-propagating probe pulses were loosely focused onto the sample by a concave mirror (FWHM ~15 μm).

A set of third-order corrected chirped mirrors (pump WL – Layertec, probe WL – Venteon) in combination with a pair of fused silica wedge prisms (Layertec) compressed the pulses to sub-10 fs at the sample, as verified by second-harmonic generation frequency-resolved optical gating (Figure **SI2**).

The transmitted probe light was collected by the same objective used to focus the pump pulses and imaged by an EMCCD camera (Qimaging Rolera Thunder, Photonmetrics). Pump light was suppressed by inserting a 650 nm long-pass filter in the detection path (Thorlabs, FELH650) in front of the camera. Additionally, a bandpass filter (Semrock/Thorlabs) was additionally placed in front of the camera to image at the desired probe wavelength in order to avoid chromatic aberration induced image artefacts. Differential imaging was achieved by modulating the pump beam at 45 Hz by a mechanical chopper. The axial focus position was maintained by an additional auto-focus line based on total internal reflection of a 405 nm continuous wave laser beam.

The TAM method builds on techniques developed by the super-resolution microscopy community where monitoring a combination of 'on' and 'off' states allows for sub-diffraction imaging When there is no time delay between pump and probe, a diffraction limited transient absorption (TA) signal is imaged onto an EMCCD camera arising only from the area of excitation. At subsequent time delays, any transport of the initial excitation energy will be observed as a spreading of the spatial pump-probe signal, outside of the initial excitation spot. Subtracting the FWHM of our spatial signals at different time delays allows read-out of the extent of transport as a function of time. Because we are subtracting spatial signals at different time delays, the only limit to our resolution is how well *different* spatial profiles can be resolved. Based on the signal to noise ratio of the measurement this is typically ~10 nm, and hence allows us to break through the diffraction limit.

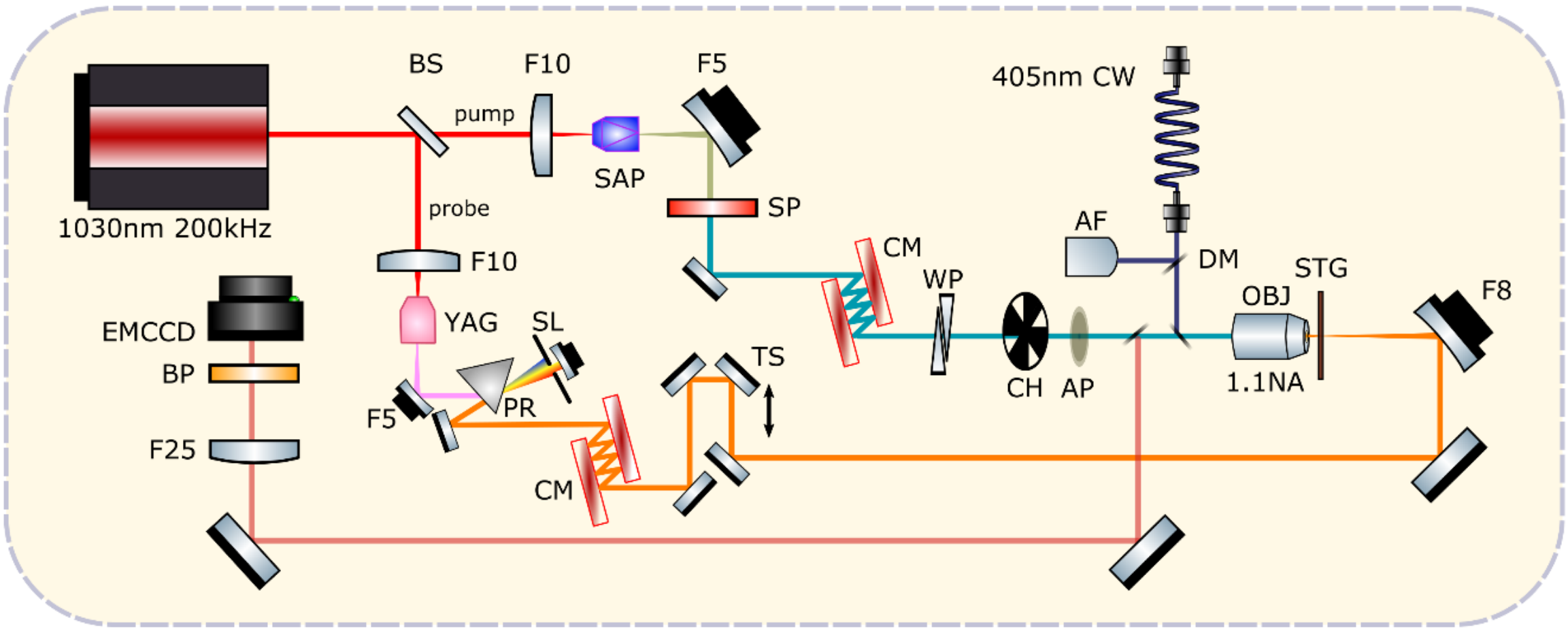

**Figure SI1:** Schematic of fs-TAM setup. Pump and probe pulses are generated by focussing the 1030 nm fundamental of the laser into YAG and sapphire crystals to generate a white light continuum. After second and third order dispersion compensation with chirp mirrors and fused silica wedge pairs, the pump is focussed onto the sample by a high NA objective whilst the probe is loosely focussed onto the sample to achieve wide field illumination. A chopper generates a series of pump on and off shots, which are subsequently imaged by the EMCCD camera. BS – beam splitter, FX – focussing or collimating lens/mirror of X focal length in cm, CM – chirp mirrors, WP – silica wedge pair, AP – pinhole aperture, OBJ – objective, BP X band pass filter of X nm width, SP – short pass filter at X nm, CH – chopper wheel, YAG – 3 mm yttrium aluminium garnet, SAP – 3 mm sapphire, DM – dichroic mirror, AF – auto focus camera, CW – continuous wave autofocus laser, PR – prism, SL – slit, TS – translation stage, EMCCD - electron multiplying charge coupled device.

Pump and probe pulse compression

The pump and probe pulse compression was confirmed using second harmonic generation frequency resolved optical gating (SHG-FROG) (*1*). The dispersion of each pulse up to the sample was compensated for with a combination of chirp mirrors and silica wedge pairs. For the pump, the SHG-FROG traces were measured following passing through the objective. Figure **SI2** shows the respective SHG-FROG traces confirming a sub-10 fs pulse compression.

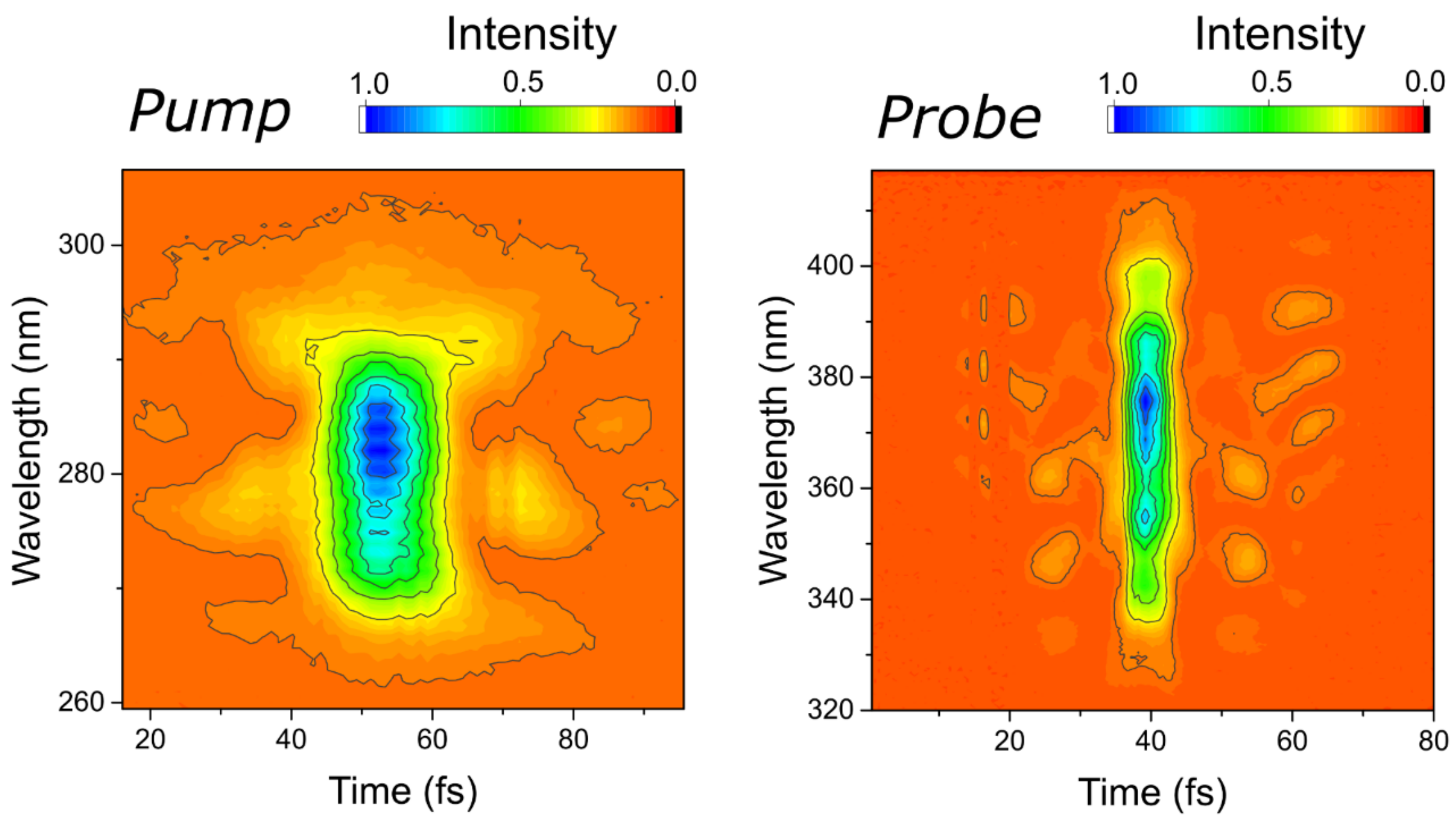

**Figure SI2: a-b.** Second harmonic generation frequency resolved optical gating (SHG-FROG) traces for pump and probe. The step size used is 1 fs and the pump FROG spectrum is taken after the beam has passed through all dispersive optics including the objective. In this case the pump has time duration of ~8 fs and probe ~6 fs.

The time resolution of our setup can also be confirmed from measurement of the samples themselves. PDA for example, is known to exhibit strong Raman bands at ~1400 $cm^{-1}$ and ~2100 $cm^{-1}$ arising from double and triple bond stretches, respectively. A sufficiently short laser pulse (<15 fs) should be able to impulsively excite these modes, resulting in an oscillatory modulation superimposed on the pump probe kinetics. As can be seen in Figure **SI3a** this is the case for the 670 nm kinetic. Subtracting the electronic background and Fourier transforming to the frequency domain shows the kinetic contains two modes: one at 1420 $cm^{-1}$ and another at 2080 $cm^{-1}$, verifying the high temporal resolution of our setup.

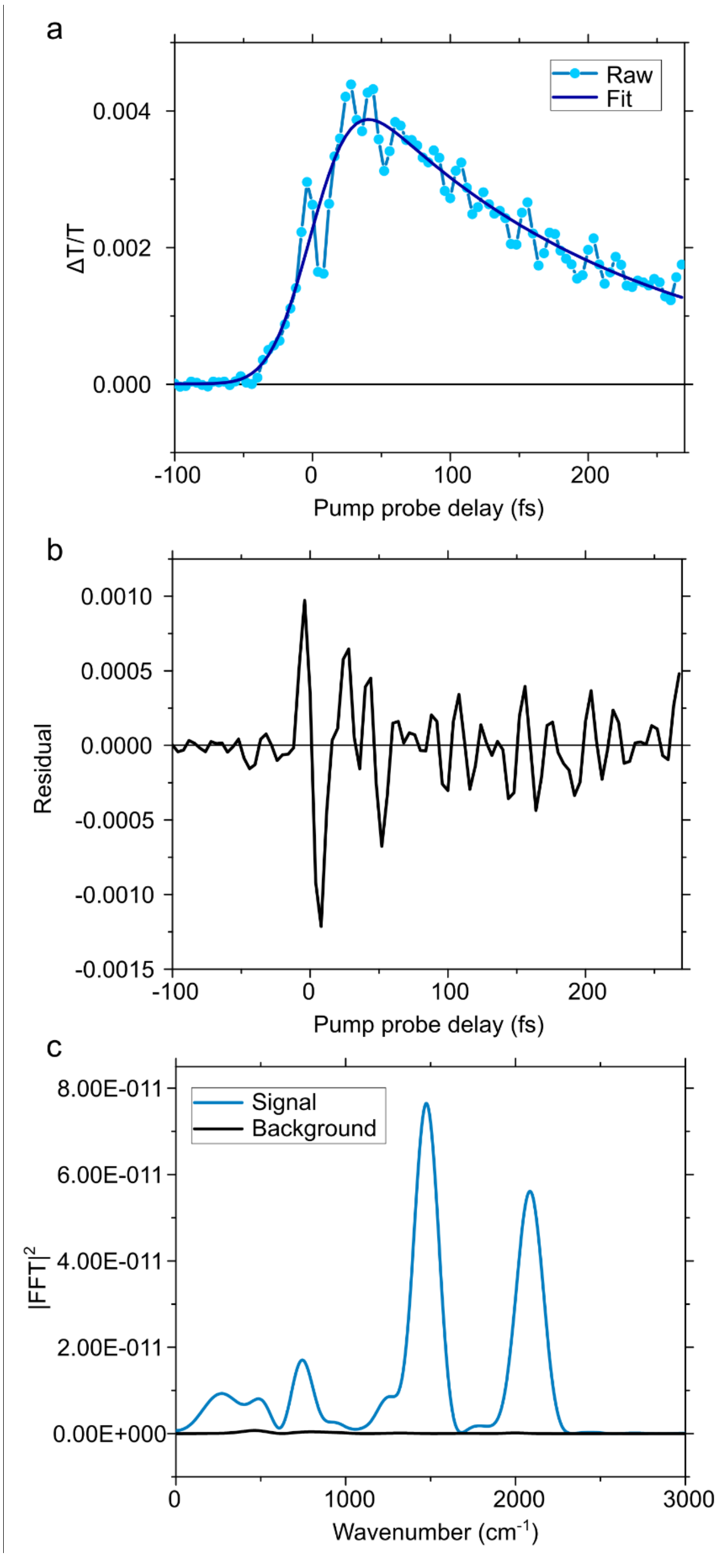

**Figure SI3: a.** PDA kinetic at 670 nm, solid dark blue line shows fit to the raw data (light blue line). **b.** Residual following subtraction of electronic decay. **c.** Fourier transform of residual (blue line) showing clear Raman modes at 1480 $cm^{-1}$ and 2080 $cm^{-1}$. Dark black line shows Fourier transform of the background.

**Femtosecond Pump Probe Spectroscopy (fs-TA)** The fs-TA experiments were performed using an Yb-based amplified system (Pharos, Light Conversion) providing 14.5 W at 1030 nm and a 38 kHz repetition rate. The probe beam was generated by focusing a portion of the fundamental in a 4 mm

YAG substrate and spanned from 520 to 1400 nm. The pump beam was generated in home-built non-collinear optical parametric amplifiers (NOPAs; 37° cut BBO, type I, 5° external angle) pumped with either the second or third harmonic of the source. The NOPAs output (~4–5 mW power) was centered between 520 nm - 860 nm depending on the experiment, and pulses were compressed using a chirped mirror wedge prism (Layterc) combination to a temporal duration of 9 fs - 25 fs for the particular experiment (upper limits determined by SHG-FROG). The white light probe was delayed using a computer-controlled piezoelectric translation stage (Physik Instrumente), and a sequence of probe pulses with and without pump was generated using a chopper wheel on the pump beam. The energy density per pulse maintained at 19 μJ/cm$^2$. After the sample, the probe pulse was split with a 950 nm dichroic mirror (ThorLabs). The visible light (520–950 nm) was then imaged with a silicon photodiode array camera (Stresing Entwicklunsbüro; visible monochromator 550 nm blazed grating) with the near-infrared proportion of the probe seeded to an IR monochromator (1200 nm blazed grating) and imaged using an InGaAs photodiode array camera (Sensors Unlimited). This technique allows simultaneous collection of the entire probe spectrum in a single shot. Offsets for the differing spectral response of the detectors was accounted for in the data post-processing.

**Picosecond Pump Probe Spectroscopy (ps-TA)** The picosecond transient absorption (ps-TA) experiments were performed using an Yb-based amplified system (PHAROS, Light Conversion) providing 14.5 W at 1030 nm and 38 kHz repetition rate. The probe beam was generated by focusing a portion of the fundamental in a 4 mm YAG substrate and spanned from 520 to 900 nm. The pump beam was generated by seeding a portion of the fundamental to a narrow band optical parametric oscillator (ORPHEUS-LYRA, Light Conversion). The pump and probe beams were focused to an area of 300 μm × 330 μm and 60 μm × 70 μm, respectively on the sample using off-axis parabolic mirrors. The energy density per pump pulse was 5 - 45 μJ/cm$^2$. The white light probe was delayed using a computer-controlled piezoelectric translation stage (Newport), and a sequence of probe pulses with and without the pump was generated using a chopper wheel (ThorLabs) on the pump beam. The probe pulse transmitted through the sample was detected by a silicon photodiode array camera (Stresing Entwicklungsbüro; visible monochromator 550 nm blazed grating).

**Transmission and Reflection Spectroscopy** Optical microscopy was performed using a Zeiss Axio.Scope optical microscope in Köhler illumination equipped with a 100× objective (Zeiss EC Epiplan-APOCHROMAT 0.95 HD DIC) coupled to a spectrometer (Avantes HS2048) via an optical fibre (Thorlabs, FC-UV50-2-SR). Five spectra were collected for each sample using an integration time of 10 ms and 20 ms for reflection and transmission measurements, respectively. The reflectance and transmittance were calculated using a silver mirror (ThorLabs, PF10-03-P01) and the glass substrate as references respectively.

**Angle resolved reflectivity** Angle dependent reflectivity measurements were performed using an Andor Shamrock SR-303i-A triple-grating imaging spectrograph, with a focal length of 0.303m. The spectra were recorded using a 300 grooves / mm grating blazed at 500 nm. The reflectivity light source was a fibre-coupled 20W tungsten halogen lamp (Ocean Optics DH-2000-BAL). The angle dependence was measured using a k-space imaging setup. The white light was focussed on to the sample at normal incidence using an Edmund Optics 20X objective with a numerical aperture (NA) = 0.63, with the reflected signal collected through the same optical path using a beam splitter. This light was then focussed into a spectrometer using a final collection lens. An additional Fourier-plane imaging lens positioned before the final lens facilitated the Fourier plane to be imaged into the spectrometer. Here, a dual axis slit was positioned at the focus of the imaging lens (before the final collection lens) which allowed the reflectivity to be spatially filtered, permitting the unwanted real space signal to be rejected.

**Power dependent photoluminescence spectroscopy** Samples were excited with ~120 fs pulses, ~10 nm full width at half maximum (FWHM), generated from a tunable optical parametric oscillator (OPO) (Spectra Physics Inspire) pumped at 820 nm with a repetition rate of 80 MHz. The power of the pulses was controlled using a variable neutral density filter (ThorLabs) mounted on a rotating stage. The attenuated pulses pass were focused by a microscope objective (Nikon 100x, numerical aperture 0.9) to excite the PDI nanobelts at 590 nm. Emission light was collected by the focusing objective and passes through tunable long pass filters (Fianum). An achromatic doublet lens (focal length f = 60 mm) focuses the filtered light onto a single photon avalanche diode (SPAD) (MPD, jitter time <50 ps). The output of the SPAD was read out and averaged over 5 seconds for each power step. In the plot shown in Figure 3c of the main manuscript, we have accounted for the transmission through the optics and the data is a representative measurement from different single nanobelts.

**Sample Preparation**

Unless otherwise stated all samples were deposited onto pre-cleaned (Acetone/Water, Isopropanol, 10 mins $O_2$ Plasma etch) 22 mm × 22 mm No. 1.5 borosilicate glass coverslips (ThorLabs). Samples were encapsulated in a $N_2$ glovebox prior to measurement by placing a second smaller coverslip atop of the sample substrate with a ~200 µm thick carbon tape spacer and sealing with epoxy resin.

**Polydiacetylene (PDA):** The 3BCMU (3-methyl-n-butoxy-carbonylmethyl-urethane) diacetylene molecules were synthetized in-house using the method previously outlined by Se *et al.* and references therein (*2*). The synthesis classically consisted of two steps: (i) oxidative coupling of 4-pentyn-1-ol (Hay's method) to produce the 4,6-decadiyn-1,10-diol and then, (ii) reaction of the diol with *n*- butylisocyanate acetate. Note that the source 4-pentyn-1-ol was not synthetized at the laboratory but purchased from Sigma-Aldrich (Merck).

Ultrathin single crystals were grown between two coverslips using a melt-processing method. The whole process was systematically carried out under a polarized optical microscope so as to be able to follow and control the sample elaboration: a very small amount of diacetylene powder is placed at one edge of the double-slides assembly; when heating above the melting temperature (~65°C) the liquid diacetylene fills the empty space by capillary action to form a thin liquid film between the two substrates. Rapid cooling leads to the formation of a highly polycrystalline film. The sample is then heated again to around the melting temperature until the melting of all the crystallites took place. When only a few crystal germs remain the sample is cooled again at a very slow cooling rate (typically < 0.1°C/mn) to induce the growth of large single monocrystalline domains from the germs.

The sample prepared in this manner is then studied as it is (without separation of the slides) after the desired amount of polymer is generated inside the film. It is important noting that submicronic thickness is obtained (*3*) corresponding to a reduction of 1 to 2 orders of magnitude in thickness as compared with diacetylene monocrystals usually considered in advanced transmission spectroscopy experiments (*4–6*). The optical density (OD) of the studied samples was kept below $10^{-2}$ (that is close to the detection limit of our CARY 5000 spectrophotometer). Taking an upper bound ~0.8 µm for the film thickness leads to a chain content by weight, $x_p$, that keeps below $\sim 2.3\ 10^{-4}$ (using $OD = (\alpha\, l\, x_p)/2.3$ with $\alpha \sim 10^6\ cm^{-1}$ at room temperature, $\alpha$ being the absorption coefficient of 3BCMU chains inside their host crystal (*7*)). It is then estimated that such a polymer content corresponds to an interchain distance, $d_{inter}$, of several tens of nm; for $x_p = x_{p,\,max} \sim 2.3\ 10^{-4}$ we find here $d_{inter} = d_{inter,\,min} \sim 30$ nm (considering the shorter interchain distance along the crystal **c** axis, $d_{inter} \sim 0.45$ nm, for $x_p = 1$ as well as homogeneous topochemical polymerization (*7*, *8*)). In these conditions interchain interactions can reasonably be discarded in the analysis.

Highly aligned regions of the crystal were selected using a transmission microscope prior to fs-TAM measurements. A 3BCMU crystal (monoclinic space group) is a birefringent medium with its optic axis lying in the surface plane and corresponding to the long axis of the aligned polymer chains. Another principal axis also lies in the surface plane of the crystal). To identify high quality crystalline domains (constant thickness, absence of strain), samples are placed between crossed polarisers and illuminated. The rotation of the sample allows to identify domains that present the most homogeneous optical extinction (Figure **SI4**). Regions exhibiting such behaviour were predefined by application of masks to the sample.

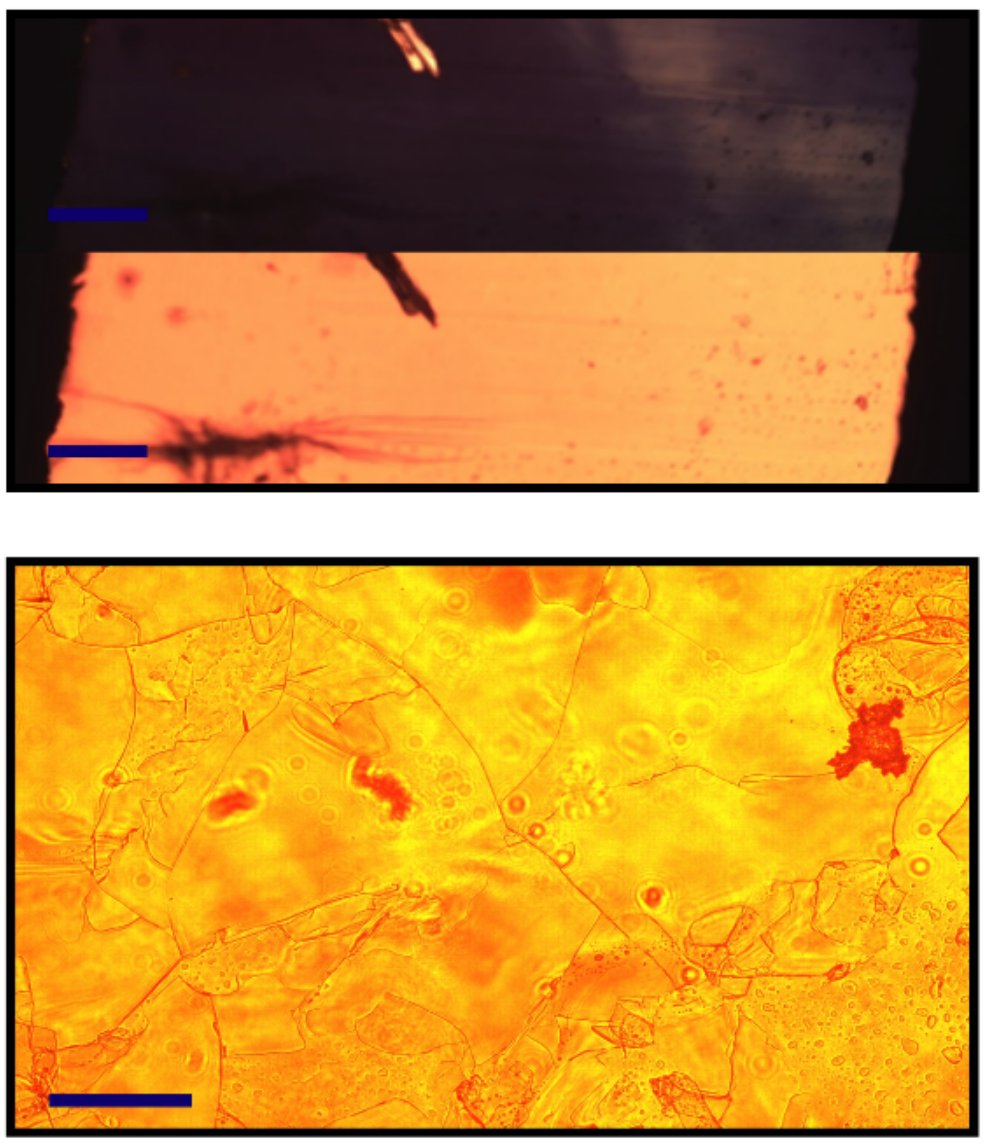

**Figure SI4:** Top panel: Optical microscope image of PDA thin film in transmission and between crossed polarizers. When the linearly polarized incident light has its polarization along one of the optic axes, complete extinction is obtained. In "dark" domains the chains present in the diacetylene matrix have their conjugated backbone either in the direction of incident polarization or perpendicular to it. In general the chains are parallel to the longest dimension of the monocrystalline domain. Adjacent domains do not have the same orientation as observed in the figure and transmitted light acquires elliptical polarization as it propagates through the film. Scale bar in the top image is 2 µm. Bottom image shows transmission microscope image through PDA crystal. Clear crystal domains can be observed. Scale bar for bottom image is 150 µm.

**PIC:** To prepare the aggregates, we modified existing procedures (*9*, *10*) to create a high-concentration sugar glass similar to that initially detailed for double-walled cyanine nanotubes by Caram *et al.* (*11*). PIC monomer (Sigma Aldrich) was dissolved in methanol by shaking overnight to produce a 10 mM solution. We then mixed 250 µL of monomer solution with 250 µL of a saturated solution of sucrose. We deposited 200 µl of the sugar-monomer mixture onto pre-cleaned glass substrates (acetone, isopropanol, $O_2$ plasma *etc*; each 10 mins) and spin coated at 2500 rpm for 2 mins. All spin coating was performed in an inert $N_2$ environment, to prevent oxidative degradation. All solvents used were purged of $O_2$ via a freeze-pump-thaw cycle and bubbling of Argon through the solvent. The samples were then dried under vacuum (0.5 atm) for 24 hrs, to form a uniform amorphous glass. Samples for electron microscopy were prepared by dropping ~3 µL of monomer-sugar solution matrix on to a TEM copper grid with a holey carbon support film in an inert atmosphere, followed by spin coating under the same conditions as above. Once dried in vacuum, bright-field TEM was carried out in an FEI Tecnai Osiris TEM operated at 200 kV. In all cases, samples were measured within 48 hrs of preparation to avoid morphological changes.

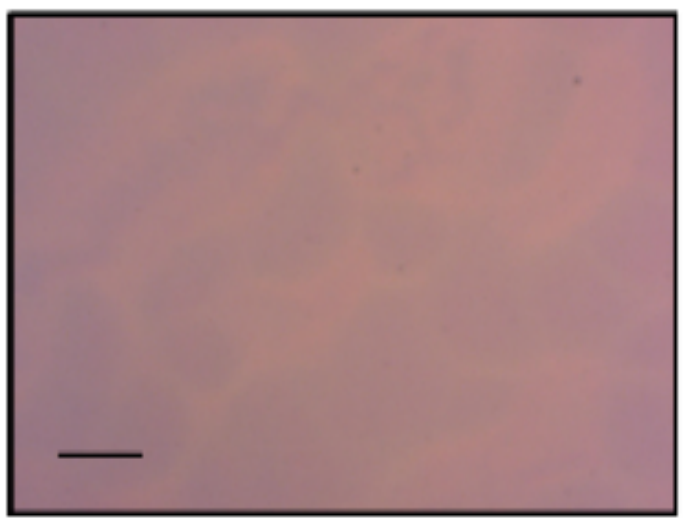

**Figure SI5:** Optical microscope image of PIC thin film. The fibrous pale yellow streaks on the pink background are suggested to be bundles of nanotubes. Scale bar is 5 μm.

**PDI:** In order to prepare PDI nanobelts, the PDI monomer was synthesised as described below.

Abbreviations

| | |
|---|---|
| CF | Chloroform |
| EtOAc | Ethyl Acetate |
| EtOH | Ethanol |
| eq. | Equivalents |
| HCl | Hydrochloric Acid |
| NMR | Nuclear Magnetic Resonance |
| PTCDA | Perylene - 3, 4, 9, 10 - Tetracarboxylic Dianhydride |
| $R_f$ | Retardation Factor |

Materials

Solvents and reagents were purchased from commercial suppliers and used as received. Flash column chromatography was performed using silica gel (60 Å, 40-60 μm, Acros Organics). Analytical thin layer chromatography was carried out with silica gel 60 F254 plates (Merck).

Methods

Reactions were carried out with oven-dried glassware. $^{1}H$ and $^{13}C$ NMR spectra were recorded on a Bruker 500 MHz DCH Cryoprobe Spectrometer. Proton chemical shifts (δ) were measured in parts per million (ppm) relative to tetramethylsilane (0 ppm) and referenced to the residual undeuterated solvent peaks. Shifts are reported to the nearest 0.01 ppm, coupling constants (J) reported to the nearest 0.1 Hz. Multiplicity is reported using: singlet (s), doublet (d), triplet (t), quartet (q), and multiplet (m), and combinations thereof. Carbon chemical shifts are also measured in ppm relative to tetramethylsilane, referenced to the solvent peaks, and are quoted to the nearest 0.1 ppm.

Synthesis

***N,N'*-dicyclohexyl -3,4,9,10-perylenedicarboximide**

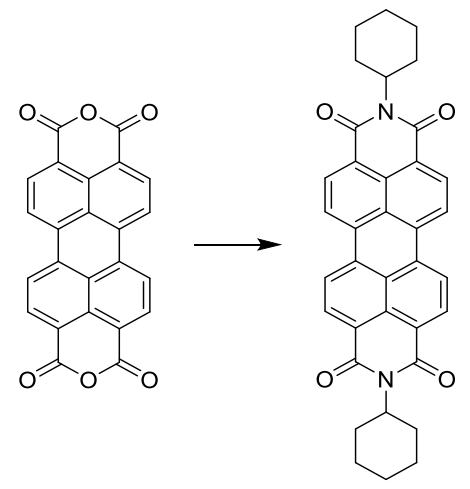

PTCDA (390 mg, 1 mmol, 1 eq.), $Zn(OAc)_2 \cdot 2H_2O$ (196 mg, 4 mmol, 4 eq.), cyclohexylamine (343 µL, 3 mmol, 3 eq.), and imidazole (4 g) were stirred under argon at 160 °C for 2 hrs. The reaction was cooled and the crude dispersed in EtOH (150 mL), 2M HCl (100 mL), and water (100 mL) and left stirred overnight in air at room temperature. The precipitate was collected by filtration, washed with water, and dried. The solid was purified by column chromatography (silica gel, CF/EtOAc, 9:1) to yield the title compound as a dark red solid (283 mg, 0.51 mmol, 51 %).

**$^1$H NMR** (500 MHz, $CDCl_3$, <0.5 mg/mL) δ 8.67 (d, *J* = 8.0 Hz, 4H), 8.61 (d, *J* = 8.0 Hz, 4H), 5.05 (tt, *J* = 12.2, 3.8 Hz; 2H), 2.62-2.53 (m, 4H), 1.94-1.91 (m, 4H), 1.80-1.73 (m, 6H), 1.50-1.33 (m, 6H).

**$^{13}$C NMR** (126 MHz, $CDCl_3$) δ 163.9, 134.5, 131.1, 129.4, 126.4, 123.9, 123.0, 54.0, 31.87, 23.5, 22.7.

**$R_f$** 0.54 (CF:EtOAc, 9:1)

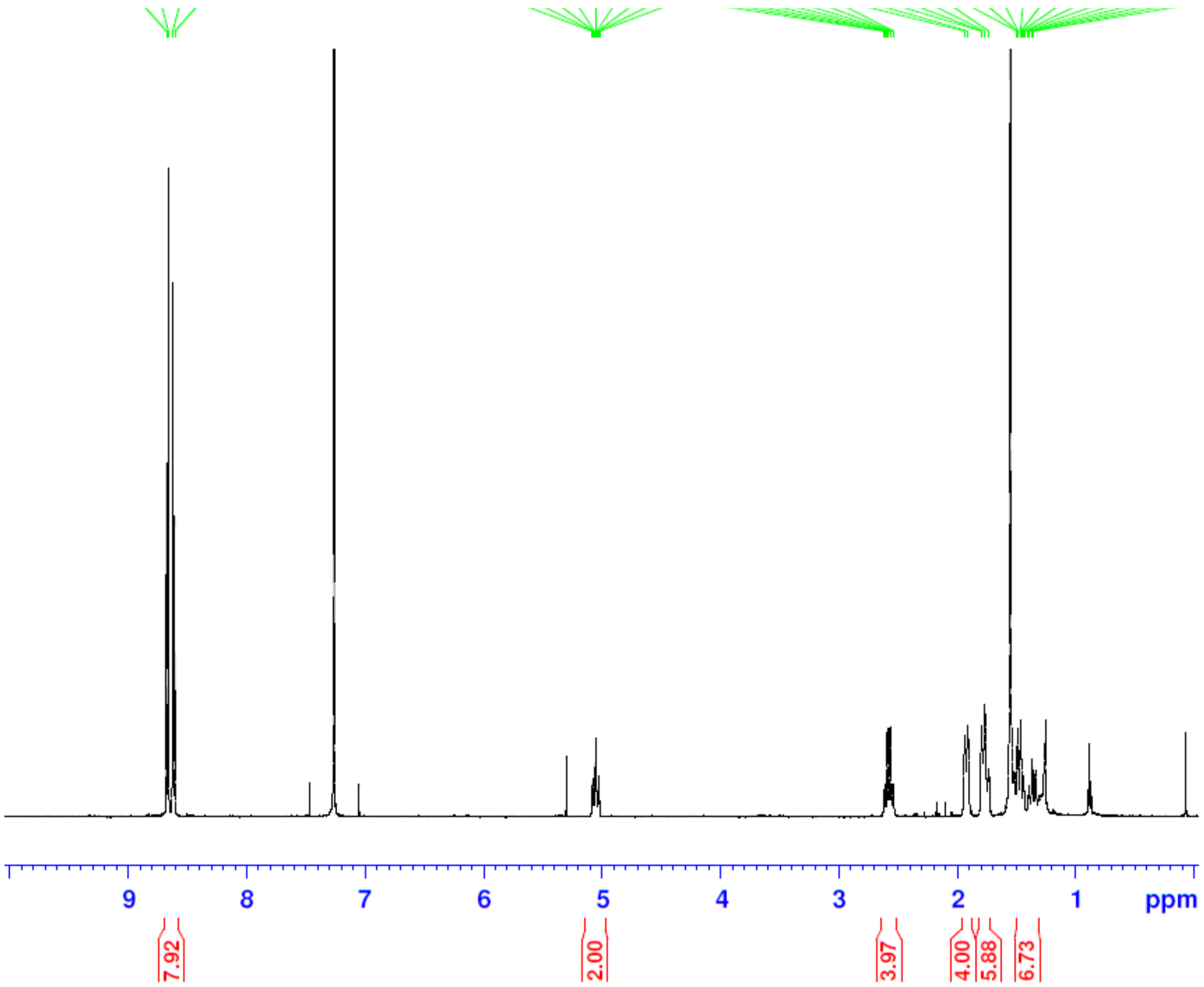

**Figure SI6**: $^{1}$H-NMR of *N*,*N′*-dicyclohexyl -3,4,9,10-perylenedicarboximide

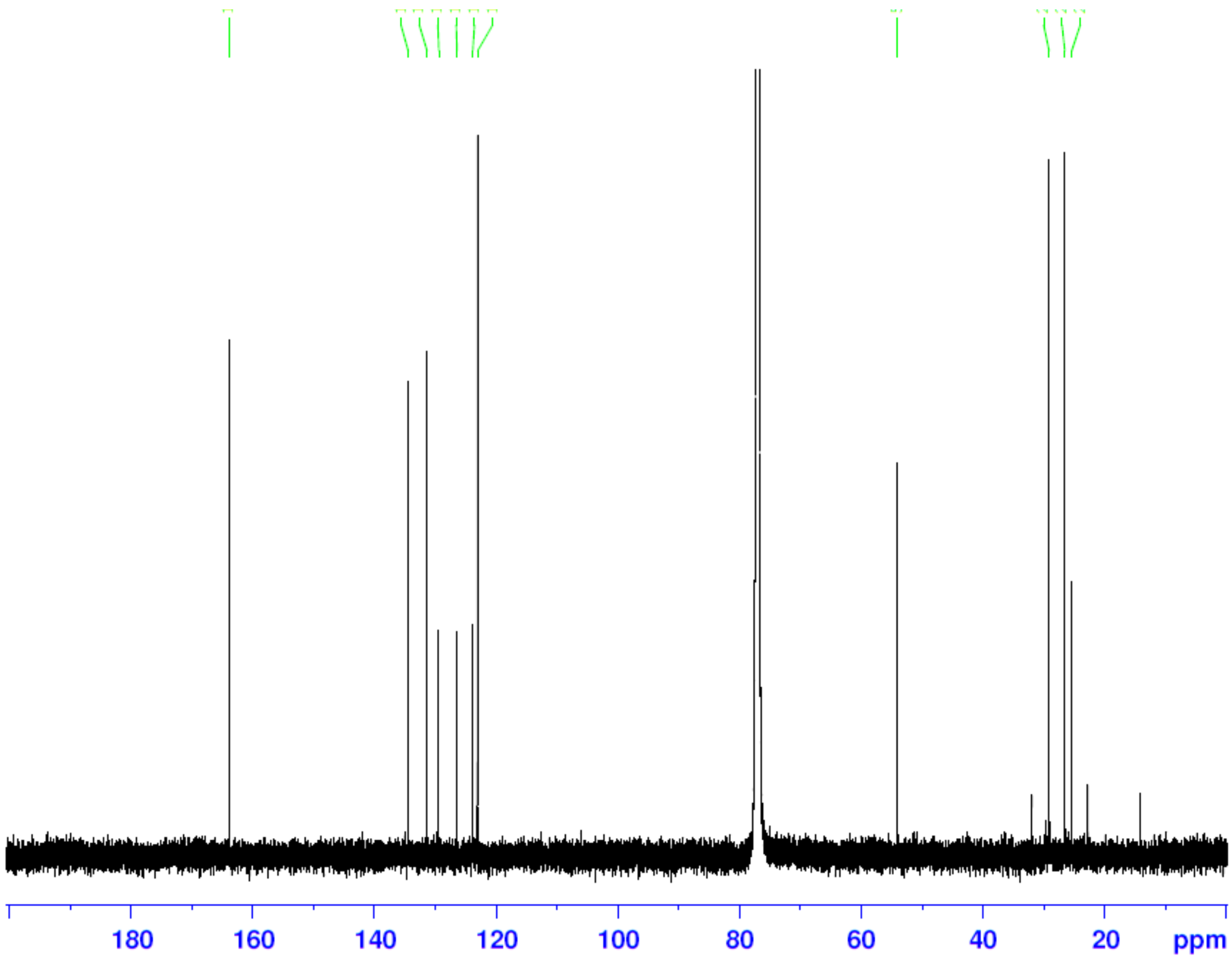

**Figure SI7:** $^{13}$C-NMR of *N*,*N'*-dicyclohexyl -3,4,9,10-perylenedicarboximide

Nanobelts were fabricated by self-assembly of CH-PTCDI molecules using a phase transfer method as outlined previously by Che *et al.* (*12*). A concentrated solution of CH-PTCDI (0.5 mL, 0.3 mM) was prepared in chloroform in a glass vial. This was then injected to the bottom of a solution of ethanol (~11:3 volume ratio, EtOH : PDI). The solution was allowed to sit in the dark at room temperature for 24 hrs and nanobelts were formed at the interface of the solvents. After mixing the two solvents, the nanobelts diffused into the whole solution phase and were transferred onto glass coverslips by pipetting (Figure **SI8**). The samples were then allowed to dry in a $N_2$ glove box before encapsulation and measurement. The solution was not shaken during the self-assembly process to prevent nanobelts with sharp ends forming.

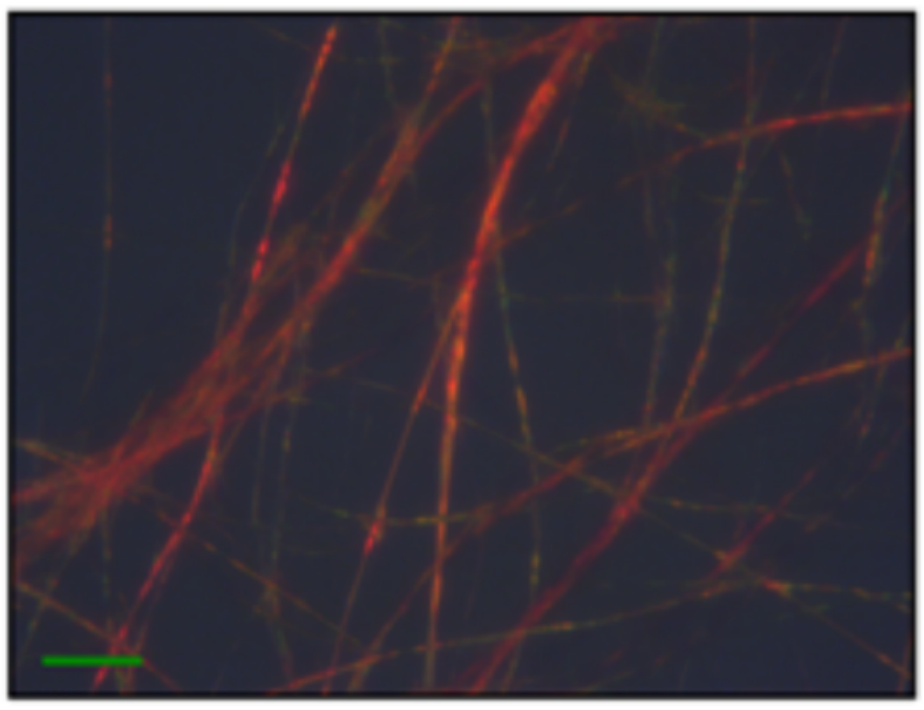

**Figure SI8:** Optical microscope image of PIC thin film in transmission. The wires are long and highly twisted. Scale bar is 10 µm.

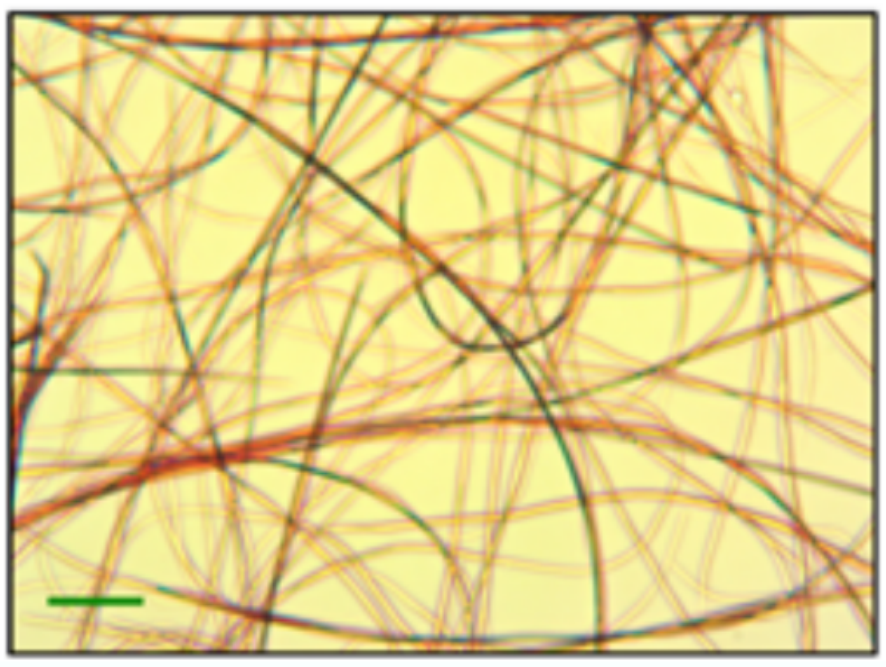

**Figure SI9:** Optical microscope image of PIC thin film in reflection. Scale bar is 10 µm.

Powder X-ray diffraction (PXRD) patterns (Figure **SI10**) were measured on a Bruker D8 Advance diffractometer equipped with a Cu $K_\alpha$ source (0.1 mm divergence slit, static) and a LynxEye XE detector. $K_\beta$ radiation was attenuated with a 0.0125 mm Ni filter.

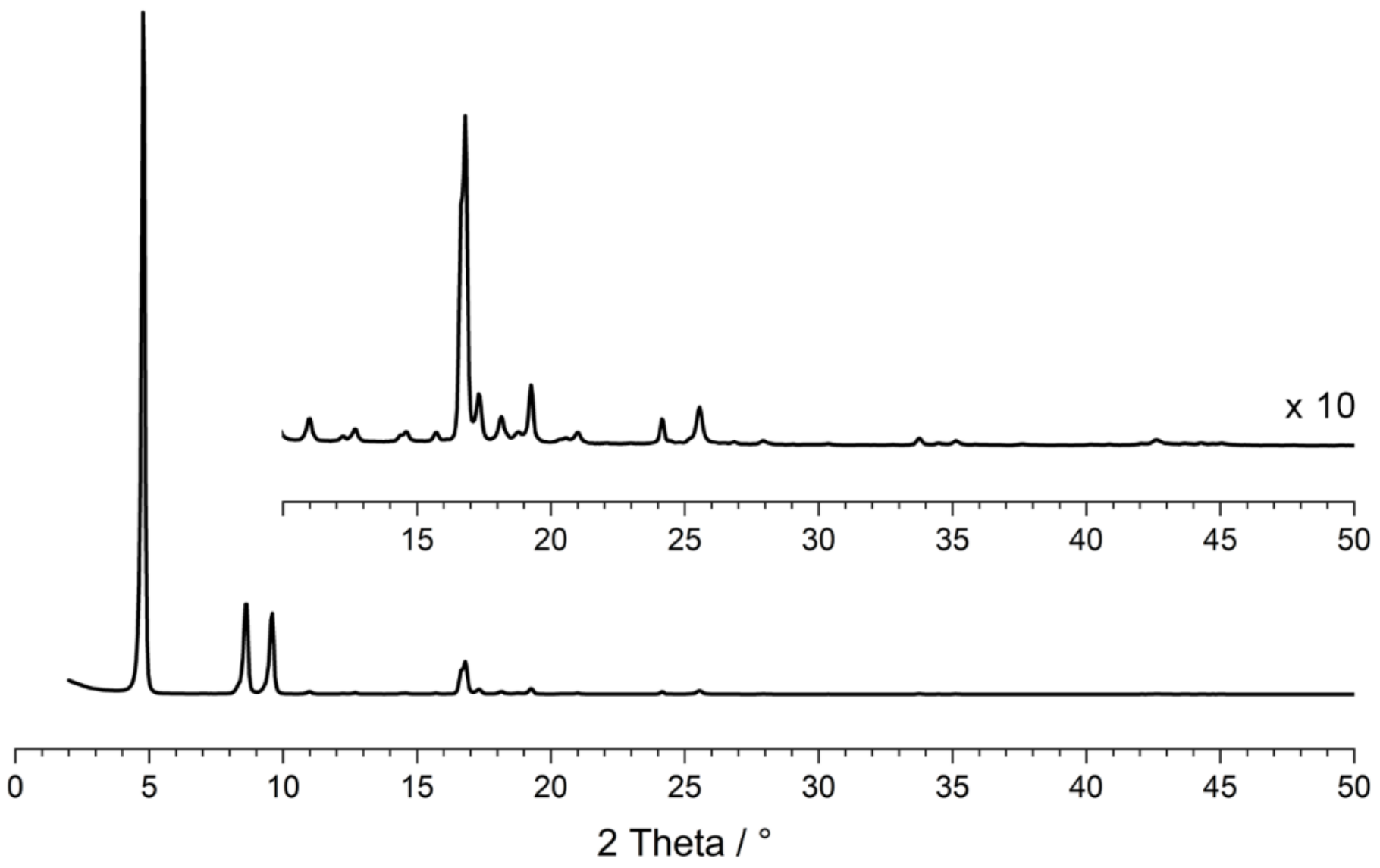

**Figure SI10:** Powder XRD pattern of the cyclohexyl PDI wires.

## S2: Sample characterization

To characterize the structure and morphology of the semiconductor thin films, a combination of techniques were employed, including; continuous rotation electron diffraction (cRED), transmission electron microscopy (TEM), scanning electron diffraction (SED), scanning electron microscopy (SEM), atomic force microscopy (AFM), and kelvin probe force microscopy (KPFM).

### Cyclohexyl PDI structure determination by continuous rotation electron diffraction

cRED data were acquired from PDI nanobelts deposited on standard holey carbon coated 200 mesh Cu TEM grids (EM Resolutions Ltd.) using a JEOL JEM-2100 TEM operated at an accelerating voltage of 200 kV. The specimen was cooled to a temperature of ~100 K during data collection and during acquisition the goniometer was rotated continuously with a speed of 0.45 °/s. The PDI nanobelts were typically single crystals and selected electron diffraction patterns, acquired near to crystallographic zone axes, during this tilt series, are shown in Figure **SI11**.

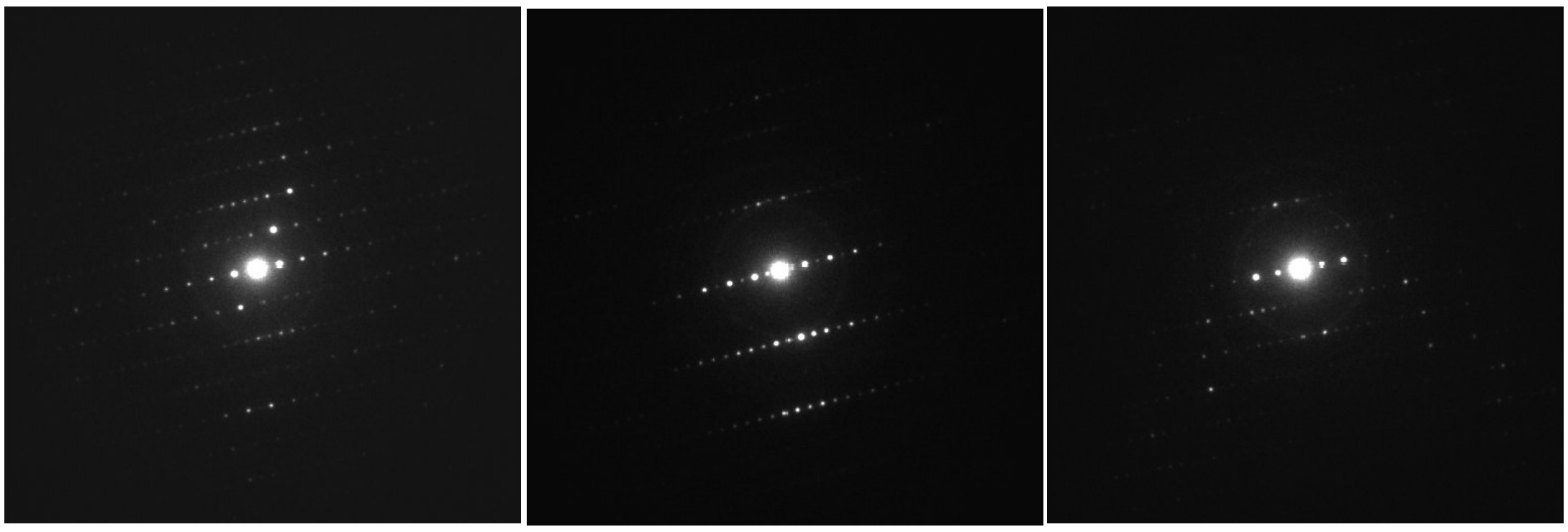

**Figure SI11:** Selected electron diffraction patterns from a continuous Rotation Electron Diffraction data set. The selected patterns are from orientations close to crystallographic zone axes.

The cRED datasets were processed using X-ray Detector Software (XDS) (*13*) where the observed reflections were indexed using a single unit cell and the corresponding intensities were extracted for structure solution and refinement. The structure was solved using the software SHELXT (*14*) and a least-squares refinement was performed in SHELXL-97 (*15*), converging with an R1 of 2879%, which is a typical value for structures solved using electron diffraction data and refined in a kinematical framework (*16*), details of the data and least-square refinement can be found in Table **T2**..

The determined structure was orthorhombic ($Pn2_1a$) with unit cell parameters (a = 22.9 Å, b = 37.6 Å, c = 7.2 Å, 90°, 90°, 90°). The cyclohexyl PDI molecules exhibited a herringbone packing with π-stacking near to the crystallographic c-axis as shown in Figure **SI12**.

crystallographic c-axis as shown in Figure **SI11**.

**Table T2.** Details of the electron diffraction data and refinement

| | |
|---|---|
| **Crystal system** | Orthorhombic |
| **Space group** | $Pn2_1a$ (No. 33) |
| **a, Å** | 22.9 |
| **b, Å** | 37.6 |
| **c, Å** | 7.2 |
| **α, °** | 90 |
| **β, °** | 90 |
| **γ, °** | 90 |
| **Volume, Å$^3$** | 6199.5 |
| **λ, Å** | 0.0251 |
| **Exposure time per frame (s)** | 0.6 |
| **Tilt range, °** | -59.71 – +77.85 |
| **Resolution, Å** | 0.97 |
| **Completeness, %** | 79.9 |
| **$R_{int}$** | 0.159 |
| **R1** | 0.288 |
| **No. of symmetry independent reflections** | 6043 |
| **Number of parameters** | 256 |
| **Number of restraints** | 7 |

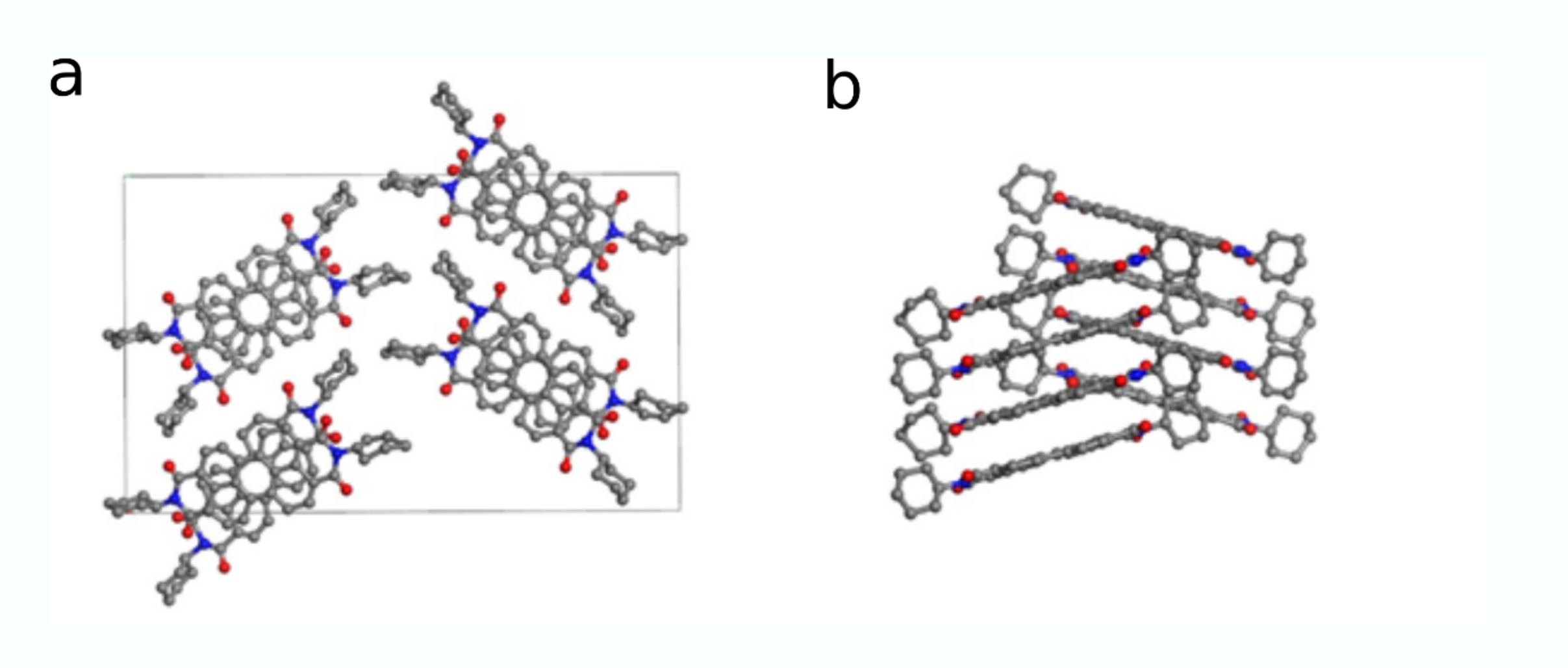

**Figure SI12:** The crystal structure of as determined from continuous Rotation Electron Diffraction data. **a.** View along the crystallographic *c*-axis **b.** shows the π-stacking of the molecules along an axis near to the (vertical) crystallographic *c*-axis..

Cyclohexyl PDI crystallographic mapping by scanning electron diffraction

Scanning electron diffraction (SED) data was acquired from the PDI nanobelts using a JEOL GrandARM 300F (S)TEM operated at an accelerating voltage of 300 kV with a probe current of 1.8 pA and a convergence semi-angle of ~0.6 mrad leading to a diffraction limited probe diameter of ~5 nm. Electron diffraction patterns were recorded using a Merlin/Medipix detector with 256×256 pixels at a camera length of 20 cm giving a calibrated pixel size of 0.010 $Å^{-1}$/px. Scans were performed with a calibrated step size of 7.73 nm/px over 256×256 probe positions. Virtual dark-field (VDF) images were formed by plotting the summed intensity in the diffraction plane, in the scattering vector magnitude range 0.75 to 1 $Å^{-1}$, as a function of probe position using the pyXem python library, as shown in Figure **SI12**. These images reveal the nanobelts to have a width of ~100 nm. Diffraction patterns from selected regions along the nanobelts illustrate that the nanobelts are approximately single crystals with a common orientation axis aligned along the long axis of the nanobelts, as shown in Figure **SI13.**

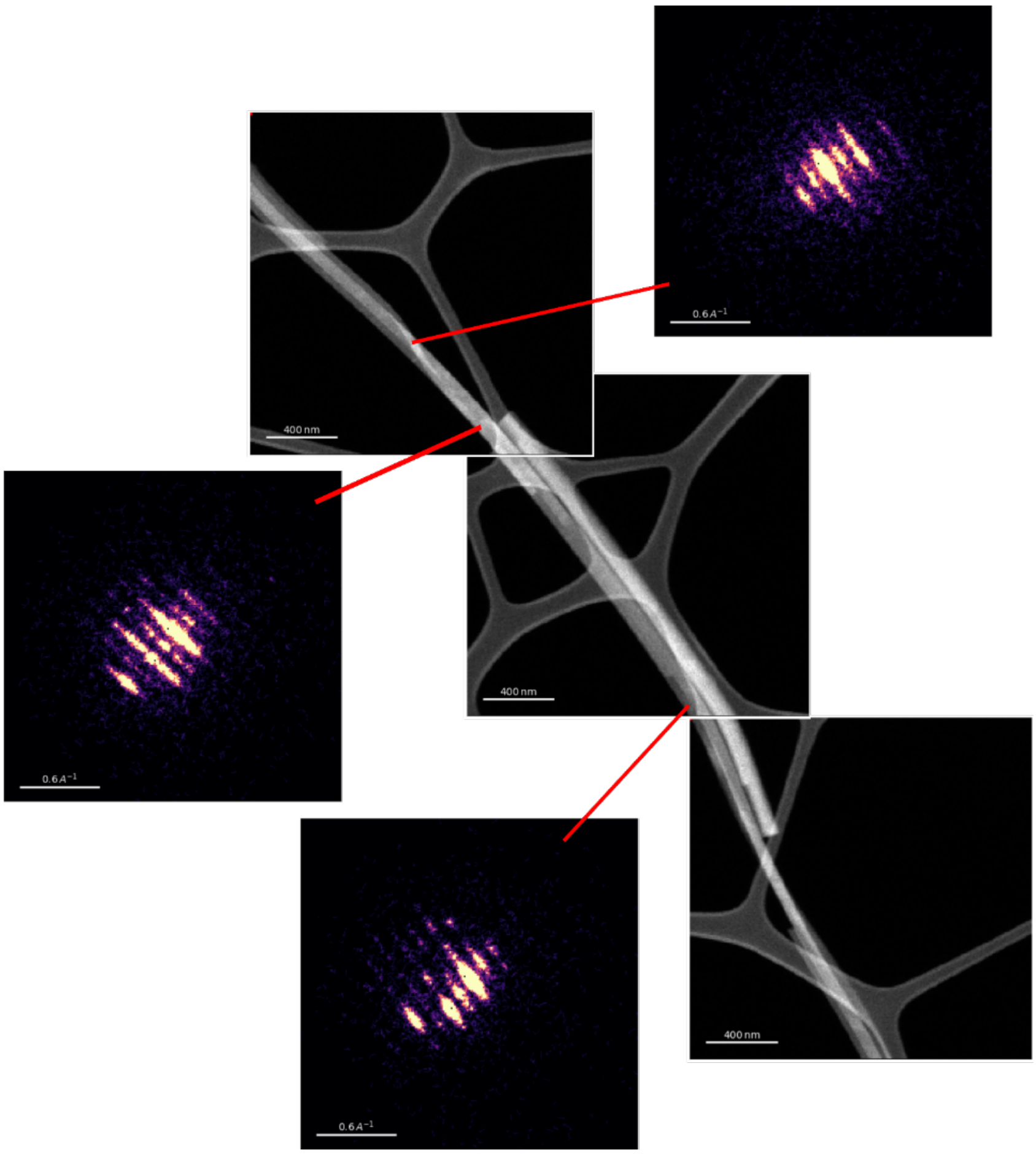

**Figure SI13:** Virtual diffraction images reconstructed from the SED microscopy data, integrating the scattering range 0.75 to 1 $Å^{-1}$, and representative electron diffraction patterns from selected regions along the nanobelts illustrating an approximately common crystal orientation.

Scanning electron microscopy of PDA and PIC thin films

To better characterise sample morphology, scanning electron microscopy (SEM) was performed on PDA and PIC thin films. In PDA (Figure **SI14**) it can be clearly seen that the crystal grows into 100 $\mu m^2$ domains with well-defined grain boundaries between the domains. Optical microscopy reveals the polymer to be highly aligned within a domain. Unpolymerized/unaligned regions of the sample (Figure **SI14c**) appear as small <1 $\mu m^2$ sized islands on the glass substrate.

SEM of PIC thin films (Figure **SI15**) shows the sample to be highly textured. In particular, there are large (>5 μm) grains in which it is presumed the nanotubes are bound. The remainder of the film appears to consist of a smooth background potentially arising from isolated nanotubes embedded within the sugar matix or isolated dye molecules.

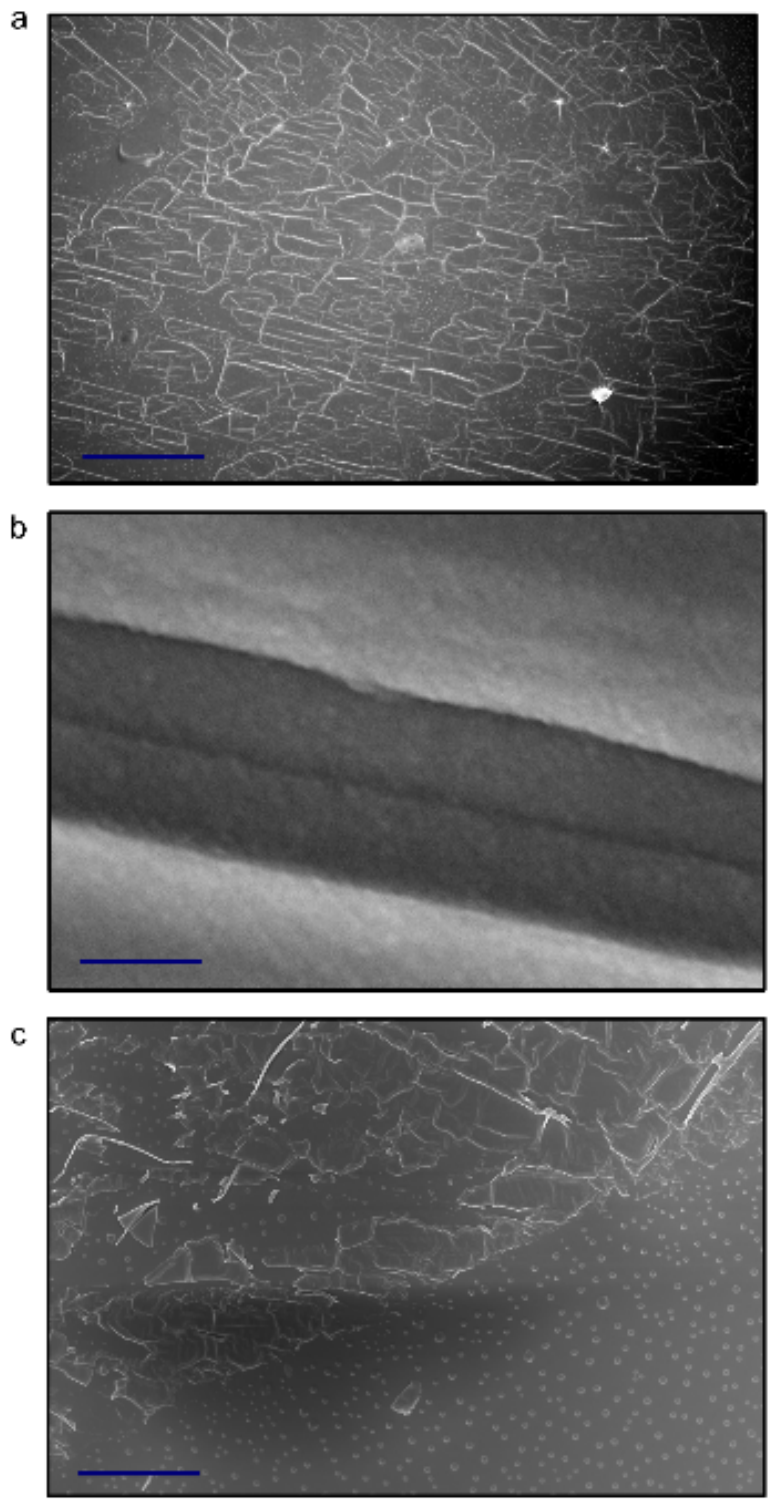

**Figure SI14: a.** Scanning electron microscopy (SEM) image of thin PDA crystal. The polymer chains are arranged into 100 $\mu m^2$ domains which themselves are highly aligned. The scale bar is 20 μm. **b.**

Grain boundary between two PDA crystal domains. Scale bar is 500 nm. The domain boundary is relatively smooth. **c.** SEM image showing unaligned regions of PDA polymer. The domains are broken into small clusters and also show regions of incomplete spherulitic crystal growth. Scale bar is 20 μm.

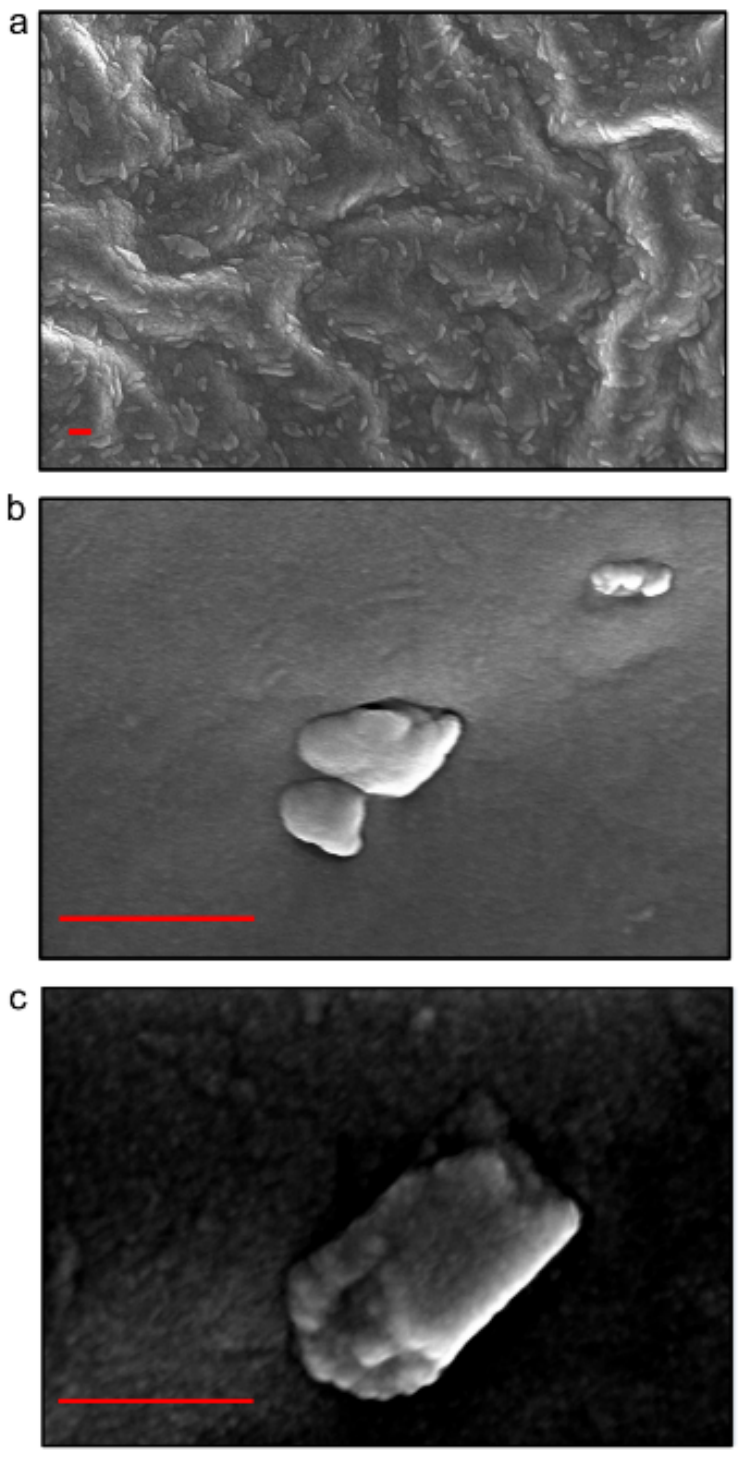

**Figure SI15: a.** Scanning electron microscopy (SEM) image of a PIC thin film. The clusters correspond to large bundles of nanotubes, with the background composed of the sugar matrix. The films charges easily even when sputtered in gold (50 nm) leading to some bright spots on the image. Scale bar is 5 μm in all the images. **b-c.** Zoom into SEM of cluster of nanotubes, which show a very rough surface structure. The background shows less aggregated nanotubes embedded into the sugar matrix.

Atomic force microscopy of PDI nanobelts

Atomic force microscopy was performed with a Veeco Dimension™ 3100 on individual PDI wires at room temperature (Figure **SI16**). The width of a wire was found to be ~130 nm and is uniform across the length of the wire shown in the image (>5 um). The surface of the wires is smooth with a root mean square (RMS) roughness of less than 1 nm, as illustrated in the height profile. This suggests highly-ordered packing of individual monomers. Consequently, the aggregates we suggest aggregates to be low in structural disorder.

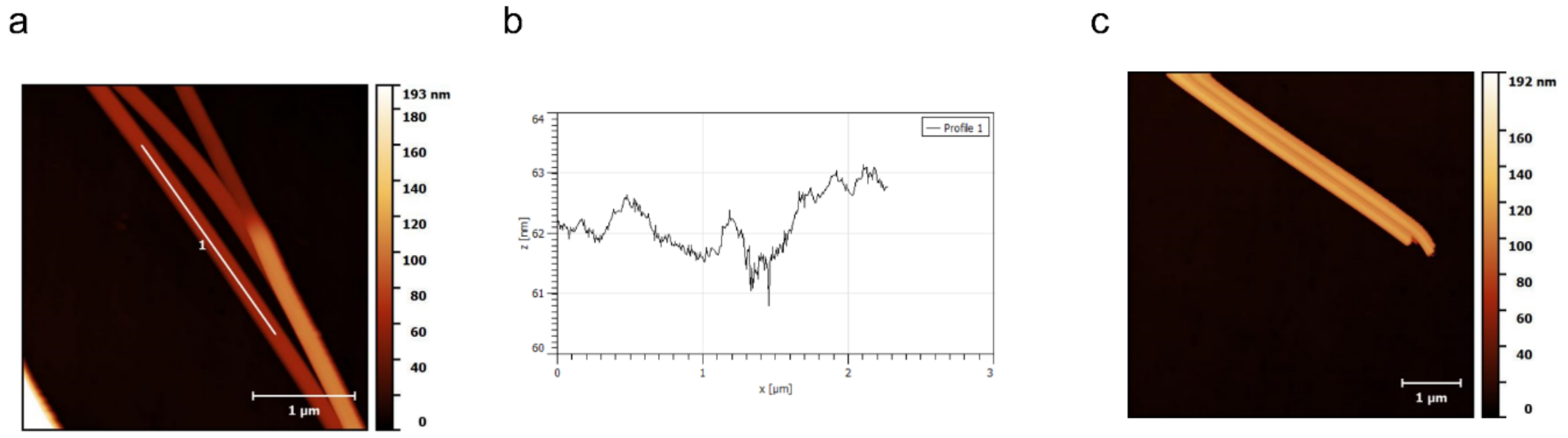

**Figure SI16: a.** AFM of PDI nanobelts showing the uniform topology of the surface. **b.** Topology profile taken along white line in a, the variation is height is less than 1 nm. **c.** The nanobelts show blunt ends. In cases where the ratio of ethanol to PDI monomer is incorrect, the wire ends can be sharp. Such wires were not considered in this work.

Transmission electron microscopy of PIC

To further characterize the PIC films, transmission electron microscopy was performed. Figure **SI17** shows the J-aggregates consist of highly twisted and knotted tubes tens of nanometres thick and hundreds of nanometres long; bundles of such tubes are presumed to make up the clusters found in SEM. The nanowires are highly unstable even to low dose electron beams, hence high resolution images could not be collected.

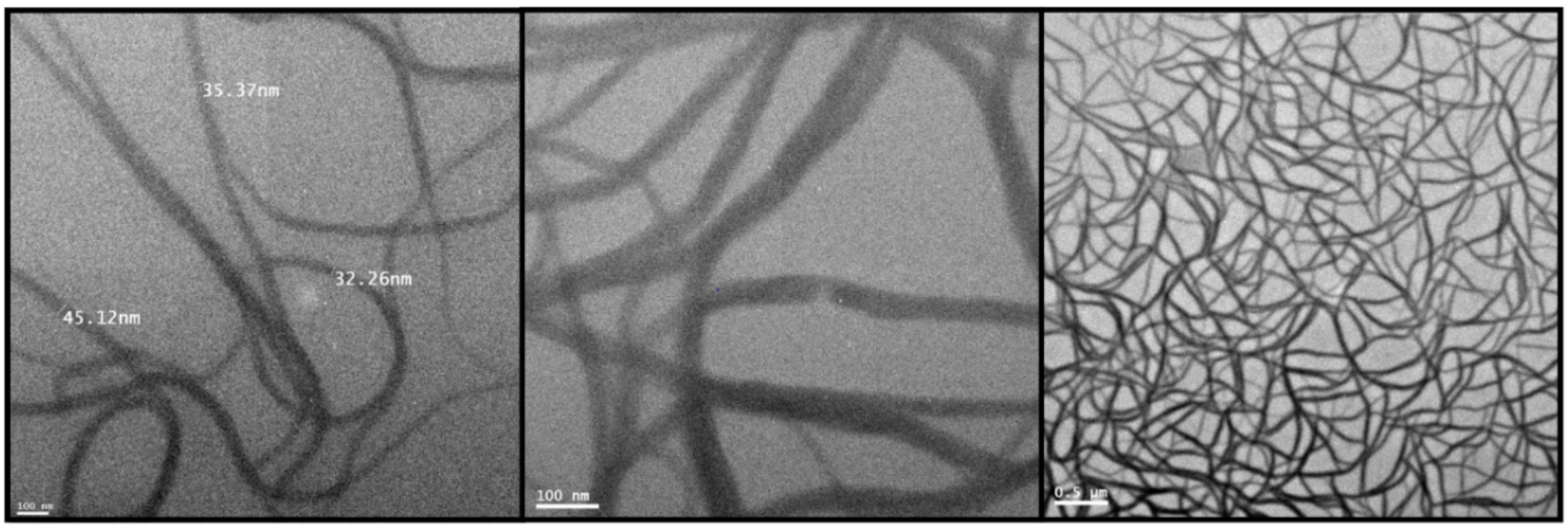

**Figure SI17:** PIC nanotubes embedded in a sugar matrix twist into threads tens of nanometres thick (left). Zooming into the left image (centre) shows each set of threads appear to consist of several nanotubes. Each of the threads is bundled into large agglomerations which are presumed to make up the clusters observed in SEM.

Kelvin probe force microscopy

Kelvin probe force microscopy was performed on eight different wires deposited on ~100 nm gold-coated glass slides. Separate locations were selected where wires were sufficiently separate from any aggregation of wire. The variation of work function along any isolated wires was less than 50 meV, suggesting low disorder.

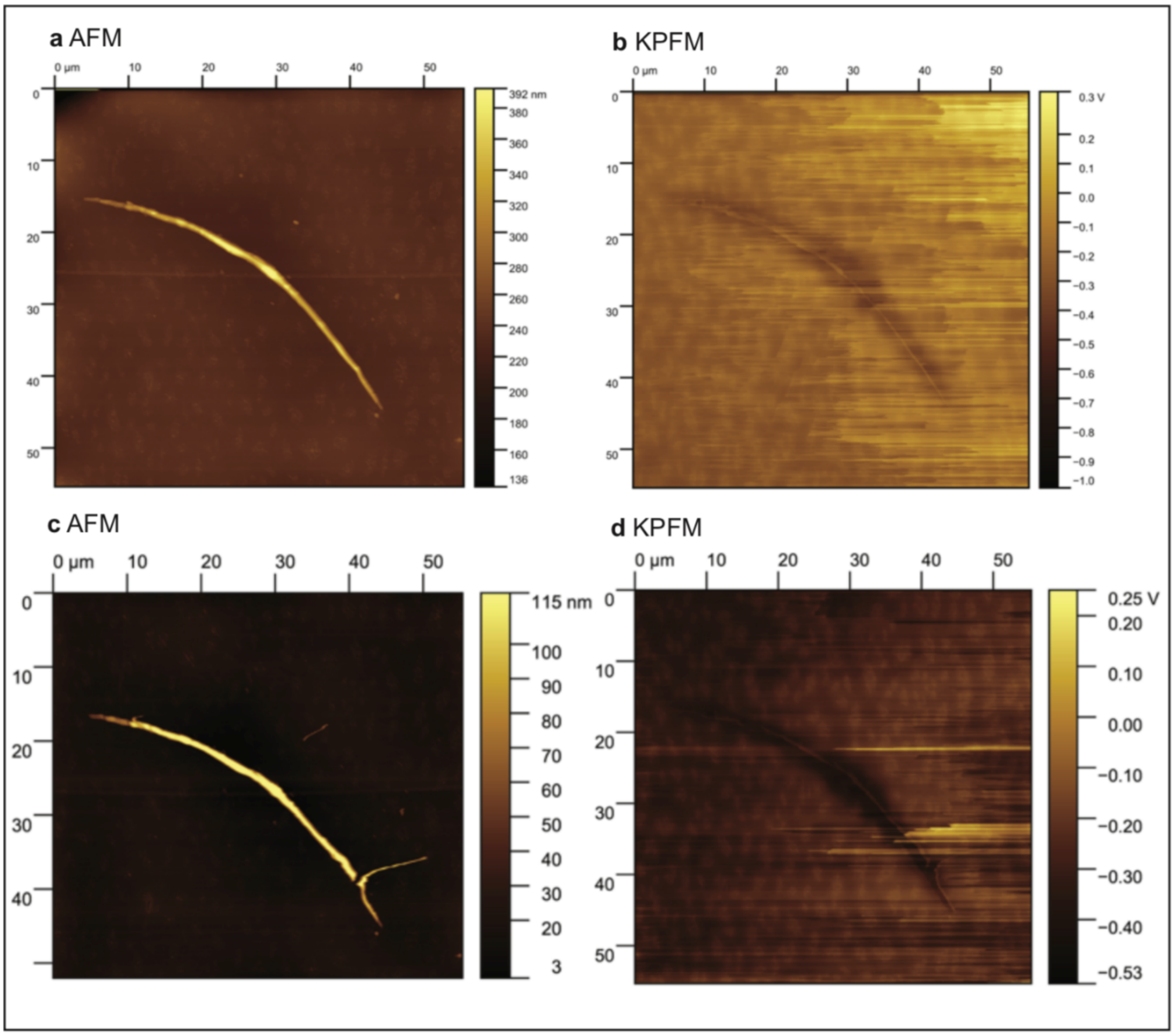

**Figure SI18:** Selected Amplitude Modulation mode Kelvin probe force microscopy (KPFM) and Atomic Force Microscopy (AFM) topology measurements taken on an Asylum Instruments KPFM-AFM using a n-type silicon tip with a platinum coating from Windsor Scientific (HQ:NSC35/PT). i) and ii) refer to the same wire, whereas iii) and iv) refer to a different wire where the end has been broken by indentation.

Confocal Microscopy

Confocal microscopy although cannot directly probe the microstructure of films is a useful tool in characterizing their microstructure. This is discussed below for PIC and PDI, 'blue' PDA is non-emissive and hence such measurements could not be used.

Photoluminescence microscopy of PDI nanobelts

Figure **SI19a** shows a confocal photoluminescence (PL) microscopy image of a PDI nanobelt. The PL intensity is relatively uniform across the nanobelt in keeping with the suggestion of the samples being low in structural defects. Measuring the spectrally resolved PL decay of nanobelts (405 nm excitation; decay captured with Andor iSTAR ICDD) shows a decay that is although asymmetric in shape, decays uniformly (23 ± 2 ns) across all wavelengths and is approximately independent of the position on the nanowire (white asterisk). This suggests there be to a single recombination pathway in nanobelts as opposed to localised defect emission.

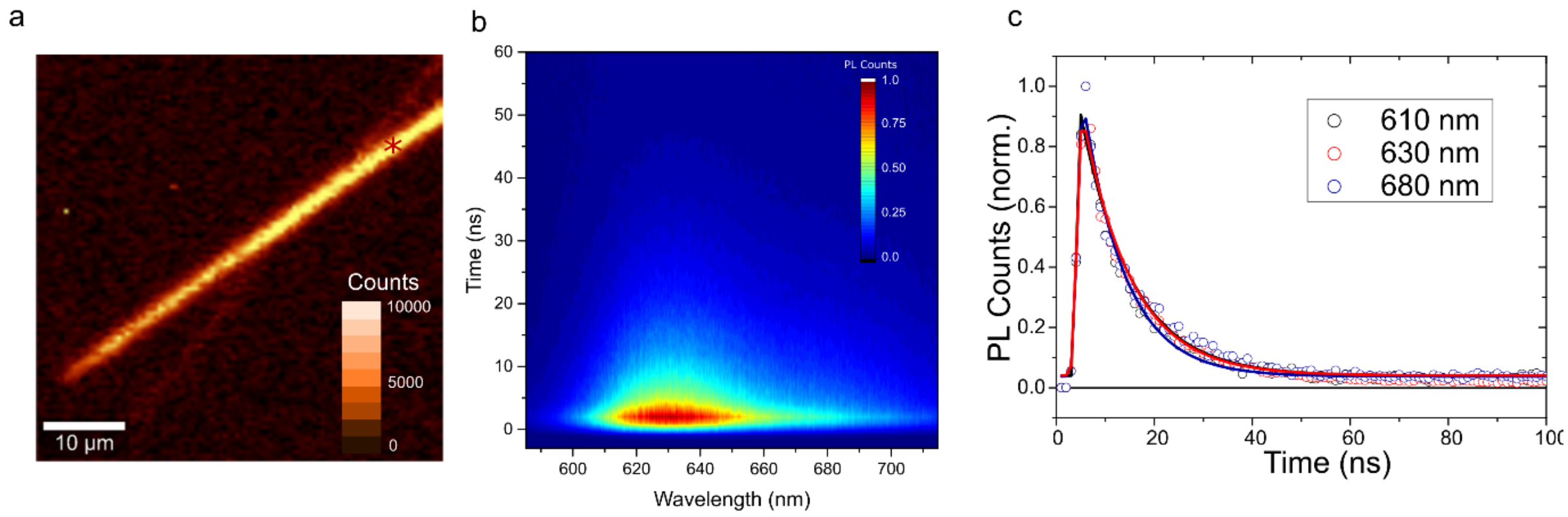

**Figure SI19: a.** Confocal photoluminescence (PL) image of two overlapped PDI wires, the bottom wire is in a different focal plane and hence appears dimmer. The PL is relatively uniform across the wire in line with a structure low in defect concentration. **b.** Spectrally resolved PL decay from PDI wire at location marked with white asterisk. **c.** The rate of decay is spectrally uniform with the normalised decay kinetics at 610 nm, 630 nm and 680 nm being nearly identical.

Photoluminescence microscopy of PIC thin films

Thin films of PIC are inhomogeneous. This is most readily demonstrated by confocal photoluminescence (PL) microscopy (405 nm excitation; ~500 nm FWHM spot size) as shown in Figure **SI20** for a typical 10 × 10 μm region of the films. The film appears to contain regions where the nanotubes are clustered together leading to bright emission, as well as regions which appear to contain little to no nanotubes within the matrix. Unfortunately, experimental limitations preclude correlation of the TAM images with confocal PL. However, we speculate that the highly bundled regions give rise to the largest expansion.

Cryogenic absorption spectra of PIC

Cryogenic absorption spectroscopy is a commonly used technique to estimate energetic disorder in molecular materials. We measured the absorbance of our PIC films at a range of temperature from 298 K down to 17 K. Figure **SI20a** shows the normalized absorbance as a function of temperature. The sharp J-band is clearly observed at 585 nm at room temperature and blue shifts to 580 nm when cooled to 17 K, accompanied by a small degree of narrowing. After a quadratic background removal, we fitted the J-band to a pseudo-Voigt profile (i.e. a linear combination of a Gaussian and a Lorentzian line shape):

$$y = A_{Gauss}\frac{e^{-(\hbar\omega - E_0)^2/2\sigma^2}}{\sigma\sqrt{2\pi}} + A_{Lorentz}\frac{\gamma}{\pi[(\hbar\omega - E_0)^2 + \gamma^2]}$$

where $\sigma$ and $2\gamma$ are the Gaussian standard deviation and Lorentzian FWHM respectively. The fitting results are shown in Figure **SI20(b-d).** We note that the absorbance spectra are asymmetrical, implying it is impossible to achieve a perfect fit to the data using a simple pseudo-Voigt line shape. However, it provides a reasonable approximation throughout the investigated temperature range, with a stronger contribution from the homogeneous Lorentzian line shape. The Lorentzian linewidth, $\gamma$, is found to decrease from 15 meV (~120 $cm^{-1}$) to 10 meV (~80 $cm^{-1}$) when cooled to 17K, which is notably smaller but on the same order of magnitude as many self-assembled J-aggregates (e.g. 50 meV in TDBC J-aggregates). This low disorder is likely to reduce exciton scattering and aid long-range polaritonic transport. Fitting $\gamma$ to a power law reveals its quadratic dependence in temperature as shown in Figure **SI20c**, which theoretical models have shown relates to pure electronic dephasing (*17*) due to elastic two-phonon scattering of the excitons. The inhomogeneous broadening is likely the result of interactions with the inhomogeneous sugar matrix.

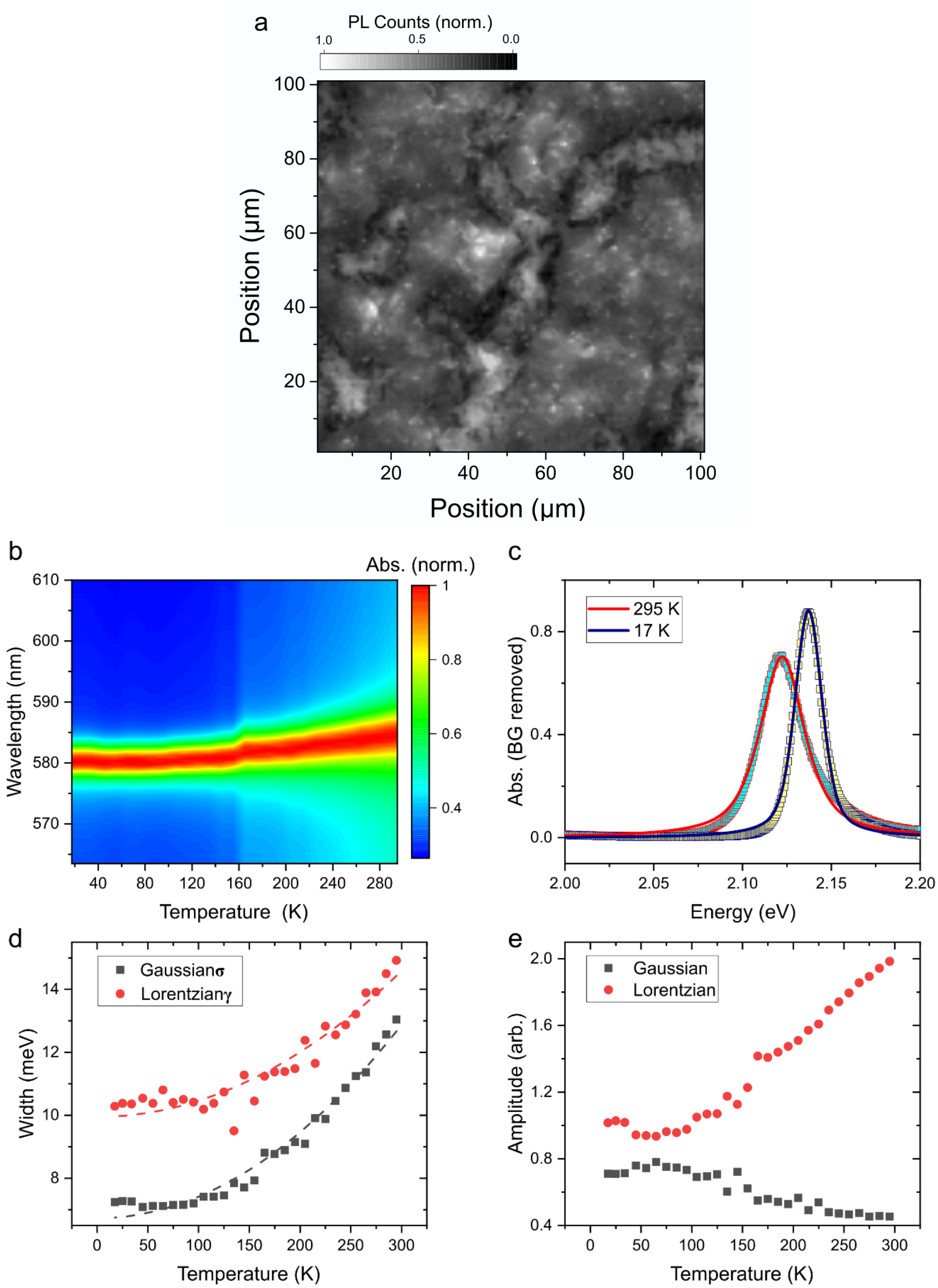

**Figure SI20: a.** Confocal photoluminescence (PL) map from a typical region of PIC thin film (normalized counts). The brighter regions are suggested to arise from tightly packed cluster of nanowires. They are also suggested to show the greatest level of polaritonic transport associated with Fresnel modes supported by the higher volume. **b.** Background-subtracted absorbance at 17 K (blue) and 295 K (red) with corresponding fits to a pseudo-Voigt function. **c.** Temperature dependent

Gaussian $\sigma$ and Lorentzian $\gamma$ and fits to a quadratic relationship, suggesting a 2-phonon pure dephasing mechanism for the observed broadening. **d.** Amplitudes (areas) of the fitted Gaussian and Lorentzian profiles. The Lorentzian line shape has a larger contribution across the whole temperature range. However, there is a considerable contribution from the Gaussian profile, suggesting inhomogeneous broadening mechanisms in our samples.

## S3: Data analysis

### Data acquisition

Consecutive $pump_{ON}$ and $pump_{OFF}$ images for each time delay were first corrected for camera background and then converted to a single $\Delta T/T$ image using the equation below:

$$\frac{\Delta T}{T} = \frac{pump_{ON}}{pump_{OFF}} - 1$$

where $pump_{ON/OFF}$ denotes the background-corrected transmission image with and without the presence of the pump pulse. The extraction of time zero ($t_0$) is discussed later.

### Self-referencing method

Due to the random nature of photon emission in a laser, fluctuations in the laser output power (i.e. shot noise) is unavoidable and may introduce significant noise to the final images. Although our system is shot-noise limited and the 15 Hz mechanical chopper should average out high-frequency fluctuations, low-frequency fluctuations should be observable through the varying background photon counts in the transmission. In order to mitigate these effects, we defined a correction factor for each pair of images, $\varepsilon$, as the ratio of averaged photon counts on the camera between the $pump_{ON}$ and the $pump_{OFF}$ images and recalculated every $\Delta T/T$ image using a revised equation:

$$\left(\frac{\Delta T}{T}\right)_{new} = \frac{1}{\varepsilon}\frac{pump_{ON}}{pump_{OFF}} - 1$$

Figure **SI21** compares the $\Delta T/T$ intensity at a user-defined signal centre as a function of the sweep number before and after applying the self-referencing correction. It is evident that the average amplitude of the signal is not affected but the spikes are supressed. Applying this correction before averaging over multiple sweeps should further improve our SNR, as shown in Figure **SI21** where we plot the $\Delta T/T$ signals before and after correction against time. The SNR can be improved by up to a factor of three depending on the experimental conditions. For example, airflow and vibrations induced body movement in the laboratory are sufficient to introduce significant fluctuations in the numbers of photons reaching the camera. This self-referencing correction was applied to all fs-TAM data presented in this manuscript.

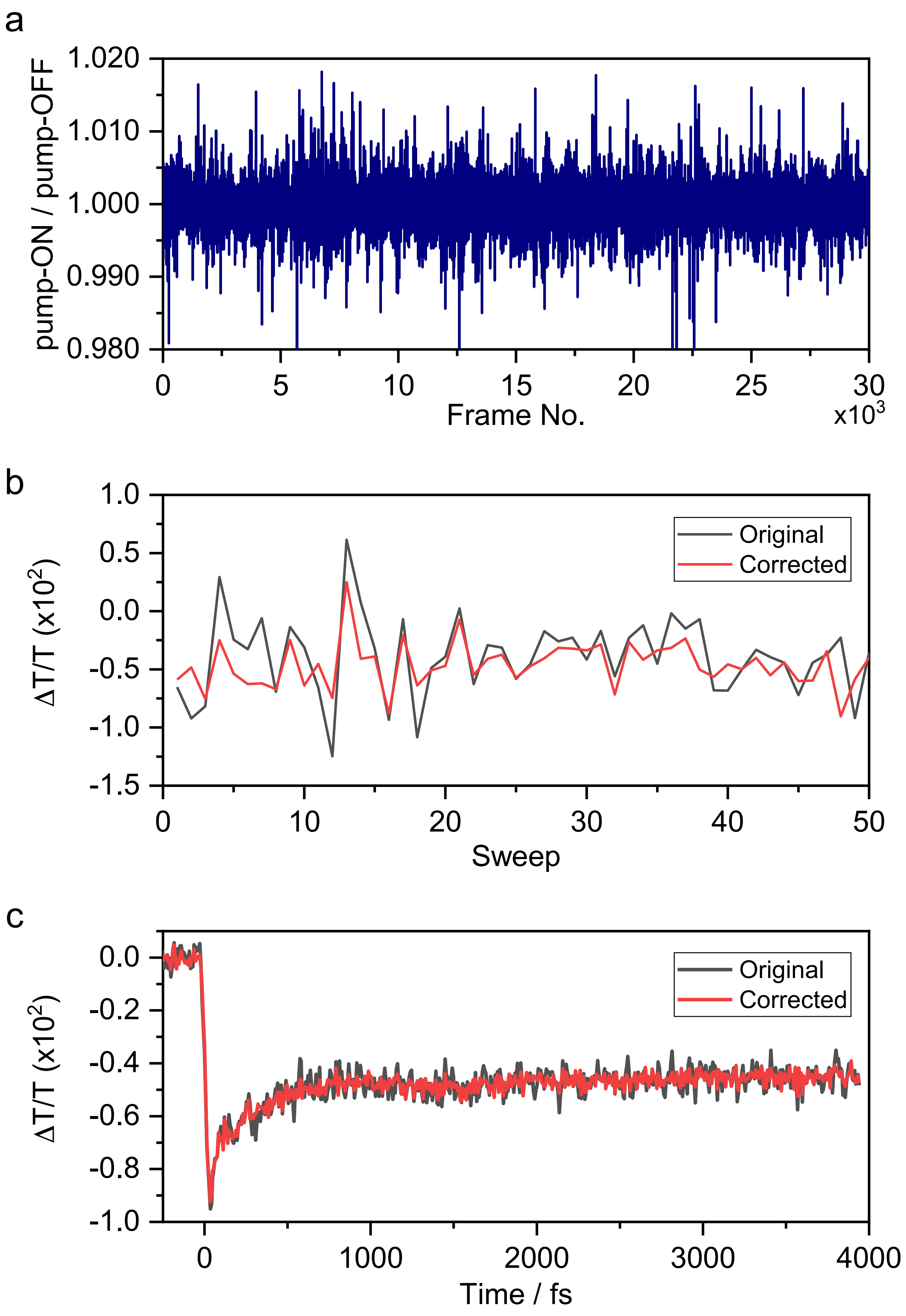

**Figure SI21: a.** The ratio of averaged photon counts between each pair of $pump_{ON}$ and $pump_{OFF}$ images, excluding all pixels affected by the actual signal and associated optical artefacts. This shows an averaged 0.4% fluctuation in the measured laser output with numerous spikes of over 1% in magnitude spanning the whole measurement window. **b.** $\Delta T/T$ response at a user-defined centre in the image at t = 300 fs, plotted against the sweep number before and after applying the self-referencing correction. **c.** Kinetics at the same spot averaged over 50 sweeps before and after applying the self-referencing correction, showing a noticeable improvement in the sign-to-noise ratio.

A further background correction was performed on the averaged $\Delta T/T$ images to make sure that the average $\Delta T/T$ is zero before time zero. In practice, we estimated the background $\Delta T/T$ by averaging the first five images (i.e. 64 × 64 × 5 = 20480 pixels), which was subsequently subtracted from all images. The estimated background $\Delta T/T$ is usually on the $0.5 \times 10^{-5}$ order of magnitude and has little effect on data analysis. Nevertheless, all TAM data presented in this work were background corrected.

Data smoothing - moving average

In order to suppress the random noise whilst preserving the real signal, we investigated the effects of applying a two-dimensional moving average filter to our data.

The moving average is the most common filter in digital signal processing. It operates by averaging several points from the input signal to produce each point in the output signal. For example, in a one-dimensional problem the output signal can be expressed mathematically as

$$y[i] = \frac{1}{N}\sum_{j=0}^{N-1} x[i+j]$$

where $x[i]$ and $y[i]$ are the i-th data point in the input and output signal respectively. This equation only uses a total of $N$ points on one side of the signal. However, it is straightforward to use an equal number of points on either side of the output data point. For a two-dimensional dataset, the equation can be modified to average over data points surrounding the target point:

$$y[i,j] = \frac{1}{N^2}\sum_{k=0}^{N-1} x[i+k, j+k]$$

We applied the moving average filter to average over $N^2$ input data points surrounding the output point and assessed its performance as a function of $N$. Unsurprisingly, increasing $N$ improves the noise reduction but the signal becomes less sharp at the edge. Similarly, applying the filter again further reduces the noise at the expense of signal strength and sharpness. We found that applying the filter once with $N = 3$ produces the ideal result with the best balance between preservation of the original signal and elimination of random noise.

Despite its simplicity, we find the 2D moving average filter consistently outperforms other conventional image analysis filter such as the 2D Fourier filter. As a result, we decided to adopt the moving average as our default post-processing filter. Unless otherwise stated, all TAM data was averaged once with the 2D moving average filter with $N = 3$.

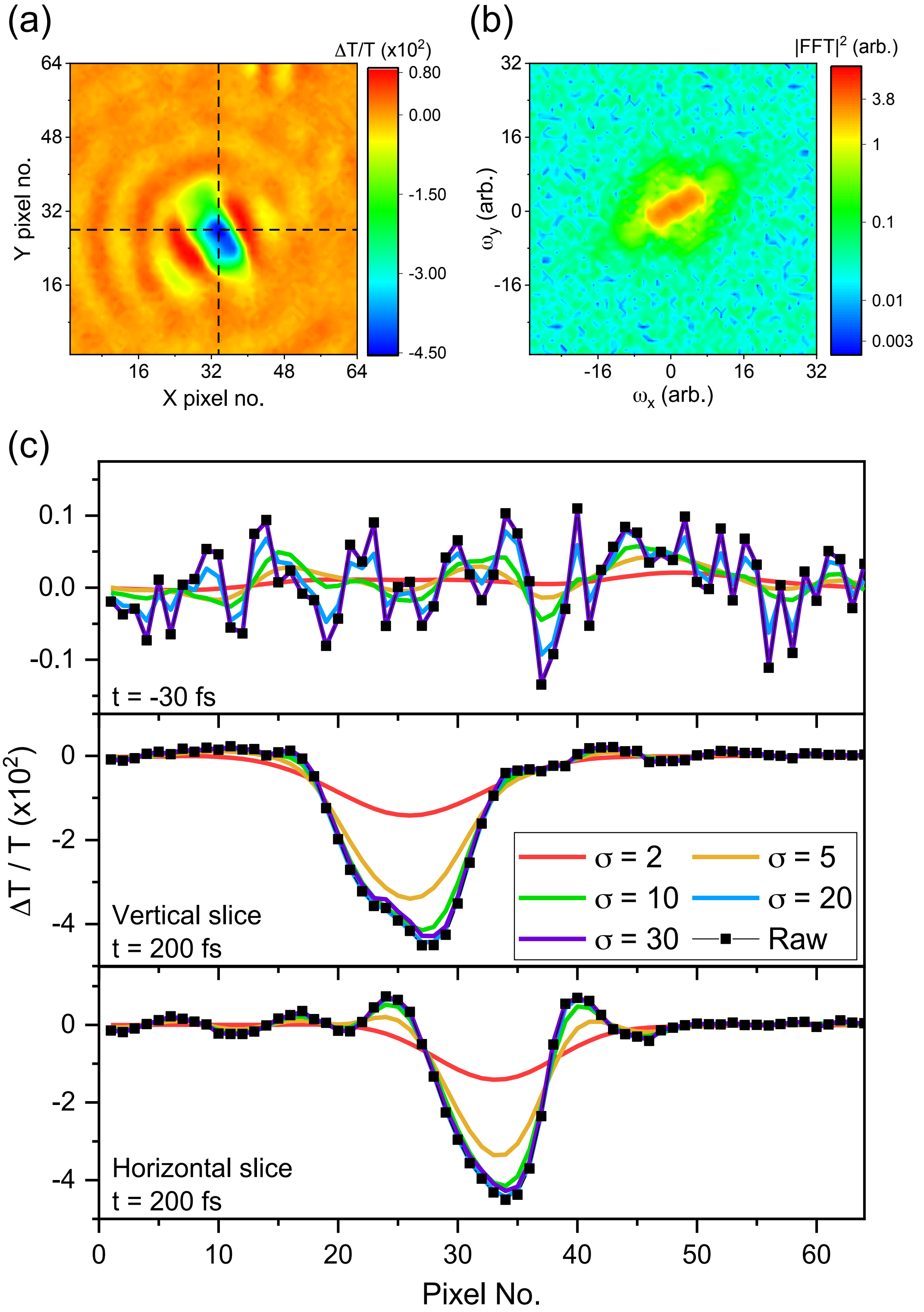

**Figure SI22: a.** A typical $\Delta T/T$ image from PDI sample with a pump-probe delay of 200 fs, showing a clear photo-induced absorption feature in the middle and the surrounding optical artefacts. The image was obtained by averaging over 50 sweeps. **b.** The same $\Delta T/T$ image in the frequency domain after a Fast Fourier Transform operation. The logarithmic colour scale helps to highlight the high-frequency noise, which is much weaker in amplitude. **c.** Effect of applying a Gaussian Fourier filter of a range of variances on the original signal. The top panel compares the ability of the different filters to remove the random noise from an image before time zero whereas the middle and the bottom panels

compare the effect of the filters on the actual signal amplitude and shape, plotted along the black dashed lines in **a**. It is clear that a filter with a large σ is able to preserve the signal better but it also lets more noise through.

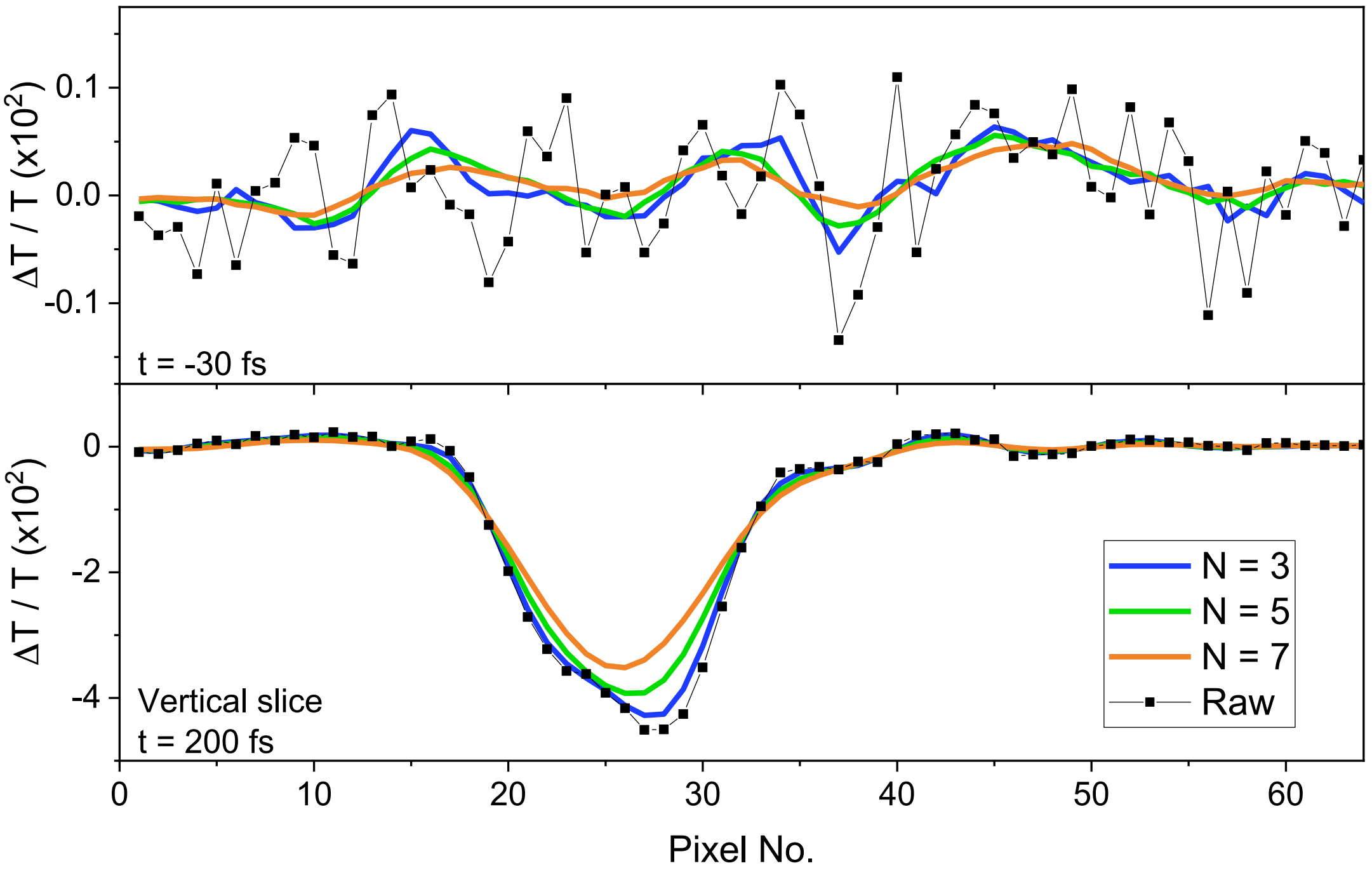

**Figure SI23:**.3Effects of applying a two-dimensional moving average filter of various window size N on the original signal (PDI). The top panel compares the ability of the different filters to remove the random noise from an image before time zero whereas the bottom panel compares the effect of the filters on the actual signal amplitude and shape. Surprisingly even with N = 3 the random noise is significantly suppressed and the improvement in noise reduction by increasing N is nullified by the loss in signal strength and shape as shown in the bottom panel.

Accurate determination of time zero, signal centre and sigma propagation

The most common method employed in analysis of TAM data is by the fitting of a two dimensional (2D) Gaussian (*18*, *19*). To extract the material's diffusion constant $D$ along a principle axis, one simply fits the diffusion length equation to the sigma-time graph. Assuming the initial population at $t = t_0$ is Gaussian in space with an initial width $\sigma(t_0)$ and undergoes anomalous diffusion, the sigma at any later time $t \geq t_0$ can be described by

$$\sigma^2(t) = \sigma^2(t_0) + D(t - t_0)^n \textbf{ (E1)}$$

where $n = 1$ for a typical diffusion process. $n < 1$ and $n > 1$ represent a sub-diffusion and super-diffusion process respectively. Sub-diffusion usually suggests the existence of carrier traps whereas super-diffusion is commonly found in active cellular transport. If $n = 2$ a particle undergoes ballistic transport with the transport length determined by its mean free path.

However, although this method produces quantitatively similar results it is not necessarily the most applicable to the quasi one-dimensional systems investigated, due to their inhomogeneous geometry. We therefore need to analyse our data differently, yet the extracted diffusion coefficient must still be comparable to those acquired from a 2D Gaussian fitting.

General method

The above equation describes how the width of the Gaussian grows in time. In a similar fashion, we can define $\sigma$ as the distance between the centre of the signal and the point in the image plane (in any direction) where the signal strength is $e^{-1/2}$ of that at the centre. Mathematically this is expressed by

$$\sigma^2(\theta, t) = \left(x_{\frac{1}{2}}(\theta, t) - x_0\right)^2 + \left(y_{\frac{1}{2}}(\theta, t) - y_0\right)^2$$

where $(x_0, y_0)$ and $(x_{\frac{1}{2}}(\theta, t), y_{\frac{1}{2}}(\theta, t))$ are the coordinates of the signal centre and the points where the signal strength at time $t$ drops to $e^{-1/2}$ of that at the centre at angle $\theta$ relative to the positive x-axis, respectively. We will refer to $(x_{\frac{1}{2}}(\theta, t), y_{\frac{1}{2}}(\theta, t))$ as the 'sigma points' from now on. We can then extract the diffusion coefficient and investigate if the carriers diffuse more in a certain direction.

Any scientific statement or comparison of parameters is valid only when it is significant compared to the relevant uncertainties. Therefore, it is important that we evaluate the uncertainties in our derived quantities and minimise them by adopting the best methods in data analysis. In the presence of random noise, one can estimate the uncertainty in the x- and y-coordinates to be 1 pixel (i.e. 55.5 nm on the image plane). Therefore, each direction in E1 will have a combined error of about 1.4 pixels (i.e. 80 nm) and the overall uncertainty $\sigma^2$ will be over 110 nm. Consequently, all the coordinates must be determined with an accuracy and precision better than our pixel resolution, or else the large uncertainties in the coordinates will render $\sigma^2$ and any other fitted parameters un-usable. Another reason for correct determination of the centre is that E1 is non-linear, which means that an inaccurate centre may skew the fitted parameters differently as angle $\theta$ varies and makes fair comparisons difficult.

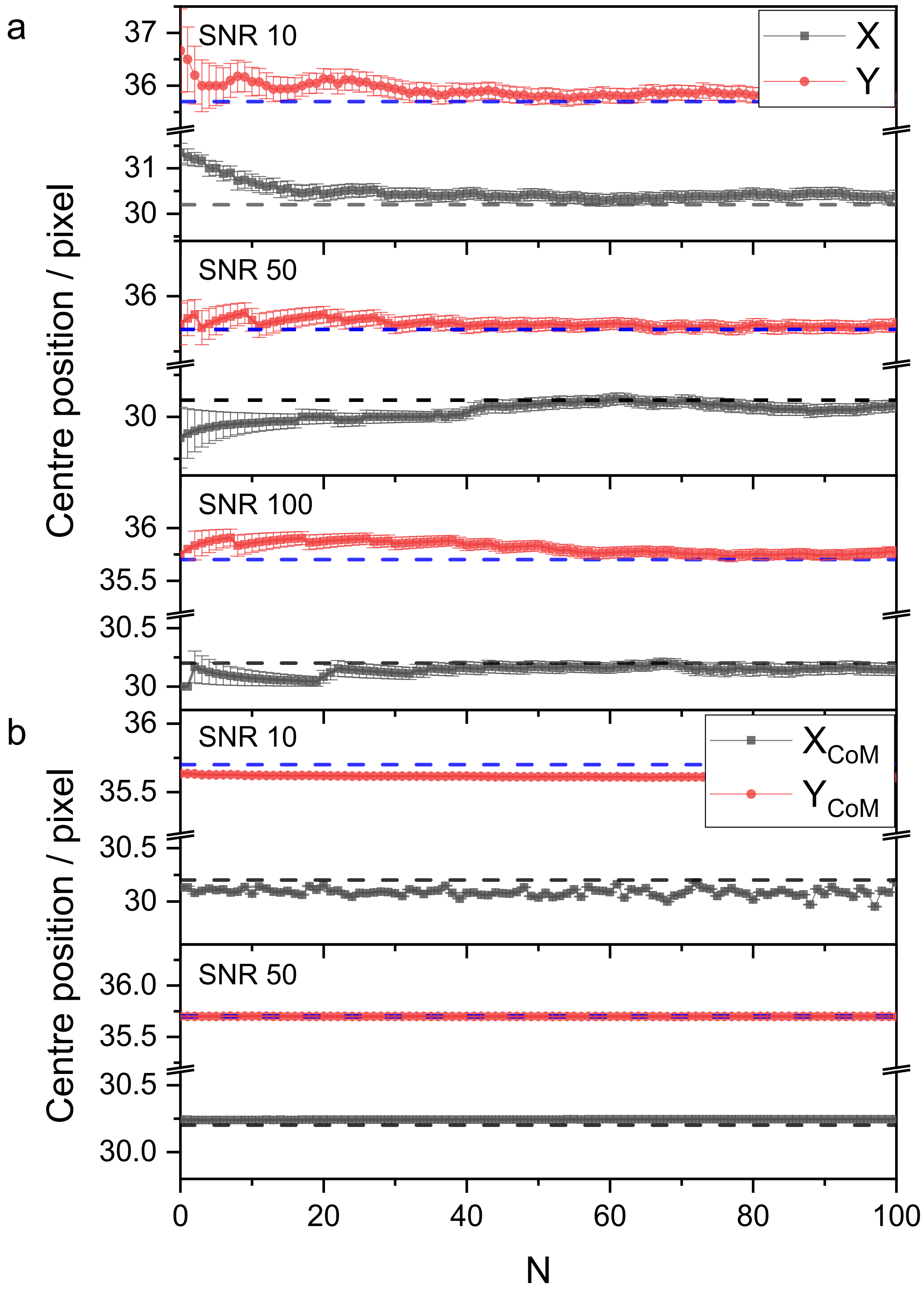

**Figure SI24:** Estimated position and uncertainties of the signal centre in datasets with various signal-to-noise ratios, using **a.** Method 1: averaging the positions of the pixel with maximum signal and **b.** Method 2: calculating the centre of mass (CoM) in each frame, then averaging over N images after time zero. The blue and black dashed lines represent the true positions of the centre: (30.2, 35.7). Method 1 is much more sensitive to N, and requires a much larger number of frames (more than 50) to achieve a stable and accurate estimate.

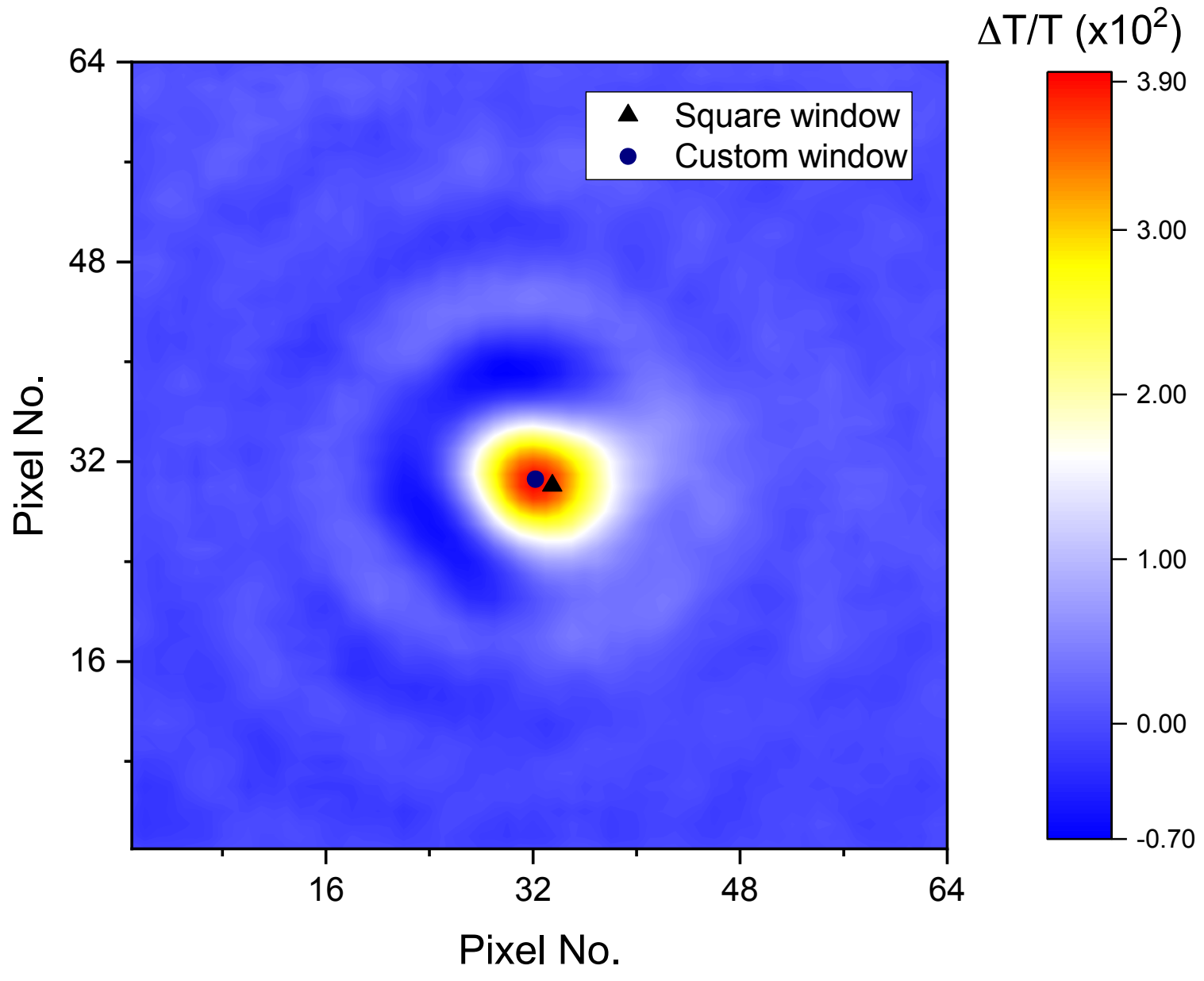

**Figure SI25:** A typical TAM image from PDA, showing the effect of using a square window (*i.e.* including pixels with a strong negative $\Delta T/T$) and manually selecting pixels in the centre-of-mass (CoM) calculations. The negative C-shaped signal on the left is probably an optical artefact due to the microscope objective or inhomogeneous surface morphology. Note that it pushes the CoM away from where we would believe the true pump focus was. By manually selecting the pixels with a strong positive $\Delta T/T$ signal, we are able to retrieve a CoM much closer to our expectations.

Determination the signal centre

It is tempting to use the pixel with the strongest signal in each frame as the centre, as one might expect its position to be constant over time. However, since one pixel on the EMCCD camera corresponds to 55.5 × 55.5 $nm^2$ in real space and the pump spot has a FWHM of ~270 nm, the central four pixels often receive similar numbers of photons. In practice, the position of the maximum signal pixel fluctuates between 1-2 pixels in either direction. This is undesirable for not only the reasons described in the previous section, but physically the centre of the signal is defined as the centre of the pump pulse, which has a well-defined position that stays constant throughout the measurement window. There are two common ways to achieve sub-pixel resolution and improve accuracy: 1) Average the coordinates of the pixel with maximum signal strength over a number of frames; 2) Determine the centre using multiple pixels within the same image. Method 1 works in a similar way to dithering in digital signal processing, where uncorrelated noise is added purposefully to an analogue signal before digitisation to reduce quantisation noise. The random noise present in our data pushes the signal maximum to neighbouring pixels in time. Therefore, averaging over many frames should yield a more precise estimate of the true position of the signal centre. We tested the method using simulated TAM data with a known centre at (30.2, 32.5) and a range of maximum SNR from 10 to 100. The estimated centre coordinates are plotted in Figure **SI24** as a function of $N$, which is the number of frames averaged. The errors on each averaged coordinates is taken as the standard error in mean given by $\sigma_{\bar{x}} = \frac{\sigma_x}{\sqrt{N}}$. We found that a large number of frames (more than 50) is required to achieve a stable and accurate average, even for a dataset with a high SNR. As a result, the averaged position will not reflect the true pump centre. This method also requires significant noise to push the signal to neighbouring pixels efficiently, which is not always available if the SNR is too high especially after smoothing. Method 2, on the other hand, utilises a number of pixels around the maximum to estimate the centre through a weighted average. It works almost identically to calculating the centre of mass (CoM) of $N$ discrete masses on a plane, which can be expressed algebraically using the following equations:

$$X_{CoM} = \frac{1}{M}\sum_{i=1} m_i x_i$$

$$Y_{CoM} = \frac{1}{M}\sum_{i=1} m_i y_i$$

where $(X, Y)_{CoM}$ is the coordinates of the centre of mass, $M$ is the total mass, $m_i$ and $(x_i, y_i)$ are the mass and coordinates of the i-th individual mass respectively. To calculate the centre of our signal, we substitute the $\Delta T/T$ data points for the individual masses in a 10 × 10 square window surrounding the maximum signal pixel (*i.e.* $i_{max} = 100$). We can then average the CoM over several frames and calculate its uncertainty in a similar fashion. Using the same set of data, it is clear from Figure **SI25** that the CoM calculated from each frame is much closer to true value and the frame-to-frame fluctuation is much reduced (less than 0.1 pixels in each direction) when compared to that of Method 1 (1 pixel in both directions). Therefore, a stable CoM was achieved using a significantly smaller number of frames. In fact, there is no need to average over multiple frames as the estimated centre is very close to the true value using a single frame. The slight drifts in the estimated centre are most likely caused by a slightly asymmetrical selection of pixels around the true maximum. However, in most cases, the magnitude of such drift is less than 0.1 pixel (around 5 nm in the sample) so this is not of serious concern.

It is important to note that if the $\Delta T/T$ value of a pixel has an opposite sign to the main signal, it acts as a 'negative' mass which pushes the centre of mass away from the true position. This can sometimes significantly skew the centre when 'Airy disk rings' are much more pronounced on one side of the signal. The solution to this problem is to exclude the pixels heavily influenced by the 'Airy disk rings'. As a result, we adopted method 2 as our default method to define our signal centre (Figure **SI25**).

Determine sigma at different time points and angles

After we have estimated the signal centre, we need to extract the sigma points (i.e. coordinates where the signal strength is $e^{-1/2}$ of that at the centre), at any angle and time with sub-pixel resolution. This is achieved by mapping the smoothed data onto a fine polar grid centred at $(X, Y)_{CoM}$, followed by selecting in each direction the point which has an amplitude closest to $e^{-1/2}$ of that at the centre. $\sigma^2(\theta, t)$ can then be calculated.

We investigated the effect of noise on the errors in the CoM and $\sigma^2(\theta, t)$. A circular Gaussian signal with a constant width and strength in time (a minimum of 250 frames) was corrupted with random noise such that the signal-to-noise ratio ranges from 1 to 100. The CoM and $\sigma^2(\theta, t)$ for each frame are calculated as above. The uncertainty in the CoM is taken as the standard deviation of the coordinates computed from 250 frames. $\sigma$ was calculated for 180 angles in each frame and the uncertainty is taken as the standard deviation of the 250 × 180 = 45000 values obtained. Figure **SI26** summarises the findings and it is clear that the uncertainties decreases rapidly as the SNR improves (almost linearly on a log-log scale plot). The CoM has an impressively low uncertainty even when the SNR is 1. On the other hand, we need a minimum SNR of 10 to achieve an uncertainty of less than 1 pixel in $\sigma$. This suggests the importance of having a good SNR if we wish to extract the diffusion parameters accurately from TAM data. It is important to note that these uncertainties are associated with individual CoM coordinates and $\sigma$ points at a particular angle. By averaging the CoM and $\sigma$ points over multiple frames and angles respectively will further reduce the errors by a factor of $\sqrt{N}$.

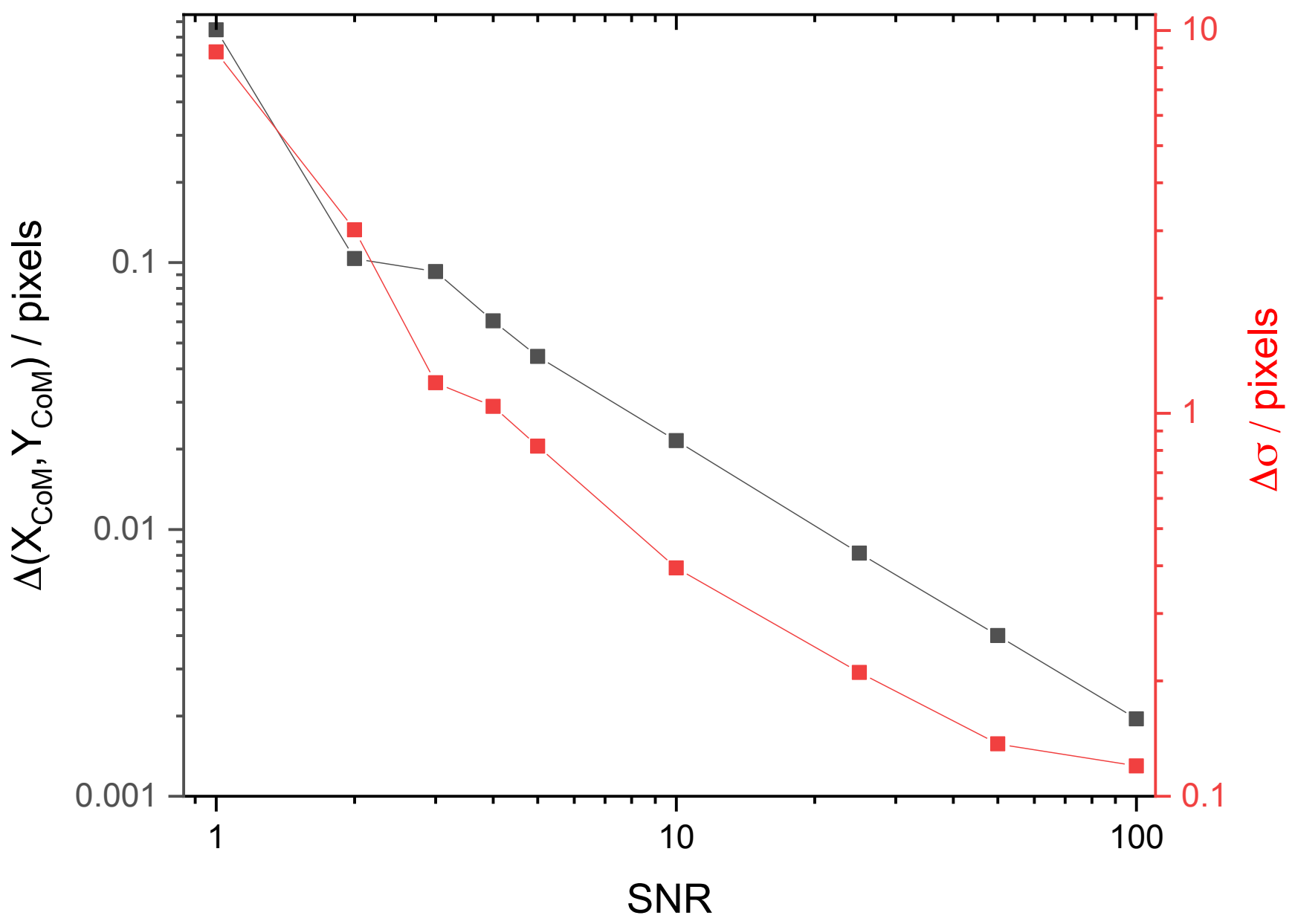

**Figure SI26**: Uncertainties in the centre of mass (CoM) and σ plotted against signal-to-noise ratio (SNR). The uncertainties fall off rapidly as the SNR improves from 1 to 3. Random noise has little effects on the CoM even for a low SNR of 1. On the other hand, a minimum SNR of 10 is required to reduce the uncertainty in σ to below 1 pixel.

Extraction of time zero ($t_0$)

In order to estimate the diffusion coefficient using E1, we need to define time zero (i.e. the moment when the pump pulse arrives at the sample), for which three common definitions exist: The moment when the $\Delta T/T$ signal 1) appears through the noise; 2) is half-way to the maximum and; 3) is at the maximum. Placing time zero too early may result in sigma increasing rapidly at early times and an illusion of ultrafast diffusion process as the signal grows through the noise. On the contrary, putting time zero too late in time would mask any ultrafast carrier processes (*e.g.* ballistic transport). Hence, it is crucial we can detect time zero with confidence. It is not always straightforward to choose one of the above definitions to find time zero. For example, the observed $\Delta T/T$ signal could keep growing in for hundreds of fs (or even longer) if we were probing a photo-induced absorption whose growth in space and time depends on the exact dynamics of two different populations (e.g. individual lifetimes, rates of interconversion and annihilation, diffusion properties, *etc.*).

One interesting observation when we plot the extracted sigma points on top of the TAM images, as illustrated in Figure **SI27**, is that the algorithm is unable to define the sigma points before time zero.

Before the signal comes in, the sigma points are scattered randomly over the whole image, giving a large average distance from the estimated centre $\bar{\sigma}$, as shown in Figure **SI28**. As the signal starts to grow in, as shown in the middle image in Figure **SI28** (Frame 43), the sigma points gradually aggregate into an elliptical ring. The smoothness improves and $\bar{\sigma}$ increases, as the signal gets stronger within a few frames (i.e. within 20 fs). We decided to define time zero as the first frame where the ring formed by the sigma points is smooth and has a $\bar{\sigma}$ closest to the diffraction-limited pump spot size ($2\bar{\sigma} \approx 1.7 \times FWHM \approx 1.7 \times \frac{\lambda}{1.1NA}$). This is crosschecked with the signal strength at the CoM and also the $t_0$ obtained by fitting the electronic response $y$ ($\Delta T/T$ at CoM pixels) with the following multi-exponential equation:

$$y = \sum_n \frac{1}{2} \cdot A_n \cdot e^{\frac{s^2}{2\tau_n{}^2}} \cdot e^{\frac{-(t-t_0)}{\tau_n}} \cdot \left(1 + erf\left(\frac{t - t_0 - \frac{s^2}{\tau_n}}{\sqrt{2}s}\right)\right)$$

here $n$ is the number of lifetimes $\tau$ to be extracted, $s$ is the instrumental response and $A_n$ is the amplitude of the $n$-th exponential decay. The error function considers the probe arriving at $t = t_0$ is Gaussian temporally. For all data considered in this manuscript $t_0$ is obtained from this method lies at or beyond the half way point in the signal rise. Consequently, we find this to be a reasonable and consist approach to find time zero.

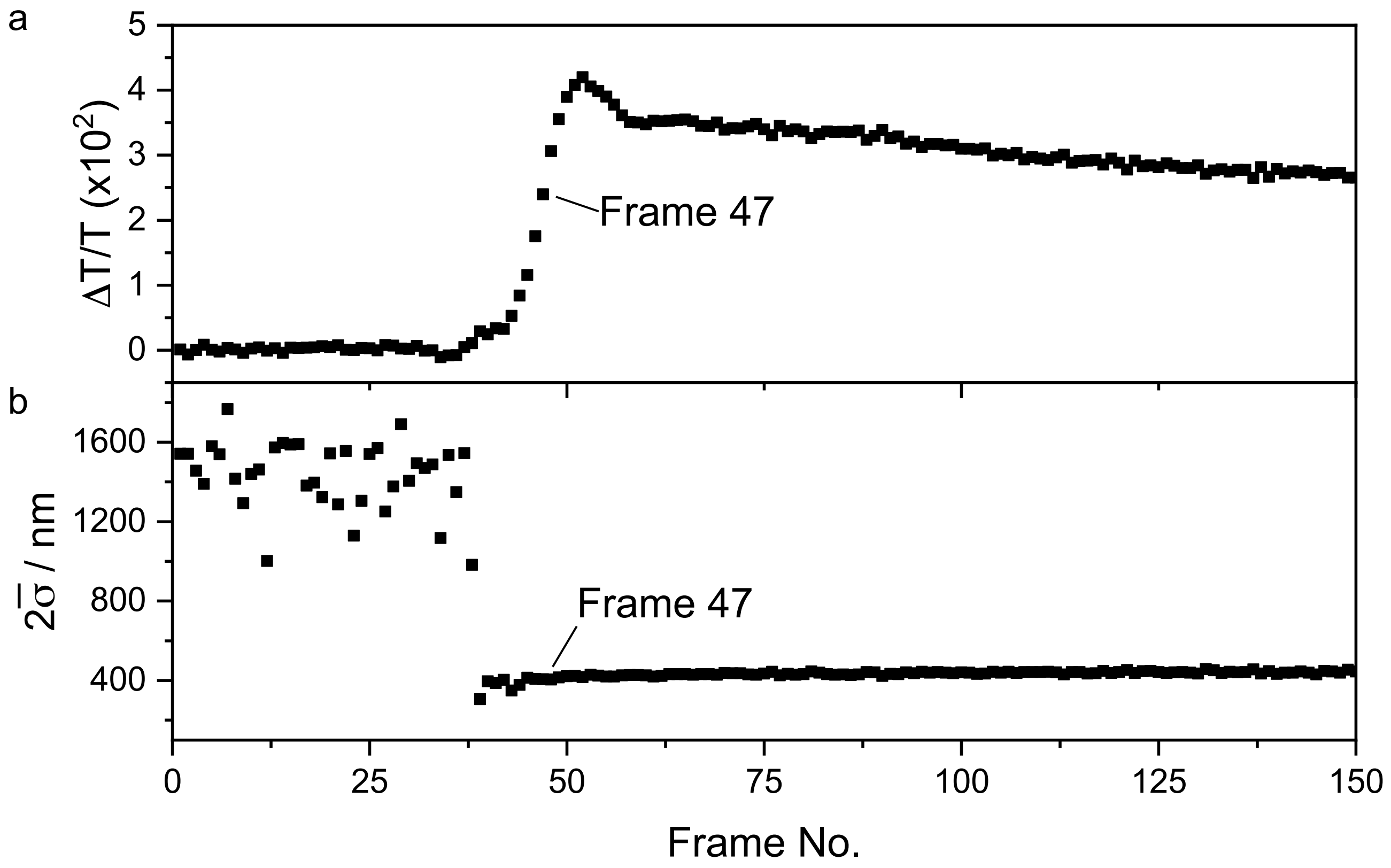

**Figure SI27: a.** Signal strength at the centre of mass. **b.** average diameter of the ring formed by the sigma points plotted against the frame number. Before time zero, the sigma points are randomly scattered across the image resulting in a large average distance from the centre. As the signal starts to appear, a ring is formed with its diameter growing with the signal. In the case with no carrier diffusion, the signal should have a constant size approximately equal to that of the pump spot ($2\bar{\sigma}{\sim}400\ nm$) throughout the measurement window.

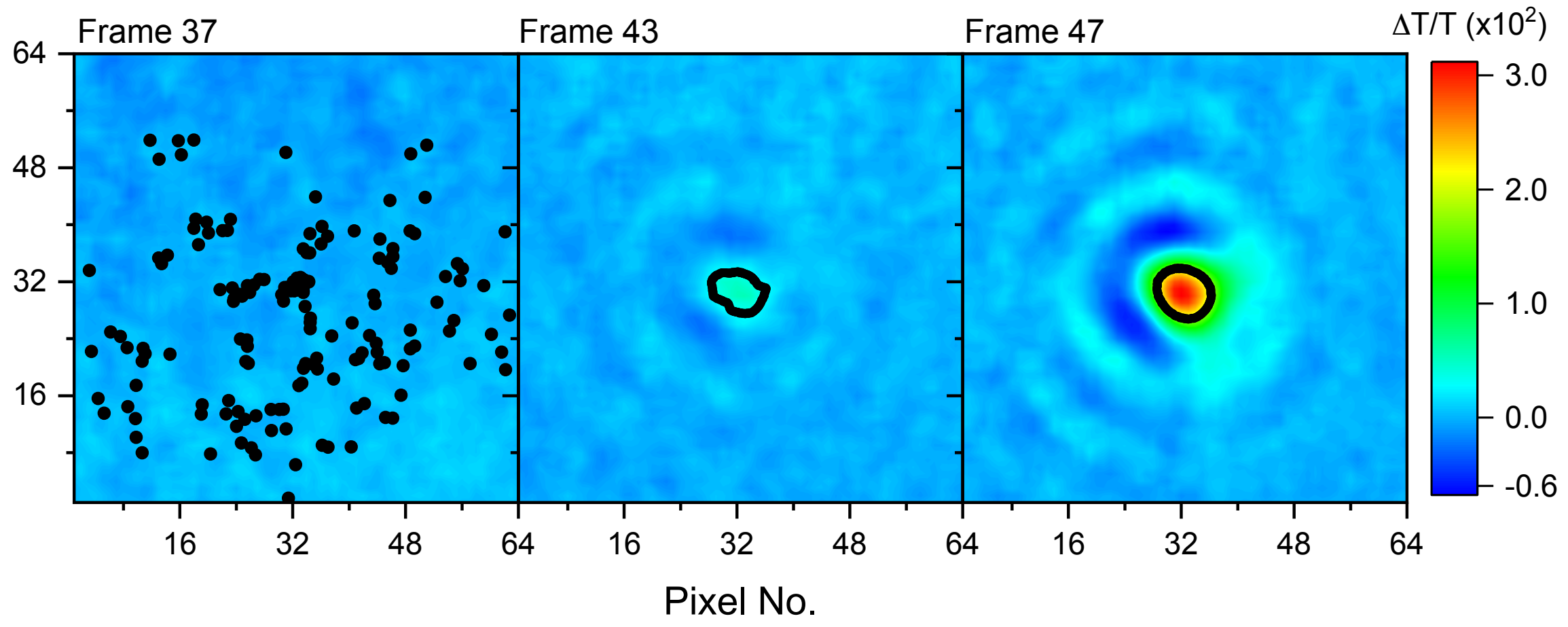

**Figure SI28:** Sigma points plotted on top of the corresponding TAM images with the same colour scale. (Left) The sigma points are scatter randomly over the whole image before time zero; (middle) a ring starts to form as the signal begins to grow; (right) the first frame when the ring is smooth and has a spot size very close to that of the diffraction-limited pump. Consequently, we define Frame 47 as our time zero.

Extraction of diffusion parameters $D$ and $n$

The most prominent advantage of using an ultra-fast TAM setup is that the spatial resolution across the image plane enables us to track movement of any pump-induced population on the femtosecond timescale. The standard data analysis procedure entails the fitting of a two-dimensional Gaussian to each image and plot the fitted sigma $\sigma(t)$ against pump-probe delay time $t$. To extract the material's diffusion constant $D$ along a principle axis, one simply fits the diffusion length equation to the sigma-time graph. Assuming the initial population at $t = t_0$ is Gaussian in space with an initial width $\sigma(t_0)$ and undergoes anomalous diffusion, the sigma at any later time $t \geq t_0$ can be described by

$$\sigma^2(t) = \sigma^2(t_0) + D(t - t_0)^n$$

where $n = 1$ for a typical diffusion process. $n < 1$ and $n > 1$ represent a sub-diffusion and super-diffusion process respectively. Sub-diffusion usually suggests the existence of carrier traps whereas super-diffusion is commonly found in the transport of active matter. If $n = 2$ a particle undergoes ballistic transport with the transport length determined by its mean free path.

The $R_1$ region observed and described in the main text arises from the ballistic transport of polaritons. In many cases the initial part of the mean free path trace shows a quadratic dependence in time, but to ease comparison between sample locations we only report a diffusion constant $D$ obtained by fitting the straight-line parts of the trace (within $R_1$). However, in order to confirm that this ballistic component is present we show in Figure **SI29** a quadratic fit to the first 20 – 25 fs of transport. We note that the behaviour is close to the instrument response and hence do not over stress its significance.

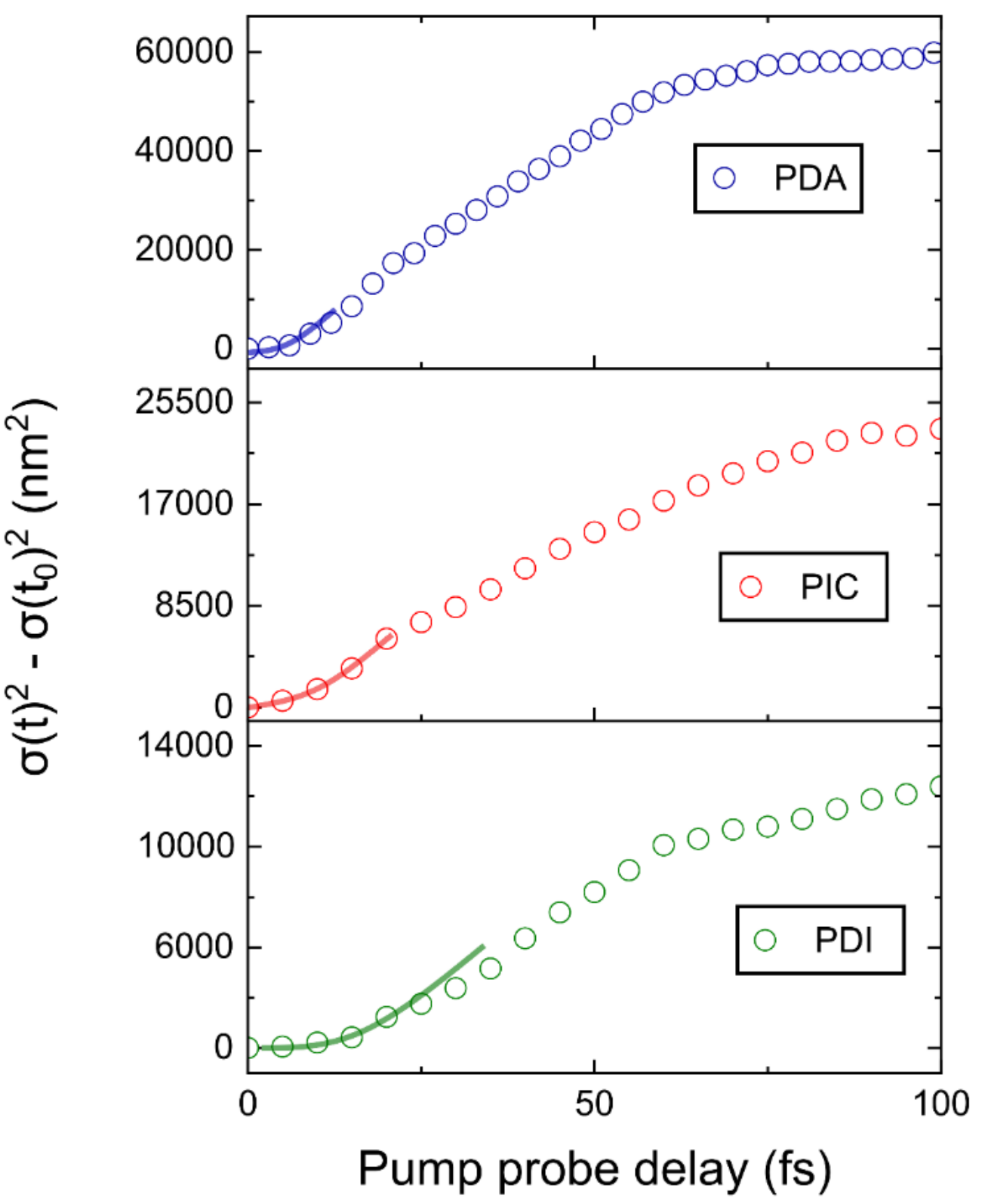

**Figure SI29:** Mean square displacement for PDA, PIC and PDI showing a quadratic fit to the initial time points. This is in keeping with the proposed ballistic transport of the initial polariton wavefuction. However, we refrain from extracting a diffusion coefficient from the data in this region as it is close to the instrument response of the setup.

Spot size at time zero

The spot size at time zero is determined from the method outlined in the above sections. The values obtained across all measurements of PDA, PIC and PDI are shown in Figure **SI30**. In all cases the FWHM at $t_0$ lies in the range 300 – 460 nm, with mean values of $\langle FWHM(t_0)_{PDA} \rangle$ = 362 ± 25 nm, $\langle FWHM(t_0)_{PIC} \rangle$ = 367 ± 25 nm and $\langle FWHM(t_0)_{PDI} \rangle = 365 \pm 25$ nm obtained. These values lie in the expected range based on the numerical aperture of the objective, pump/probe wavelength and calibration of the pump spot size using fluorescent microspheres, showing this method for determining $FWHM(t_0)$ to be robust.

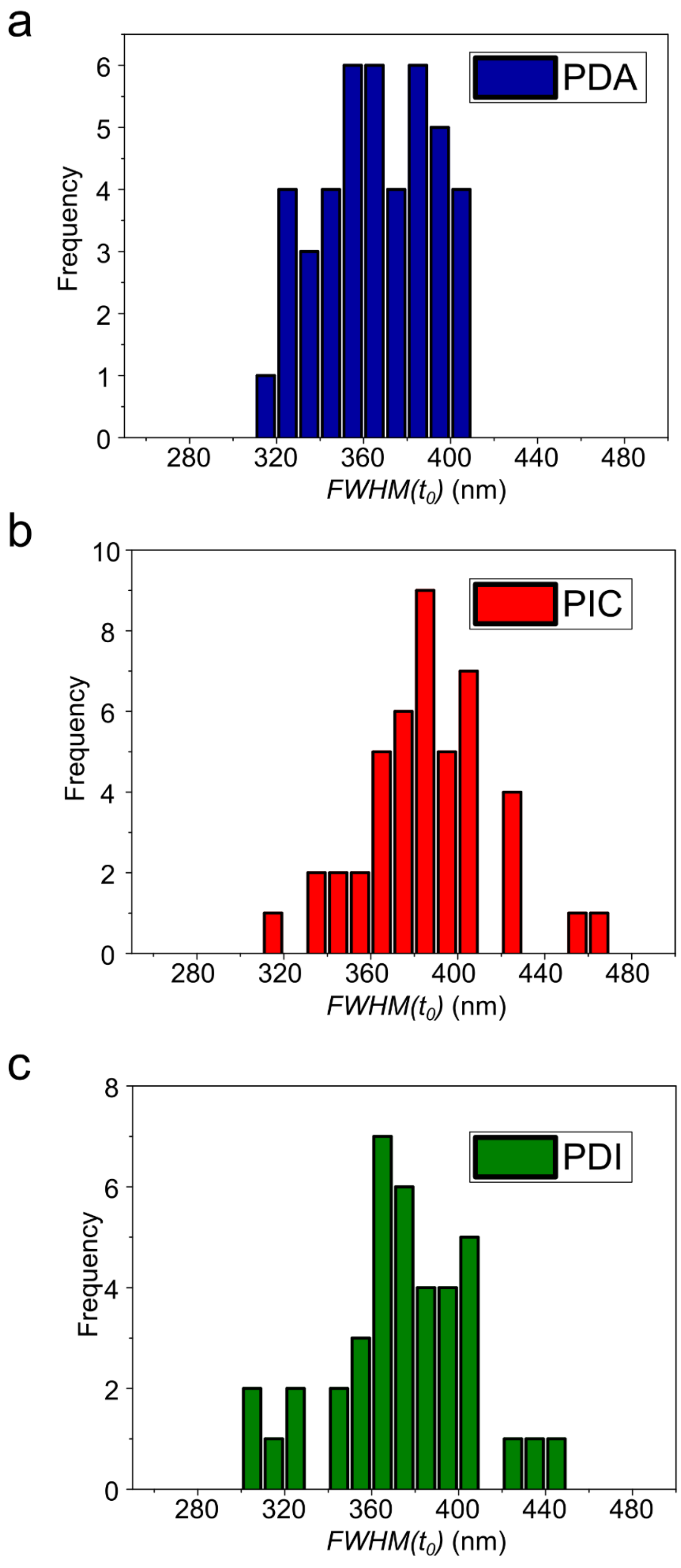

**Figure SI30: a-c.** Histograms showing distribution of pump spot size at $t_0$ for all sample locations reported in the main text. In all cases, the mean size lies close to the diffraction limit based on the numerical aperture of the objective and pump wavelength.

Radial analysis of TAM kinetics

To track the rise and decay of spatially migrating population it is possible to take a ring cut through a stack of fs-TAM images. Because the signals arising from the structures considered in this manuscript are typically highly anisotropic, ellipsoidal as opposed to circular cuts were taken. The results of such a fitting method are shown in Figure **SI31-33**. The line thickness of ellipses in each reflect the number of pixels averaged over to obtain kinetic for that particular ellipse, typically 0.5 - 1.5 pixels radially . The radius of the ellipse reported in the legend is that for the semi-major axis, as determined from the centre of mass pixel.

For PDA, PIC and PDI the growth of the kinetic from the outermost ellipse is delayed compared to that obtained by averaging pixels around the centre, and reflects the time taken for spatial transport of population. For PDA specifically, the signal at 500 nm or 240 nm (Figure **SI31**) shows a significantly slower rise ($\tau$ = 70 ± 5 fs - a; 100 ± 5 fs - b) compared to that at the centre, suggesting that in this material there are intermediate electronic relaxation processes involved in the transport of energy. The decay rate of the signal at 500 nm or 240 nm is also much slower than that at the centre ($\tau$ = 230 ± 10 fs - a; 620 ± 5 fs - b), where the kinetic of the decay is similar to that obtained from ensemble spectroscopy. We note simulations (Figure **SI34**) show that annihilation effects play little role in the observed dynamics and constantly cannot be responsible for the fast decay in the central region. Consequently, it is suggested that in PDA the transported signal at the edge pixels of the profile could arise from a state that is different to that which is within the pump excitation area.

For PIC the signal from the edge pixels decays at a similar rate to that in the centre, with the rise off-set due to energy migration. However, the decay of the kinetic, as discussed later, is much slower that obtained from ensemble spectroscopy. In PDI depending on the exact sample location very different kinetic profiles can be obtained. For example, in Figure **S33a** the pump-probe signal at 10 nm and 190 nm are off-set but decay at a similar rate ($\tau$ = 220 ± 2 fs), with the rise of the former marginally slower ($\tau$ = 15 ± 2 fs - 10 nm; 28 ± 2 fs – 190 nm). However, for the sample location examined in Figure **S33b** the signal at 250 nm decays far slower than that at 10 nm ($\tau$ = 95 ± 10 fs - 10 nm; 930 ± 10 fs – 250 nm) suggesting either a decay from a different electronic state at the edge pixels or a slow underlying transport of population to the ends of the wire that is convoluted with decay of the signal.

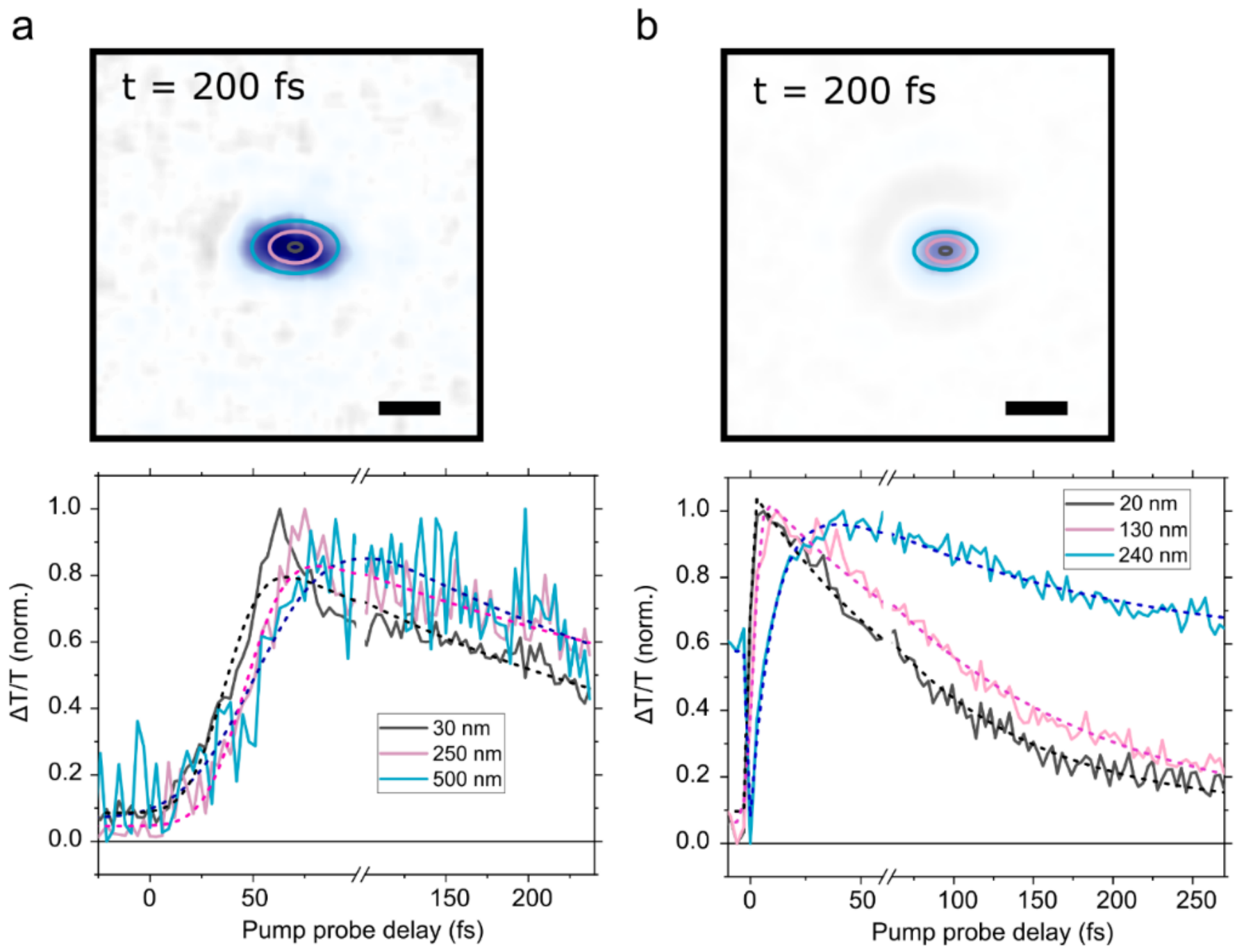

**Figure SI31: a-b.** Radial analysis of fs-TAM data for PDA at two different sample locations. Images are shown for a constant time delay of 200 fs. Ellipsoidal ring cuts are taken through the image and pixels on the ellipse averaged to generate the kinetics shown beneath. We average pixels along the lines shown in the top images. The distance shown in legend corresponds to the radius along the semi-major axis ellipse. The rise and decay of the kinetic at increasing distances from the centre is delayed with respect to that at the centre. In some cases, the strong Airy disks prevent absolute extraction of the signal (*e.g.* the signal before 0 fs in the 240 nm kinetic of panel b normalizes to 0.6). Solid lines show raw data whereas dotted lines are fits. Scale bars in both top images corresponds to 500 nm.

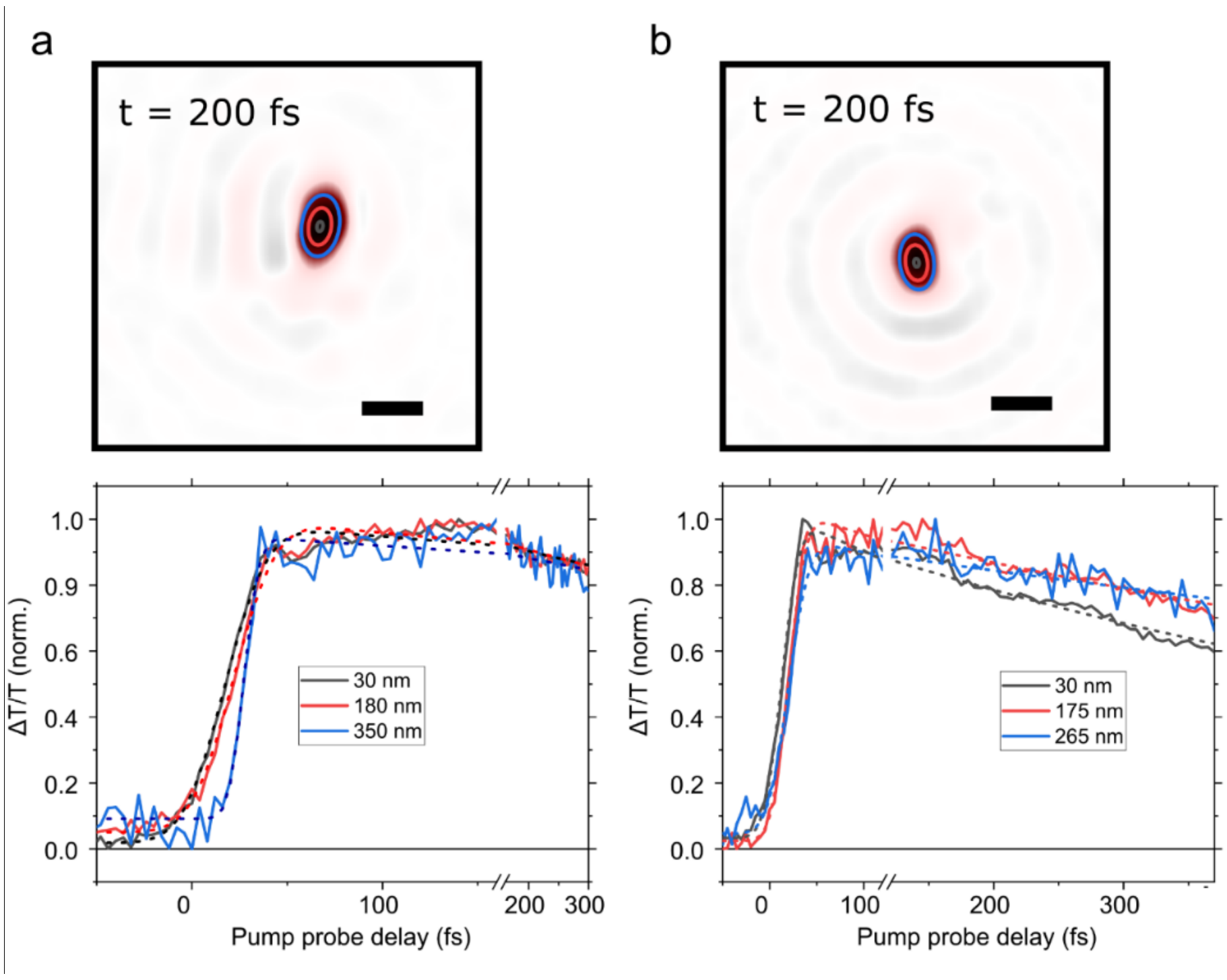

**Figure SI32: a-b.** Radial analysis of fs-TAM data for PIC at two different sample locations. Images are shown for a constant time delay of 200 fs. Ellipsoidal ring cuts are taken through the image and pixels on the ellipse averaged to generate the kinetic shown beneath. We average pixels along the lines shown in the top images. The distance shown in legend corresponds to the radius along the semi major axis of the ellipse. The decay of the kinetics is approximately the same at different distances from the centre but the rise of the signal away from the centre is delayed, indicative of ultrafast energy migration. Scale bars correspond to 500 nm.

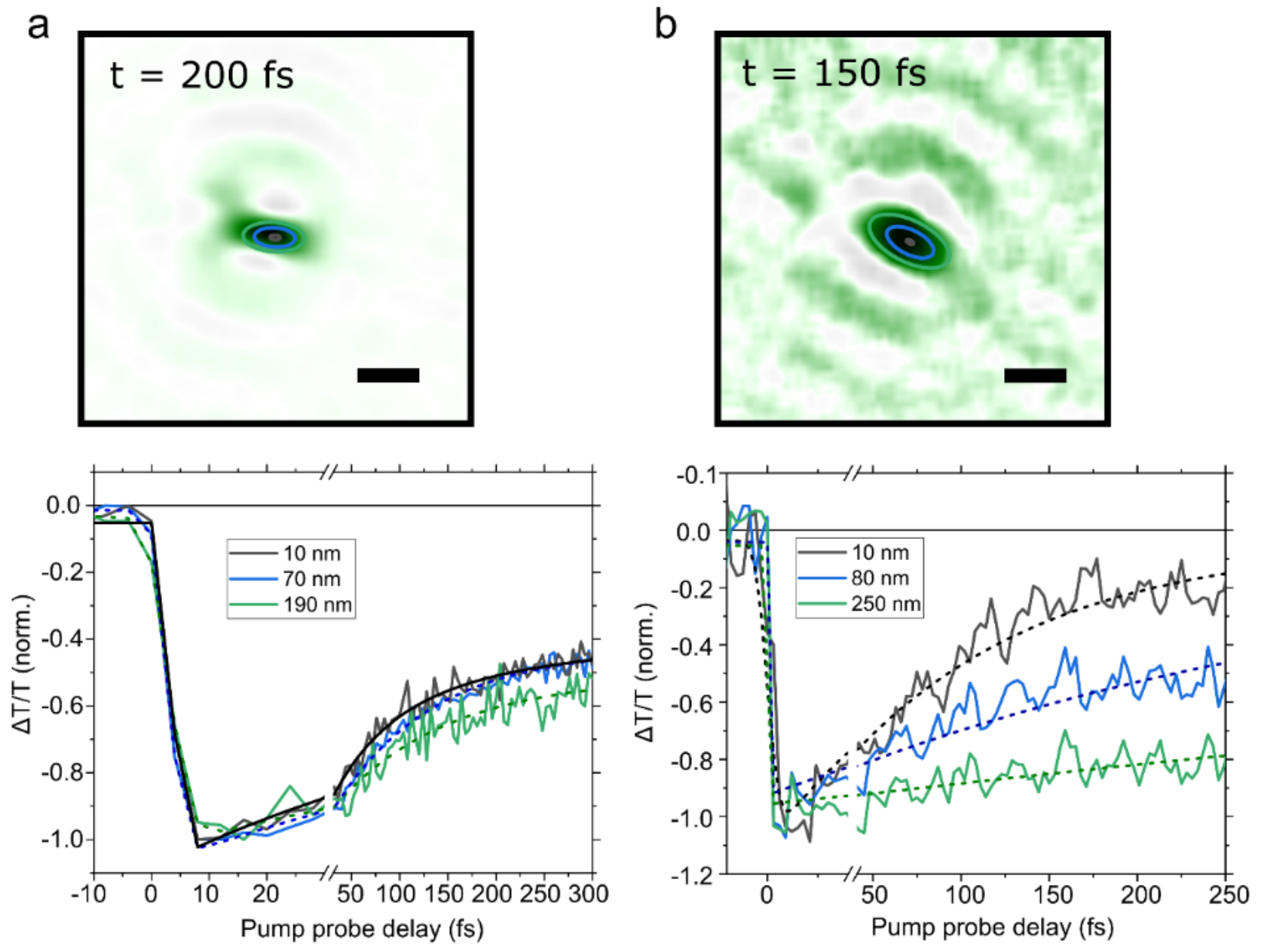

**Figure SI33: a-b.** Radial analysis of fs-TAM data for PDI. Ellipsoidal ring cuts are taken through the image and pixels on the ellipse averaged to generate the kinetic shown beneath. We average pixels along lines shown in the top image. The distance shown in legend corresponds to the radius of the ellipse along the semi major axis. The decay of the kinetics can be approximately the same (**a**) at different major radii or very different (**b**) depending on the sample location, but the rise of the signal is delayed in all cases, indicative of ultrafast energy migration. Scale bars correspond to 500 nm. The strong oscillations atop of the kinetic in (**b**) are Raman modes. The analysis and discussion of this is however beyond the scope of this work.

Empirical Model Fitting for PIC

In order to unravel the relative contributions of diffusion and population decay in the observed TAM spatial and temporal dynamics, we attempted to fit the following second-order differential equation to our fs-TAM data:

$$\frac{dp(x,y,t)}{dt} = D\,\nabla^2 p(x,y,t) - \lambda_1 p(x,y,t) - \lambda_2\, p(x,y,t)^2$$

where $p(x,y,t)$ is the population (or ΔT/T signal strength) at time $t$, and $D$ is the diffusion coefficient, $\lambda_1$ is the one-excitonic decay rate and $\lambda_2$ is associated with the rate of bi-excitonic annihilation. The raw data was smoothed using a bilateral filter based on a Gaussian weighting. D was fixed using the value obtained from experiments in $R_2$. Background offset was corrected, and a plane was subtracted from each image to ensure the background is flat. The image with maximum signal strength was then used as the initial condition for fitting with the Dirichlet boundary condition $p(\pm\infty, \pm\infty, t) = 0$. Strictly speaking, this boundary condition is only valid when the pump spot size (200 nm FWHM) is much smaller than the sample dimensions (at least 1 um) and hence the exciton population does not 'feel' the crystal boundary as they diffuse. However, this is the best approximation we have, since detailed morphological studies on our sample is beyond the scope of this work. The fit converges as residual sum of squares is minimized. As illustrated in Figure SI34, the model produces a good fit to our fs-TAM data with an outstanding goodness of fit ($R^2 > 0.95$). The table below summarizes the fitted parameters averaged over 10 data sets for PIC, where the uncertainties are the 90% confidence interval. .

| D [$cm^2$ $s^{-1}$] | $\lambda_1$ [$fs^{-1}$] | $\lambda_2$ [$cm^2$ $s^{-1}$] | $\lambda_1 p_{max}$ [$cm^{-2}$ $s^{-1}$] | $\lambda_2 p_{max}^2$ [$cm^{-2}$ $s^{-1}$] |
|---|---|---|---|---|
| 300 | $1/(337 \pm 9)$ | $11 \pm 50$ | $3.4 \times 10^{-1}$ | $1.6 \times 10^{-4}$ |

$\lambda_1 p_{max}$ and $\lambda_2 p_{max}^2$ denote the maximum contribution of one-excitonic and bi-exciton decay to the observed $\Delta T/T$ signal dynamics. Although there is a high fractional uncertainty in $\lambda_2$ due to its small amplitude, it is clear that exciton-exciton annihilation does not play a significant role in our fs-TAM experiments. This further demonstrates that the expansion in our signal is a result of carrier transport, not an 'artefact' due to annihilation effects at the centre of the pump spot. We chose to apply the model to PIC because of its relatively simple electronic structure and decay kinetics *i.e.* no triplets, excimer states *etc.*. However, we expect calculations on PDA and PDI to show similar results.

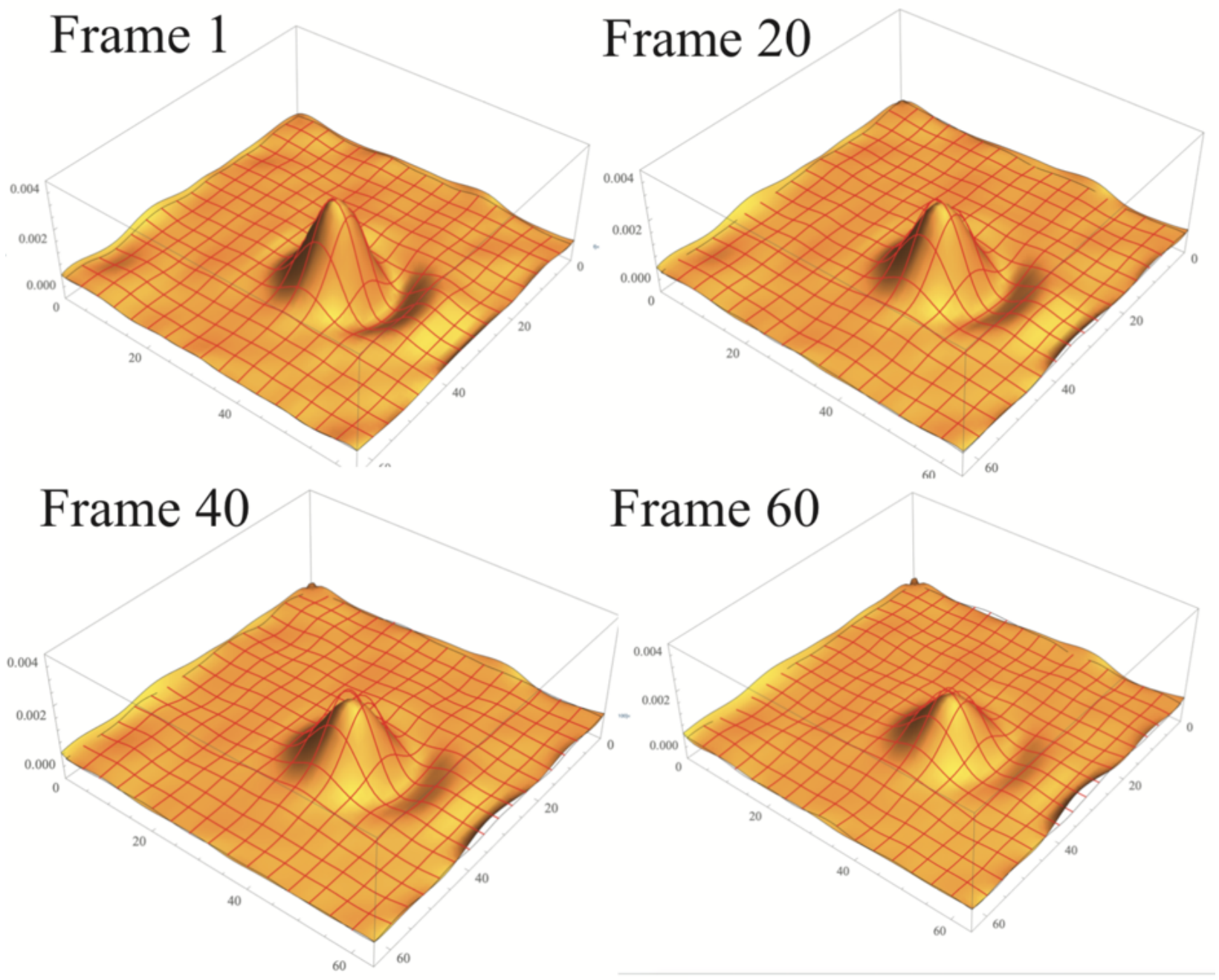

**Figure SI34:** The red mesh corresponds to experimental data and the filled yellow surface is the model fit. We find a good fit between the model and the measured data, with an R-squared value in excess of 0.95.

Fitting with two-dimensional Gaussian function

A two-dimensional (2D) rotational Gaussian function can alternatively be fit to each frame and used to analyse the spatial expansion of the $\Delta T/T$ signal. Extracting the Gaussian standard deviation ($\sigma$) along the two principle components of the 2D Gaussian, nominally referred to as $x$ and $y$ and plotting the mean square displacement, $\sigma(t)^2 - \sigma(t_0)^2$, against time allows us to determine the transport velocities and diffusion coefficients, $D_{x,y}$, in the two principle directions. In Figure **SI35** we compare the $D$ values obtained from this analysis, $D_{2D\ Guass}$, to the radial method $D_{Radial}$ (along the wire axis), detailed previously in the main text. As the growth of the signals is typically anisotropic, we take the $x$ direction as that in which there is maximal growth. The solid straight line denotes $D_{2D\ Guass} = D_{Radial}$. It is clear that the $D$ values obtained from the two methods are roughly equivalent with some small variation in the exact nature of the MSD plot.

We emphasise here that order to fit a 2D Gaussian the excited species must not see boundaries which might affect the distribution of population in space. As we are dealing with 1D nanostructures whose boundaries inherently lie within the area of the pump spot the boundaries of the material can skew the 2D Gaussian and hence it is not an appropriate function to capture the physics, despite the reasonable fit.

The error on the fits is not shown on the plots for clarity, but was estimated using the MATLAB curve fitting tool box; fits were considered acceptable when the error was less than 5%.

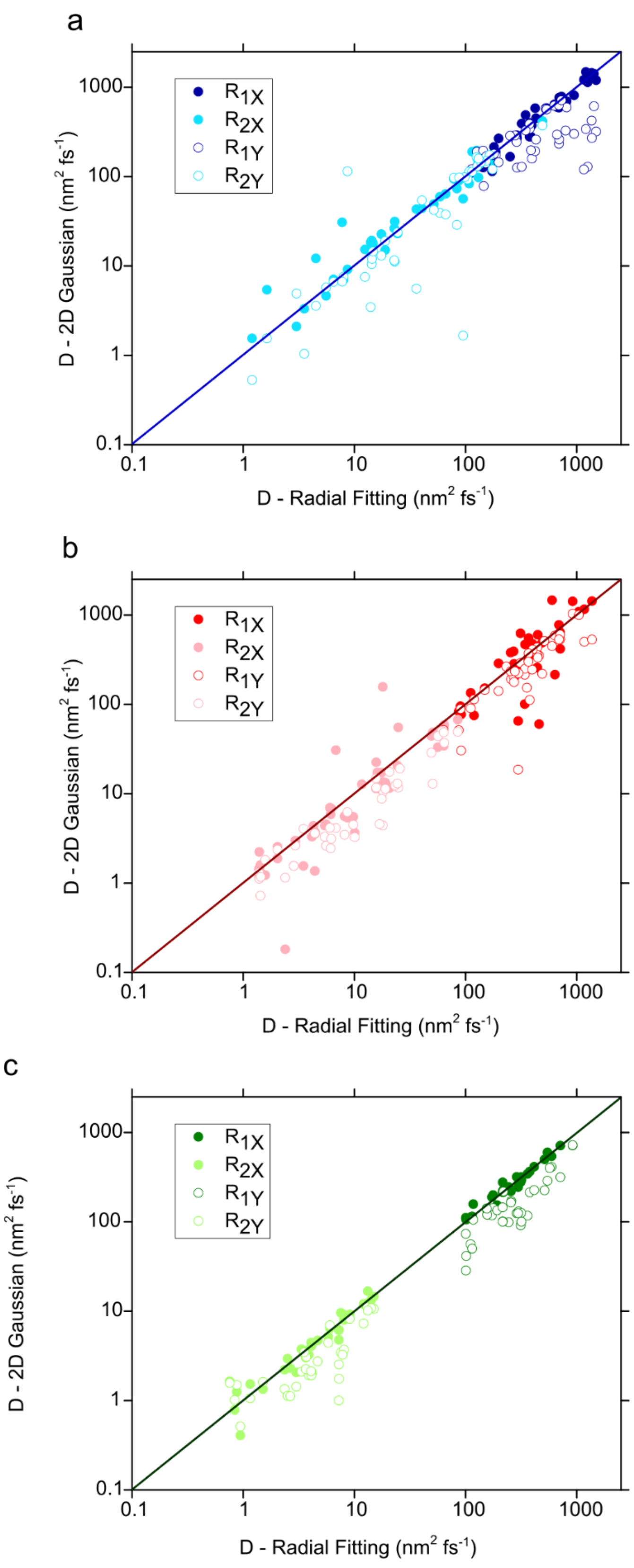

**Figure SI35:** Comparison of diffusion coefficient obtained from radial and 2D Gaussian fitting, for **a.** PDA (blue), **b.** PIC, (red) and **c.** PDI (green). Diffusion constants are obtained for both the $R_1$ and $R_2$

region. The solid line across the frame shows the one-to-one comparison between D obtained via the two methods.

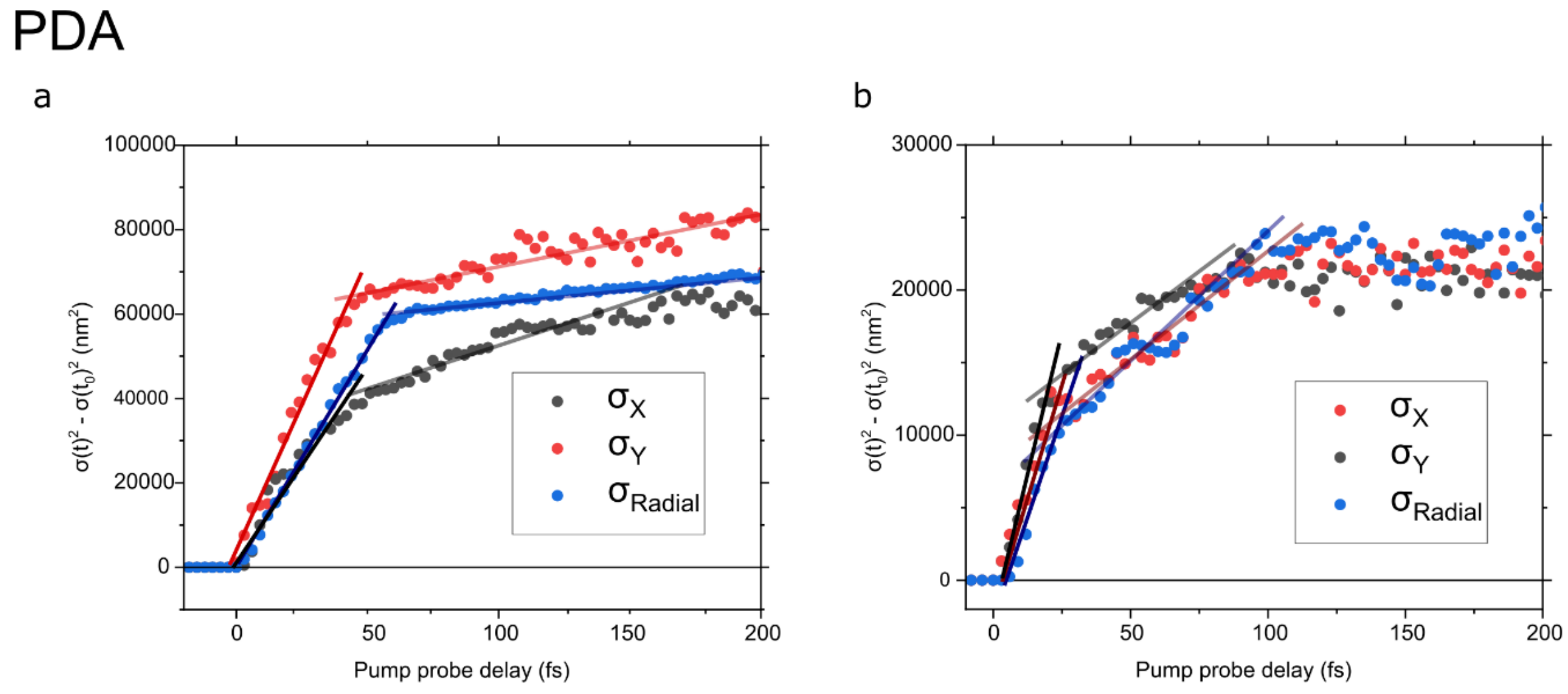

**Figure SI36: a-b.** Mean square displacement for the two different sample locations in PDA shown above. There is a good agreement between $\sigma_X$ and $\sigma_{radial}$ (value obtained from fitting method detailed above and used in the main text).

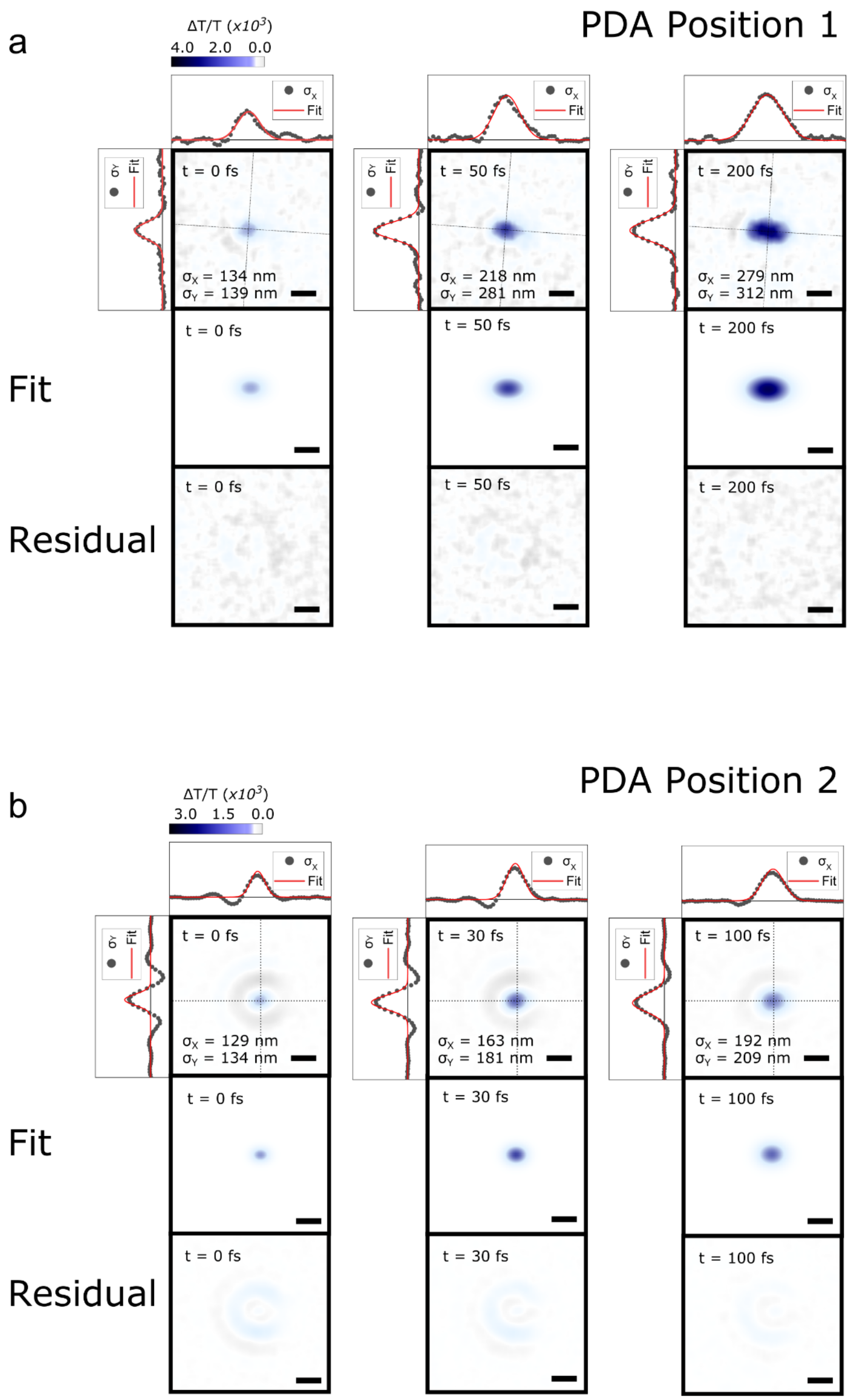

**Figure SI37: a-b.** fs-TAM images of PDA at three time delays, overlaid with the principle axes . Side panels show data along the principle axes (black dots) and the corresponding fits (red line). The fits

and residuals are shown in the bottom half; the residuals are dominated by strong Airy disks. The scale bar is 500 nm in all images.

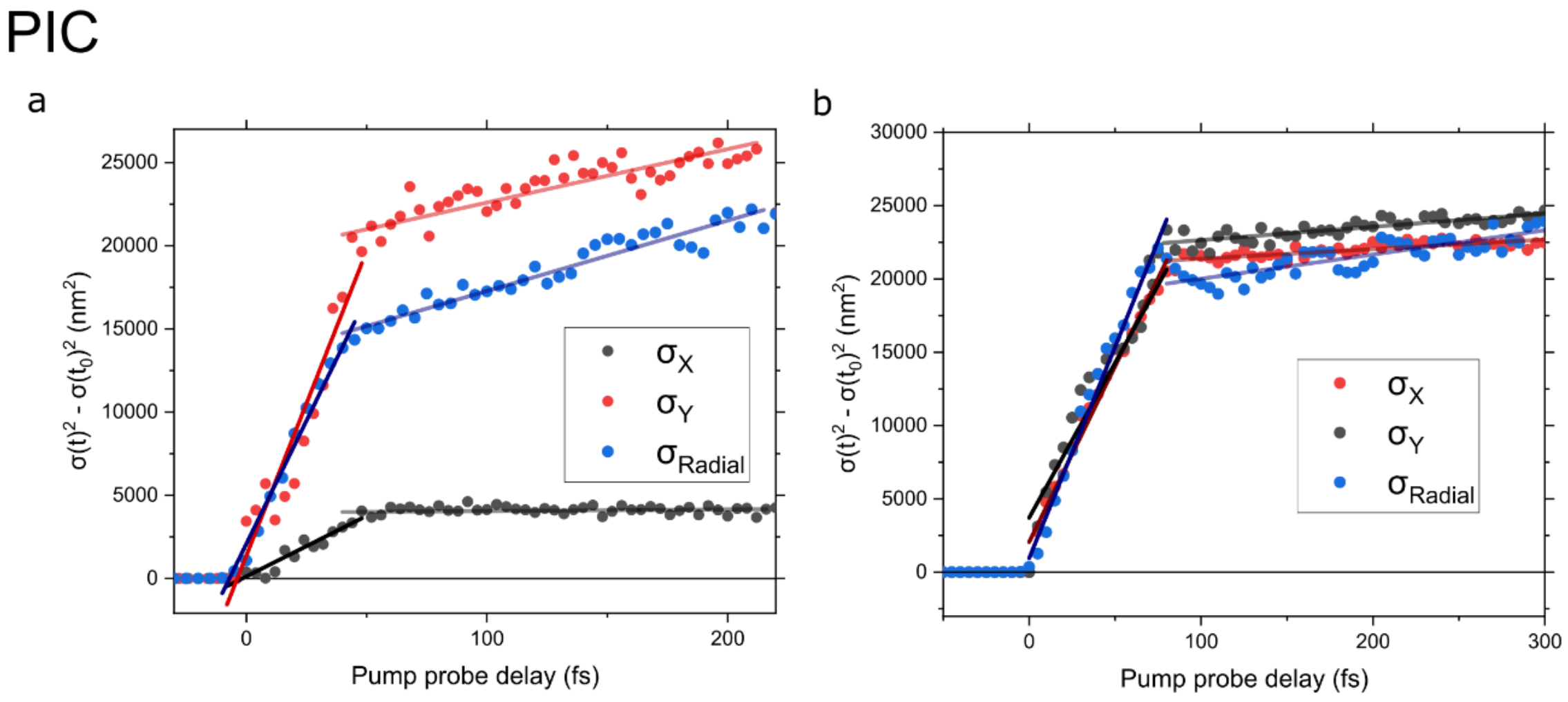

**Figure SI38: a-b.** Mean square displacement for the two different sample locations in PIC shown above. The x and y directions also correspond to those in the above. There is a good agreement between $\sigma_X$ and $\sigma_{radial}$ (value obtained from fitting method detailed above and used in the main text), with a small degree of overestimation in the case of the 2D Gaussian. When the motion is particular anisotropic displacement in the direction orthogonal to transport is small, in this case $\sigma_Y$.

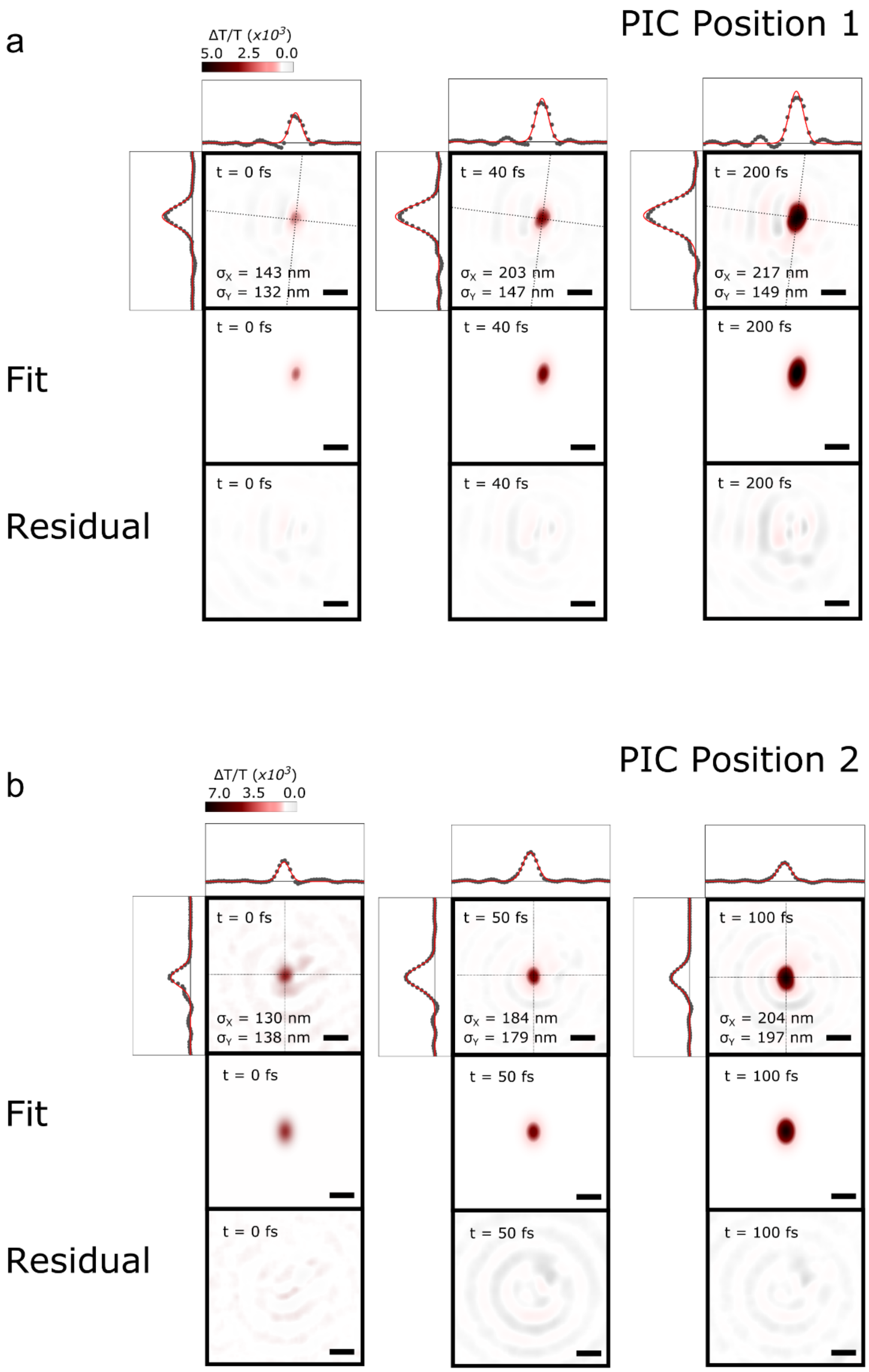
a
PIC Position 1
ΔT/T (x10³)
5.0 2.5 0.0
t = 0 fs
σx = 143 nm
σy = 132 nm
t = 40 fs
σx = 203 nm
σy = 147 nm
t = 200 fs
σx = 217 nm
σy = 149 nm
Fit
t = 0 fs
t = 40 fs
t = 200 fs
Residual
t = 0 fs
t = 40 fs
t = 200 fs
b
PIC Position 2
ΔT/T (x10³)
7.0 3.5 0.0
t = 0 fs
σx = 130 nm
σy = 138 nm
t = 50 fs
σx = 184 nm
σy = 179 nm
t = 100 fs
σx = 204 nm
σy = 197 nm
Fit
t = 0 fs
t = 50 fs
t = 100 fs
Residual
t = 0 fs
t = 50 fs
t = 100 fs

**Figure SI39: a-b.** fs-TAM images of PIC at three time delays and two different positions. Dotted lines show principle axis of the 2D Gaussian fit to data. Side panels show data along the principle axes (black dots) and the corresponding fits (red line). The scale bar is 500 nm in all images.

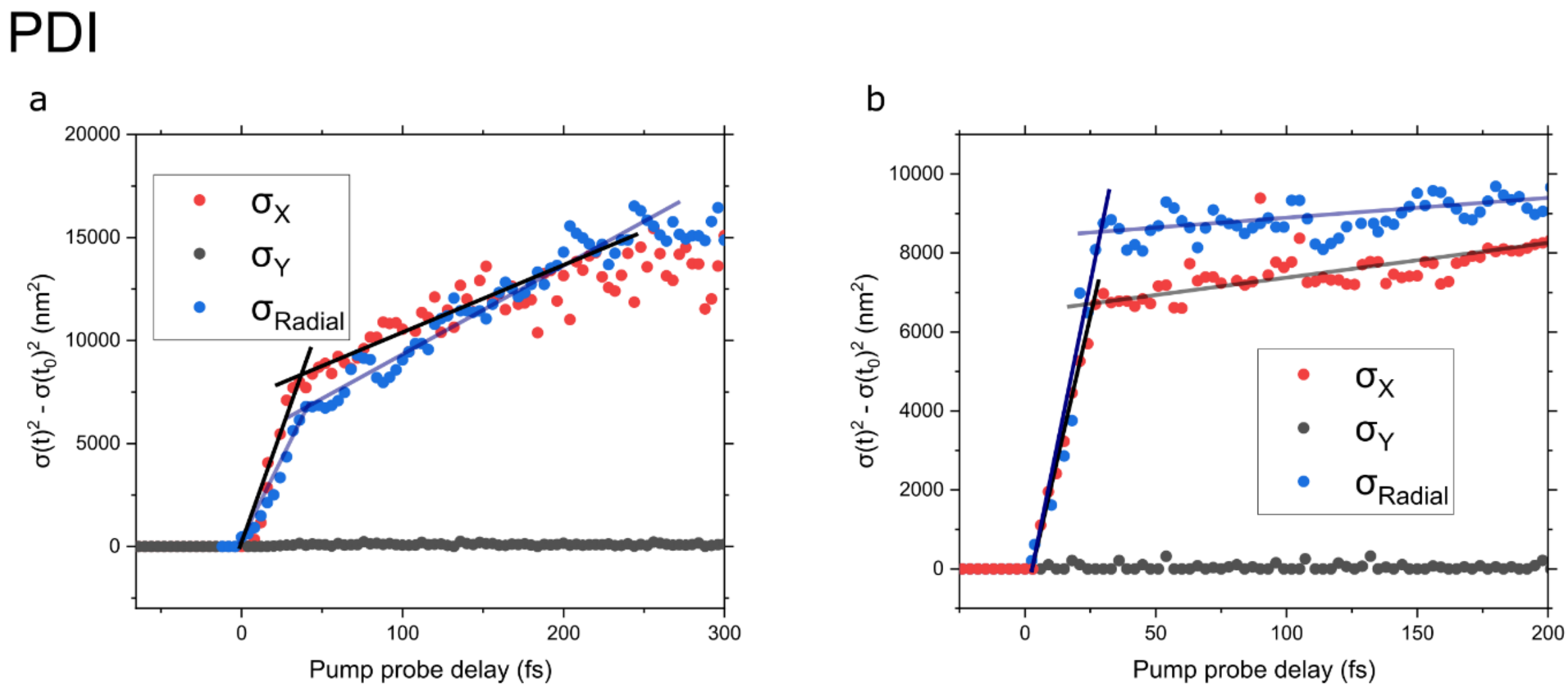

**Figure SI40: a-b.** Mean square displacement for the two different sample locations in PDI shown above. There is a good agreement between $\sigma_X$ and $\sigma_{radial}$, with a small degree of underestimation in the case of the 2D Gaussian. When the motion is particularly anisotropic displacement in the direction orthogonal to transport is small.

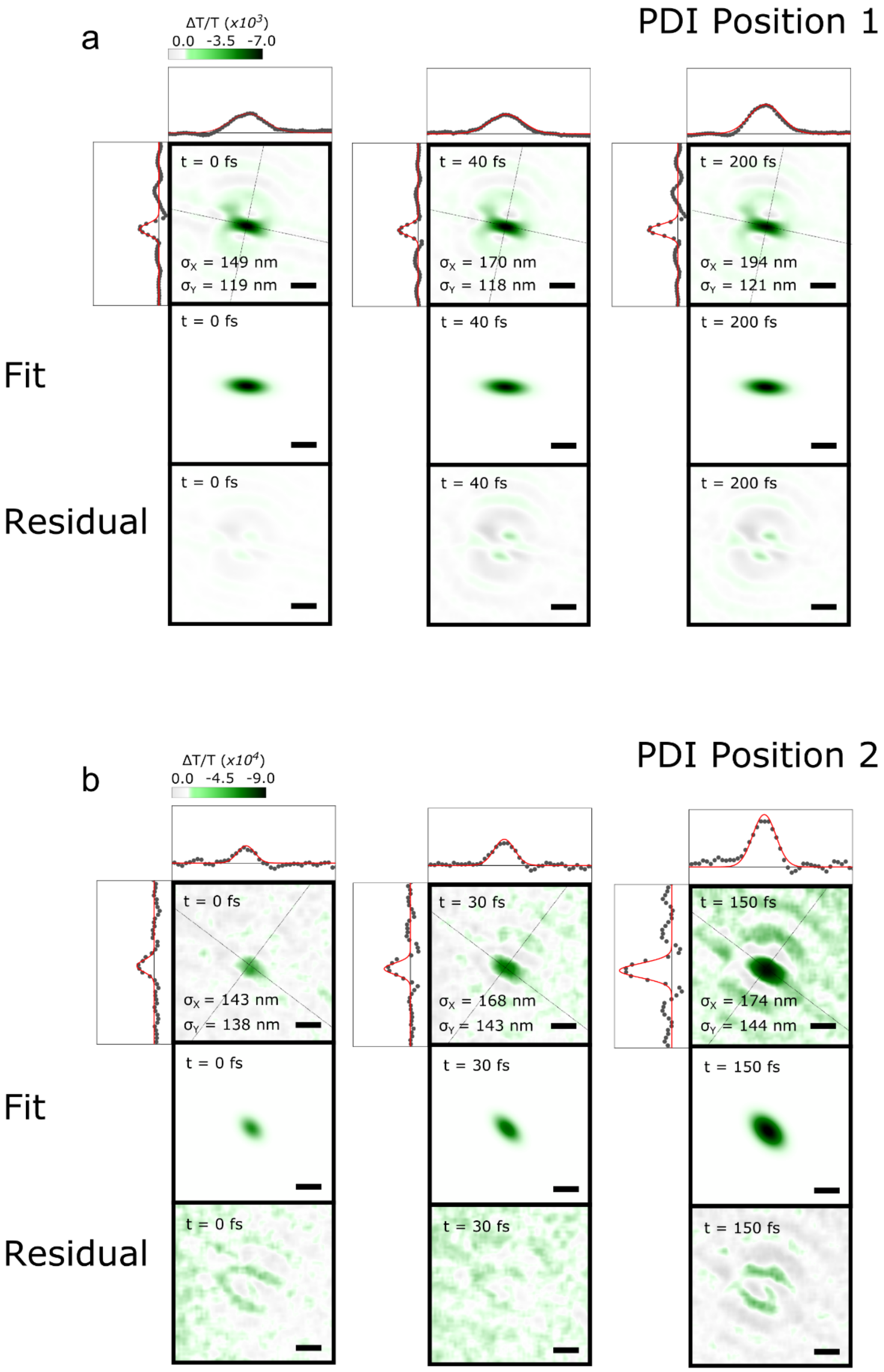

a
ΔT/T (x10³)
0.0 -3.5 -7.0
PDI Position 1
t = 0 fs
σX = 149 nm
σY = 119 nm
t = 40 fs
σX = 170 nm
σY = 118 nm
t = 200 fs
σX = 194 nm
σY = 121 nm
Fit
t = 0 fs
t = 40 fs
t = 200 fs
Residual
t = 0 fs
t = 40 fs
t = 200 fs
b
ΔT/T (x10⁴)
0.0 -4.5 -9.0
PDI Position 2
t = 0 fs
σX = 143 nm
σY = 138 nm
t = 30 fs
σX = 168 nm
σY = 143 nm
t = 150 fs
σX = 174 nm
σY = 144 nm
Fit
t = 0 fs
t = 30 fs
t = 150 fs
Residual
t = 0 fs
t = 30 fs
t = 150 fs

**Figure SI41: a-b.** fs-TAM images of PDI at three time delays and two different positions. The scale bar is 500 nm in all images.

## S4 Probe wavelength dependence

For PDA and PIC the spectrally sharp stimulated emission bands represents the only transitions we could probe because of the limited probe wavelength range and the pump wavelength. However, in PDI where we probed the broad PIA feature, we measured the dynamics of a different wavelength (780 nm), in order to check whether the probe wavelength had any effects on the reported signal diffusion. As can be seen from Figure **SI42** there appears to be little dependence on probe

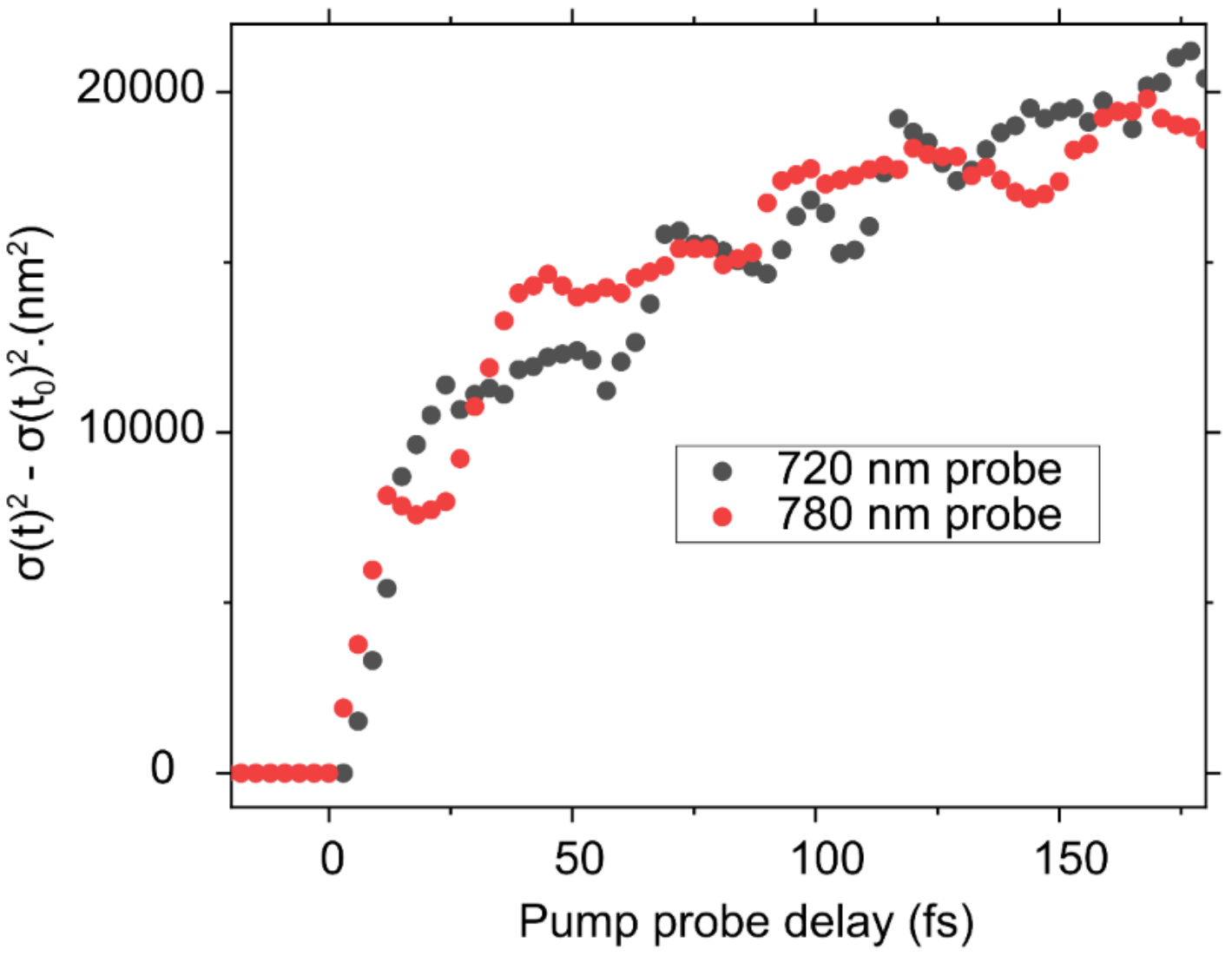

wavelength.

**Figure SI42:** Mean square displacement in PDI wires at 720 nm (black) and 780 nm (red) probe wavelengths. No difference can be observed and hence we conclude that within the same band the probe wavelength has little effect on the observed dynamics.

<u>Femtosecond pump-probe spectra of PDA</u>

PDA is an archetypal conjugated polymer which has been extensively studied via time-resolved spectroscopies. In Figure **SI43** we present the pump-probe spectra of PDA; all experiments are carried out with the pump and probe polarized parallel to the long axis of the PDA chains. There are three distinct species in the spectra whose decays can all be fitted with a single exponential. Between 650 – 800 nm, we observe a rapid (~320 ± 5 fs) decay of the PDA SE from the $1B_u$ state. The SE consists of a vibrational progression of peaks. Additionally, between 820 – 900 nm there is an energetically broad band that has been previously assigned to the $2A_g$ photo-induced absorption (PIA). Vibronic relaxation from $1B_u$ to $2A_g$ means the growth of the $2A_g$ PIA is delayed with respect to that of the SE; $2A_g$ then decays with an exponential time constant of ~380 ± 5 fs. The results above are fully consistent with previous pump-probe data and state assignments on a range of 'blue' PDA. A prominent third species between 675 – 750 nm grows from the $1B_u$ and $2A_g$ states ($\tau_{rise}$ ~350 ± 10 fs). This state has been recently assigned to arise from a vibrationally 'hot' ground state. TAM measurements were performed at a probe wavelength of 670 nm on the edge of the stimulated emission band of PDA. Discussion of results at other probe wavelengths *e.g.* within the $2A_g$ band or 'hot' ground state are beyond the scope of this work.

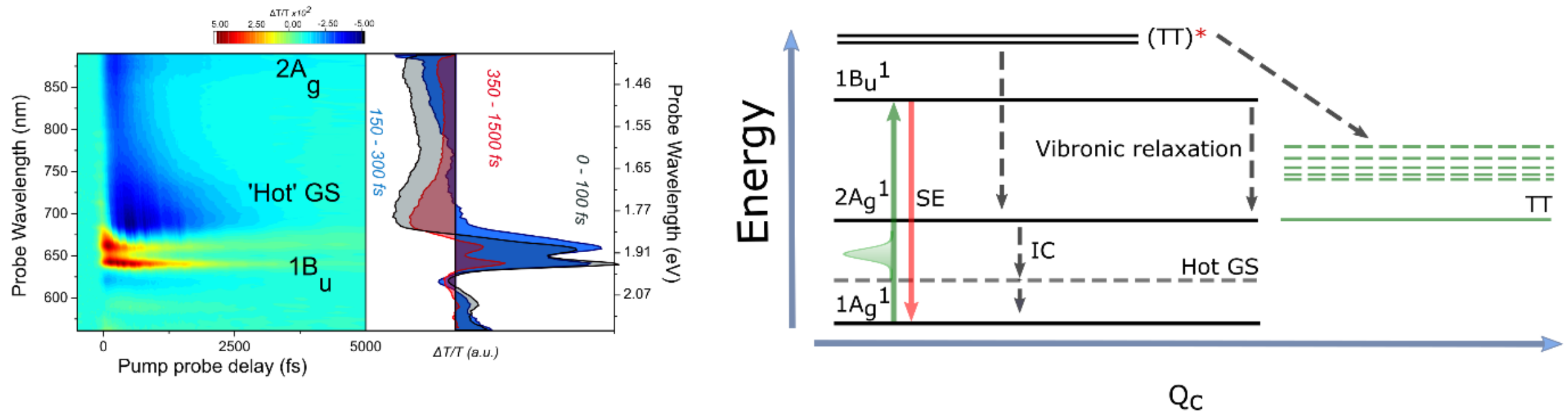

**Figure SI43:** Pump-probe spectrum and corresponding spectral cuts at indicated time delays of PDA crystal. The positive (red) regions at 640 nm and 670 nm correspond to stimulated emission from the first vibronic and zero phonon transitions. Additionally, at 720 nm and 870 nm there are two photo-induced absorption features (PIA), which we assign to the vibrationally 'hot' ground state (GS) and the $2A_g$ state respectively. The PIA below 630 nm are not assigned. The right hand figure shows an energy diagram summarizing the electronic states of PDA. The x-axis $Q_c$ represents the reaction coordinate. The pump pulse (green Gaussian) directly populates $1B_u$, which is followed by a rapid (~150 fs in PDA) vibronic relaxation to $2A_{g.}$ $1B_u$ and $2A_g$ then relax back to $1A_g$ via the 'hot' ground state (Hot GS).

<u>Femtosecond pump-probe spectra of PIC</u>

The pump-probe response of PIC thin films has been described extensively in the literature, hence here we only provide a brief description. Thin films were excited with a broadband pump pulse (520 nm centre, ~60 nm FWHM, 10 fs) with the polarization of the white light continuum probe set to 'magic angle' to avoid photoselection effects. Figure **SI44** shows the pump probe spectrum contains two sharp transitions: a positive band at ~583 nm associated with the overlapping ground state bleach (GSB)/stimulated emission (SE) of the nanotubes and a negative photo-induced absorption (PIA) band at 575 nm resulting from $S_1$-$S_2$ transitions. The congested spectrum precludes absolute determination of the rate constants for individual relaxation processes but we note the decay appears approximately uniform across the spectrum with a fast component $\tau$ = 130 ± 20 fs and a slow component $\tau$ = 5.2 ± 0.2 ps. Within the range of excitation densities considered here, $(\Delta T/T)_{max}$ =

0.008, the kinetic dynamics is independent of fluence. In fs-TAM, to achieve the required bandwidth for pump and probe pulse compression we are limited to probing solely the stimulated emission band at ~600 nm which is strongly overlapped with GSB albeit on the low energy red edge.

For completeness, transient reflectivity measurements were performed on sample locations that showed a splitting in their reflection spectra and that those did not. The pulse configuration used was similar to that detailed in (S1), however the sample was tilted slightly with respect to the incident beams in order to collect the reflected signal shown in Figure SI44a. A high resolution 2700 lines per mm (550 nm blaze) grating (Andor) was used. The resulting ΔR/R spectrum contains similar features to that obtained in transmission measurements but for regions that do show a splitting the GSB/SE peak around 580 nm is divided into two branches, assigned to an upper and lower polariton respectively. The congested spectrum prevents any splitting within the PIAs from being observed.

In Figure **SI44d** we compare the kinetic at an integrated probe wavelength 590 – 610 nm (GSB/SE) for a sample location that does show splitting and one that does not. In the former case the decay and rise of the signal is much slower than where there is no splitting ($\tau_{rise}$ = 120 ± 10, τ = 280 ± 20, τ = 5.9 ± 0.3 ps – no splitting; $\tau_{rise}$ = 170 ± 10, τ = 375 ± 10, τ = 7.2 ± 0.2 ps – splitting), as was observed in fs-TAM measurements. The origin of this is unknown and investigation beyond the scope of this work, but for example may represent some relaxation time between the upper and lower polariton branches or the lower polariton branch and the exciton reservoir.

Similar experiments were performed on PDA thin films. However, the smaller splitting precluded observation of split bands in transient reflection experiments within the signal-to-noise limit.

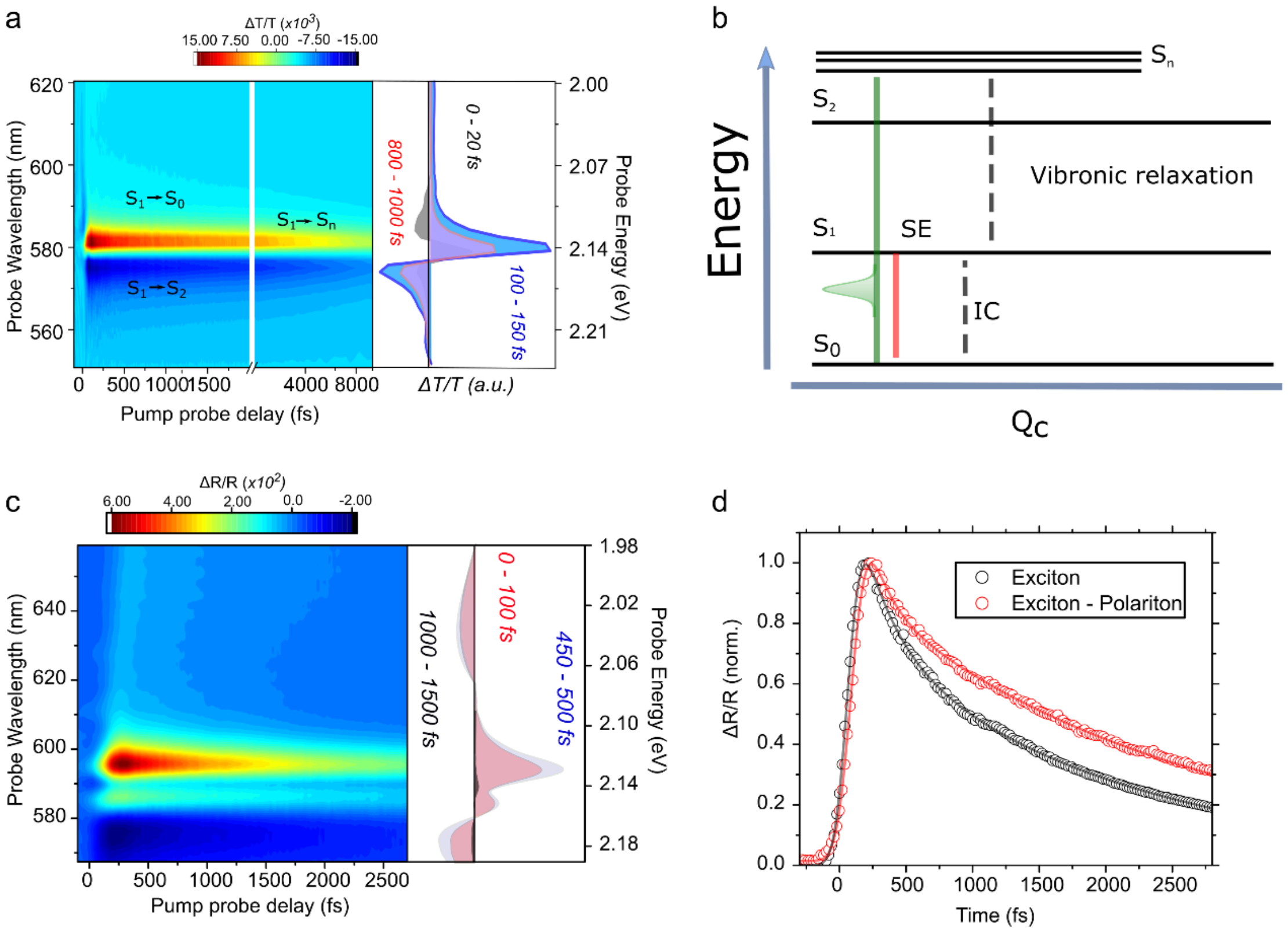

**Figure SI44: a.** Pump probe spectrum and corresponding spectral cuts at indicated time delays of PIC thin film. The positive (red) regions correspond to the highly overlapped ground state bleach (GSB)/stimulated emission (SE), with a sharp photo-induced absorption band centred around 577 nm. **b.** The right hand figure shows a simplified cartoon of the energy levels in the system. Excitation occurs into a $S_n$ manifold from which there is ultrafast vibronic relaxation to $S_1$. Stimulated emission, which is slightly red shifted from the GSB, occurs from this state. The PIA corresponds to transitions between $S_2 - S_1$. **c.** Transient reflection spectrum of PIC in a sample location that shows a splitting of the main J-band in steady state reflectivity. The GSB/SE transition is split into bands assigned to the upper (UP) and lower polariton (LP). **d.** The decays of the UP and LP are bi-exponential with the fast component having a higher amplitude. The faster component of UP decays significantly shorter than the LP; $\tau_{UP} = 220 \pm 10$ fs versus $\tau_{LP} = 375 \pm 10$ fs.

<u>Femtosecond pump-probe spectra of PDI nanobelts</u>

Extensive description and analysis of the pump probe spectra of PDI nanobelts is beyond the scope of this work, hence we limit the following to a brief description of the spectra. Thin films consisting of large clusters of PDI nanowires were excited with a broadband pump pulse (530 nm centre, ~55 nm FWHM, 12 fs). Within the probe range considered the spectrum shows a broad positive feature 550 nm – 650 nm which is well overlapped with emission of the nanowires and hence assign to stimulated emission (SE). Between 650 nm – 800 nm there is a broad negative photo-induced absorption band which we tentatively ascribe to an $S_1$-$S_2$ transition. At later time delays there may be relaxation into excimer sub-levels, however based on previous reports this would be expected to be on nanosecond time scales. Furthermore, although there have been several reports of intersystem crossing and singlet fission in perylene dimiide based semiconductors, we believe on the timescales considered here the effects of crossover into a triplet manifold are negligible. Consequently, we feel confident that in fs-TAM measurements that at 720 nm we are probing a singlet exciton transition.

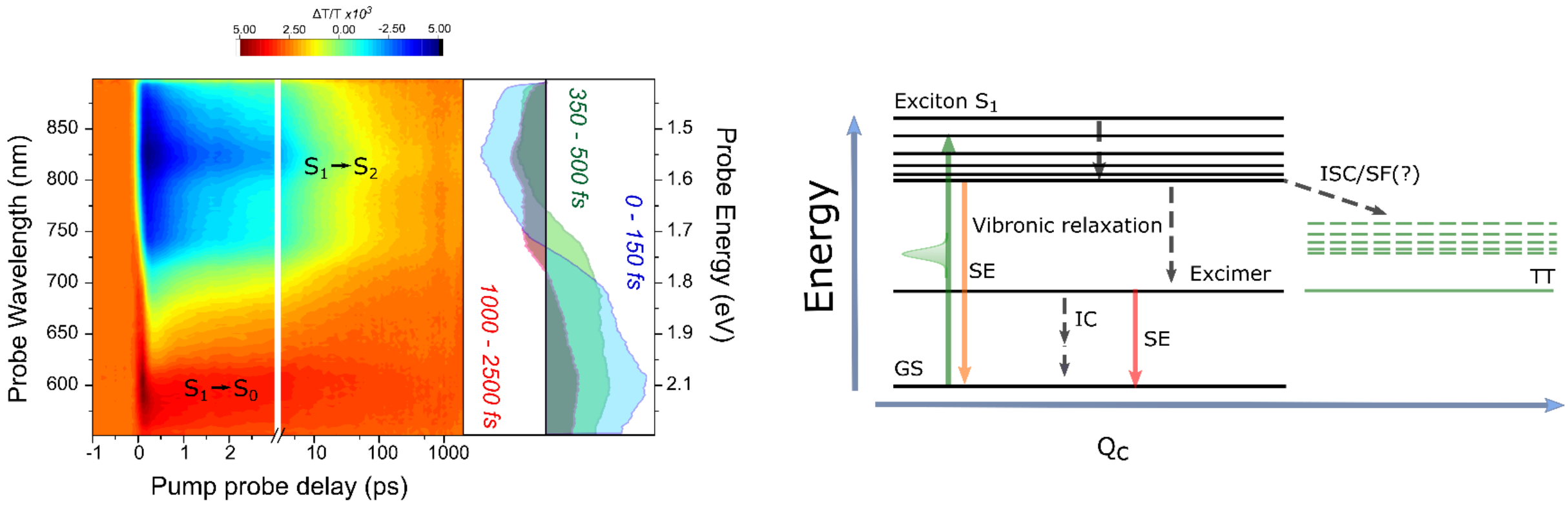

**Figure SI45:** Pump probe spectrum and corresponding spectral cuts at indicated time delays of PDI nanobelts. The positive (red) regions correspond to stimulated emission (SE) and the broad negative region (probed in TAM) is the $S_1$-$S_2$ photo-induced absorption band (PIA). On the right is shown a cartoon of the corresponding energy levels ($Q_c$ – indicates reaction coordinate). For the timescales studied it is assumed there is no crossing to the triplet manifold. Furthermore, any vibronic relaxation or internal conversion (IC) to low energy excimer states is presumed to be relatively inefficient.

**S5 Transport Kinetics**

Kinetics of fs-TAM signal

As discussed in the main text, the fs-TAM response shows a large variation depending on the spot on the sample. This is particularly evident in the kinetics of response (averaged across the entire spatial signal) as shown in Figure **SI46**. For example, in PDA there is sometimes a slow ~70 fs growth to the signal (**a** and **c**) where in other positions the signal rise is instrument response limited (**b**). In all cases there is an initial fast ~250 fs decay component to the SE but for several positions an additional 1.2 – 1.8 ps lifetime is required to fit the decay. As seen in Figure **SI46c** there is also the potential for small coherent artefact contributions to the signals at some locations in the sample.

In PIC and PDI the response is slightly more homogenous. For PIC the signal shows a slow ~180 fs rise, followed by a bi-exponential decay ($\tau_1$~350 fs and $\tau_2$~7.6 ps). In PDI the rise of the signal is often instantaneous with an overall multi-exponential decay.

Understanding the origin of this behaviour goes beyond the scope of the present work. However, we highlight in all three samples that the spatial growth is independent of the signal strength, especially in the region of ultrafast expansion. Indeed, the signal expands outwards throughout the initial rise and the subsequent decay, proving that the observations described do not arise from an incorrect assignment of $t_0$.

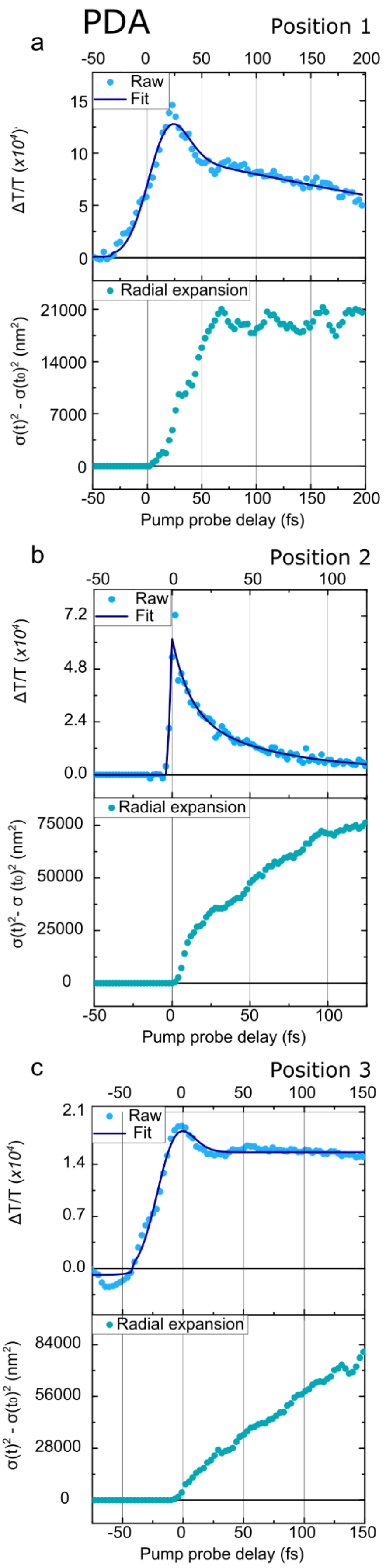

a
PDA
Position 1
Raw
Fit
ΔT/T (x10$^4$)
Radial expansion
$\sigma(t)^2 - \sigma(t_0)^2$ (nm$^2$)
Pump probe delay (fs)
b
Position 2
Raw
Fit
ΔT/T (x10$^4$)
Radial expansion
$\sigma(t)^2 - \sigma(t_0)^2$ (nm$^2$)
Pump probe delay (fs)
c
Position 3
Raw
Fit
ΔT/T (x10$^4$)
Radial expansion
$\sigma(t)^2 - \sigma(t_0)^2$ (nm$^2$)
Pump probe delay (fs)

**Figure SI46: a-c.** Kinetics (dots - raw data; solid lines - fit) and mean square displacement traces at three different spots in PDA. The $\sigma(t)^2 - \sigma(t_0)^2$ trace consistently shows a biphasic response but the kinetics vary significantly between different sample locations. In some cases, (**c**), a small coherent artefact can be observed before time zero. The vertical grid lines are added to allow for comparison between the two plots.

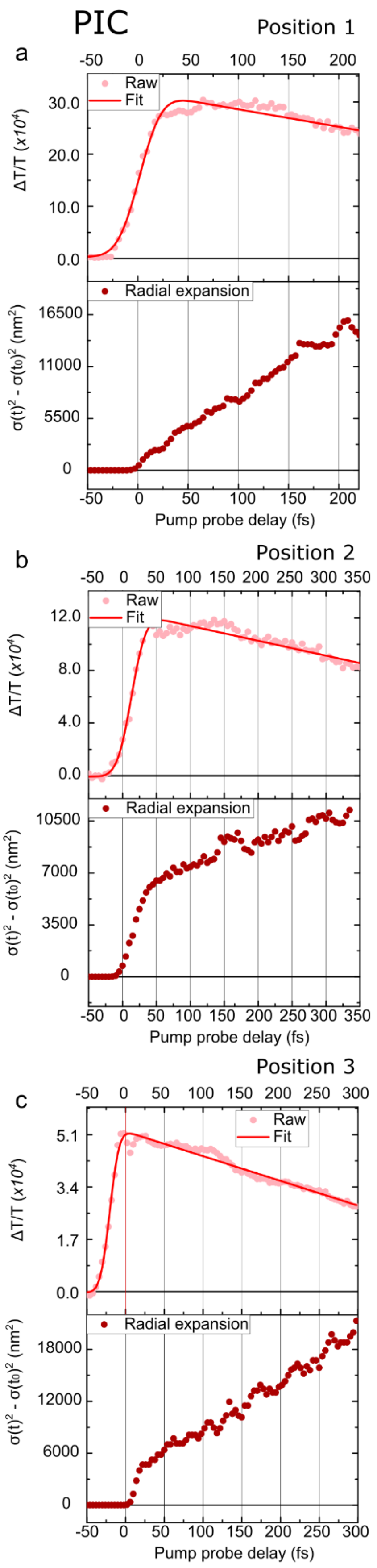
PIC
a
Position 1
Raw
Fit
ΔT/T (x10⁴)
30.0
20.0
10.0
0.0
Radial expansion
σ(t)² - σ(t₀)² (nm²)
16500
11000
5500
0
-50 0 50 100 150 200
Pump probe delay (fs)
b
Position 2
-50 0 50 100 150 200 250 300 350
Raw
Fit
ΔT/T (x10⁴)
12.0
8.0
4.0
0.0
Radial expansion
σ(t)² - σ(t₀)² (nm²)
10500
7000
3500
0
Pump probe delay (fs)
c
Position 3
-50 0 50 100 150 200 250 300
Raw
Fit
ΔT/T (x10⁴)
5.1
3.4
1.7
0.0
Radial expansion
σ(t)² - σ(t₀)² (nm²)
18000
12000
6000
0
Pump probe delay (fs)

**Figure SI47: a-c.** Kinetics (dots - raw data; solid lines - fit) and mean square displacement traces at three different spots in PIC. The $\sigma(t)^2 - \sigma(t_0)^2$ trace consistently shows a biphasic response and in this sample the kinetics are relatively consistent between sample locations.

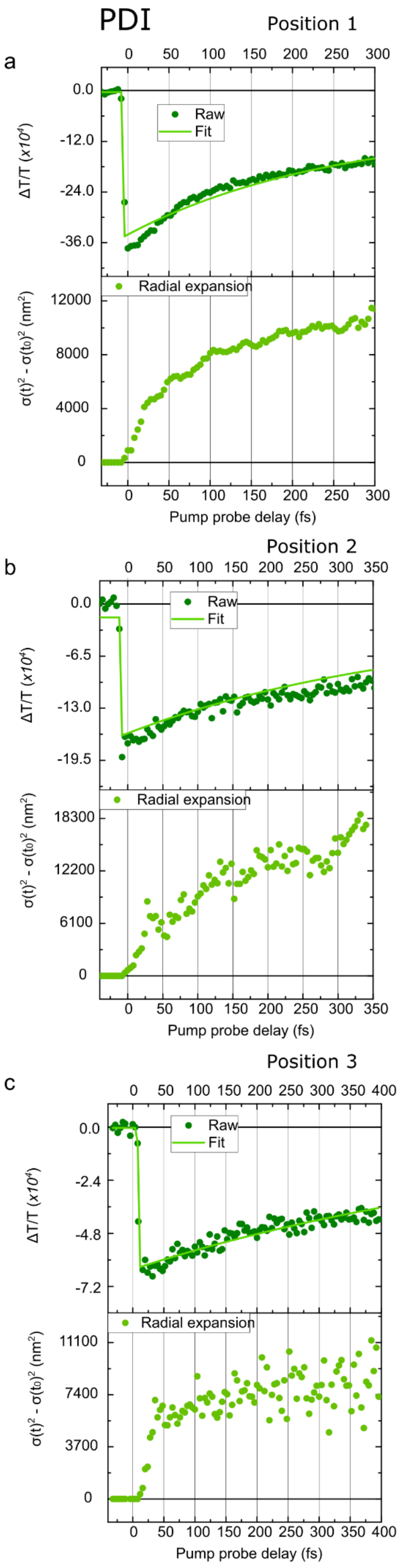
PDI
Position 1
a
Raw
Fit
ΔT/T (x10⁴)
0.0
-12.0
-24.0
-36.0
Radial expansion
σ(t)² - σ(t₀)² (nm²)
12000
8000
4000
0
0 50 100 150 200 250 300
Pump probe delay (fs)
Position 2
b
Raw
Fit
ΔT/T (x10⁴)
0.0
-6.5
-13.0
-19.5
Radial expansion
σ(t)² - σ(t₀)² (nm²)
18300
12200
6100
0
0 50 100 150 200 250 300 350
Pump probe delay (fs)
Position 3
c
Raw
Fit
ΔT/T (x10⁴)
0.0
-2.4
-4.8
-7.2
Radial expansion
σ(t)² - σ(t₀)² (nm²)
11100
7400
3700
0
0 50 100 150 200 250 300 350 400
Pump probe delay (fs)

**Figure SI48: a-c.** Kinetics (dots - raw data; solid lines - fit) and mean square displacement traces at three different spots in PIC. The $\sigma(t)^2 - \sigma(t_0)^2$ trace consistently shows a biphasic response and in this sample the kinetics are relatively consistent between sample locations.

## S6 Transport velocities

In addition to the diffusion constant, transport in PDA, PIC and PDI can be characterized by the transport velocity ( $v = \frac{\sqrt{\sigma(t)^2 - \sigma(t_o)^2}}{t}$), as displayed in Figure **SI49**. The following mean velocities are obtained from the data: PDA $\langle v_{R_1} \rangle$ = 4.8 × $10^6$ m $s^{-1}$, $\langle v_{R_2} \rangle$ = 0.28 × $10^6$ m $s^{-1}$; PDA $\langle v_{R_1} \rangle$ = 4.1 × $10^6$ m $s^{-1}$, $\langle v_{R_2} \rangle$ = 0.19 × $10^6$ m $s^{-1}$; PDA $\langle v_{R_1} \rangle$ = 3.4 × $10^6$ m $s^{-1}$, $\langle v_{R_2} \rangle$ = 0.09 × $10^6$ m $s^{-1}$. The spread in the velocities is estimated from the standard deviation μ as follows: PDA: $\mu_{R1}$ = 0.652 × $10^6$ m $s^{-1}$; PIC: $\mu_{R1}$ = 0.509 × $10^6$ m $s^{-1}$; PDI: $\mu_{R1}$ = 0.430 × $10^6$ m $s^{-1}$. In R2 it is as follows: PDA: $\mu_{R2}$ = 0.303 × $10^6$ m $s^{-1}$; PIC: $\mu_{R2}$ = 0.355 × $10^6$ m $s^{-1}$; PDI: $\mu_{R2}$ = 0.271 × $10^6$ m $s^{-1}$.

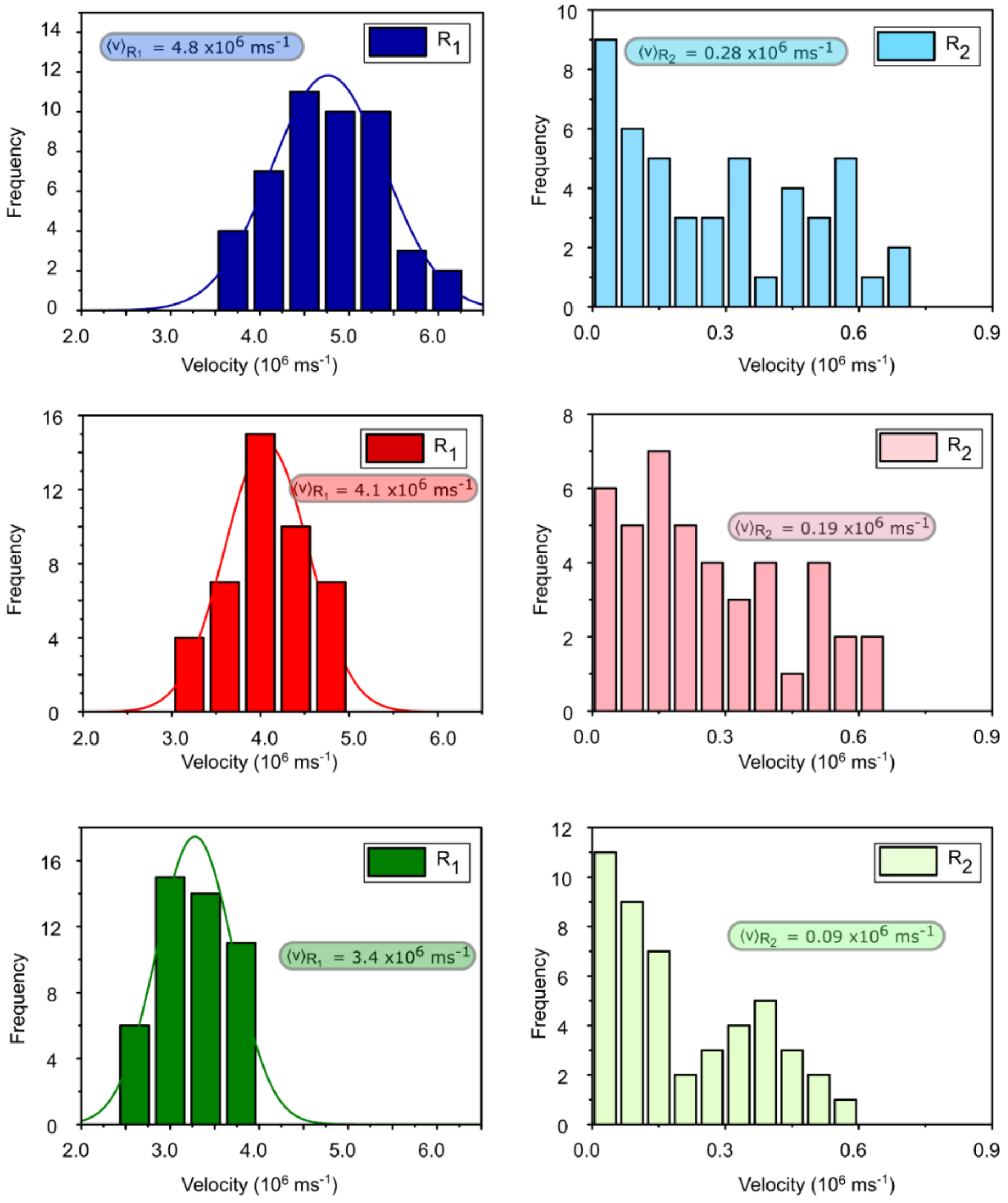

**Figure SI49:** Distribution of transport velocities in $R_1$ and $R_2$ for PDA (blue), PIC (red) and PDI (green). In $R_1$ the velocities follow a Gaussian distribution with a tight range whereas for $R_2$ the transport velocity is skewed towards lower values with a larger more random spread of values.

Pump fluence dependence

For all three materials the effect of varying the fluence of the pump pulse was investigated. The mean square displacement traces in Figure **SI50** show that the spatial dynamics of PDA, PIC and PDI have little dependence on the pump intensity. This is the case for both $R_1$ and $R_2$. In the former, this is expected because exciton-polaritons are formed from interaction of the optical dipoles with the vacuum electromagnetic field. Consequently, the polaritonic states are present and accessible even in the absence of excitation. In the low-excitation-density regime, these states and the transport they mediate should be independent of photon flux. In $R_2$ this is likely the case because we are in the low fluence regime of excitation where thermal and annihilation effects do not play a significant role. The independence of the diffusion coefficient in $R_2$ suggests that exciton-exciton annihilation effects play a negligible role in the observed behaviour. The differences in the absolute distance travelled are small do not follow a systematic trend and instigation of their origin is beyond the scope of the work.

Repeating the fluence series at 5 separate sample locations in addition to those detailed in the main text allows estimation of the fluence dependence of the mean diffusion coefficient. As shown in Figure **SI50**, there is only small variation in the values obtained, with $\langle D \rangle$ unvarying.

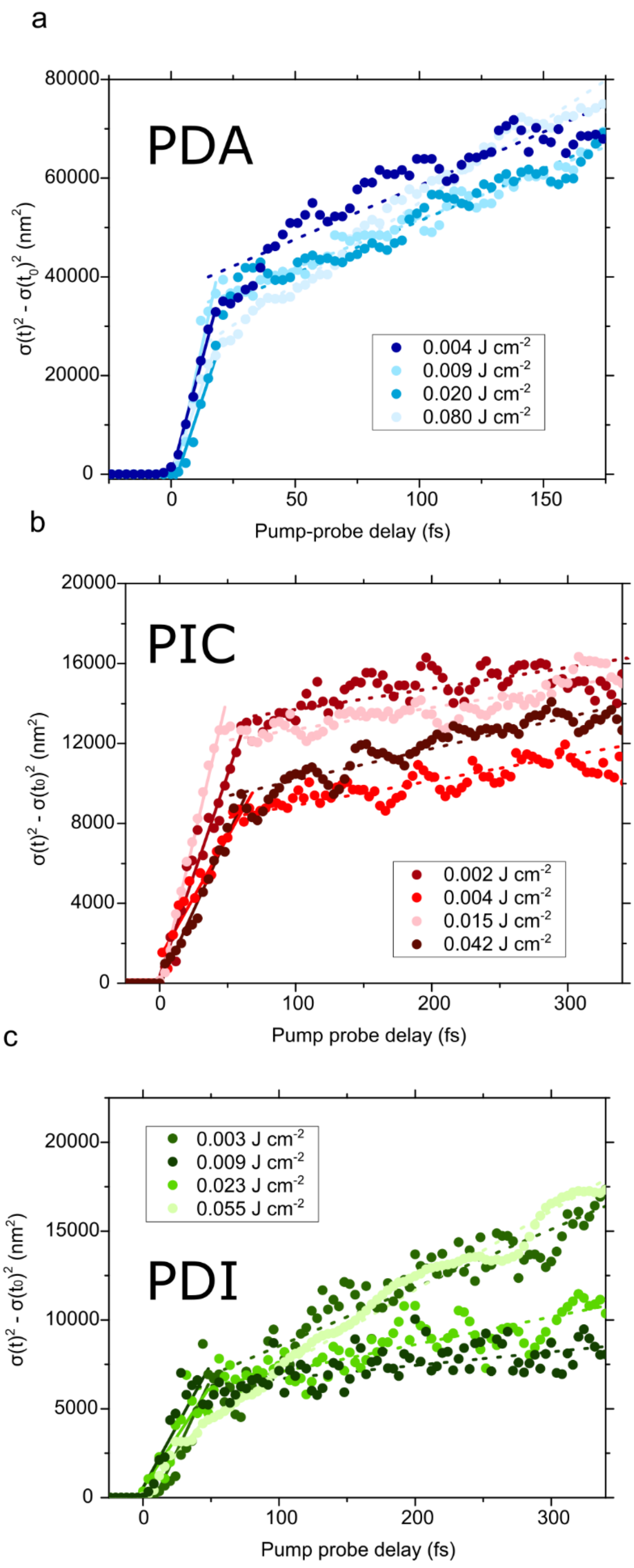

**Figure SI50:** Mean square displacement as a function of time at different pump fluences for **a.** PDA, **b.** PIC and **c.** PDI . The straight lines indicate a fit to the data in $R_1$ (solid) and $R_2$ (dashed). The absolute distance travelled varies to a small degree between the fluences but there appears to be no systematic trend and the diffusion coefficient (gradient of straight lines) is independent of the pump fluence.

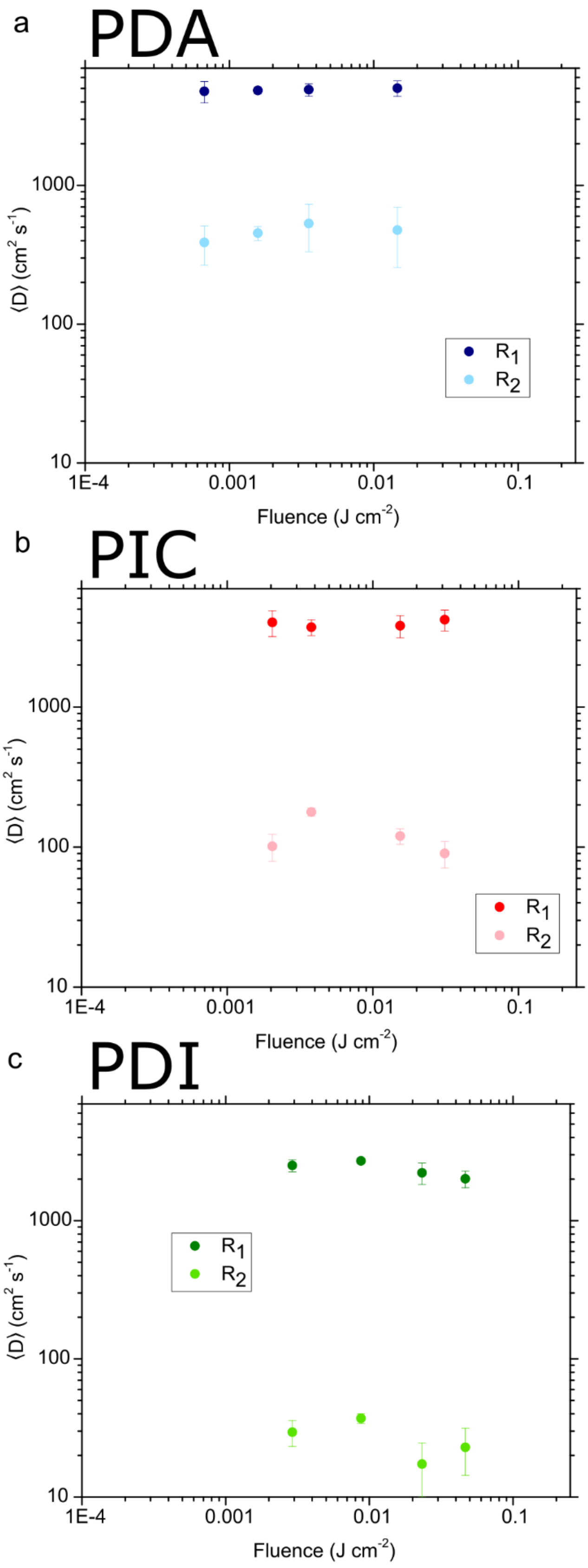

**Figure SI51: a-c.** Mean diffusion coefficient ($\langle D \rangle$) in $R_1$ (dark circles) and $R_2$ (light circles) regions at different laser fluences. At each fluence a minimum of at least 5 different mean square displacement traces are used to obtain the average diffusion coefficient. In general $\langle D \rangle$ appears to be independent of fluence.

**S7: Spread in D**

In $R_1$ the range of log of the diffusion coefficients shows a normal distribution (Figure 2d-f of main text) hence we estimate the spread in values from the standard deviation (of Log(D)), obtaining values as follows. PDA: $\mu_{R1} = 0.334$; PIC: $\mu_{R1} = 0.316$; PDI: $\mu_{R1} = 0.258$.

In $R_2$ due the larger and more random spread we simply calculate the standard deviation in the D values. PDA: $\mu_{R2} = 0.532$; PIC: $\mu_{R2} = 0.524$; PDI: $\mu_{R2} = 0.398$.

## S8: Theoretical modelling of exciton diffusion in PDI

*Ab initio* calculations of lowest electronic transitions and transition densities of PDI molecules

Structure of a single PDI molecule, Figure **SI52a**, was optimized using DFT as implemented in Turbomole 6.0 (*20*). For the structure optimization, we used the B3LYP hybrid functional (*21*) and the triple-ζ def2-TZVP basis set (*22*) that provide a reasonable balance between the precision of the results and computation time. Five lowest singlet electronic excitations of PDI molecule were computed using TDDFT with the same functional and the basis set. The lowest electronic excitation is optically active with the corresponding transition dipole about $\mu = 9.2$ Debye and the transition frequency $\hbar\omega = 2.37\text{eV}$. The computed frequency agrees well with the frequency measured experimentally in a solution of PDI molecules, see Figure **SI52b-c**. This transition is dominated by the HOMO → LUMO. The next four electronic transitions are separated from the lowest one by the gap of the order of 0.8 eV and are optically dark, see dashed lines in Figure **SI52b**.

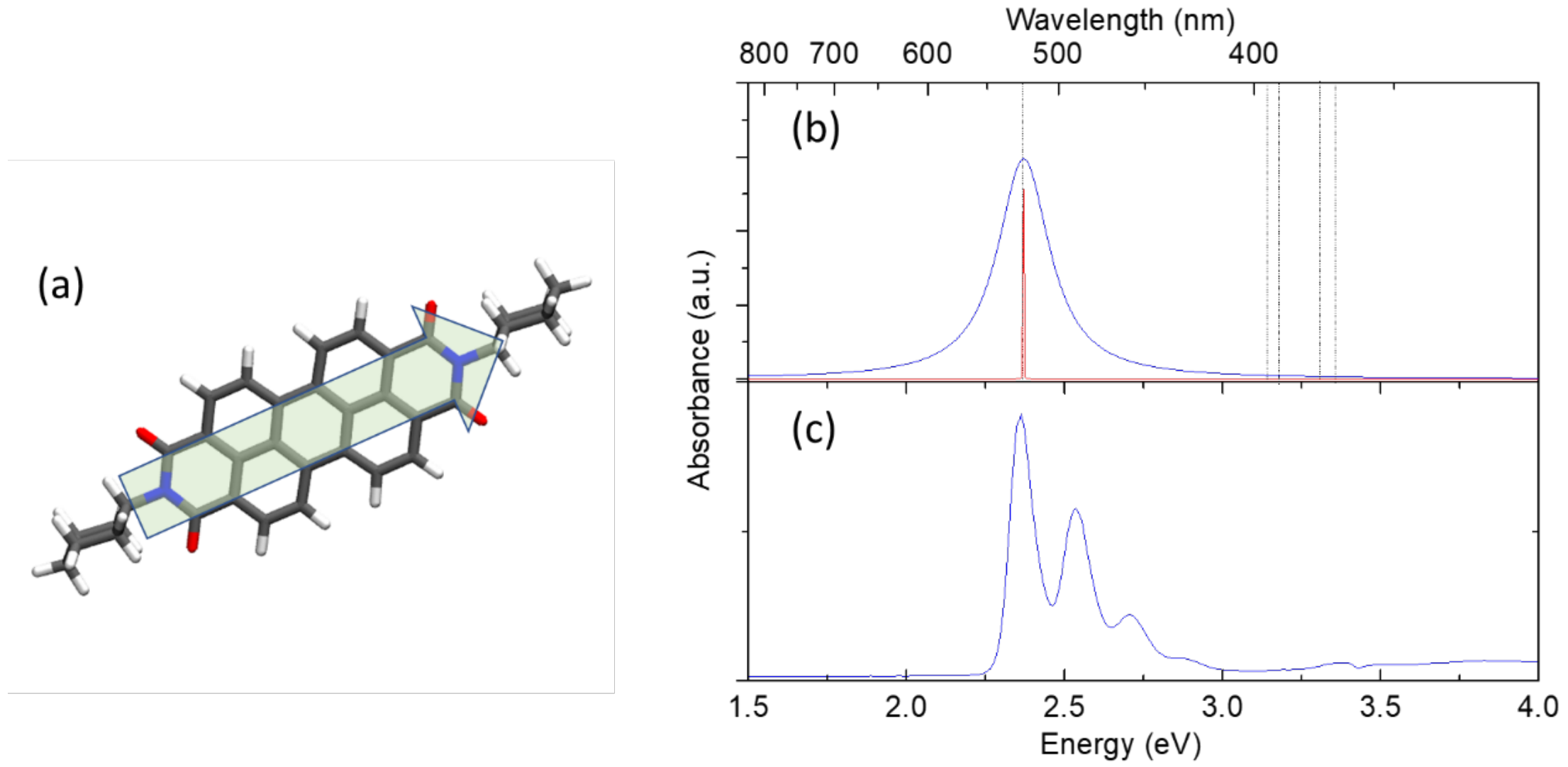

**Figure SI52: a.** Structure of PDI molecule and the orientation of the transition dipole associated with the lowest electronic excitation. **b.** Computed electronic excitation spectrum of PDI molecule. Frequencies of the lowest five transitions are marked by dashed lines. Only the lowest electronic transition is optically active. The red line corresponds to the computed frequency and the blue line is for the Lorentz-broadened transition with $\Gamma = 50$ meV. **c.** Measured absorption spectrum of PDI molecules in solution showing vibronic progression.

Intermolecular couplings and exciton transport in PDI nanobelts

The transport of singlet excitons in PDI nanocrystals is controlled by the Förster coupling between the intramolecular electronic transitions (*23*). We compute the Förster coupling between PDI molecules using the electron transition densities obtained from DFT calculations. Figure **SI53a** shows the coupling strength in a dimer of PDI molecules as a function of the relative angle between two molecules displaced axially by the Van der Waals distance, which is equal to 3.62 A. Figure **SI53b** shows the distance dependence of the interaction while the relative angle is fixed at $\varphi = 30°$. In order to efficiently compute excitonic Hamiltonian of nanocrystals, we fit the a extended dipole model (*24*) to the intermolecular coupling dependencies obtained using transition densities. The resulting parameters of extended dipoles are $q = 0.25e$ and $l = 7.6$ Å.

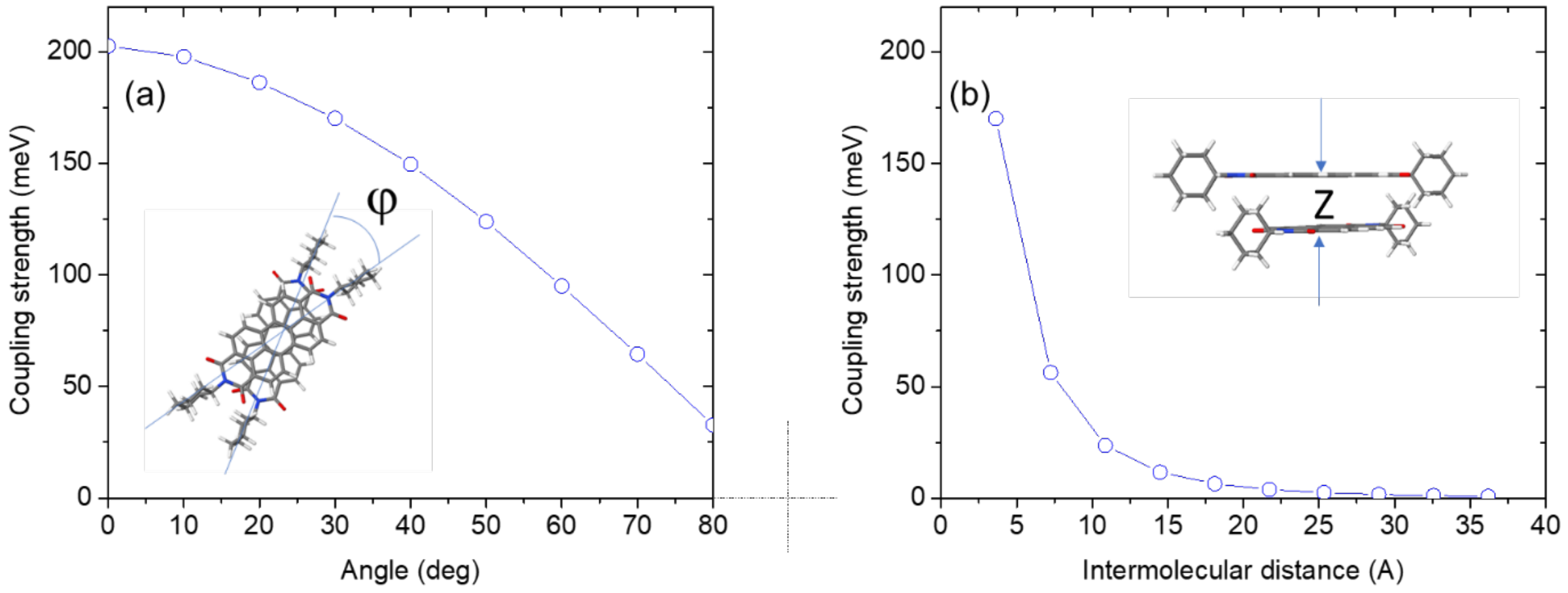

**Figure SI53:** Computed excitonic coupling strength in a dimer of PDI molecules as a function of the relative angle **a** and the intermolecular distance **b**.

To model absorption spectra and exciton dynamics in PDI nanocrystals, we built three slabs that represent 3-dimensional, 2-dimensional, and 1-dimensional domains. The slabs are composed correspondingly of 3×3×17 lattice cells (1224 molecules), 13×13×1 lattice cells (1352 molecules), and 1×1×151 lattice cells (1208 molecules). Then, the excitonic Hamiltonians were constructed with the values of extended transition dipoles obtained from the fitting procedure and the electronic excitation spectra were computed. The theoretical spectra of the slabs, Figure **S54a-c**, are consistent with the measured absorption spectrum of PDI nanobelts, Figure **SI54d**. Firstly, in all computed spectra, the majority of the bright excitonic transitions are shifted to shorter wavelengths (blue part of the spectrum) as compared to the transition in monomers. Secondly, the ranges of the shift in the theory and experiment are comparable. However, in our computational model the values of the shifts depend sufficiently on the shape of the slab. This is due to the saturation of the Förster interaction in stacks of PDI molecules. Finally, all spectra show a feature red-shifted from the monomer peak by about 100-150 meV. It is likely that the experimental spectrum shows responses from multiple domains of different shapes. We conclude that the derived Hamiltonian can be used for further modelling of exciton transport.

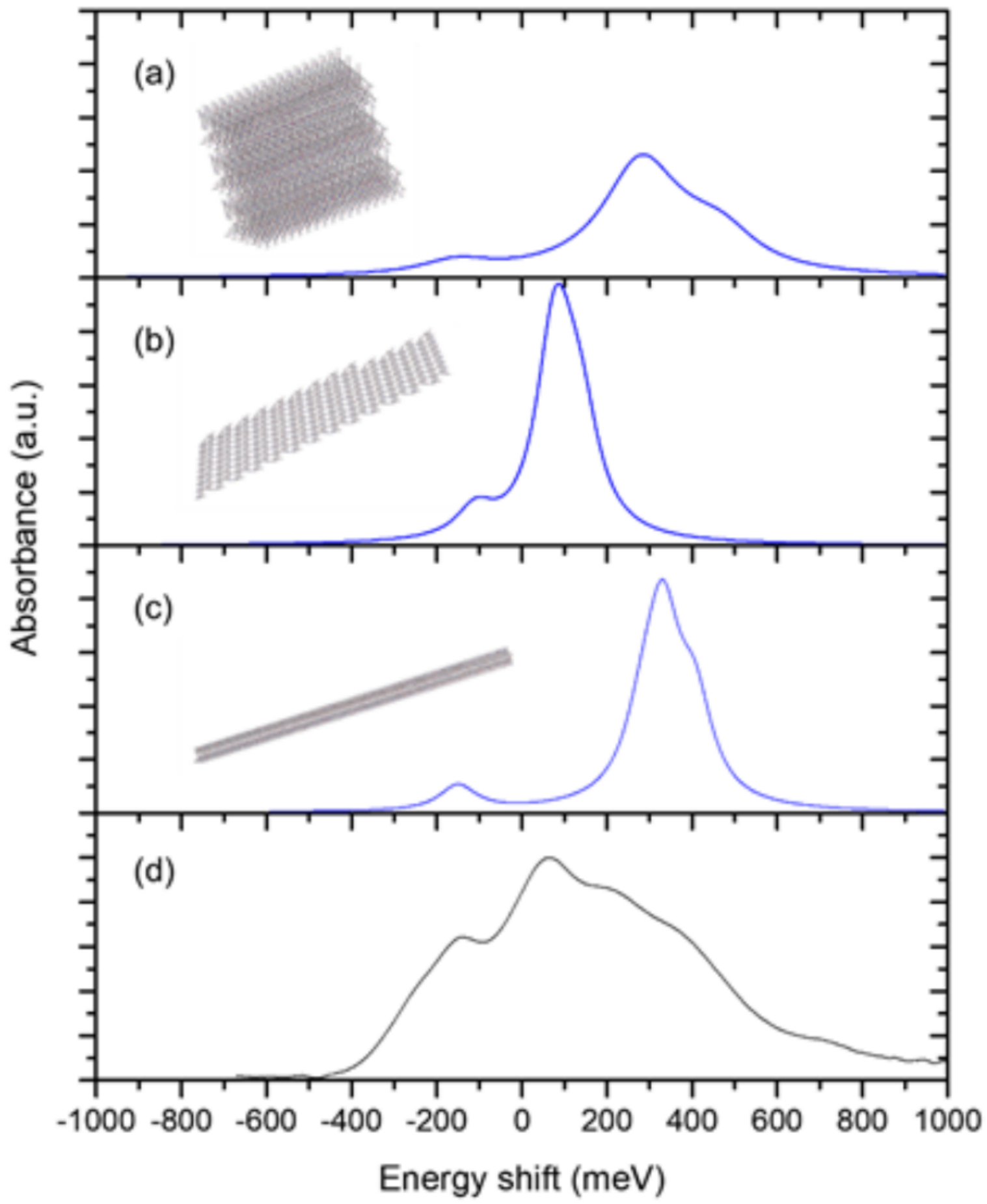

**Figure SI54: a-c**. Computed spectra of 3D, 2D, and 1D PDI slabs. The relative intensities are normalized to the number of molecules. The transition lines are broadened using a Lorentzian function with $\Gamma = 50$ meV. **d.** Measured absorption spectrum of PDI nanobelts.

The diffusion coefficient for singlet excitons in PDI crystals has been estimated using the model developed by Haken, Reineker, and Strobl (*25*, *26*), where the exciton scattering is modelled using a phenomenological source of white noise. This model provides realistic estimates of transport properties, both ballistic and diffusive, at high temperatures keeping the number of model parameters minimal. Specifically, in the model we include a single phenomenological parameter, Γ, which characterizes the exciton dephasing rate due to the fluctuation of intramolecular electronic transition frequencies. For perfect molecular crystals, the second moment of the exciton wave function can be written as (*27*):

$$M_2(t) = \frac{\sum_{m>0}(x_m V_m)^2}{\Gamma}\left\{\frac{e^{-2\Gamma t/\hbar}-1}{\Gamma} + \frac{2t}{\hbar}\right\},$$

where $V_m$ is the strength of excitonic couplings between two molecules, $x_m$ is the intermolecular distance, and the sum is taken over only positive direction. At short times $t \ll \hbar/\Gamma$ the exciton transport is ballistic with the velocity $v \propto \sum_{m>0} x_m V_m/\hbar$, and it becomes diffusive at longer time with the diffusion coefficient $D = \sum_{m>0}(x_m V_m)^2/(\hbar\Gamma)$. In the derived model, the intermolecular couplings along the $\pi - \pi$ stacked chains are several times stronger than the interactions between molecules belonging to different chains. Therefore, for the sake of simplicity we estimate 1D transport along single chains.

The computed ballistic velocity of excitons $v \propto 10$ m/sec, which is about 30 times smaller than the value estimated from the experiments in the $R_1$ region. In order to have the diffusion coefficient the same as the experimentally measured in $R_2$ region, the dephasing rate should be about $\Gamma = 5$ meV. While this value is realistic, it disagrees with the measured line widths that are of the order of $\Gamma = 50$ meV. One potential reason for this discrepancy is that in the $R_2$ region excitons still have some small admixture of a photon mode. In this case, the diffusion can be fast even for large values of $\Gamma$. Another potential reason is that a large fraction of the line width in the measured absorption spectra is due to the inhomogeneous broadening.

Finally, we compute exciton dynamics in the three slabs shown in Figure **SI54a-c**. In the model we assume the dephasing rate $\Gamma = 5$ meV. Additionally, we include a disorder in molecular transition frequencies with a Gaussian distribution of $\Delta\omega = 50$ meV width. Figure **SI55** shows the second moments of the exciton wave function as a function of time for three slabs. Each simulation is averaged over 100 different realizations of structural disorder. At time $t = 0$ the excitation was localized on a single molecule in the middle of the slab. The initial dynamics of the exciton wave packet is coherent with a ballistic propagation that is reflected as a quadratic time dependence of the second moment. At longer times, the exciton dynamics becomes diffusive with a linear time dependence. The slab boundaries limit the transport of excitons which can be seen for the all three slabs. Only 1D slab is sufficiently long that the boundaries do not affect the exciton dynamics on the time scales of 100 fs. In this case the diffusion coefficient along the stack of PDI molecules, z-axis, is equal to $D = 370$ Å$^2$/fs, see Figure **SI55c**, that is comparable to the value obtained in the experiments.

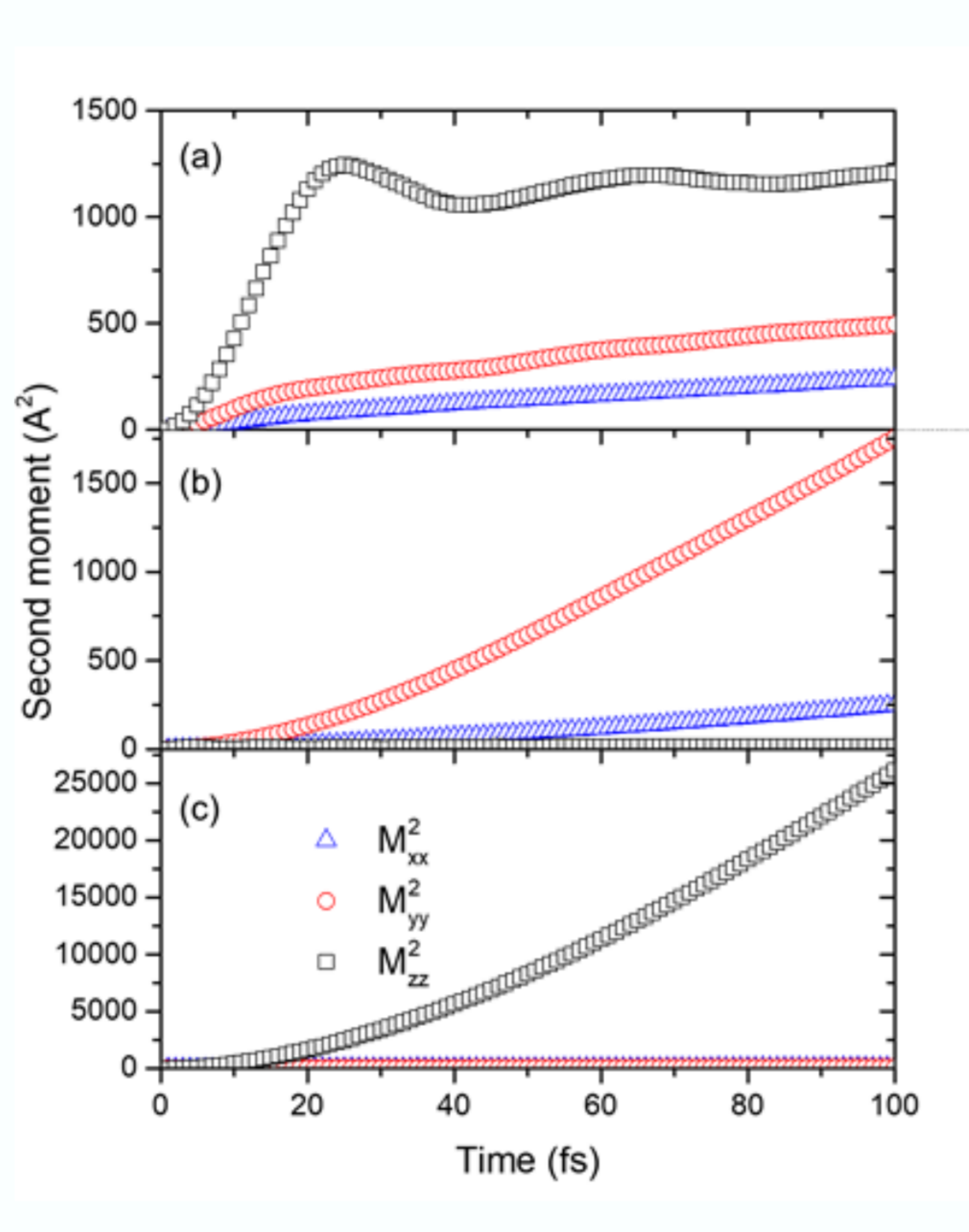

**Figure SI55: a-c.** Computed dynamics of excitons in **a.** 3D, **b.** 2D, and **c.** 1D slabs of PDI molecules. The plots show second moments of the exciton wave function (z-axis is along the stacks of PDI molecules) as a function of time for the first 100 femtoseconds. The dephasing rate is $\Gamma = 5$ meV.

## S9: Specular reflection

### Specular reflection measurements on PIC

Specular reflection measurements were performed on ~70 randomly selected locations of a PIC film. The reflection spectra obtained fall into two categories: those that do not show a splitting and those that do (Figure **SI56**). In the former case the spectra appear as approximately the inverse of the absorption with a sharp single excitonic dip at 583 nm (2.13 eV) and as well as a broad tail out to 550 nm, which we suggest to be associated with residual monomers. In the case of regions of the sample that do show a splitting, the J-band clearly shows two pronounced dips in reflectivity at ~579 nm and ~587 nm, whose relative intensities vary based on the precise sample location. In all cases, the separation between the peaks (splitting) is between 39 – 40 meV (Figure **SI57**).

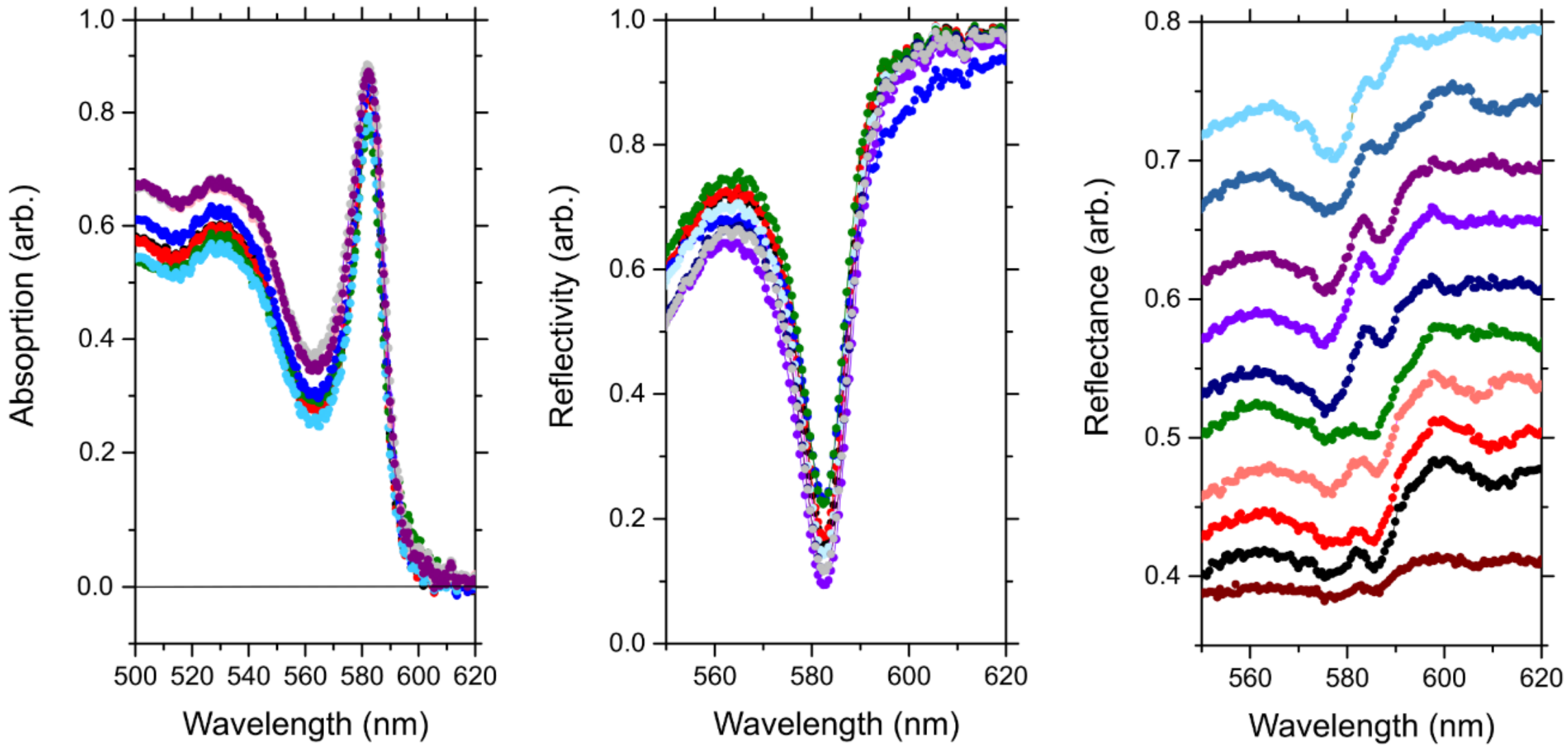

**Figure SI56: a.** Absorption spectra of PIC film in eight different locations. **b.** Reflection spectra in PIC film in regions that do not show a splitting in the excitonic peak. **c.** Reflectivity spectra of PIC film in regions of the sample showing a line splitting of ~ 40 meV. The off-set between spectra has been added to ease viewing.

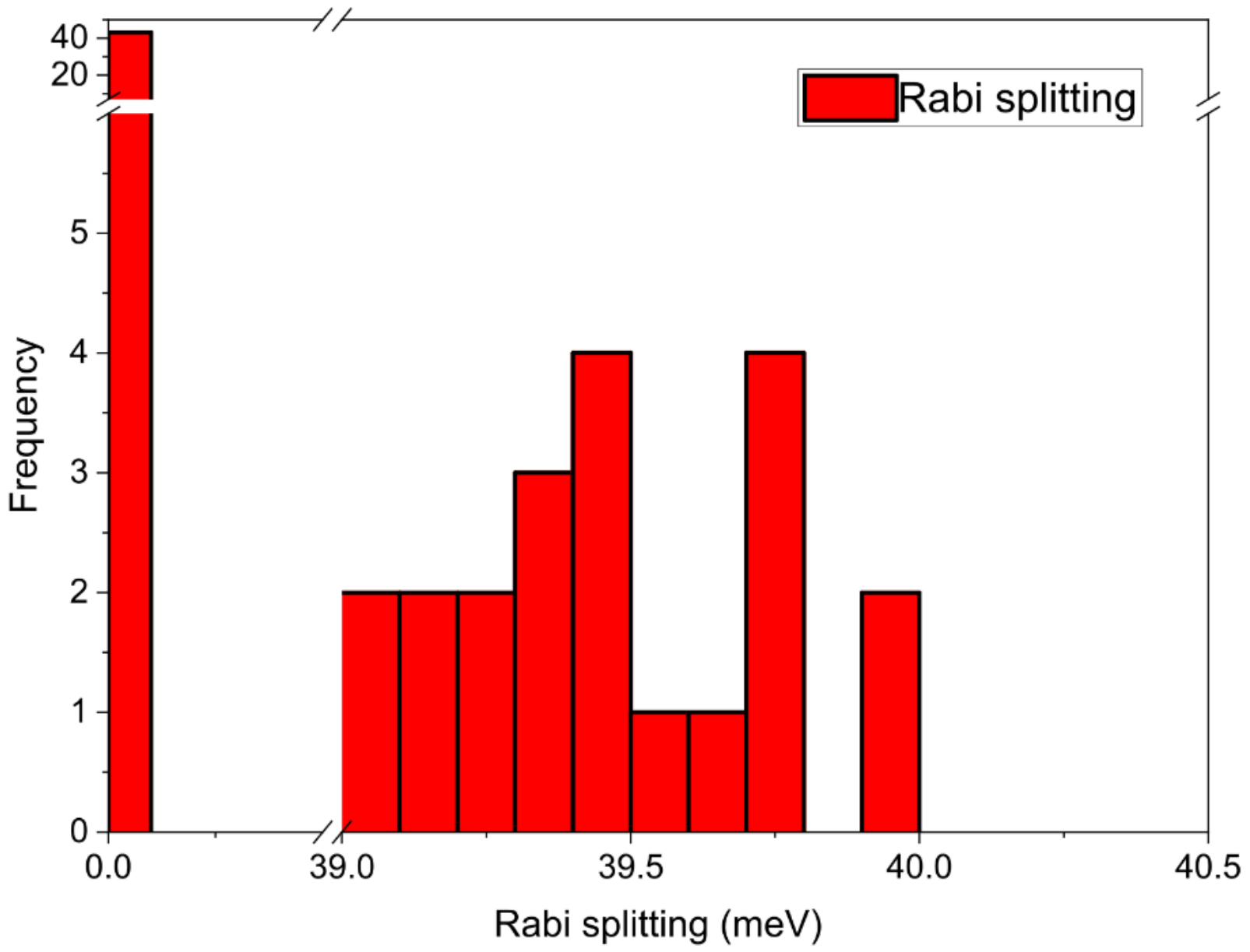

**Figure SI57:** Distribution of Rabi splittings measured in PIC thin films over ~70 spots. In a large number of locations (~40) no splitting is observed. In some places however, a splitting between 39 – 40 meV is observed.

Specular reflection measurements on PDA

Specular reflection measurements were performed on ~70 randomly selected locations of a PDA crystal. In all cases, the incoming light was polarized parallel to the long axis of the polymer chains. The reflection spectra obtained fall into two categories those that do not show a splitting and those that do (Figure **SI58**). In the former case the spectra appear as approximately the inverse of the absorption with a sharp zero-phonon and first vibronic peaks at 630 nm (1.97 eV) and 581 nm (2.13 eV) respectively. In the case of regions that do showing a splitting the spectral shape is similar, but the peaks associated with the zero-phonon line and first vibronic transition are clearly split. For the former the splitting is higher around ~18 meV, whereas in the latter case the splitting is slightly smaller at ~17 meV, in-line with the lower absorbance. The values are obtained from the peak to peak separation, but we do not stress their relative magnitude too strongly due the small difference which is also close to the noise level of our experiment.

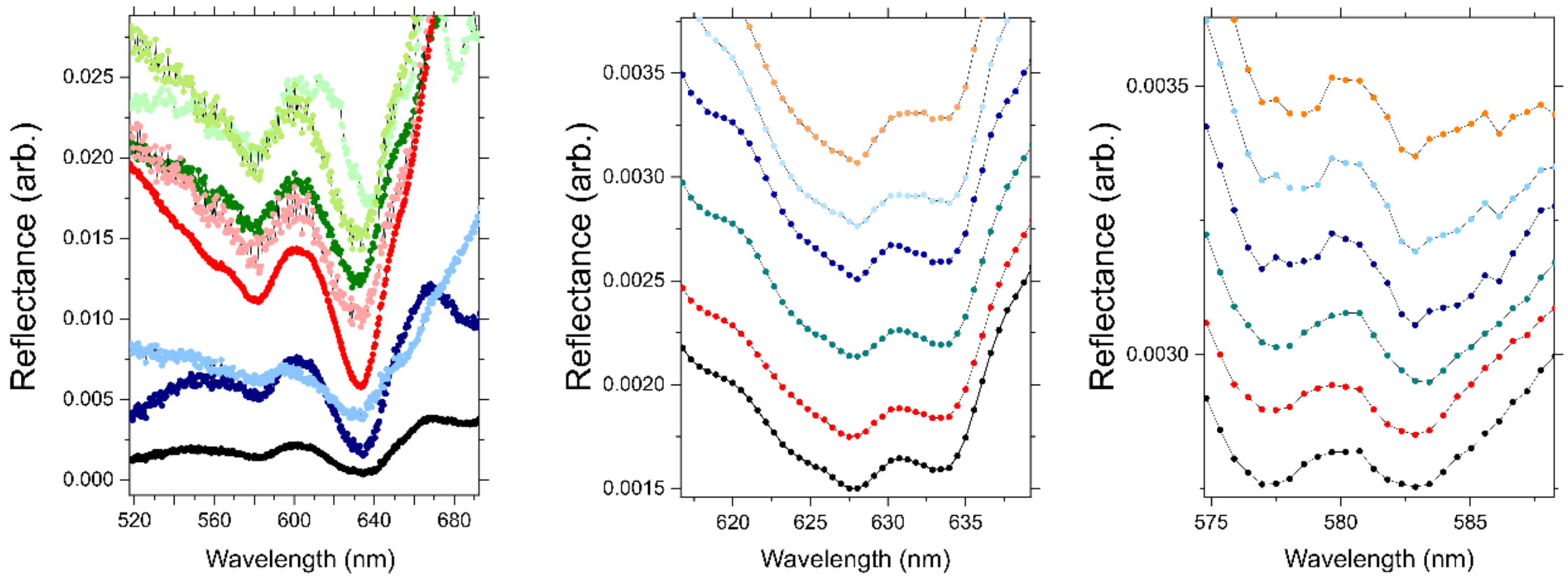

**Figure SI58: a.** Reflectance spectra from a region of PDA crystal that does not show a splitting. The curves are offset from one another for ease of viewing hence the absolute intensity is arbitrary. **b.** **S**plitting around the zero-phonon peak for several different sample locations in PDA, the lower polariton branch is typically more intense than that of the upper polariton. **c.** Splitting around first vibrational peak in PDA crystal. We note that in **b** and **c** the scale is also arbitrary with curves offset from one another.

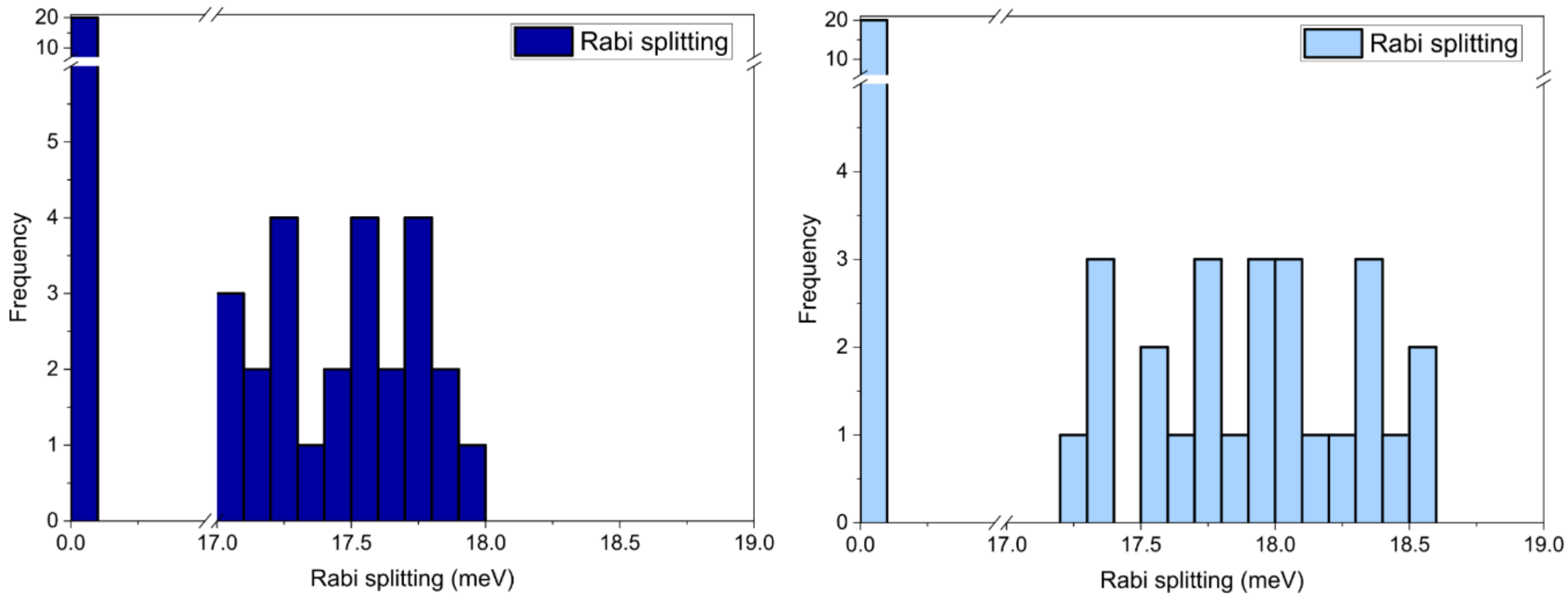

**Figure SI59:** Distribution of Rabi splittings measured in PDA crystalline films measured over ~70 spots. As for PIC a large number of sample locations do not show any splitting. However, in some cases the zero-phonon and first vibronic peaks at 630 nm (1.965 eV) and 581 nm (2.134 eV) are split. The splitting around the zero phonon peak (17.2 – 18.5 meV) is empirically larger than at the first vibronic transition where it is 17 -18 meV.

Absorption and Reflection spectra of PDI

Figure **SI60** shows the absorption spectrum of a dilute solution of the PDI monomer, an evaporated film and the PDI nanobelts. The absorption spectrum blue shifts to higher energies between the solution and evaporated films or nanobelts, with a broadening of the vibronic peaks. Such behaviour is characteristic of the formation of an H-aggregate.

Microscopic specular reflectance measurements were additionally performed on PDI nanobelts. The reflectance spectrum is highly sensitive to polarization as can be seen from the red and black curves in **SI60b** where the incoming light is parallel and perpendicular to the nanowire direction. The multitude of inhomogeneously broadened vibronic peaks precludes the observation of any Rabi splitting as in PDA or PIC. Under unpolarised light the reflectance spectrum (Figure **SI45c**) appears approximately as the inverse of the absorption.

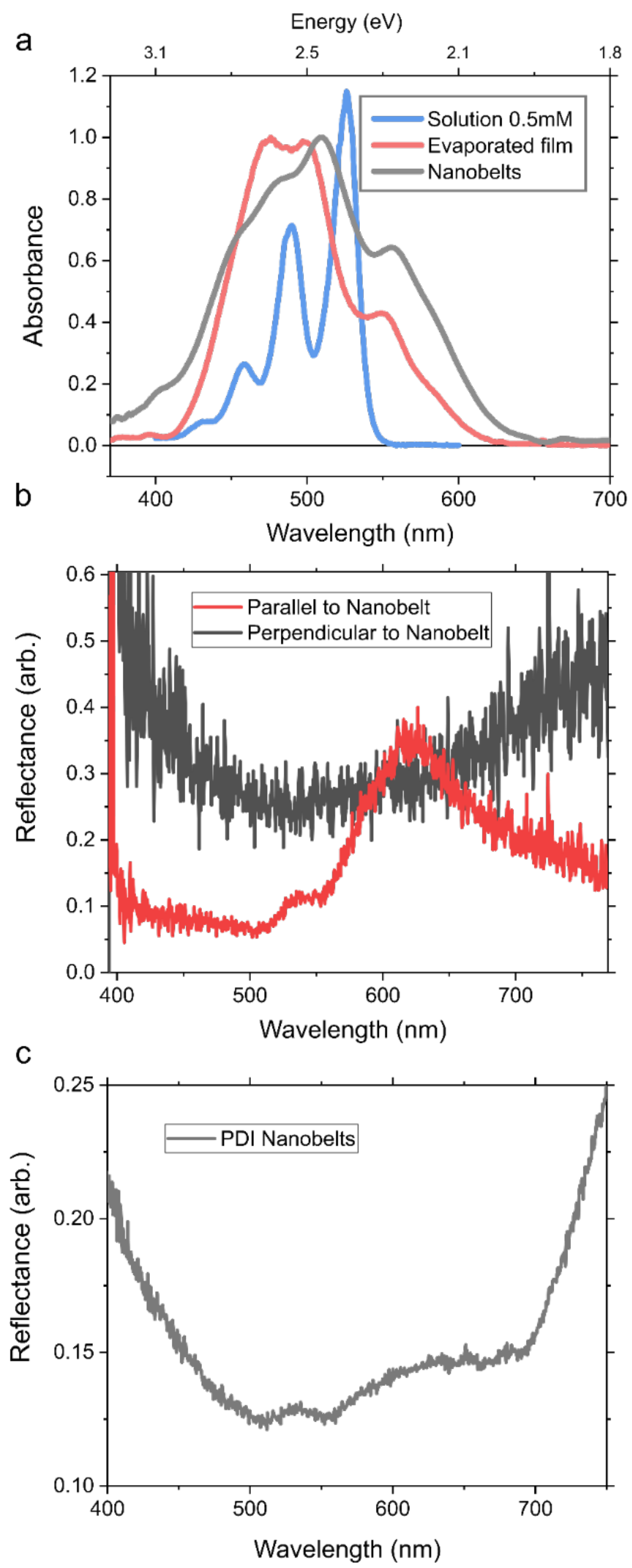

**Figure SI60: a.** Absorption spectrum of PDI monomer molecules diluted in chloroform (blue spectrum; 0.5 mM; 1 mm path length cuvette Hellma), a clear progression of vibronic peaks can be observed. The red and black curves show the absorption spectrum of an evaporated thin film of PDI and PDI nanobelts. The red shift is indicative of H-aggregation. **b.** Reflection spectrum of PDI nanobelts with light polarized parallel (red) and perpendicular to the long axis of the nanobelts. **c.** Reflectance spectrum of PDI nanowires with unpolarised light, the broad featureless spectrum precludes the observation of any Rabi splitting.

## S10: Transfer Matrix Simulations

The optical properties of thin films of PIC and PDA were explored in detail using transfer matrix simulations as previously reported (*28*), parameterised from the measured absorption of the organic films. The absorption spectra were fitted to a series of Lorentzian peaks, with the oscillator strength of each individually tuned to correctly reproduce the extinction of the measured film. Because the main absorption band in these materials is particularly strong and narrow, we found that we could satisfactorily describe the optical properties with consideration of only that band. We simulated a uniform slab of this absorber, on top of a 200 nm layer of $SiO_2$ and covered by 500 nm of air, to approximate the experimental situation (using tabulated values for the $SiO_2$ index of refraction, 1.0008 for air). The optical properties of the entire system (transmission, absorption, reflectivity) were calculated as a function of angle. Our aim was to determine whether and in what regimes these absorbers could reproduce the splitting in reflectivity noted in main-text Figure 3, for which purpose we focus only on the modelled normal-incidence reflectivity. On the length scales relevant to our experiments, it is clear that the PIC and PDA films are highly inhomogeneous (see *e.g.* Figure **SI20**). The bulk thin-film absorption measurement reflects the contribution of highly concentrated bundles of active material, within a much more weakly absorbing matrix. In order to describe the effects of this, we have incorporated linear scaling factor to the oscillator strength in our model (effectively equivalent to scaling the concentration). Setting the scaling factor to zero (Figure **SI61a** below) allows us to directly visualise the Fresnel modes of the material slab in the absence of any light-matter interactions.

As summarised in Figure **SI61**, we find that when PIC is treated as uniformly distributed (scaling factor = 1) or weakly bundled (scaling factor = 2) there is no reflectivity splitting at the main absorption peak (dashed line) for any slab thickness. However, on increasing the bundling (scaling factor = 5) we find a slight splitting for slabs of 2.3 and 2.5 μm. Further increasing the scaling factor results in the splitting becoming evident for still thinner slabs. The onset of the splitting is a product of both 'concentration' (scaling factor) and thickness. For a 1.7 μm film (Figure **SI61g**) we require a 15x local variation in absorption to account for our observed splitting, while at 2.3 μm thickness (Figure **SI61h**) even a 5x factor results in detectable splitting. To confirm the origin of this splitting as a polaritonic (*i.e.* light-matter coupling) effect, we have extracted the separation between peaks for a 2.3 μm film (arrows in panel h) as a function of scaling factor. Figure **SI61i** shows that this splitting closely follows the square root of the scaling factor, equivalent to the sqrt(N) dependence typical of the light-matter coupling regime. We thus consider these peaks to be a signature of bulk polaritons(*29–31*), formed as a consequence of the extremely high oscillator strength in these aggregates.

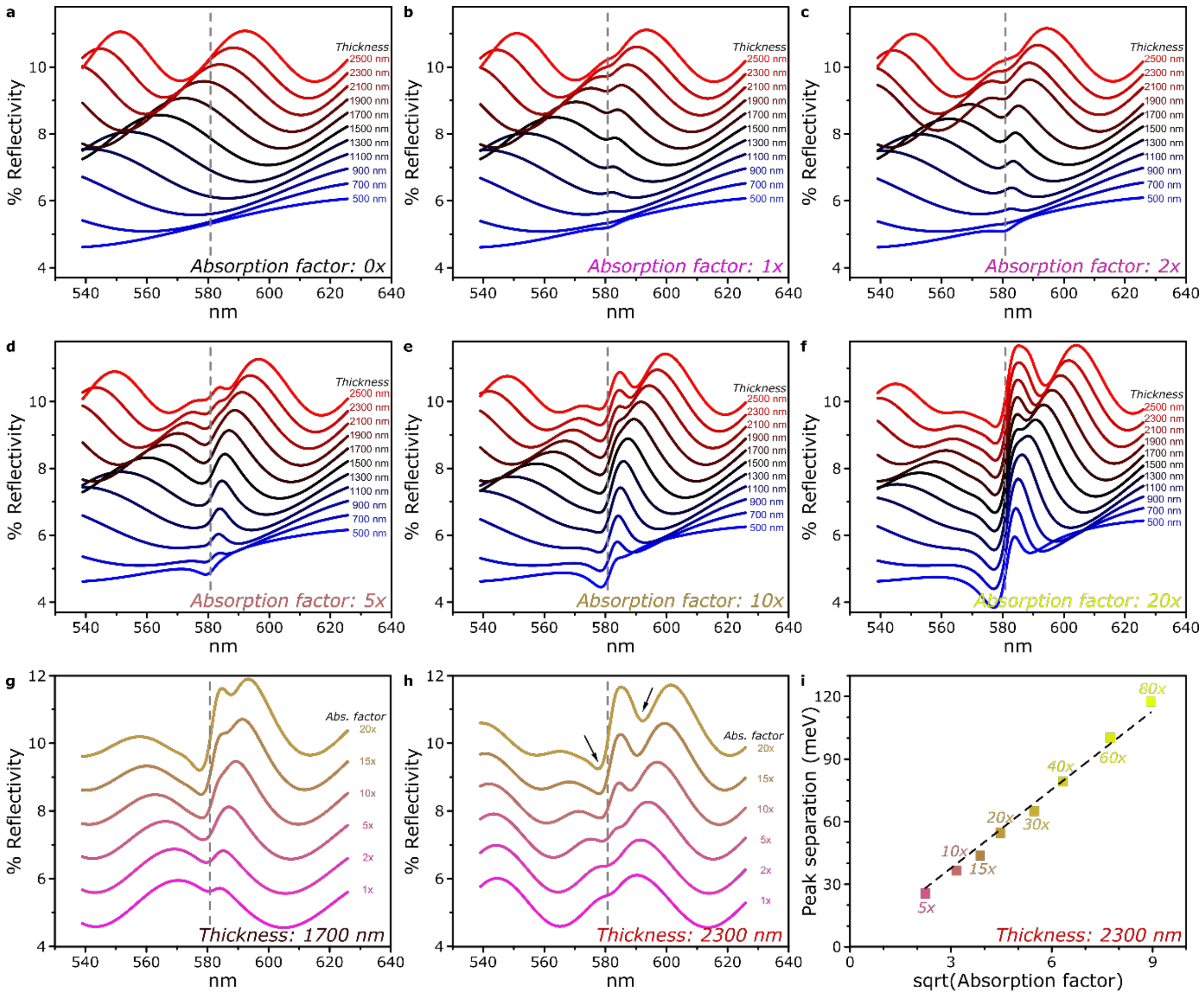

**Figure SI61:** Transfer-matrix simulations of normal-incidence reflectivity for PIC film. **a** Thickness-dependent Fresnel modes are evident when the dye oscillator strength is set to zero. **b,c** Weak absorption/bundling results in a slight dip at the exciton energy (dashed line), but only at a single transition. **d,e,f** At still higher oscillator strength, for sufficiently thick films, the dip in reflectivity splits into two peaks on either side of the exciton energy, a sign of exciton-polariton formation. **g,h** . Reflectivity at fixed film thickness as a function of scaling factor, showing the onset of splitting is both thickness- and absorption-dependent. **i.** Degree of peak separation, plotted versus the square root of the absorption scaling factor (equivalent to concentration). The linearity (dashed line) confirms that the origin of these peaks is polaritonic, *i.e.* light-matter interactions.

Our simulations do not depend on the particular chemical identity of the material, and we thus expect the result from PIC to be generalizable: for a given transition linewidth, there should be some region in the thickness/scaling factor space where a splitting is observable (though whether that regime is practically achievable depends very much on material). Thus we have obtained comparable splitting using our model for the strongest absorption peak of PDA. However, to reproduce the small splitting suggested in the main text, we had to consider a region of reduced oscillator strength and narrowed transition linewidth (Figure **SI62**), suggesting that our measurement might report on a subset of particularly well-ordered regions within the bulk.

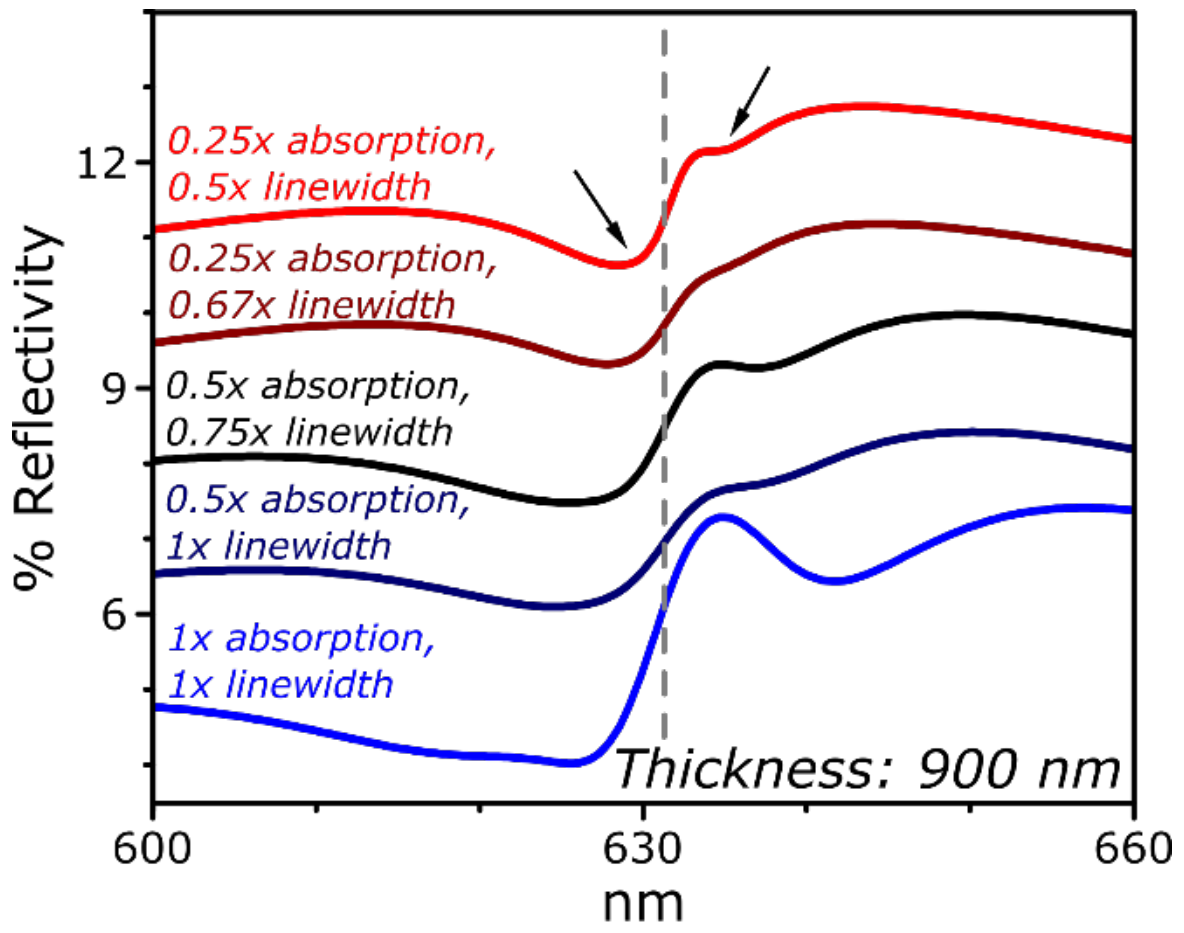

**Figure SI62:** Transfer-matrix simulations of normal-incidence reflectivity for PDA thin film. Only for a significant reduction in the oscillator strength and linewidth can we reproduce a narrow splitting (arrows) comparable to that tentatively assigned in the main text.

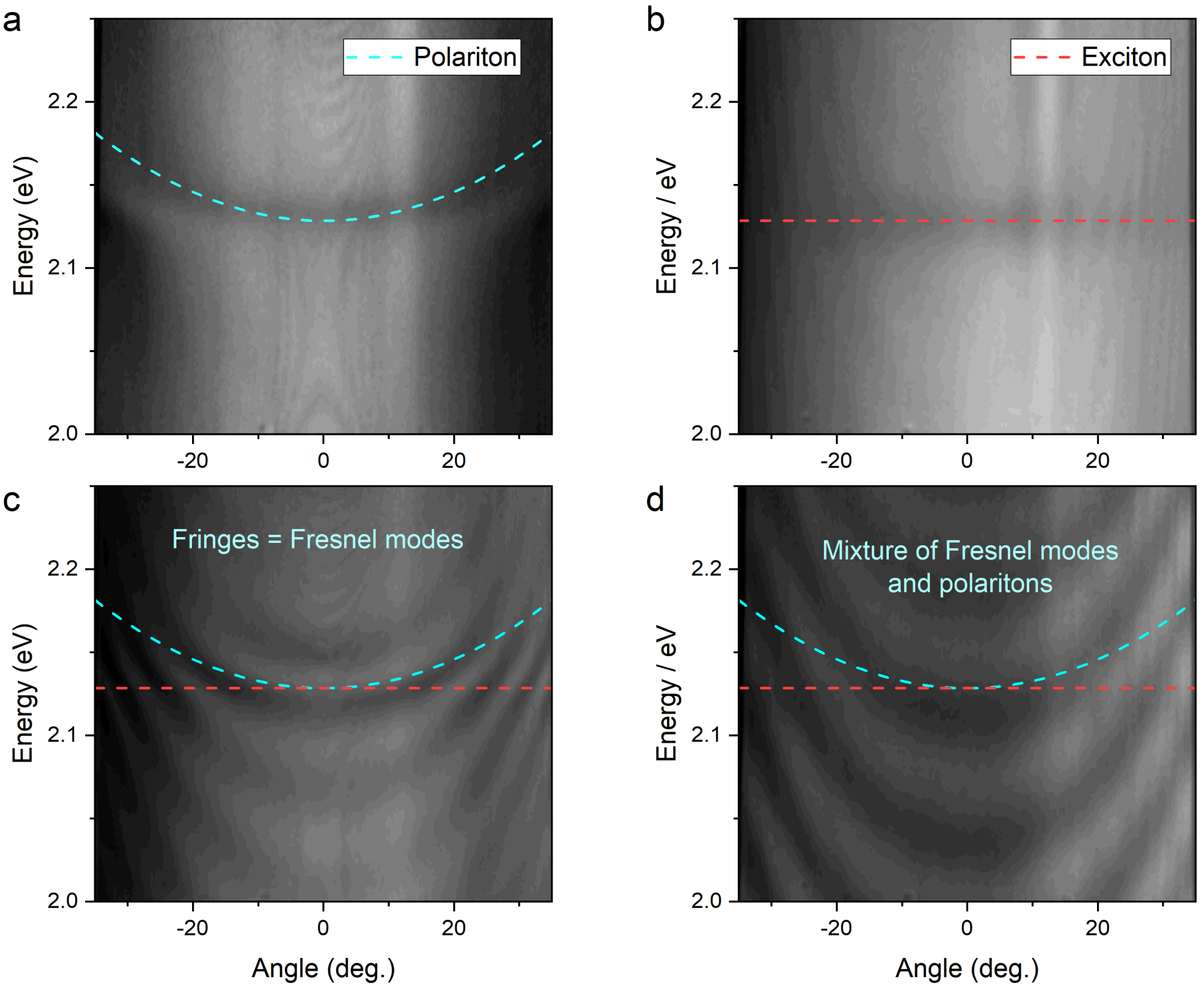

**Figure SI63.** Angle-dependent white-light reflectivity of PIC film, measured in k-space using a high-NA lens. **a.** In isolated areas of the film we primarily resolve a single clean dispersion which closely approaches the bare exciton energy (2.13 eV) at 0 degrees. Very faint signatures of underlying Fabry-Perot modes with steeper dispersion can be identified at high angles and/or low energies. **b** On other areas of the same sample where no optical modes can be supported, we detect the dispersionless signature of the exciton. **c** In other regions of the film we primarily detect a ladder of intrinsic Fabry-

Perot modes with relatively steep dispersion. These cross directly over the exciton absorption energy. **d.** Complex mixtures of these features can also be detected.

To probe the nature of the splitting of the exciton transition identified with normal-incidence reflectivity measurements, we investigated the angular dependence of the reflectivity. These measurements were performed on PIC films, where the splitting is most pronounced, using the k-space setup described in Methods. Spatial filtering was used to select areas ranging from 10-50 µm, and thus the signal is highly sensitive to the inhomogeneity of the film. Accordingly, we observe a range of different behaviours as we scan across the film surface. In some locations (Figure **SI63a**) we recover a single shallow dispersion, positioned at slightly higher energy than the exciton transition (panel b). This dispersion never crosses the energy of the exciton transition, and from the analysis below we can assign it to the upper polariton. In many locations, where the oscillator strength is too low to support polaritons, we instead detect a ladder of photonic modes. These have distinctly steeper dispersion than the band in panel a, confirming they have a different physical origin.

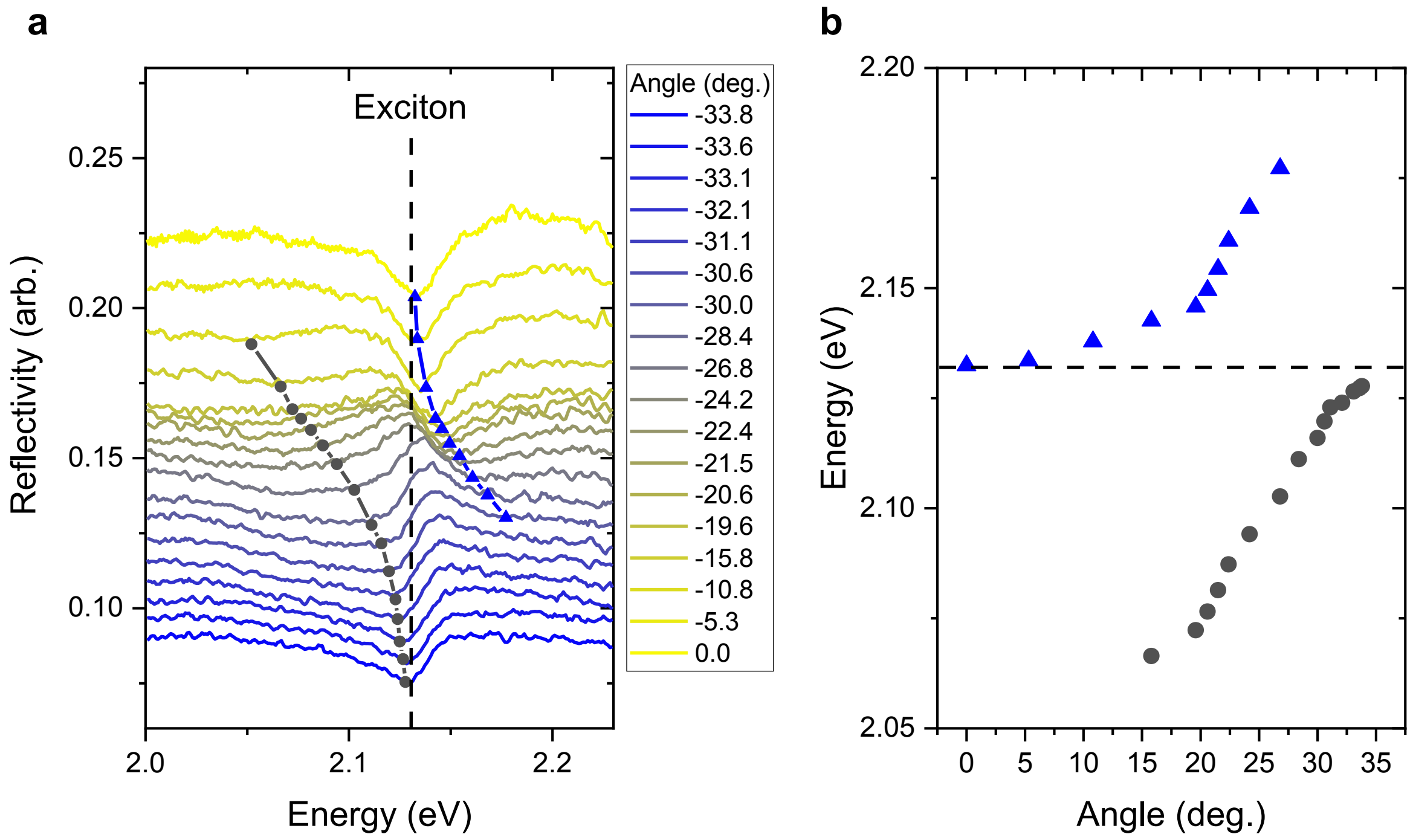

**Figure SI64: Reflectivity as a function of photon energy at different incident angles is shown in panel a**. Each curve has been displaced vertically for clarity. The solid black and blue lines are only guides to the eye to highlight the angle-dependent dips in reflectivity below and above the excitonic transition respectively. The energies of these dips are also plotted against their angles in panel **b**. resembling a typical anti-crossing of a cavity photon mode and an excitonic transition.

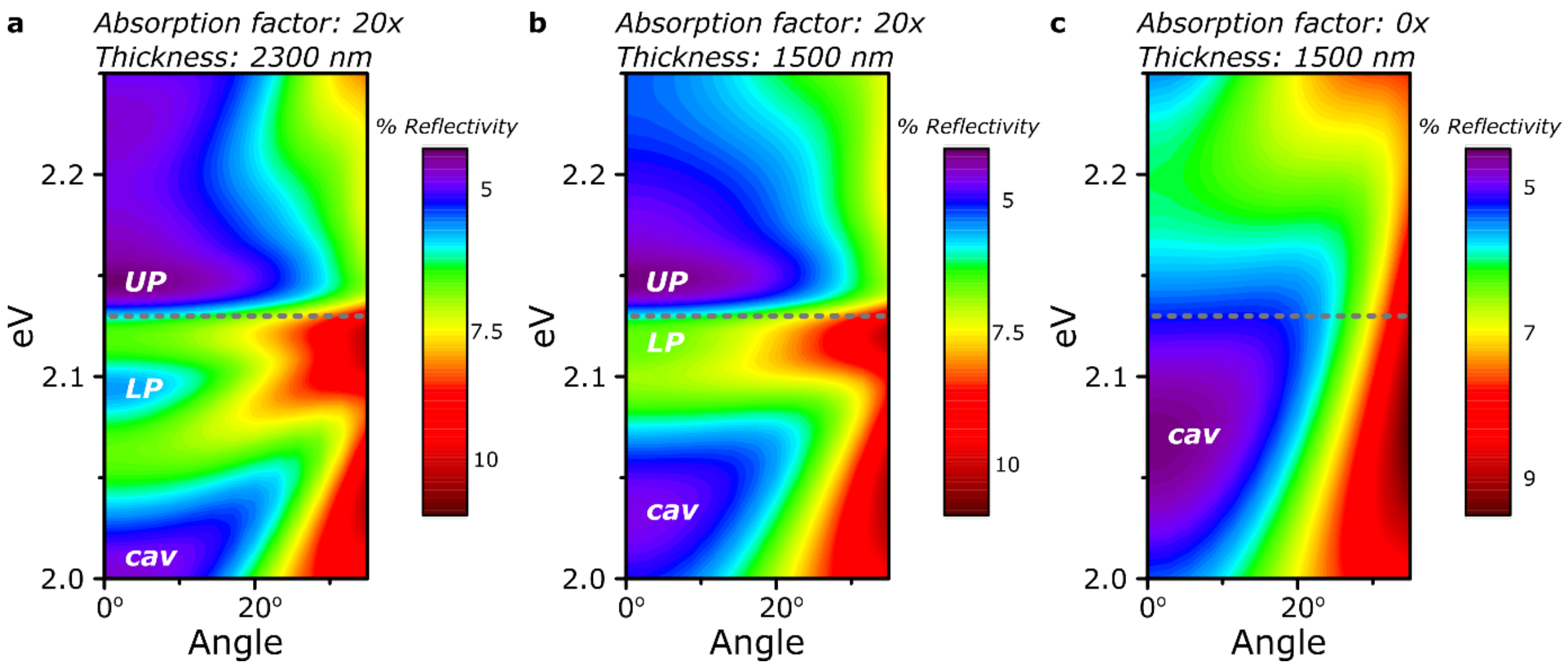

**Figure SI65: Reflectivity dispersion for PIC films, calculated using transfer matrix model.** Varying film thickness in **a** and **b** shifts the energetic positions of polariton (UP, LP) and photonic (cav) modes, as in normal-incidence calculations in Figure **SI61**. **c** When the PIC oscillator strength is set to zero, the steeper angular dispersion of the intrinsic Fabry-Perot modes can be discerned. Upper and lower polariton are labelled based on their identification in oscillator strength-dependent calculations in Figure **SI61i**. Energy of the primary PIC exciton transition is shown in grey dash.

In order to understand the appearance of these features we have performed further transfer matrix modelling (Figure **SI65**). We have used the same sample structure as in the normal-incidence reflectivity calculations above (Figure **SI61**). Here we focus on the regime of strong 'bundling' with an absorption scaling factor of 20, but the same conclusions apply to many other combinations of parameters. In panel **a** we show the angle-dependent reflectivity over the same range measured above, for a film of 2300 nm thickness where the upper and lower polariton peaks can be readily identified from normal-incidence calculations (Figure **SI61h**). We compare to calculations for a 1500 nm film, where the lower polariton can only be weakly discerned at normal incidence (Figure **SI61f**). In both structures the simulations show a low-energy photonic mode with relatively steep dispersion, labelled cav. By contrast, the upper polariton (UP) band exhibits very weak angle dependence. The lower polariton (LP) has a similarly shallow dispersion and additionally exhibits a much weaker modulation in overall reflectivity. Indeed, depending on the film thickness (i.e. detuning) the lower polariton can be almost impossible to detect despite the upper polariton branch remaining prominent. By comparison, if the oscillator strength of the material is set to zero (panel c) the simulation returns a single well-defined optical mode with steep dispersion. In light of these simulations, we find that the behaviour evident in Figure **SI65a** – a single prominent dispersion slightly above the exciton energy, with much slighter angle dependence than the Fabry-Perot modes – is most consistent with light-matter interactions, leading to the formation of distinct polariton states.

## S11: MEEP Calculations

The characteristic spectral signature of light-matter coupling between a photonic mode and a resonant emitter excitation is the presence of two peaks that are respectively redshifted and blueshifted from the interacting resonances. This distinctive peak splitting is observed in the reflectance spectra for nanostructures of PDA (Figure 3a) and PIC (Figure 3b) but not of PDI (Figure **SI60b**). However, the experiments probing long-range energy transport (Figure 1) suggest that light-matter coupling is present in all three materials; presumably, the polariton states cannot be spectrally resolved for PDI due to its progression of substantially damped vibronic transitions (Figure **SI60a**, blue curve). In this section, we report calculated reflectance and absorbance spectra for a model system resembling the PDI nanobelts. These simulations are carried out with the finite-difference time-domain (FDTD) method using the software package MEEP (*32*). Assuming one absorptive resonance to represent the first few PDI vibronic transitions, as well as considering only the major geometric features of the nanobelts, our computational treatment shows that light-matter coupling is possible in the PDI nanobelts. Specifically, the results provide insight on the nature of the photonic modes and their coupling to PDI.

The model system consists of a single nanobelt, treated as a rectangular prism that extends infinitely and possesses translational invariance along the $z$-axis. In light of this size assumption, which is motivated by the fact that the actual nanobelt length is much larger than any dimension of the beam spot, we restrict our computational cell to the $xy$-plane, *i.e.*, the plane containing the short-axes cross section (width × height = 50 nm × 800 nm) of the structure; the essential spectral features are expected to be preserved upon this dimensional reduction (*33*). The nanobelt material is modelled with the orientationally isotropic Lorentzian dielectric function (e.g., see (*34*); in this section, $\hbar$ = 1 hereafter)

$$\epsilon(\omega) = \epsilon_\infty + \frac{f\omega_0^2}{\omega_0^2 - \omega^2 - i\omega\gamma_0},$$

where $\epsilon_\infty = n_\infty^2 = 1.62^2$ is the high-frequency dielectric constant, $\omega_0$ = 2.422 eV (corresponding to wavelength of 512 nm) is the frequency of the molecular transition, $\gamma_0$ = 40 meV is the damping, and $f = \frac{2\mu^2\rho}{\epsilon_\infty\epsilon_0\omega_0}$ is the dimensionless oscillator strength. In calculating $f$, we actually use $\mu^2\rho$ = 60 $D^2/nm^3$ (see Section S8), where $\mu$ is the magnitude of the transition dipole moment along either $x$ or $z$ and $\rho$ is the number of molecules per unit volume. Furthermore, since our simulations only consider light polarized along the nanobelt long-axis, we would obtain the same results if we use an anisotropic dielectric tensor polarized along $z$. All points below the nanobelt lie in borosilicate glass ($n$ = 1.517), while the remainder of the surrounding medium lies in air ($n$ = 1). The absorbing boundaries of the computational cell are of the perfectly matched layer (PML) type. For all calculations (*i.e.*, near-field intensity distributions and optical spectra), we do the following once with all materials present (sample run) and once with only vacuum (background run): first propagate the fields produced by a Gaussian line source—which only has field component $E_z$—placed above the nanobelt until all fields have sufficiently decayed, and then Fourier transform the fields. Using the frequency-domain fields, the total flux passing through reflectance and absorbance power monitors is calculated. At each frequency, reflectance and absorbance are calculated by dividing the total flux through their corresponding power monitors for the sample run by the total flux through the reflectance power monitor for the background run. Lastly, the employed setup and fixed electric polarization of $E_z$ cause the simulated fields to be symmetric with respect to the *yz*-plane; specifying this reflection symmetry in MEEP shrinks the simulation cell even more, leading to additional gains in computational efficiency. See Figure **SI66** for other geometric parameters and a summary of the above simulation details.

To understand the nature of the electromagnetic modes coupled to PDI, we first analyze the reflectance spectrum (Figure **SI67a**) when the single molecular transition at 512 nm is not coupled to light, i.e., $f$ = 0 and $\epsilon(\omega) = \epsilon_\infty$. There are four peaks, two redshifted and two blueshifted with respect to 512 nm. Upon calculating the near-field intensity distributions at each of the peak frequencies (Figure **SI68**), we find that the photonic eigenmodes correspond to standing waves with $m$ = 2, 3, 4, and 5 antinodes confined along the nanobelt height. Observed in the simulated reflectance spectra for PIC film (Figure **SI61**) and also referred to as Fresnel modes (see Section S10), these electromagnetic resonances are well-known in photolithography (*35*): for a stack of three dielectric slabs, where the middle slab has a different refractive index than the other slabs, light traveling from first to third slab will produce standing waves in the middle slab. The standing waves arise from interference between the wave transmitted from the first slab to the second slab and the wave reflected from the third slab to the second slab.

As a control, we also calculate the absorbance spectrum (Figure **SI67b**) of the system when the oscillator strength of the molecular transition is low, namely at 1% of its full value. The single observed peak is just a Lorentzian curve centred at the molecular transition energy. This spectral profile is exactly what we expect for weak light-matter coupling. Indeed, the observed full width at half maximum (FWHM) is 44 meV, only a little more than the linewidth of 40 meV for the excitation without any coupling to light, i.e., zero oscillator strength.

We now determine how strongly the standing-wave electromagnetic modes interact with the effective PDI transition at 512 nm when the oscillator strength is set to its full value. The computed reflectance spectrum (Figure **SI67c)** displays four peaks that resemble those of the spectrum in the absence of light-matter coupling, as well as a peak around 512 nm with nontrivial lineshape. In particular, each of the former four peaks is either redshifted or blueshifted if the corresponding peak in the bare spectrum is higher or lower in energy, respectively, than the molecular transition. These observations suggest that each shifted peak represents a polariton state with mostly photonic character, formed from the off-resonant interaction between a standing-wave mode and a molecular excitation of the same spatial symmetry.

To verify this claim, we first confirm that the energy shifts are due to light-matter coupling. For each photonic eigenmode $m$ corresponding to the peak with energy $\omega_{c,m}$ in the reflectance spectrum with no light-matter coupling (Figure **SI67a**), we fit the appropriate eigenvalue of

$$H_m = \begin{pmatrix} \omega_{c,m} & g_m\sqrt{N} \\ g_m\sqrt{N} & \omega_{e,m} \end{pmatrix}$$

to the energy of the corresponding shifted peak (Figure **SI67c**), with the light-matter coupling strength

$$g_m\sqrt{N} = \sqrt{\frac{\omega_{c,m}\rho}{2\epsilon_\infty\epsilon_0 V_m/V_{sample}}}$$

as the only fitting parameter. $N$ is the number of coupled molecules, $V_m/V_{sample}$ is the ratio of the volume of photonic mode $m$ to the volume occupied by these molecules, and $\omega_{e,m} = \omega_0$ is the energy of the molecular transition interacting with electromagnetic resonance $m$. The other eigenvalues of the fitted Hamiltonians $H_m$ are well-aligned with the peak and shoulders around 512 nm (Figure **SI67c**). Therefore, these spectral contributions are those of the matter states in the presence of the same off-resonant light-matter interaction.

We now discuss the achievement of light-matter coupling. Summarized in Table **T2**, the fitted $g_m\sqrt{N}$ are on the order of 100 meV and the associated $V_m/V_{sample}$ are order 1-10. The latter values indicate moderate field confinement, a typical prerequisite for achieving light-matter coupling (*36*). To more accurately assess if this interaction regime has been reached, we must also consider damping processes. This necessity rests on the following intuition (*36*): hybrid light-matter states form when the constituent states decay slow enough to allow coherent energy exchange. Thus, we also determine an effective strength $\kappa_m/2$ of coupling between each electromagnetic resonance and decay-inducing bath modes (Table **T2**). We evaluate $\kappa_m/2$ as the half width at half maximum (HWHM) of the reflectance peak for mode $m$ in the absence of light-matter coupling (Figure **SI67a**). The HWHM for each peak is calculated using the side (left or right) that reaches lowest in signal. That is, we use the right, left, and left sides of the peaks at 442, 572, and 833 nm. Since neither side of the peak at 368 nm has points whose signal is less than the maximum signal, we do not determine $\kappa_m/2$ for $m$ = 5. Comparing $g_m\sqrt{N}$ and $\kappa_m/2$, we conclude that photonic modes $m$ = 3 and 4 couple to PDI almost as much, if not more, than they do to their respective damping degrees of freedom. Moreover, $g_m\sqrt{N}$ is much larger than the strength $\gamma_0/2$ at which PDI couples to its bath. Therefore, the light-matter subsystems for $m$ = 3, 4 are near or at the limit of light-matter coupling. Further evidence that this interaction regime has been reached is the absorbance spectrum (Figure **SI67d**), which demonstrates splitting of the peak from the spectrum with relatively weak light-matter coupling (Figure **SI67b**) into a peak and shoulder that are shifted in energy from the original molecular transition. As an aside, note that there is a lack of a definitive Rabi splitting around the molecular transition in both simulated and experimental (Figure **SI60b**, red curve) spectra for PDI nanobelts; this result suggests that the characteristic signature of polariton formation is precluded not only by the vibronic progression of PDI, but also the significant detuning between photonic and molecular states. In addition, while coupling between PDI and standing-wave modes $m \neq 2, 3, 4, 5$ can also contribute to optical spectra, these contributions are expected to be insignificant compared to the more resonant light-matter interactions of the modes considered above.

We now discuss the design parameters that enable light-matter interactions in the PDI nanobelt system. If we increase the nanobelt width from 50 to 800 nm, standing-wave modes and their large coupling to the molecular transition are still obtained. However, if instead the nanobelt height is reduced from 800 to 50 nm, eigenmodes in the visible and near-IR regions lose standing-wave character, and light-matter coupling is apparent from neither the reflectance nor the absorbance spectra. These observations indicate that a minimum height of approximately half a wavelength of incident light is required for light-matter coupling to be observed in simulated reflectance spectra. Therefore, as stated in Section S10, light-matter coupling and long-range energy transport should be a general feature of dielectric media with absorptive transitions. Indeed, the light-matter interaction in both PDI nanobelts and PIC film (Figure **SI57**) depends on nanostructure height in the same way that such interaction in microcavity exciton-polaritons depends on cavity transverse length: the nanostructure height should be chosen to allow formation of standing-wave modes that are at or near resonance with the molecular transition to be coupled. In summary, the realization of light-matter coupling relies on the high oscillator strength inherent in J-aggregate dyes, a difference in refractive index of the dye from those of the surrounding, and optimal height of the dye layer (*36*).

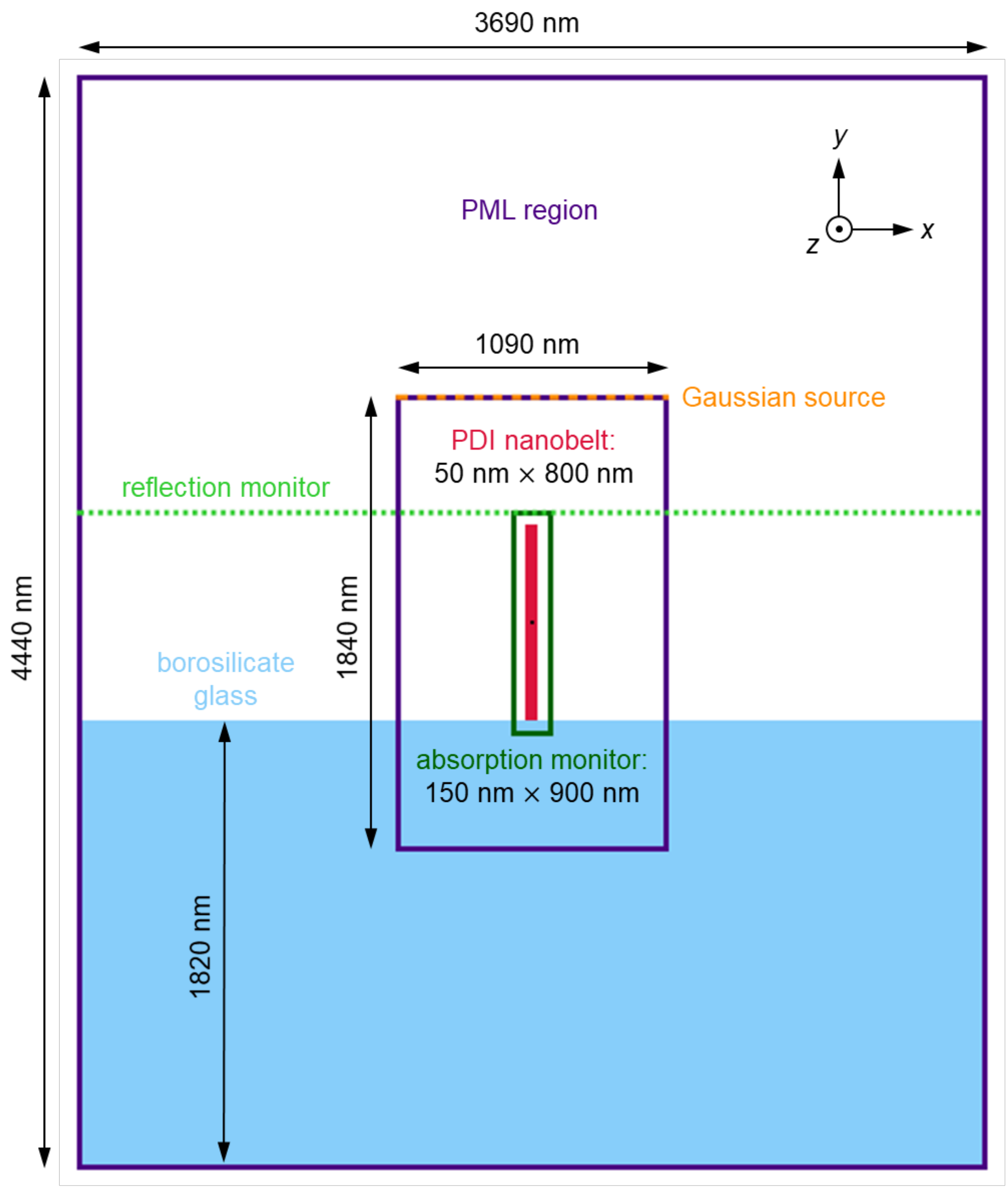

**Figure SI66:** FDTD simulation cell. All rectangles shown, except for the rectangle representing borosilicate glass, are centred at the origin (black dot in the middle). The grid resolution in the simulation is 0.1 pixels/nm. Upon indicating *yz*-mirror symmetry in MEEP, the simulation cell is halved, and additional pixels may be added to facilitate this halving. For other software packages, the user may have to specify different and/or additional parameters pertaining to object lengths, numerical mesh, etc.

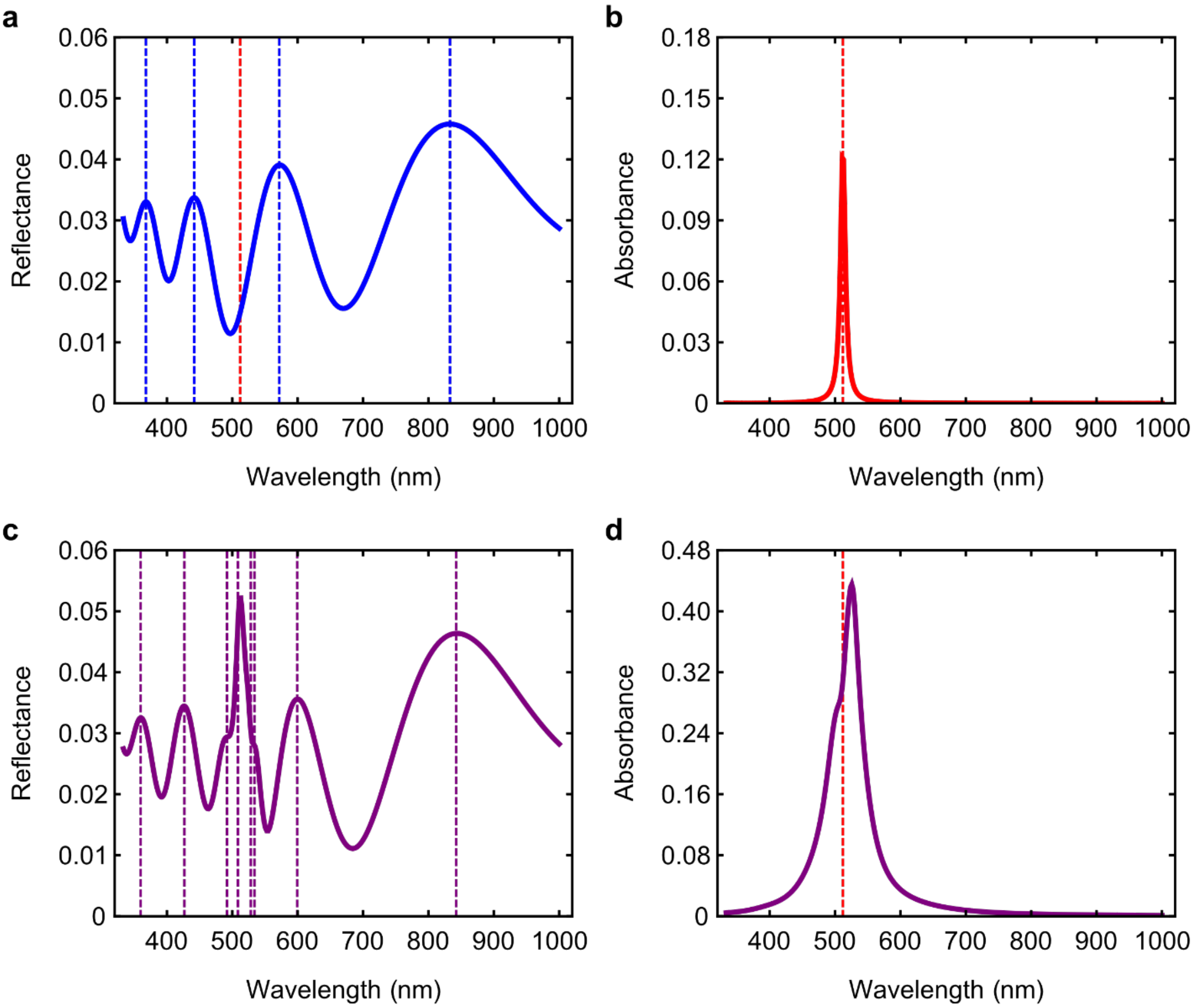

**Figure SI67:** FDTD-simulated spectra. (a) Reflectance spectrum for J-aggregate material on glass without light-matter interaction between the electromagnetic modes of the material and the molecular transition at 512 nm. The blue dashed lines indicate the peak energies of the electromagnetic modes. The red dashed line indicates the energy of the molecular transition at 512 nm. (b) Absorbance spectrum for system with relatively weak light-matter interaction, namely with the oscillator strength of the molecular transition being 1% of that in the system characterized in (c) and (d). (c) Reflectance spectrum for the same system but with light-matter interaction between the electromagnetic modes of the material and the molecular transition at 512 nm. The purple dashed lines indicate the fitted polariton energies (see SI text). (d) Absorbance spectrum for the same system as (c). The red dashed line indicates the energy of the molecular transition at 512 nm.

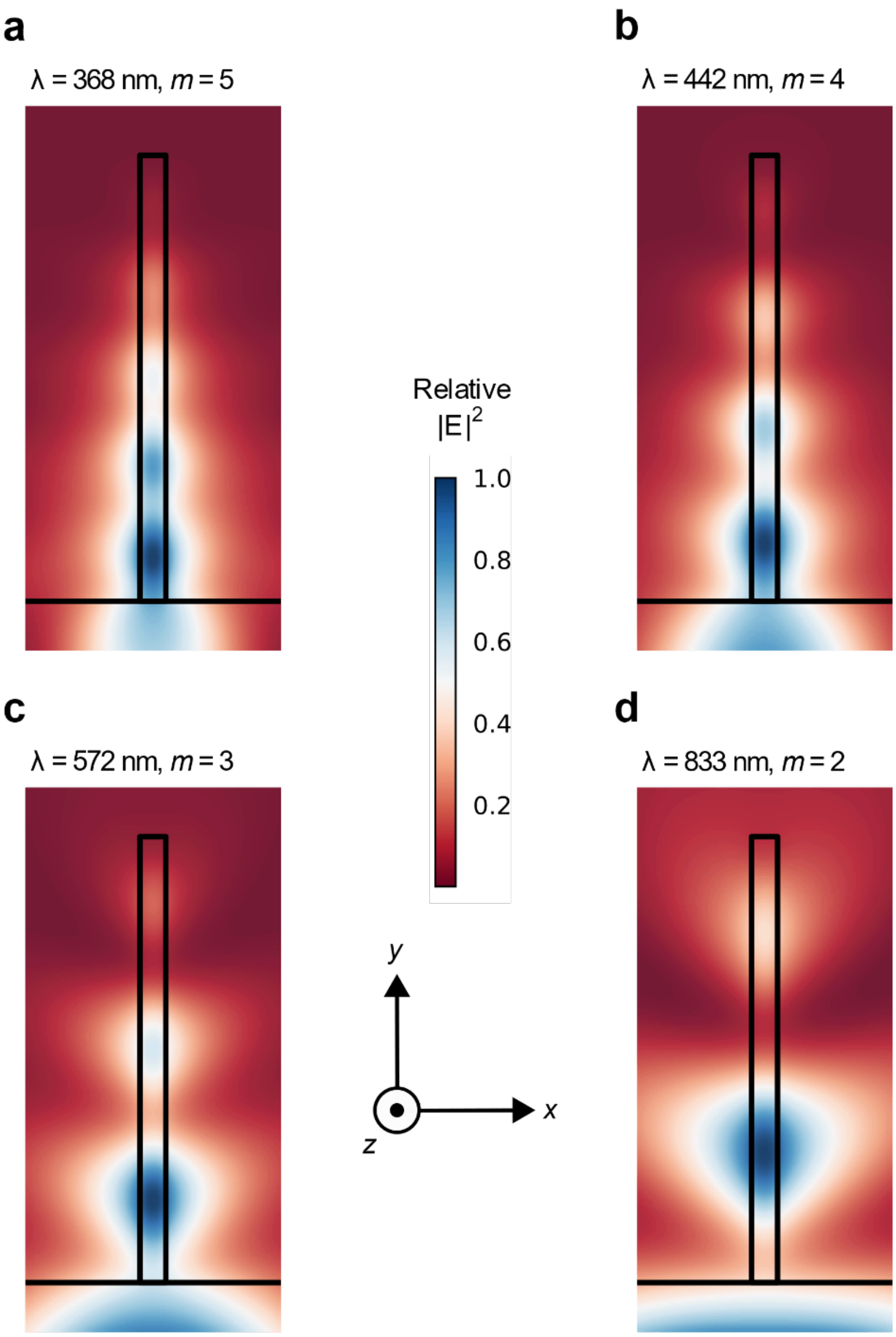

**Figure SI68:** Near-field distributions. Intensity of electric-field Fourier component with wavelength (a) 368 nm, (b) 442 nm, (c) 572 nm, and (d) 833 nm. The standing-wave index $m$ characterizes the mode excited at each wavelength. For each plot, the intensities are normalized to the maximum, and the simulation was carried out using a Gaussian source whose centre and width equal the energy corresponding to the specified wavelength. A black rectangle denotes the PDI nanobelt, and all points below a spanning horizontal line lie in borosilicate glass. All points above the line but not in the black rectangle lie in vacuum.

| Mode index $m$ | Wavelength (nm) | Light-matter coupling strength $g_m\sqrt{N}$ (meV) | Mode volume $V_m$ ($V_{sample}$) | Strength of coupling to bath modes $\kappa_m/2$ (meV) |
|---|---|---|---|---|
| 2 | 833 | 125 | 8.47 | 248 |
| 3 | 572 | 187 | 5.54 | 209 |
| 4 | 442 | 220 | 5.21 | 214 |
| 5 | 368 | 276 | 3.95 | not determined [a] |

[a]See SI text for details.

**Table T2:** Light-matter coupling strengths, mode volumes, and strength of coupling to bath modes for photonic modes in system simulated in Figure **SI67**.

## S12: Waveguiding

The rapid initial growth described in the main text could be argued to result from the waveguiding properties of aggregate 1D crystals, which propagate light efficiently away from the initial excitation spot. Such light can be absorbed along the fiber- or wire-shaped organic waveguides and produce an excited population, which becomes visible in our TAM data.

Balzer *et al.* studied the propagating light modes in self-assembled organic needle-like aggregates which have rectangular cross sections, on a dielectric substrate (*37*, *38*). They demonstrated that such waveguides only support transverse magnetic modes (TM modes) when the fibres are assumed to be optically uniaxial with the dielectric tensor components $\varepsilon_{/\!/}$ (along the fibre) and $\varepsilon_{\perp}$ (perpendicular to the fibre). The number of possible modes, m, is restricted by:

$$m < \frac{2a}{\lambda}\sqrt{\frac{\epsilon_\perp}{\epsilon_\parallel}}\sqrt{\epsilon_\parallel - \epsilon_S}$$

where $a$ is the width of the waveguide, $\lambda$ is the wavelength of the propagating light, and $\epsilon_S$ is the dielectric constant of the substrate. From our ellipsometry measurements and published data, $n_\perp$ was determined to be ~1.60 for PIC (*39*) and PDA (*40*), and 1.62 for PDI (*41*). By further assuming that all three systems are optically isotropic, we adopted $\epsilon_{iso} = n_\perp^2$ as the value for the isotropic dielectric constant. The calculated values of m for each of the three systems are summarised in the table below, using the value $\epsilon_S = n_S^2 = 1.53^2 = 2.34$ for the glass coverslip used in our TAM measurements.

| | PIC | PDA | PDI |
|---|---|---|---|
| $a$ / μm | 1 | 2 | 0.13 |
| $\epsilon_{iso}$ | 2.56 | 2.56 | 2.62 |
| $\lambda_{pump}$ / nm | 565 | 590 | 590 |
| $m$ | 1.66 | 3.18 | 0.23 |

**Table T3:** Table summarising refractive index and supported waveguiding modes for three materials considered.

This mode analysis suggests that only m = 1 is possible in PIC whereas m up to 3 is supported in PDA. PDI, on the other hand, is unable to support any waveguides due to its dimensions and this has been shown to be the case at room temperature experimentally (*41*).

It can be shown that the wavevector $k$ and the group velocity $v_g$ of the supported waveguiding mode are given by:

$$k = \sqrt{\frac{\omega^2}{c^2}\epsilon_\parallel - \frac{\epsilon_\parallel}{\epsilon_\perp}\left(\frac{m\pi}{a}\right)^2}$$

$$v_g = \frac{d\omega}{dk} = \frac{c^2 k}{\omega \epsilon_{\parallel}}$$

The calculated group velocities plotted in Figure **SI69** show a minimum velocity of 0.5c for any supported modes in PIC and PDA, which is at least 60 times faster than the maximum propagation speed observed in our experiments (i.e. 1% of light speed in vacuum). This suggests that the observed movement is not the result of waveguiding of the pump pulse and subsequent absorption along the length of the crystals.

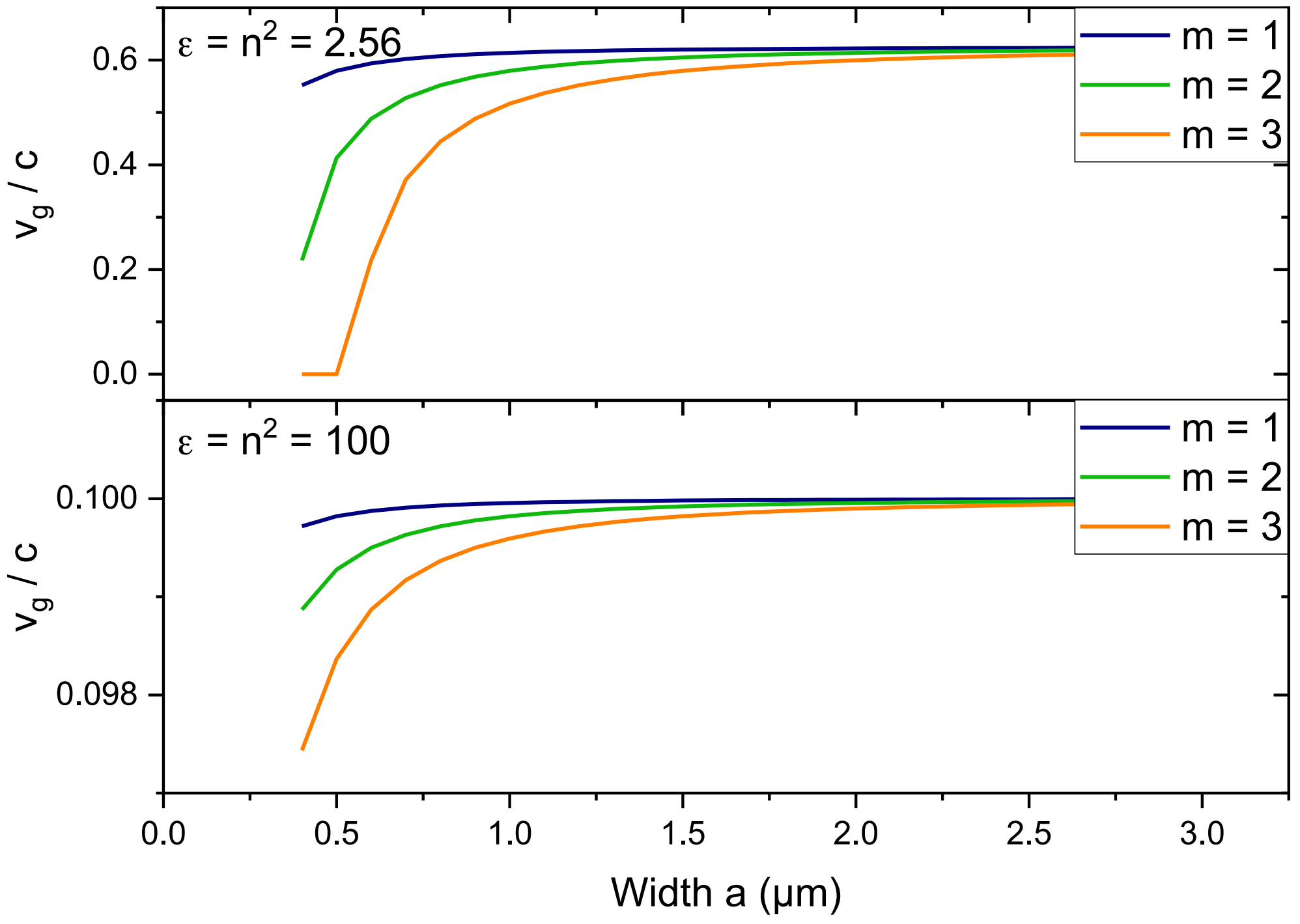

**Figure SI69:** Calculated group velocity for indicated waveguiding modes for PIC (top) and PDA (bottom).

It has been reported that exciton-polaritons (EP) could be formed by coupling between propagating fluorescence and the excitons within nanofibers. The authors observed a refractive index parallel to the nanofibers exceeding 10 at energies close to the anti-crossing point of the lower EP branch. This unusually high refractive index would result in a much slower group velocity. However, even with n = 10, the calculated group velocities (0.1c) is still about an order of magnitude higher than our observations. Therefore, it seems reasonable to rule out the possibility of waveguiding and argue that the observed expansion in our TAM signal is the result of ultrafast electronic population transport in our nanostructures. Nonetheless, weak EP coupling to the supported wave guiding modes could theoretically increase the group velocity. Further studies, theoretically and experimentally, is required to elucidate the nature of the transport mechanism.

**S13: Control measurements**

In order to confirm that the ultrafast, ultra-long energy propagation described in the main text arises from the specific chemical and morphological nature of the samples control measurements were performed on the isolated PIC, PDI dye monomers as well as amorphous poorly polymerised regions of the PDA sample. Additional measurements were also performed on diluted films of CdSe nanocrystals. In all cases no (ultrafast) energy propagation could be observed, confirming the behaviour observed in the main text is unique to the chemical structure of the materials and does not arise from a systematic measurement artefact.

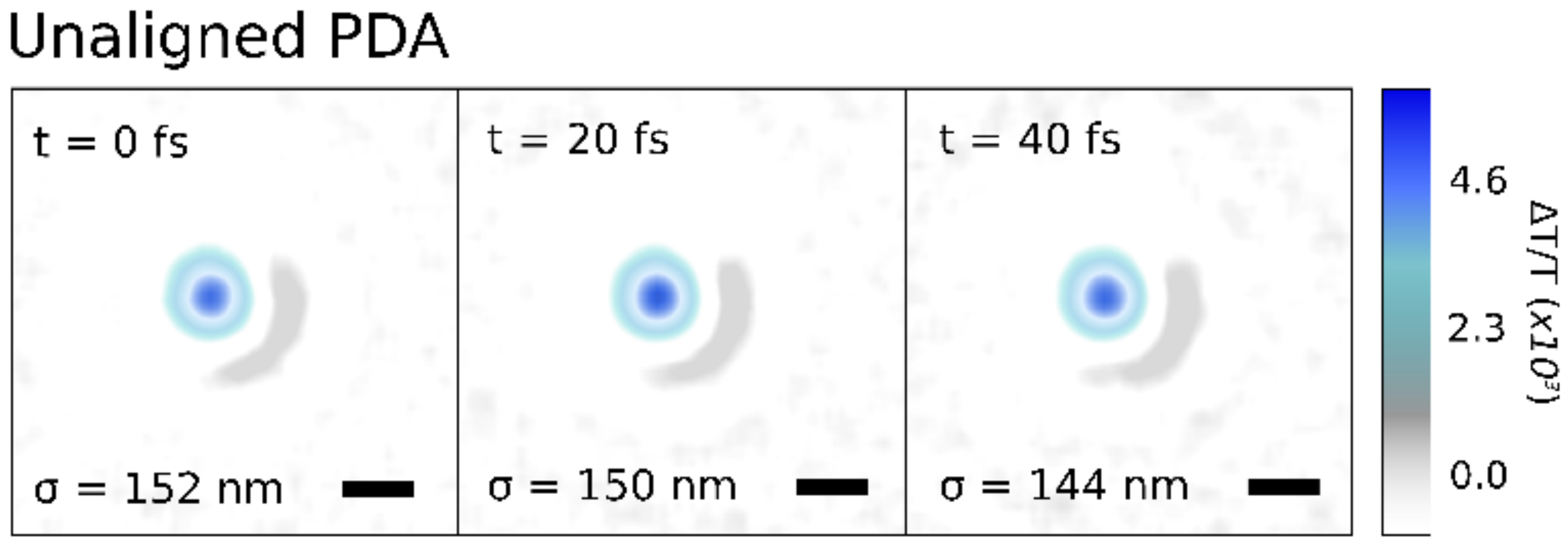

**Figure SI70:** fs-TAM images unaligned poorly polymerized PDA (Figure **SI14c**) following photoexcitation. The 10-fs pump pulse is at 520-650 nm, and probing is carried out at 670 nm at the edge of the presumed stimulated emission band. The scale bar is 500 nm and the Gaussian standard deviation (σ) is indicated in the bottom left corner of the images.

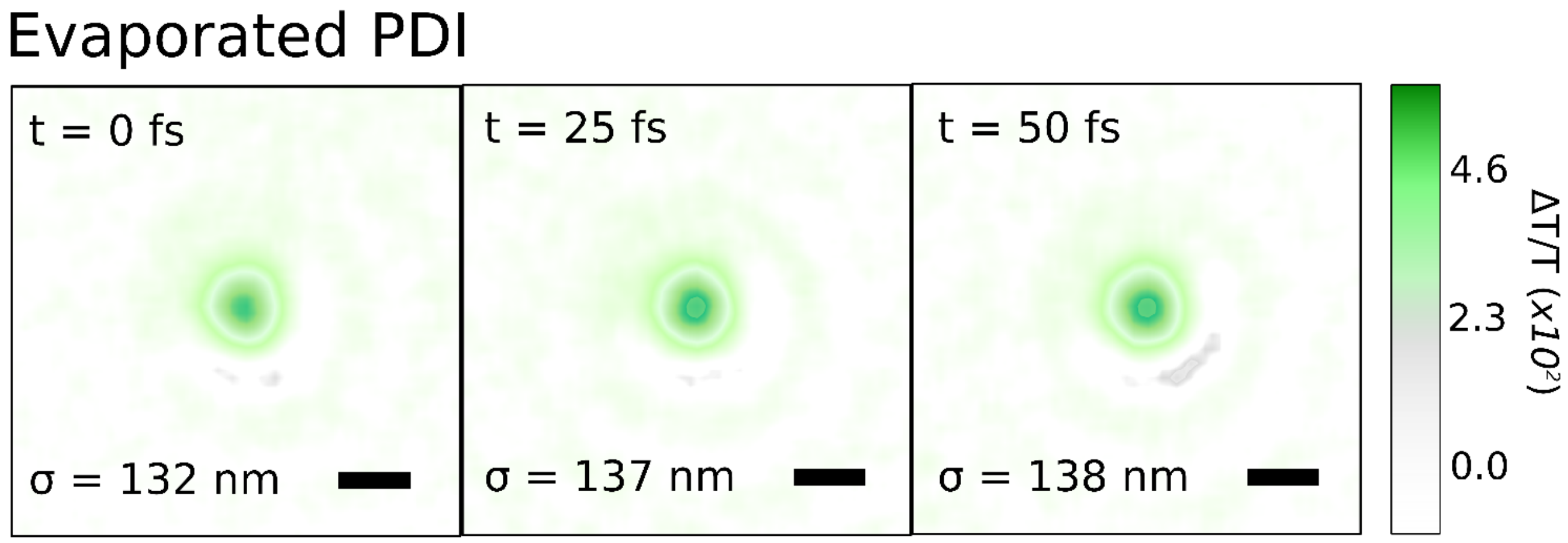

**Figure SI71:** fs-TAM images of an evaporated thin film (100 nm thickness) of PDI following photoexcitation. The 10-fs pump pulse is at 520-650 nm, and probing is carried out at 780 nm in the singlet photo-induced absorption band. The scale bar is 500 nm.

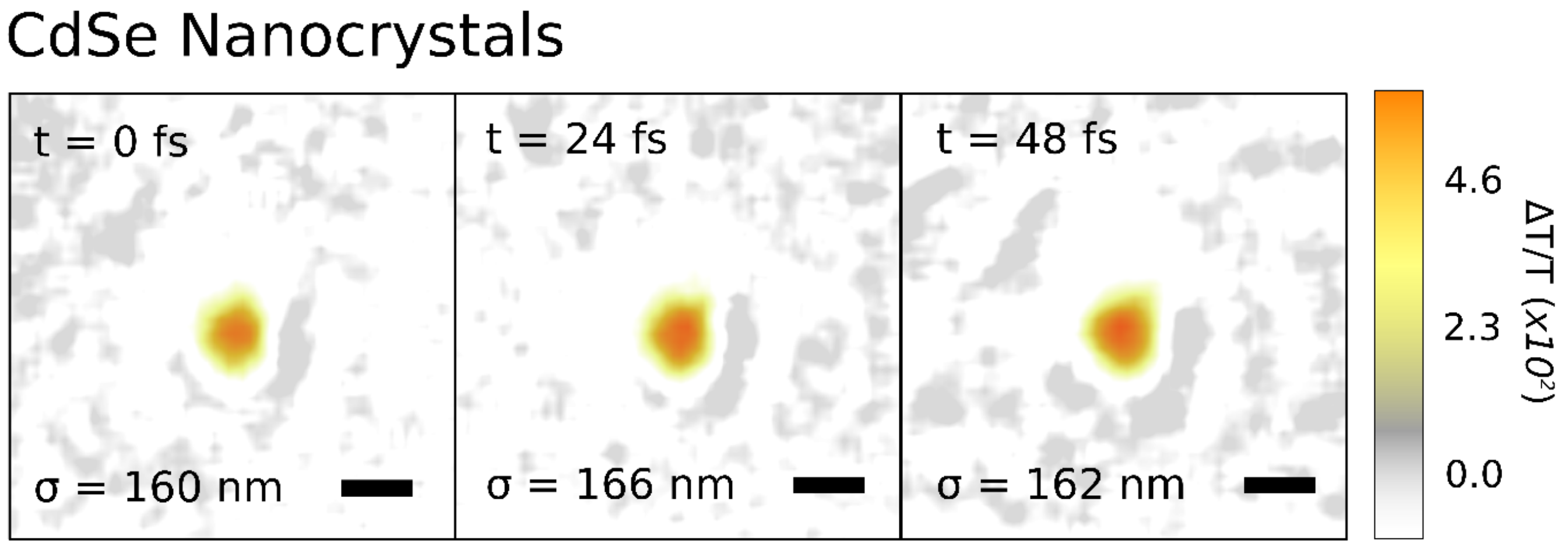

**Figure SI72:** fs-TAM images of CdSe nanocrystals dispersed 2% w/v in a polystyrene matrix ($M_w$ ~250, 000; film thickness 150 nm) following photoexcitation. The 10-fs pump pulse is at 520-650 nm, and probing is carried out at 700 nm in the broad stimulated emission band. The scale bar is 500 nm.

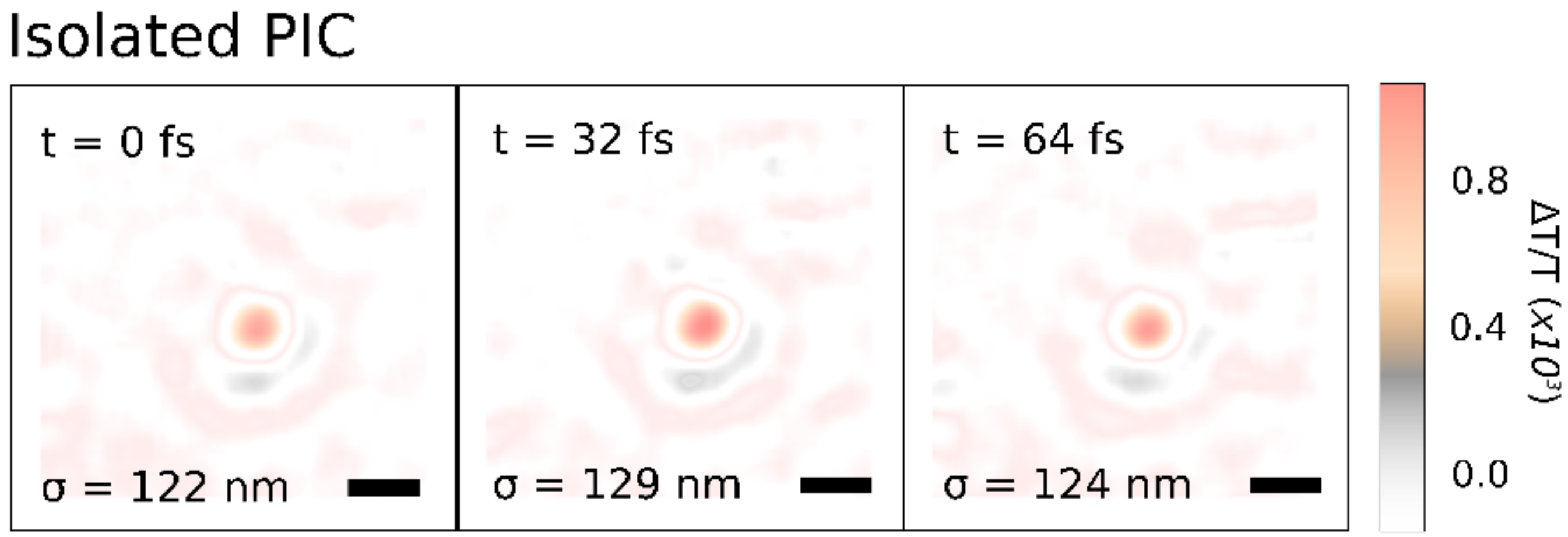

**Figure SI73:** fs-TAM images of PIC dye monomers dispersed 1% w/v in a polystyrene matrix (~250, 000 $M_w$; film thickness 130 nm) following photoexcitation. The 10 fs pump pulse is at 520-650 nm, and probing is carried out at 660 nm in the photo-induced absorption band. The scale bar is 500 nm.

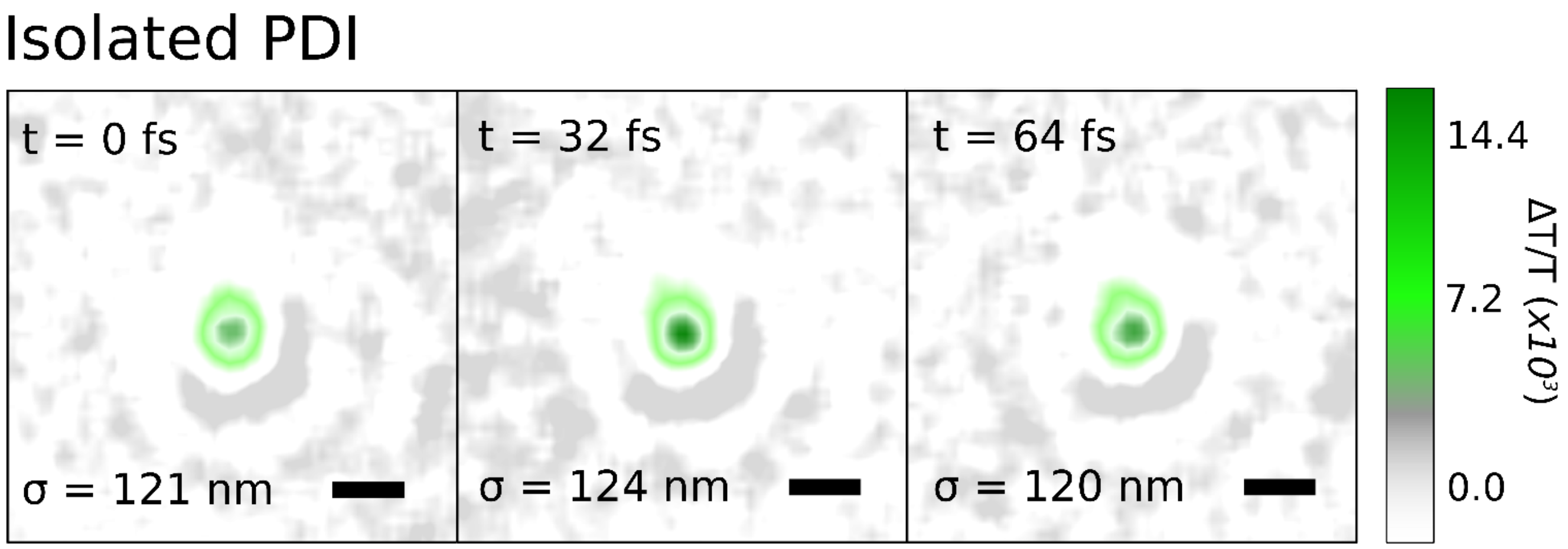

**Figure SI74:** fs-TAM images of PDI dye monomers dispersed 1% w/v in a PMMA matrix ($M_w$ ~500, 000; film thickness 200 nm) following photoexcitation. The 10-fs pump pulse is at 520-650 nm, and probing is carried out at 770 nm in the photo-induced absorption band. The scale bar is 500 nm.

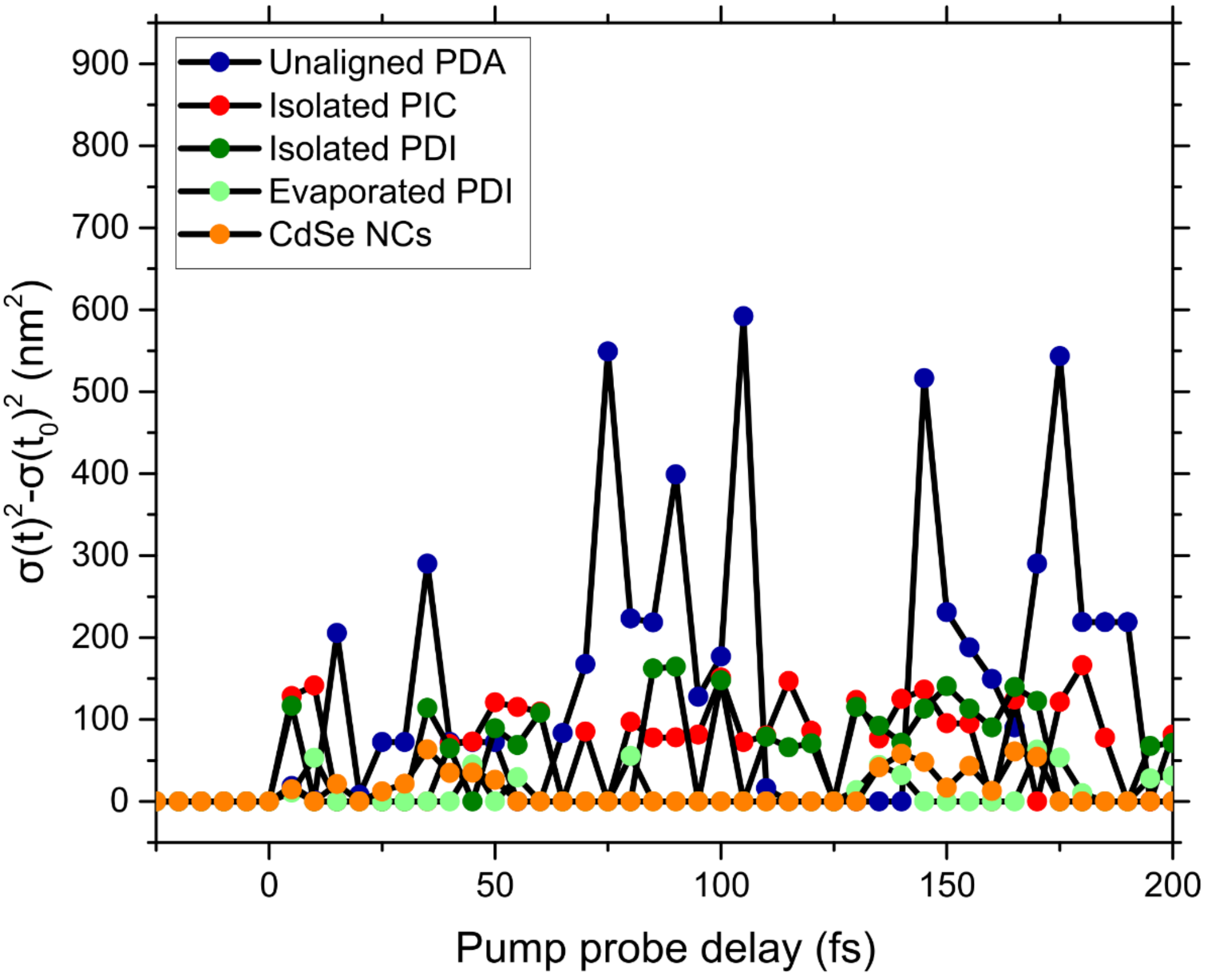

**Figure SI75:** Mean square displacement for control samples shown above. In all cases there appears to be no expansion and the radial extent of the spatial $\Delta T/T$ fluctuates about the value obtained at $t_0$. In the case when, $\sigma(t)^2$ drops below $\sigma(t_0)^2$ we set by convention set the mean square to 0 for ease of viewing. Any spikes in growth are assigned to random noise fluctuations in the relatively low signals.

## S14: Rabi Flopping measurements

In Figure **SI76-SI78** we show the results of laser fluence dependent measurements on the PL intensity of PDI nanobelts and isolated PDI monomers. In the case of multiple PDI wires no Rabi flopping can be observed due to ensemble averaging effects. Isolated PDI molecules in solution no Rabi flopping either because of rapid competing decay processes which are not mitigated by light-matter coupling.

To access the frequency content of the Rabi flopping, we first truncated the data to exclude field strengths < $9 \times 10^8$ $Vm^{-1}$ and subtracted the slowly varying intensity with a 4-point B-spline fit. The residual coherence was smoothed with a Savitzky-Golay filter (4 side points, third order polynomial) and zero-padded to 128 data points before Fourier-transformation (Figure 3c, right side). The same approach was carried out for the model.

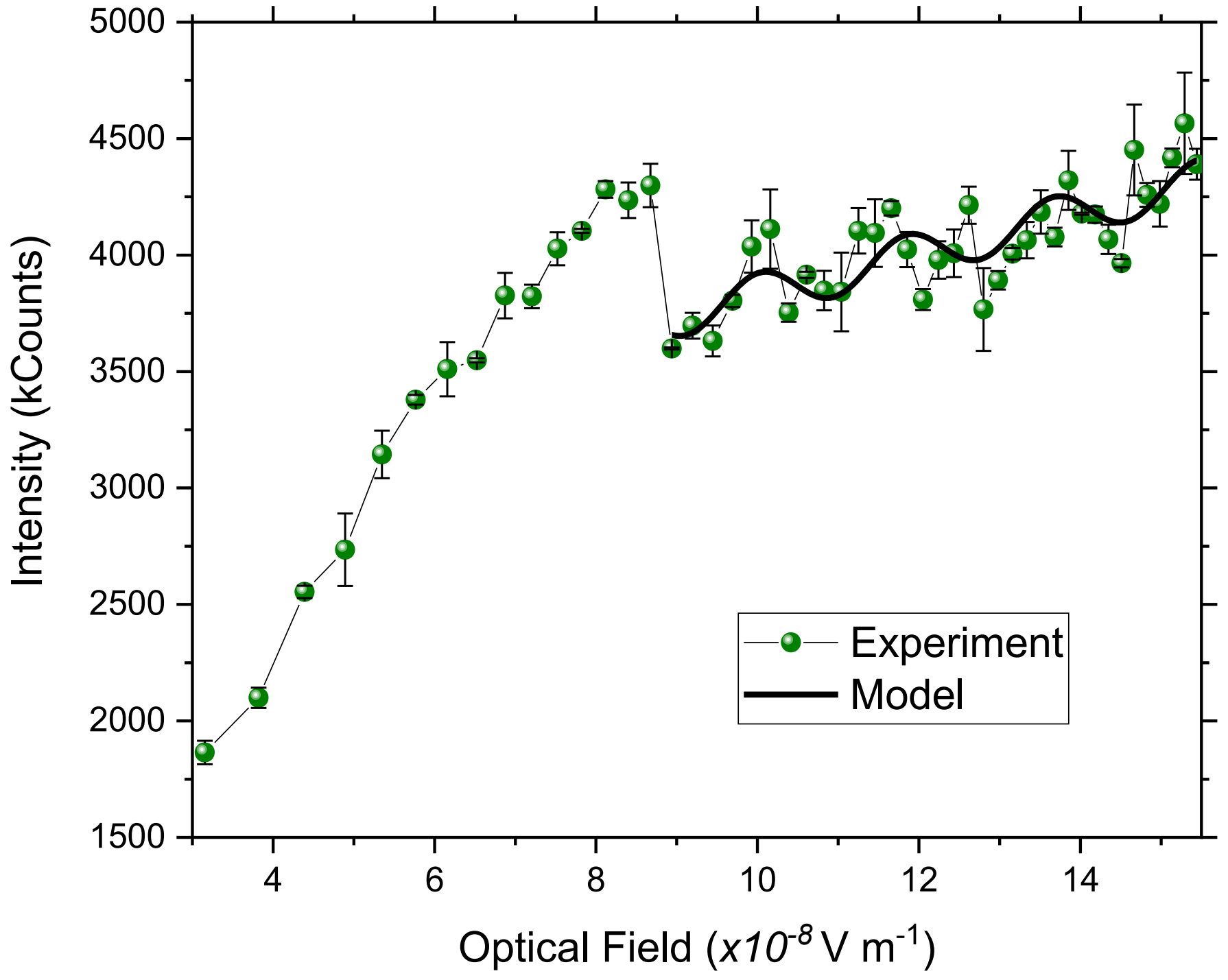

**Figure SI76:** Photoluminescence counts as a function of optical field for isolated PDI nanobelts as shown in the main text Figure 3c. The PL intensity initially increases linearly before oscillating as described in the main text. The fitted model black overlay shows good agreement with the experimental data

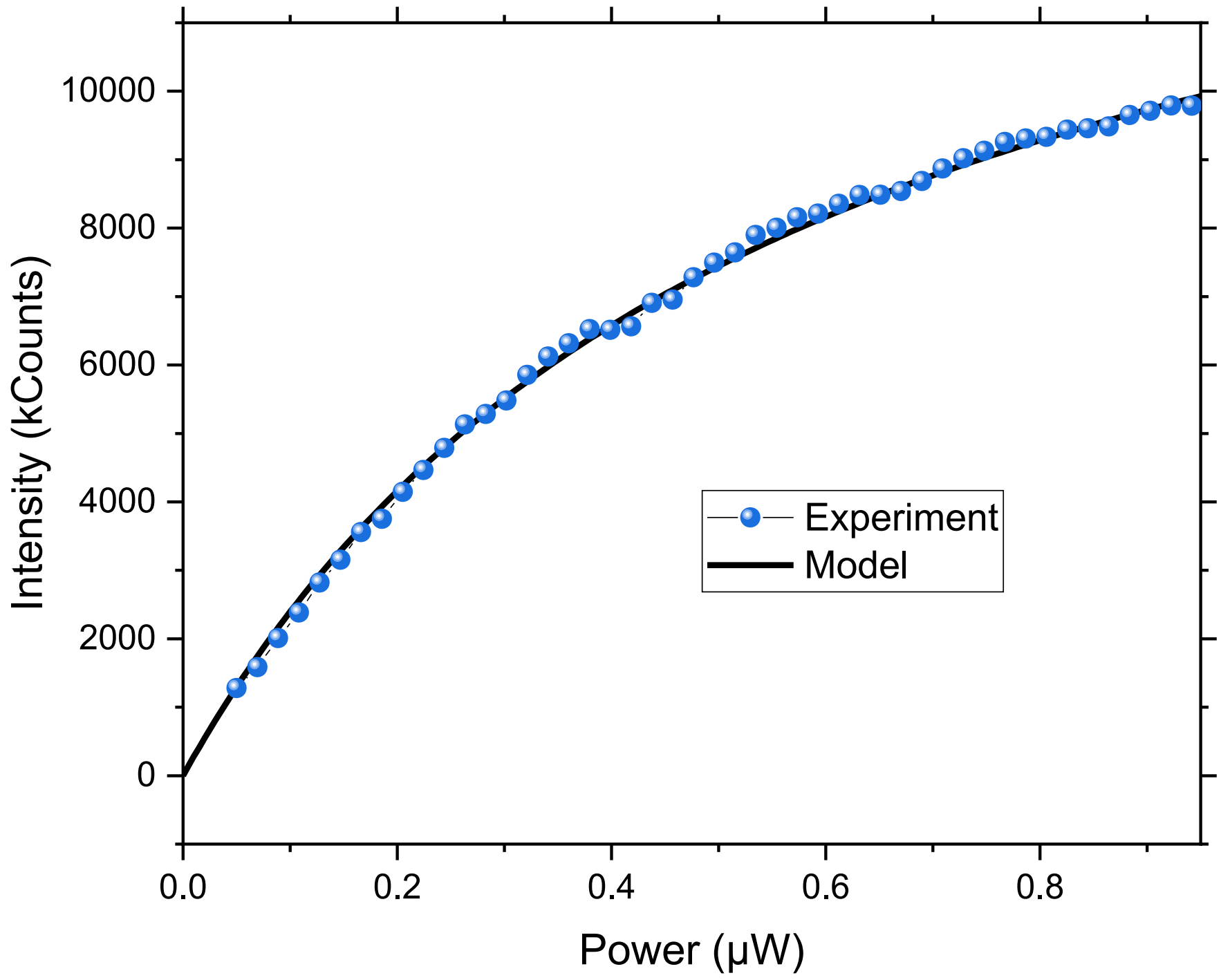

**Figure SI77:** Photoluminescence counts as a function of power for bundles of PDI nanobelts. Measurements were performed using a manner to that described in S1. The PL intensity initially shows quadratic increase (as expected from the fitted model – solid black line), before saturating beyond 0.8 μW.

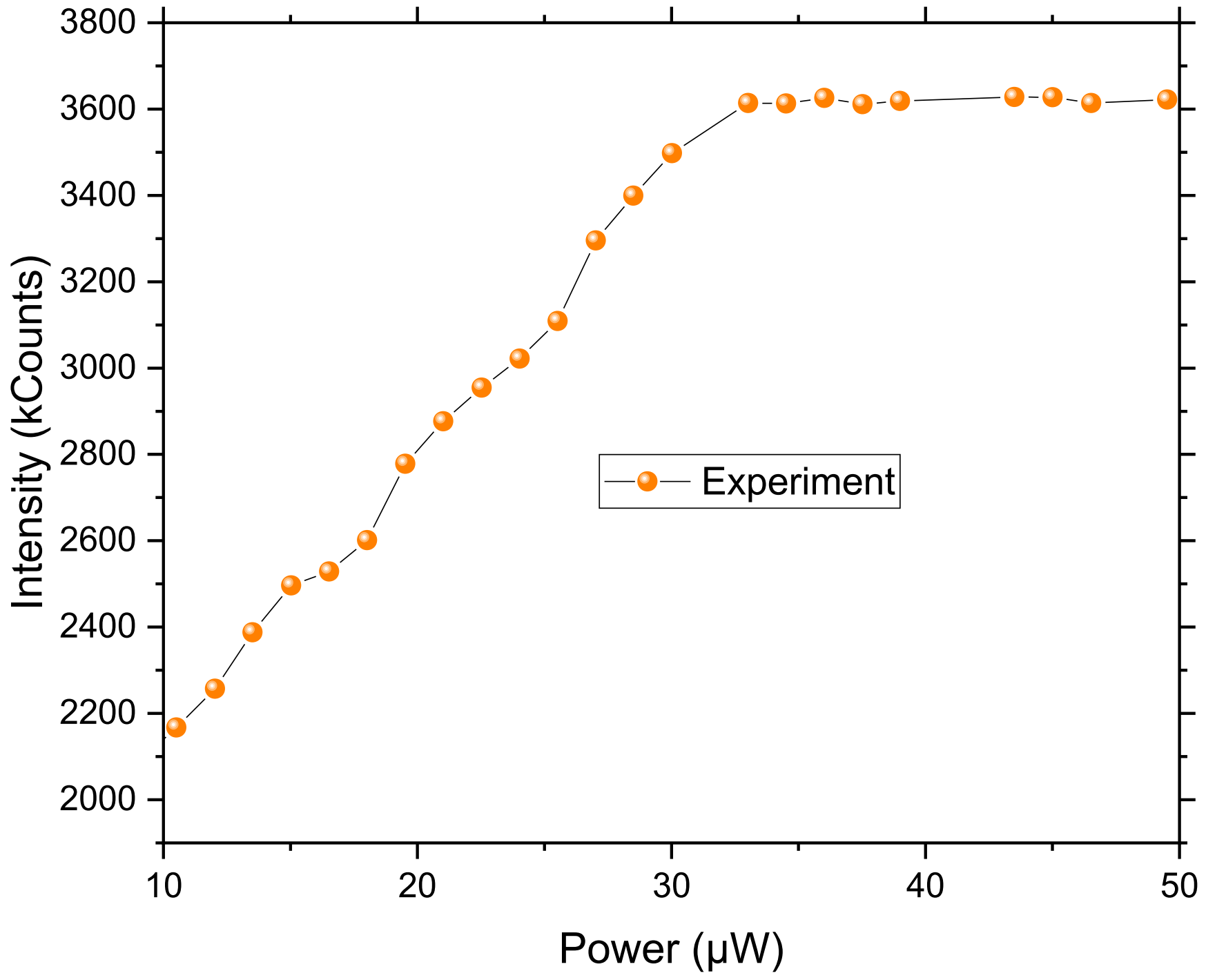

**Figure SI78:** Photoluminescence counts as a function of power for isolated PDI molecules. Measurements were performed using a manner to that described in S1but with focussing via f=50 mm focussing lens and dye in a 1 mm cuvette (Hellma). The PL intensity initially increases linearly before saturating after 30 μW.

## References

1. R. Trebino *et al.*, Measuring ultrashort laser pulses in the time-frequency domain using

frequency-resolved optical gating. *Rev. Sci. Instrum.* **68**, 3277–3295 (1997).

2. S. Kazunori, H. Ohnuma, T. Kotaka, Urethane Substituted Polydiacetylene: Synthesis and Characterization of Poly[4,6-decadiyn-1,10-diol-bis(n-butoxy-carbonyl-methyl-urethane)]. *Polym. J.* **14**, 895–905 (1982).

3. J.-C. Ribierre *et al.*, A solvent-free and vacuum-free melt-processing method to fabricate organic semiconducting layers with large crystal size for organic electronic applications. *J. Mater. Chem. C.* **7**, 3190–3198 (2019).

4. A. Vierheilig *et al.*, Femtosecond dynamics of ground-state vibrational motion and energy flow: Polymers of diacetylene. *Chem. Phys. Lett.* **312**, 349–356 (1999).

5. A. Horvath, G. Weiser, C. Lapersonne-Meyer, M. Schott, S. Spagnoli, Wannier excitons and Franz-Keldysh effect of polydiacetylene chains diluted in their single crystal monomer matrix. *Phys. Rev. B - Condens. Matter Mater. Phys.* **53**, 13507–13514 (1996).

6. B. Kraabel, M. Joffre, C. Lapersonne-Meyer, M. Schott, Singlet exciton relaxation in isolated polydiacetylene chains studied by subpicosecond pump-probe experiments. *Phys. Rev. B - Condens. Matter Mater. Phys.* **58**, 15777–15788 (1998).

7. G. Lanzani, *Photophysics of Molecular Materials: From Single Molecules to Single Crystals* (John Wiley & Sons, 2006).

8. S. Spagnoli, M. Schott, M. Johnson, L. Toupet, Structural study of two reactive diacetylenes. *Chem. Phys.* **333**, 236–245 (2007).

9. R. Brito de Barros, L. M. Ilharco, The role of cellulose acetate as a matrix for aggregation of pseudoisocyanine iodide: Absorption and emission studies. *Spectrochim. Acta - Part A Mol. Biomol. Spectrosc.* **57**, 1809–1817 (2001).

10. H. von Berlepsch, C. Böttcher, L. Dähne, Structure of J-Aggregates of Pseudoisocyanine Dye in Aqueous Solution. *J. Phys. Chem. B* (2002), doi:10.1021/jp000085q.

11. J. R. Caram *et al.*, Room-Temperature Micron-Scale Exciton Migration in a Stabilized Emissive Molecular Aggregate. *Nano Lett.* **16**, 6808–6815 (2016).

12. Y. Che, X. Yang, K. Balakrishnan, J. Zuo, L. Zang, Highly polarized and self-waveguided emission from single-crystalline organic nanobelts. *Chem. Mater.* **21**, 2930–2934 (2009).

13. W. Kabsch, *XDS. Acta Crystallogr. Sect. D Biol. Crystallogr.* **55**, 125–132 (2010).

14. G. M. Sheldrick, Crystal structure refinement with SHELXL. *Acta Crystallogr. Sect. C, Struct. Chem.* **71**, 3–8 (2015).

15. G. M. Sheldrick, SHELX. *Acta Cryst. A*. **64**, 112–122 (2008).

16. L. Palatinus *et al.*, Structure refinement using precession electron diffraction tomography and dynamical diffraction: Tests on experimental data. *Acta Crystallogr. Sect. B Struct. Sci. Cryst. Eng. Mater.* **71**, 740–751 (2015).

17. D. J. Heijs, V. A. Malyshev, J. Knoester, Thermal broadening of the J-band in disordered linear molecular aggregates: A theoretical study. *J. Chem. Phys.* **123**, 1445071–1445083 (2005).

18. T. Zhu, Y. Wan, L. Huang, Direct Imaging of Frenkel Exciton Transport by Ultrafast Microscopy. *Acc. Chem. Res.* **50**, 1725–1733 (2017).

19. Y. Wan, G. P. Wiederrecht, R. D. Schaller, J. C. Johnson, L. Huang, Transport of Spin-Entangled Triplet Excitons Generated by Singlet Fission. *J. Phys. Chem. Lett.*, 6731–6738 (2018).

20. R. Ahlrichs, M. Bär, M. Häser, H. Horn, C. Kölmel, Electronic structure calculations on workstation computers: The program system turbomole. *Chem. Phys. Lett.* **162**, 165–169 (1989).

21. A. D. Becke, Density-functional thermochemistry. III. The role of exact exchange. *J. Chem. Phys.* **98**, 5648–5652 (1993).

22. A. Schäfer, H. Horn, R. Ahlrichs, Fully optimized contracted Gaussian basis sets for atoms Li to Kr. *J. Chem. Phys.* **97**, 2571–2577 (1992).

23. T. Förster, *Chapter III: Modern Quantum Chemistry* (Dover Scientific, 1965).

24. T. Kobayashi, *J-Aggregates* (World Scientific, 1996).

25. H. Haken, G. Strobl, An exactly solvable model for coherent and incoherent exciton motion. *Zeitschrift für Phys.* **262**, 135–148 (1973).

26. H. Haken, P. Reineker, The coupled coherent and incoherent motion of excitons and its influence on the line shape of optical absorption. *Zeitschrift für Phys.* **249**, 253–268 (1972).

27. P. Reineker, Equations of motion for the moments of the coupled coherent and incoherent motion of triplet and singlet excitons. *Zeitschrift für Phys.* **261**, 187–190 (1973).

28. R. T. Grant *et al.*, Strong coupling in a microcavity containing β-carotene. *Opt. Express*. **26**, 3320 (2018).

29. J. J. Hopfield, D. G. Thomas, Polariton Absorption Lines. *Phys. Rev. Lett.* **15**, 22–25 (1965).

30. G. C. Morris, M. G. Sceats, The 4000 Å transition of crystal anthracene. *Chem. Phys.* **3**, 164–179 (1974).

31. J. M. Turlet, P. Kottis, M. R. Philpott, in *Polariton and Surface Exciton State Effects in the Photodynamics of Organic Molecular Crystals* (John Wiley & Sons, 2007), pp. 303–468.

32. A. F. Oskooi *et al.*, Meep: A flexible free-software package for electromagnetic simulations by the FDTD method. *Comput. Phys. Commun.* **181**, 687–702 (2010).

33. A. Chutinan, S. Noda, Waveguides and waveguide bends in two-dimensional photonic crystal slabs. *Phys. Rev. B - Condens. Matter Mater. Phys.* **62**, 4488–4492 (2000).

34. A. Cacciola *et al.*, Subdiffraction Light Concentration by J-Aggregate Nanostructures. *ACS Photonics*. **2**, 971–979 (2015).

35. C. Mack, *Fundamental Principles of Optical Lithography: The Science of Microfabrication* (John Wiley & Sons, 2007).

36. R. F. Ribeiro, L. A. Martínez-Martínez, M. Du, J. Campos-Gonzalez-Angulo, J. Yuen-Zhou, Polariton chemistry: controlling molecular dynamics with optical cavities. *Chem. Sci.* **9**, 6325–6339 (2018).

37. F. Balzer, V. G. Bordo, A. C. Simonsen, H.-G. Rubahn, Optical waveguiding in individual nanometer-scale organic fibers. *Phys. Rev. B*. **82** (2003).

38. F. Balzer, V. G. Bordo, A. C. Simonsen, H.-G. Rubahn, Isolated hexaphenyl nanofibers as optical waveguides. *Appl. Phys. Lett.* **67**, 115408 (2003).

39. H. Von Berlepsch, S. Möller, L. Dähne, Optical properties of crystalline pseudoisocyanine (PIC). *J. Phys. Chem. B*. **105**, 5689–5699 (2001).

40. C. Sauteret *et al.*, Optical nonlinearities in one-dimensional-conjugated polymer crystals. *Phys. Rev. Lett.* **36**, 956–959 (1976).

41. D. Chaudhuri *et al.*, Enhancing long-range exciton guiding in molecular nanowires by H-aggregation lifetime engineering. *Nano Lett.* **11**, 488–492 (2011).